\pdfoutput=1  

\documentclass[aps,prd,superscriptaddress,floatfix,nofootinbib,preprintnumbers,eqsecnum,twocolumn]{revtex4-1}

\usepackage{amsmath,float,graphicx,placeins,amssymb,multirow,pgfplots,pgfplotstable}
\usepackage{tikz}
\usepackage{comment}
\usepackage{enumerate,slashed,cancel,comment,color,bbm,graphicx,rotating,placeins,mathtools,multirow,hhline}
\usepackage[normalem]{ulem}
\usepackage{array}

\usepackage{hyperref}
\usepackage{color}
 \hypersetup
 {
   colorlinks,
   linkcolor={blue!80!black},
   citecolor={blue!70!black},
   urlcolor={blue!70!black}
 }

\usepackage{lipsum}
\usepackage{diagbox}
\usetikzlibrary{shapes.geometric,arrows}
\graphicspath{}

\def\msmall#1 {\mbox{\small{$#1$}}}
\def\qbar{{\overline{q}}}

\def\beq{\begin{equation}}
\def\eeq{\end{equation}}
\def\beqn{\begin{eqnarray}}
\def\eeqn{\end{eqnarray}}
\def\apo{\mbox{\small ${\frac{\alpha'}{2}}$}}
\def\half{\mbox{\small ${\frac{1}{2}}$}}
\def\third{\mbox{\small ${\frac{1}{3}}$}}
\def\quarter{\mbox{\small ${\frac{1}{4}}$}}
\def\sqapo{\mbox{\tiny $\sqrt{\frac{\alpha'}{2}}$}}
\def\sqap{\mbox{\tiny $\sqrt{{\alpha'}}$}}
\def\sqapxtwo{\mbox{\tiny $\sqrt{2{\alpha'}}$}}
\def\aptwo{\mbox{\tiny ${\frac{\alpha'}{2}}$}}
\def\apofour{\mbox{\tiny ${\frac{\alpha'}{4}}$}}
\def\bosqtwo{\mbox{\tiny ${\frac{\beta}{\sqrt{2}}}$}}
\def\btosqtwo{\mbox{\tiny ${\frac{\tilde{\beta}}{\sqrt{2}}}$}}
\def\apofour{\mbox{\tiny ${\frac{\alpha'}{4}}$}}
\def\sqaptwo{\mbox{\tiny $\sqrt{\frac{\alpha'}{2}}$}  }
\def\apoeight{\mbox{\tiny ${\frac{\alpha'}{8}}$}}
\def\sapoeight{\mbox{\tiny ${\frac{\sqrt{\alpha'}}{8}}$}}

\newcommand{\newc}{\newcommand}
\def\calZ{{\cal Z}}
\def\calM{{\cal M}}
\def\calV{{\cal V}}
\def\calF{{\cal F}}
\def\calG{{\cal G}}
\def\calS{{\cal S}}
\def\calX{{\cal X}}
\def\calK{{\cal K}}
\def\mathbbX{{\mathbb{X}}}
\def\mathbbY{{\mathbb{Y}}}
\def\mathbbK{{\mathbb{K}}}
\def\calY{{\cal Y}}
\def\bQ{{\bf Q}}
\def\bT{{\bf T}}
\def\bt{{\bf t}}
\def\Qs{{\bf q}}
\def\bz{{\bf 0}}

\def\ibar{{\overline{\imath}}}

\def\half{{\textstyle{1\over 2}}}
\def\quarter{{\textstyle{1\over 4}}}
\def\ie{{\it i.e.}\/}
\def\eg{{\it e.g.}\/}
\def\etc{{\it etc}.\/}
\def\inbar{\,\vrule height1.5ex width.4pt depth0pt}
\def\IR{\relax{\rm I\kern-.18em R}}
 \font\cmss=cmss10 \font\cmsss=cmss10 at 7pt
\def\IQ{\relax{\rm I\kern-.18em Q}}
\def\IZ{\relax\ifmmode\mathchoice
 {\hbox{\cmss Z\kern-.4em Z}}{\hbox{\cmss Z\kern-.4em Z}}
 {\lower.9pt\hbox{\cmsss Z\kern-.4em Z}}
 {\lower1.2pt\hbox{\cmsss Z\kern-.4em Z}}\else{\cmss Z\kern-.4em Z}\fi}

\def\oneRes{{       
     \underset{s=1}{\rm Res}
  }}
\def\Str{{\rm Str}}
\def\zStr{{         
      \underset{M=0}{\rm Str}
       \,}}
\def\newzStr{{         
      \underset{M\leq \mu}{\rm Str}
       \,}}
\def\pStr{{         
      \underset{M>0}{\rm Str}
       \,}}
\def\effStr{{         
      \underset{\small 0<M\lesssim \mu}{\rm Str}
       \,}}
\def\zeffStr{{         
      \underset{\small 0\leq M\lesssim \mu}{\rm Str}
       \,}}
\def\antieffStr{{         
      \underset{\small M\gtrsim \mu}{\rm Str}
       \,}}

\def\bpStr{{         
      \underset{\beta_0>0}{\rm Str}
       \,}}
\def\bzStr{{         
      \underset{\beta_0=0}{\rm Str}
       \,}}

\def\eftStr{{         
      \underset{{\rm eff}}{\rm Str}
       \,}}
\def\aMeffStr{{         
      \underset{\small 0\leq M \ll a \calM}{\rm Str}
       \,}}

\def\dmu{{ \frac{d^2\tau}{\tau_2^2} }}  

\def\arg{{  \left( \frac{r M}{a \calM} \right) }}

\begin{document}

\title{Calculating the Higgs Mass in String Theory} 

\def\andname{\hspace*{-0.5em}} 
\author{Steven Abel,}
\email[Email address: ]{s.a.abel@durham.ac.uk}
\affiliation{IPPP, Durham University, Durham, DH1 3LE, United Kingdom}
\author{Keith R. Dienes}
\email[Email address: ]{dienes@arizona.edu}
\affiliation{Department of Physics, University of Arizona, Tucson, AZ 85721 USA}
\affiliation{Department of Physics, University of Maryland, College Park, MD 20742 USA}

\begin{abstract}
In this paper, we establish a fully string-theoretic framework for calculating 
one-loop Higgs masses directly from first principles in perturbative closed string theories.
Our framework makes no assumptions other than worldsheet modular invariance  
and is therefore applicable to all closed strings, regardless of the specific string 
construction utilized.  This framework can also be employed even when spacetime 
supersymmetry is broken (and even when this breaking occurs at the  Planck scale),  
and can be utilized for all scalar Higgs fields, regardless of the particular gauge 
symmetries they break.  This therefore includes the Higgs field responsible for 
electroweak symmetry breaking in the Standard Model.  Notably, using our framework, 
we demonstrate that a gravitational modular anomaly generically relates the Higgs 
mass to the one-loop cosmological constant, thereby yielding a string-theoretic connection
between the two fundamental quantities which are known to suffer from hierarchy problems in 
the absence of spacetime supersymmetry.   We also discuss a number of crucial issues 
involving the use and interpretation of regulators in UV/IR-mixed theories such as 
string theory, and the manner in which one can extract an EFT description from such theories.
Finally, we analyze the running of the Higgs mass within such an EFT description, 
and uncover the existence of a ``dual IR'' region  which emerges at high energies as 
the consequence of an intriguing scale-inversion duality symmetry.    We also identify
a generic stringy effective potential for the Higgs fields in such theories.  
Our results can therefore serve as the launching point for a rigorous investigation of  
gauge hierarchy problems in string theory.
\end{abstract}
\maketitle
  \tableofcontents

\def\ie{{\it i.e.}\/}
\def\eg{{\it e.g.}\/}
\def\etc{{\it etc}.\/}
\def\taubar{{\overline{\tau}}}
\def\qbar{{\overline{q}}}
\def\kbar{{\overline{k}}}
\def\bQ{{\bf Q}}
\def\calT{{\cal T}}
\def\calN{{\cal N}}
\def\calF{{\cal F}}
\def\calM{{\cal M}}
\def\calZ{{\cal Z}}

\def\beq{\begin{equation}}
\def\eeq{\end{equation}}
\def\beqn{\begin{eqnarray}}
\def\eeqn{\end{eqnarray}}
\def\apo{\mbox{\small ${\frac{\alpha'}{2}}$}}
\def\half{\mbox{\small ${\frac{1}{2}}$}}
\def\sqapo{\mbox{\tiny $\sqrt{\frac{\alpha'}{2}}$}}
\def\sqap{\mbox{\tiny $\sqrt{{\alpha'}}$}}
\def\sqapxtwo{\mbox{\tiny $\sqrt{2{\alpha'}}$}}
\def\aptwo{\mbox{\tiny ${\frac{\alpha'}{2}}$}}
\def\apofour{\mbox{\tiny ${\frac{\alpha'}{4}}$}}
\def\bosqtwo{\mbox{\tiny ${\frac{\beta}{\sqrt{2}}}$}}
\def\btosqtwo{\mbox{\tiny ${\frac{\tilde{\beta}}{\sqrt{2}}}$}}
\def\apofour{\mbox{\tiny ${\frac{\alpha'}{4}}$}}
\def\sqaptwo{\mbox{\tiny $\sqrt{\frac{\alpha'}{2}}$}  }
\def\apoeight{\mbox{\tiny ${\frac{\alpha'}{8}}$}}
\def\sapoeight{\mbox{\tiny ${\frac{\sqrt{\alpha'}}{8}}$}}

\newc{\gsim}{\lower.7ex\hbox{{\mbox{$\;\stackrel{\textstyle>}{\sim}\;$}}}}
\newc{\lsim}{\lower.7ex\hbox{{\mbox{$\;\stackrel{\textstyle<}{\sim}\;$}}}}
\def\calM{{\cal M}}
\def\calV{{\cal V}}
\def\calF{{\cal F}}
\def\bQ{{\bf Q}}
\def\bT{{\bf T}}
\def\Qs{{\bf q}}

\def\half{{\textstyle{1\over 2}}}
\def\quarter{{\textstyle{1\over 4}}}
\def\ie{{\it i.e.}\/}
\def\eg{{\it e.g.}\/}
\def\etc{{\it etc}.\/}
\def\inbar{\,\vrule height1.5ex width.4pt depth0pt}
\def\IR{\relax{\rm I\kern-.18em R}}
 \font\cmss=cmss10 \font\cmsss=cmss10 at 7pt
\def\IQ{\relax{\rm I\kern-.18em Q}}
\def\IZ{\relax\ifmmode\mathchoice
 {\hbox{\cmss Z\kern-.4em Z}}{\hbox{\cmss Z\kern-.4em Z}}
 {\lower.9pt\hbox{\cmsss Z\kern-.4em Z}}
 {\lower1.2pt\hbox{\cmsss Z\kern-.4em Z}}\else{\cmss Z\kern-.4em Z}\fi}

\section{Introduction}

Extracting phenomenological predictions from string theory is a subtle task.
Chief among the complications is the question of 
finding a suitable vacuum.
Without solving this problem, one is limited to making generic statements that might hold across
broad classes of string theories.
But even within the context of specific string models with certain favorable characteristics,
most attempts at extracting the corresponding phenomenological predictions follow a common path.   First, one tallies the massless states that arise in such models.  Then, one constructs a field-theoretic Lagrangian which describes the dynamics of these states.   Finally, one proceeds to analyze this Lagrangian using all of the regular tools of quantum field theory
without further regard for the origins of these states within string theory.

Although such a treatment may be sufficient for certain purposes, calculations performed in this manner have a serious shortcoming:   
by disregarding the infinite towers of string states that necessarily accompany these low-lying modes within the 
full string theory, 
such calculations implicitly disregard many of the underlying string symmetries that ultimately 
endow string theory with a plethora of remarkable properties that transcend our field-theoretic expectations.   
At first glance, it may seem that these extra towers of states cannot play an important role for 
low-energy physics because these states typically have masses which are set by 
the string scale (generically assumed near the Planck scale) or by the scales associated with the compactification geometry.     
For this reason it would seem that these heavy states can legitimately be integrated out of the theory, thereby 
justifying a treatment based on a Lagrangian description of the low-lying modes alone, along with possible 
higher-order operators suppressed by powers of these heavier scales.   However, it is 
difficult to justify integrating out {\it infinite}\/ towers of states, 
much less towers whose state degeneracies at each mass level grow {\it exponentially}\/ with mass.  
Yet this is precisely the situation we face in string theory.
Indeed, these infinite towers of states particularly affect 
those operators (such as those associated with the Higgs mass 
and the cosmological constant) which have positive
dimension and are therefore sensitive to all mass scales in the theory.

Many of the string symmetries that rely on these infinite towers of states go beyond what can be 
incorporated within the framework of an effective field theory (EFT).~   
For example, strong/weak coupling duality relations intrinsically rely on the presence of 
the full towers of string states, both perturbative and non-perturbative.   
But there also exist stringy symmetries that operate purely within the perturbative weak-coupling regime.   
A prime example of this is T-duality, under which the physics of closed strings compactified 
on small compactification volumes is indistinguishable from the physics associated with strings 
compactified on large compactification volumes.   This sort of 
equivalence between ultraviolet (UV) and 
infrared (IR) physics cannot be incorporated within an
EFT-based approach in which we integrate out heavy states while treating light states as dynamical.

Both strong/weak coupling duality and T-duality are spacetime symmetries.
As such, like all spacetime physics, they are merely the {\it consequences}\/ of an underlying string theory.
But closed string theories have another symmetry of this sort which is even more fundamental and which 
must be imposed for consistency 
directly on the worldsheet.
This is worldsheet modular invariance, which will be the focus of this
paper. 
Worldsheet modular invariance is crucial
since it lies at the heart of many of the finiteness properties for which string theory is famous.
Moreover, since modular invariance is an exact symmetry of all perturbative closed-string vacua, it provides 
tight constraints on the spectrum of string states at all mass scales as well as on their interactions.  
Indeed, this symmetry is the ultimate ``UV-IR mixer'',
operating over all scales and enforcing a delicate balancing between low-scale and high-scale physics.  
There is no sense in which its breaking can be confined to low energies, and likewise
there is no sense in which it can be broken by a small amount. 
As an exact symmetry governing string dynamics,
worldsheet modular invariance is preserved even as the theory passes through phase transitions such as the 
Standard-Model electroweak or QCD phase transitions, as might occur 
under cosmological evolution.  Indeed, any shifts in the low-energy degrees of freedom induced by such phase transitions 
are automatically accompanied by corresponding shifts in the high-scale theory such that modular invariance is maintained 
and finiteness is preserved.  Yet this entire structure is missed if we integrate out the heavy states and 
concentrate on the light states alone.   

While certain phenomenological questions are not likely to depend on such symmetries, this need not always be the case.
For example, these symmetries are likely to be critical for addressing fundamental questions connected with finiteness and/or 
the stability of (or even the coexistence of) 
different scales under radiative corrections.    Chief among these questions are hierarchy problems,  
which provide clues as to the UV theory 
and its potential connections to IR physics.  Indeed, two of the most pressing mysteries in physics are the hierarchy problems 
associated with the cosmological constant and with the masses of scalar fields such as the Higgs field.
However, integrating out the heavy string states eliminates all of the stringy physics that may 
provide alternative ways of addressing such problems.  
The lesson, then, is clear:  If we are to take string theory literally as a theory of physics, 
then we should perform our calculations within the full framework of string theory, incorporating all of 
the relevant symmetries and infinite towers of states that string theory provides.

With this goal in mind, 
we begin this paper by 
establishing  a fully string-theoretic framework for calculating
one-loop Higgs masses directly from first principles in perturbative closed string theories.
This is the subject of Sect.~\ref{sec2}.~
Our framework will make no assumptions other than worldsheet modular invariance  
and will therefore be applicable to
all closed strings, regardless of the specific string 
construction utilized. 
Our results will thus have a generality that extends beyond individual string models.
As we shall see, this framework operates independently of spacetime supersymmetry, and can 
be employed even when spacetime supersymmetry is broken 
(or even when the string model has no spacetime supersymmetry to begin with at any scale).
Likewise, our framework can be utilized for all scalar Higgs fields, regardless of the particular gauge symmetries they break. 
This therefore includes the Higgs field responsible for electroweak symmetry breaking in the Standard Model.  

One of the central results emerging from our framework is a relationship 
between the Higgs mass and the one-loop cosmological constant.   
This connection arises as the result of a gravitational modular anomaly,
and is thus generic for all closed string theories.
This then provides a string-theoretic connection between the two fundamental quantities which are 
known to suffer from hierarchy problems in the absence of spacetime supersymmetry.   
From the perspective of ordinary quantum field theory, such a relation 
between the Higgs mass and the cosmological constant would be entirely unexpected.   
Indeed, quantum field theories are insensitive to the zero of energy.
String theory, by contrast, unifies
gauge theories with gravity.
Thus, it is only within a string context that such a relation could ever arise. 
As we shall see, this relationship does not require supersymmetry in any form. 
It holds to one-loop order, but its direct emergence as 
the result of a fundamental string symmetry leads us to believe that 
it actually extends more generally.  We stress that it is not the purpose of this paper to 
actually solve either of these hierarchy problems
(although we shall return to this issue briefly in Sect.~\ref{sec:Conclusions}). 
However, we now see that these two hierarchies are connected in a deep way within a string context.

As we shall find, the Higgs mass receives contributions from all of the states 
throughout the string spectrum which couple to the Higgs in specified ways.
This includes the physical (level-matched) string states as well as the unphysical (non-level-matched) string states.
Depending on the string model in question, 
we shall also find that our expression for the total Higgs mass can be divergent;  ultimately this will depend on the charges carried by the massless states.
Accordingly,
we shall then proceed to develop a set of regulators  
which can tame the Higgs-mass divergences while at the same time allowing
us to express the Higgs mass as a weighted supertrace over only the physical string states.
Developing these regulators is the subject of Sect.~\ref{sec3}.~
To do this, 
we shall begin by reviewing prior results in the mathematics literature which will form the basis for
our work.
Building on these results, we will then proceed to develop
a set of regulators which are completely general, 
which preserve modular invariance, and which can be used in a wide variety of contexts even beyond
their role in regulating the Higgs mass.

In Sect.~\ref{sec4}, we shall then use these modular-invariant regulators in order to 
recast our results for the Higgs mass in a form that is closer to what we might expect in field theory.
This will also allow us to  develop
an understanding of how the Higgs mass ``runs'' in string theory
and to develop a physical ``renormalization'' prescription that can operate at all scales.
Towards this end, we begin in Sect.~\ref{UVIRequivalence}
with a general discussion of how (and to what extent) one can meaningfully extract an effective field theory 
from UV/IR-mixed theories such as modular-invariant string theories.
This issue is surprisingly subtle, since modular invariance relates UV and IR divergences to each other
while at the same time softening both. 
For example, we shall demonstrate that while the Higgs mass is quadratically divergent
in field theory, modular invariance renders the Higgs mass at most logarithmically divergent in string theory.
We shall then apply our regulators from Sect.~\ref{sec3} to our Higgs-mass
results in Sect.~\ref{sec2}
and thereby demonstrate how the Higgs mass ``runs''
as a function of an energy scale $\mu$.
The results of our analysis are highlighted in Fig.~\ref{anatomy}, which 
not only exhibits features which might be expected in 
an ordinary effective field theory but also includes features which clearly transcend traditional quantum field-theoretic
expectations.  The latter include the existence of a ``dual'' infrared region at high energy scales as well as an invariance under 
an intriguing ``scale duality'' transformation {\mbox{$\mu\to M_s^2/\mu$}}, where $M_s$ denotes the string scale.
This scale-inversion duality symmetry in turn implies the existence of a fundamental limit on the extent to which a modular-invariant theory such as string
theory can exhibit UV-like behavior.

All of our results in Sects.~\ref{sec2} through \ref{sec4} are formulated in a fashion that assumes that our 
modular-invariant string theories can be described through charge lattices.
However, it turns out that our results can be recast in a completely general fashion
that does not require the existence of a charge lattice.
This is the subject of Sect.~\ref{sec5}.~
Moreover, we shall find that this reformulation has an added benefit, allowing us to extract a modular-invariant stringy effective
potential for the Higgs from which the Higgs mass can be obtained through a modular-covariant double derivative with respect
to fluctuations of the Higgs field. 
This potential therefore sits at the core of our string-theoretic calculations
and allows us to understand not only the behavior of the Higgs mass but also the overall stability of the string theory
in a very compact form.
Indeed, in some regions this potential exhibits explicitly string-theoretic behavior.   However, in other regions,
this potential --- despite its string-theoretic origins --- exhibits a number of features which are
reminiscent of the traditional Coleman-Weinberg Higgs potential.

Finally, in Sect.~\ref{sec:Conclusions},
we provide an overall discussion of our results and outline some possibilities for future research. 
We also provide an additional bird's-eye perspective on the 
manner in which modular invariance induces UV/IR mixing 
and the reason why the passage from a full string-theoretic result to an EFT description 
necessarily breaks the modular symmetry. 
We will also discuss some of the possible implications of our results for addressing
the hierarchy problems associated with the cosmological constant and the Higgs mass.
This paper also has two Appendices which provide the details of calculations  
whose results are quoted in Sects.~\ref{chargeinsertions} and \ref{Lambdasect} respectively.

Our overarching goal in this paper is to provide a fully string-theoretic framework for the calculation of the Higgs
mass --- a framework in which modular invariance is baked into the formalism from the very beginning.
Our results can therefore potentially serve as the launching point for a rigorous investigation of 
the gauge hierarchy problem in string theory.
However, our methods are quite general and can easily be adapted to other quantities of phenomenological interest,
including not only the masses of all particles in the theory but
also the gauge couplings, quartic couplings, and indeed the couplings associated with all allowed interactions.

As already noted, much of the inspiration for this work stems from our conviction that 
it is not an accident or phenomenological irrelevancy 
that string theories contain not only low-lying modes but also infinite towers of massive states.
Together, all of these states conspire to enforce many of the unique symmetries for which
string theory is famous, and thus 
their effects are an intrinsic part of the predictions of string theory.
In this spirit, one might even view our work as a continuation of the line
  originally begun in the classic 1987 paper of Kaplunovsky~\cite{Kaplunovsky:1987rp}
which established a framework for calculating string threshold corrections in which 
the contributions of the infinite towers of string states were included.
Indeed, as discussed in Sect.~\ref{higgsmin},
some of our results for the Higgs mass 
even resemble results obtained in Ref.~\cite{Kaplunovsky:1987rp} for threshold corrections.
One chief difference in our work,
however, is our insistence on maintaining modular invariance at all steps 
in the calculation, {\it including the regulators}\/,
especially when seeking to understand the behavior of dimensionful operators.
It is this extra ingredient which is critical for ensuring consistency with the underlying
string symmetries, and which allows us to probe the unique effects 
of such symmetries (such as those induced by UV/IR mixing) in a rigorous manner.

\section{Modular invariance and the Higgs mass:  A general framework\label{sec2}}

In this section we develop a framework for calculating the Higgs mass in any four-dimensional
modular-invariant string theory.
Our framework incorporates modular invariance in a fundamental way, and ultimately leads
to a completely general expression for one-loop Higgs mass.
Our results can therefore easily be applied to any 
four-dimensional 
closed-string model.
Throughout most of this paper, our analysis will focus on heterotic string models and will proceed under the assumption 
that the string model in question can be described through a corresponding charge lattice.
As we shall see, the existence of a charge lattice provides a very direct way of performing
our calculations and illustrating our main points.
However, as we shall discuss in Sect.~\ref{sec5}, our results 
are ultimately more general than this, and apply even for closed-string models
that transcend a specific charge-lattice construction.

\subsection{Preliminaries:  String partition functions, charge lattices,  and modular invariance}

We begin by reviewing basic facts about string partition
functions, charge lattices, and modular invariance,
establishing our notation and normalizations along the way.
The one-loop partition function for any closed heterotic string in four spacetime dimensions is
a statistics-weighted trace over the Fock space of closed-string states, and thus takes the general form
\beq
     {\cal Z} (\tau,\taubar) ~\equiv~ \tau_2^{-1} \frac{1}{\overline{\eta}^{12} \eta^{24}} 
       \, \sum_{m,n} \, (-1)^F \, \qbar^m q^n~.
\label{Zform}
\eeq
Here $\tau$ is the one-loop (torus) modular parameter, 
{\mbox{$\tau_2\equiv {\rm Im}\,\tau$}}, 
{\mbox{$q\equiv \exp(2\pi i\tau)$}}, 
$F$ is the spacetime fermion number,  
and the Dedekind eta-function is {\mbox{$\eta(\tau)\equiv q^{1/24}\prod_{n=1}^\infty (1-q^n)$}}.
In this expression, the $\overline{\eta}$- and $\eta$-functions represent the 
contributions from the string oscillator states [which include
appropriate right- and left-moving vacuum energies $(-1/2,-1)$ respectively],
while the $(m,n)$ sum tallies the contributions from
the Kaluza-Klein (KK) and winding excitations of the heterotic-string worldsheet fields --- 
excitations which result from the compactification
of the heterotic string to four dimensions from its critical spacetime dimensions ($=10$ for the 
right-movers and $26$ for the left-movers), 
with $(m,n)$ representing the corresponding right- and left-moving worldsheet energies.
These KK/winding contributions can be written in terms of the charge vectors {\mbox{$\bQ\equiv \lbrace \bQ_R,\bQ_L\rbrace $}} of 
a $(10,22)$-dimensional Lorentzian charge lattice ---
or equivalently the KK/winding momenta $\lbrace {\bf p}_R,{\bf p}_L\rbrace$ of a corresponding momentum lattice of the
same dimensionality ---
via
\beq
               m= {\bQ_R^2\over 2} = {\alpha' {\bf p}_R^2\over 2} ~,
           ~~~~~ n= {\bQ_L^2\over 2} = {\alpha' {\bf p}_L^2\over 2} ~,
\label{mnQ}
\eeq
where {\mbox{$\alpha'\equiv 1/M_s^2$}} with $M_s$ denoting the string scale.
Thus the partition function in Eq.~(\ref{Zform}) can be written as a sum over charge vectors $\bQ_L, \bQ_R$:
\beq
        {\cal Z}(\tau,\taubar) ~=~ \tau_2^{-1} {1\over \overline{\eta}^{12} \eta^{24}} \, \sum_{\bQ_L,\bQ_R} 
             (-1)^F \qbar^{\bQ_R^2/2} 
             q^{\bQ_L^2/2}~.
\label{preZQdef}
\eeq

In general, the spacetime mass $M$ of the resulting string state is given by 
{\mbox{$\alpha' M^2 = 2(m+n) + 2(\Delta_L+\Delta_R) + 2(a_L+a_R)$}}
where $\Delta_{R,L}$ are the contributions from the oscillator excitations and {\mbox{$(a_R,a_L)=(-1/2, -1)$}} are the corresponding
vacuum energies.
Identifying individual left- and right-moving contributions
to $M^2$ through the convention 
\beq
         M^2 ~=~ {1\over 2} (M_L^2 + M_R^2)
\label{masssum}
\eeq
then yields {\mbox{$\alpha' M_R^2  = 4 (m+ \Delta_R +a_R) $}} 
and         {\mbox{$\alpha' M_L^2  = 4 (n+ \Delta_L +a_L) $}}.
Writing these masses in terms of the lattice charge vectors then yields
\beqn
          {\alpha'\over 2} M_R^2 ~&=&~ \bQ_R^2  + 2 \Delta_R + 2 a_R~,\nonumber\\
          {\alpha'\over 2} M_L^2 ~&=&~ \bQ_L^2  + 2 \Delta_L + 2 a_L~.
\label{eq:massRL}
\eeqn
States are level-matched (physical) if {\mbox{$M_R^2 = M_L^2$}} and unphysical otherwise.
Indeed, with these conventions, 
gauge bosons in the left-moving non-Cartan algebra are massless, with {\mbox{$\bQ_L^2=2$}} and {\mbox{$\Delta_L=0$}},
while those in the left-moving Cartan algebra are massless, with {\mbox{$\bQ_L^2 =0$}} and {\mbox{$\Delta_L=1$}}.
(Indeed, such results apply to all left-moving simply-laced gauge groups with level-one affine realizations;  more
complicated situations, such as necessarily arise for the right-moving gauge groups,  are discussed in Ref.~\cite{Dienes:1996yh}.)
Note that the CPT conjugate 
of any state with charge vector $\lbrace \bQ_R,\bQ_L\rbrace$ has charge vector  
$-\lbrace \bQ_R,\bQ_L\rbrace$.
Thus CPT invariance requires that all states in the string spectrum come
in $\pm\lbrace \bQ_R,\bQ_L\rbrace$ pairs.
By contrast, 
since the right-moving gauge group is necessarily non-chiral as a result of superconformality constraints,
the chiral conjugate of 
any state with charge vector $\lbrace \bQ_R,\bQ_L\rbrace$ has charge vector  
$\lbrace \bQ_R,-\bQ_L\rbrace$.
 
One important general property of the partition functions in Eq.~(\ref{Zform}) ---
and indeed the partition functions of {\it all}\/ closed strings in any spacetime dimension --- 
is that they must be {\it modular invariant}\/, \ie, invariant
under all transformations of the form
{\mbox{$\tau \to (a\tau+b)/(c\tau+d)$}} where {\mbox{$a,b,c,d\in \mathbb{Z}$}} and {\mbox{$ad-bc=1$}}
(with the same transformation for $\taubar$).
Modular invariance is thus an exact symmetry underpinning all heterotic strings,
and in this paper we shall be exploring its consequences for the masses of the Higgs fields in such theories.
For these purposes, it will be important to understand the manner in which these
partition functions achieve their modular invariance.
In general, the partition functions for heterotic strings in four dimensions
can be rewritten in the form
\beq
   {\cal Z} (\tau,\taubar) ~\equiv~ \tau_2^{-1} \frac{1}{\overline{\eta}^{12} \eta^{24}}\, 
 \sum_{\ibar,i} N_{\ibar i} ~\overline{g_\ibar (\tau)}  f_i(\tau)~
\label{partfZ}
\eeq
where each $(\ibar,i)$ term represents the contribution from a different sector of the theory
and where the left-moving holomorphic $f_i$ functions (and the corresponding right-moving antiholomorphic
$g_\ibar$ functions) 
transform covariantly under modular transformations according to
relations of the form
\beq 
        f\left( \frac{a\tau+b}{c\tau+d}\right) ~\sim~ (c\tau+d)^{k} \,f(\tau)
\label{fis}
\eeq
where $k$ is the so-called modular weight of the $f_i$ functions
(with an analogous weight $\kbar$ for the $g_\ibar$ functions)
and where the $\sim$ notation allows for the possibility of overall 
$\tau$-independent phases which
will play no future role in our arguments.
We likewise have 
\beq
        \eta\left( \frac{a\tau+b}{c\tau+d}\right) ~\sim~ (c\tau+d)^{1/2} \, \eta(\tau)~.
\label{modfis}
\eeq
Thus, since {\mbox{$\tau_2\to \tau_2/|c\tau+d|^2$}} as {\mbox{$\tau\to (a\tau+b)/(c\tau+d)$}}, we immediately
see that modular invariance of the entire partition function in Eq.~(\ref{partfZ}) requires not only that
the $N_{\ibar i}$ coefficients in Eq.~(\ref{partfZ}) be chosen correctly but also that
{\mbox{$k=11$}} and {\mbox{$\kbar =5$}}.  
In general, 
for strings realizable through free-field constructions, 
these $f_i$ and $g_\ibar$ functions produce the lattice sum in Eq.~(\ref{preZQdef}) because
they can be written in the factorized forms 
\beq
    f_i~\sim~ \prod_{\ell=1}^{22} \vartheta
       \begin{bmatrix} \alpha_\ell^{(i)} \\ \beta_\ell^{(i)} \end{bmatrix}~,~~~~~
    g_\ibar~\sim~ \prod_{\ell=1}^{10} \vartheta
       \begin{bmatrix} \alpha_\ell^{(\ibar)} \\ \beta_\ell^{(\ibar)} \end{bmatrix}~,~~
\label{partft}
\eeq
where each $\vartheta$-function factor 
is the trace over the $\ell^{\rm th}$ direction $Q_\ell$ of the charge lattice:
\beq
        \vartheta_\ell(\tau) 
     ~\equiv~ \vartheta \begin{bmatrix} \alpha_\ell \\ \beta_\ell \end{bmatrix}(\tau)
              ~\equiv~
              \sum_{Q_\ell\in \mathbb{Z}+\alpha_\ell} e^{2\pi i \beta_\ell Q_\ell} \,q^{Q_\ell^2/2}~.~~
\label{thetadef}
\eeq
Indeed, the $\vartheta_\ell$-functions transform 
under modular transformations as in Eq.~(\ref{fis}),
with modular weight $1/2$.
The modular invariance of the underlying string theory
then ensures that there exists a special
$(10,22)$-dimensional ``spin-statistics vector''  ${\bf S}$ 
such that we may identify 
the spacetime fermion number $F$ within Eq.~(\ref{Zform})
as {\mbox{${F = 2 \bQ\cdot {\bf S}}$}}~(mod~2)
for any state with charge $\bQ$,
where the dot notation `$\cdot$' signifies the Lorentzian 
(left-moving minus right-moving) dot product.
Modular invariance 
also implies 
that the shifted charges $\bQ -{\bf S}$ 
associated with the allowed string states
together form a Lorentzian lattice which is both odd and self-dual.
It is with this understanding that 
we refer to the charges $\bQ$ themselves as populating a ``lattice''.
Indeed, it is the self-duality property of the 
shifted charge lattice $\lbrace \bQ-{\bf S}\rbrace$ which guarantees  
that the $f_i$ and $g_\ibar$ functions in Eq.~(\ref{fis})
transform covariantly 
under the modular group,
as in Eq.~(\ref{fis}).

For later purposes, we simply observe that the
general structure given in Eq.~(\ref{partfZ}) is 
typical of the modular-invariant quantities that arise
as heterotic-string Fock-space traces.
Indeed, a general quantity of the form
\beq
        \tau_2^\kappa~
     \frac{1}{\overline{\eta}^{12} \eta^{24}}\, 
 \sum_{\ibar,i} N_{\ibar i} ~\overline{g_\ibar (\tau)}  f_i(\tau)~
\label{genexp}
\eeq
cannot be modular invariant unless 
the $N_{\ibar i}$ are chosen correctly
and the corresponding $f_i$ and $g_\ibar$ functions transform
as in Eq.~(\ref{fis})
with 
\beq
         k-12 ~=~  \kbar-6 ~=~  \kappa~.
\label{genexp2}
\eeq
While {\mbox{$\kappa = -1$}} for the partition 
functions of four-dimensional heterotic strings, as
described above,
we shall see that other important Fock-space traces can have different
values of $\kappa$.
For example, the partition functions of heterotic strings
in $D$ spacetime dimensions have {\mbox{$\kappa = 1-D/2$}}, with
corresponding changes to the dimensionalities of 
their associated charge lattices.

\subsection{Higgsing and charge-lattice deformations} \label{latdeform}

In general, different string models exhibit different spectra and thus have different charge lattices.
However, Higgsing a theory changes its spectrum in certain dramatic ways, 
such as by giving mass to formerly massless gauge bosons and thereby breaking the associated gauge symmetries.
Thus, in string theory, Higgsing 
can ultimately be viewed as a process of 
transforming the charge 
lattice from one configuration to another.  

Of course, 
modular invariance must 
be maintained throughout the Higgsing process. 
Indeed, it is only in this way that
we can regard the Higgsing process as a fully string-theoretic 
operation that 
shifts the string vacuum state within the space of self-consistent string vacua.
However, modular invariance then implies that
the charge-lattice transformations induced by Higgsing are not arbitrary. 
Instead, they must preserve those charge-lattice properties, 
as described above, which guarantee the modular invariance of the theory.

This in turn tells us that the process of Higgsing is likely to be far more complicated
in string theory than it is in ordinary quantum field theory.
In general, 
the charge lattice receives contributions from all sectors of the theory,
and modular transformations mix these different contributions in highly non-trivial ways.
Thus the process of Higgsing a given gauge symmetry within a given sector of a string
model generally involves not only the physics associated with that gauge symmetry 
but also the physics of all of the {\it other}\/ sectors of the theory as well, both twisted and untwisted,
and the properties of the {\it other}\/ gauge symmetries, including gravity, that might also be present 
in the string model ---  even if these other gauge symmetries are apparently completely disjoint from 
the symmetry being Higgsed.  
We shall see explicit examples of this below.
Moreover, in string theory the dynamics of the Higgs VEV --- and indeed the dynamics of all string moduli ---
is generally governed by an effective potential  
which is nothing but the vacuum energy of the theory, expressed as a function of this VEV.~
Thus the overall dynamics associated with Higgsing can be rather subtle: 
the Higgs VEV determines the deformations of the charge lattice, and these deformations
alter the vacuum energy which in turn determines the VEV.

In this paper, our goal is to calculate the mass of the physical Higgs scalar field that emerges 
in the Higgsed phase (\ie, after the 
theory has already been Higgsed).  We shall therefore assume that our theory contains a scalar Higgs field
which has already settled into the new minimum of its potential.
This will allow us to 
sidestep the (rather complex) model-dependent issue 
concerning the manner in which the Higgsing itself occurs, and instead
focus on the {\it perturbations}\/ of the field around this new minimum.
In this way we will be able to  determine 
the curvature of the scalar potential at this local minimum, and thereby obtain the corresponding Higgs mass. 

To do this, we shall begin by 
exploring the manner in which a general charge lattice is deformed
as we vary a scalar Higgs field away from the minimum of its potential.
Our discussion will be completely general, and we shall defer to Sect.~\ref{sec:EWHiggs}  
any assumptions that
might be specific to the particular Higgs field responsible for electroweak symmetry breaking.
For concreteness, we shall let
our scalar field have
a value $\langle \phi \rangle + \phi$, where $\langle \phi\rangle$ is the Higgs VEV 
at the minimum of its potential and where $\phi$ describes the fluctuations away from this point.
If $\lbrace \bQ_L,\bQ_R\rbrace$ are the charge vectors associated with a
given string state in the Higgsed phase (\ie, at the minimum of the potential, when {\mbox{$\phi=0$}}),
then turning on $\phi$ corresponds to deforming these charge vectors.  
In general, for {\mbox{$\phi/\langle \phi\rangle\ll 1$}}, we shall parametrize these deformations
according to
\beqn  
          \bQ_L &\to&  ~\bQ_L + \sqrt{\alpha'} \phi \bQ_a + {1\over 2} \alpha' \phi^2 \bQ_b~+~... ~~~~~   \nonumber\\
          \bQ_R &\to&  ~\bQ_R + \sqrt{\alpha'} \phi \tilde \bQ_a + {1\over 2} \alpha' \phi^2 \tilde \bQ_b~+~...~,
\label{Qdeform}
\eeqn
where $\bQ_a$, $\bQ_b$, $\tilde\bQ_a$, and $\tilde \bQ_b$ are deformation charge vectors of 
dimensionalities $22$, $22$, $10$, and $10$ respectively.
Indeed, the forms of these vectors are closely correlated with the specific gauge symmetries
broken by the Higgsing process,
and as such these vectors continue to 
govern the fluctuations
of the Higgs scalar around this Higgsed minimum.

In this paper, we shall keep our analysis as general as possible. 
As such, we shall not make any specific assumptions regarding
the forms of these vectors.
However, as discussed above, we know that the Higgsing process --- and even the fluctuations around the Higgsed minimum 
of the potential --- should not break modular invariance.  In particular,
the corresponding charge-lattice deformations in Eq.~(\ref{Qdeform}) should not 
disturb level-matching.   This means that  the value of the difference $\bQ_L^2 - \bQ_R^2$ should not be 
disturbed when $\phi$ is taken to non-zero values, which in turn means that 
this difference should be independent of $\phi$.
This then constrains the choices for 
the vectors $\bQ_a$, $\bQ_b$, $\tilde\bQ_a$, and $\tilde \bQ_b$.

To help simplify the notation, let us assemble a single 32-dimensional charge vector 
{\mbox{$\bQ \equiv (\bQ_L,\bQ_R)^t$}} (where `$t$' signifies the transpose). 
Recalling that the dot notation `$\cdot$' signifies 
the Lorentzian (left-moving minus right-moving) contraction of vector indices,
as appropriate for a Lorentzian charge lattice,
we therefore require that {\mbox{$\bQ^2 \equiv \bQ^t\cdot \bQ$}} be $\phi$-independent for all $\phi$.
Given the above shifts, we find that terms within {\mbox{$\bQ^t\cdot \bQ$}} 
which are respectively linear and quadratic in $\phi$ will cancel provided 
\beqn
                    (\bQ^t_a, \tilde \bQ^t_a) \cdot \bQ ~&=&~ 0~,~~~~~\nonumber\\
                 (\bQ_b^t, \tilde \bQ^t_b) \cdot \bQ + (\bQ_a^t, \tilde \bQ_a^t ) \cdot 
           \begin{pmatrix}
              \bQ_a\\
              \tilde \bQ_a
           \end{pmatrix}
            ~&=&~0~.
\label{levelmatching}
\eeqn
These are thus modular-invariance
constraints on the allowed choices for the shift vectors $\bQ_a$, $\bQ_b$, $\tilde\bQ_a$, and $\tilde \bQ_b$.

We can push these constraints one step further if we write these shift vectors in terms of $\bf Q$ itself via relations of the form
\beq
  \begin{pmatrix}
     \bQ_a\\
        \tilde \bQ_a
     \end{pmatrix}
    = {\cal T} \cdot \bQ ~,~~~~~
     \begin{pmatrix}
          \bQ_b\\ \tilde \bQ_b
         \end{pmatrix}
   = {\cal N} \cdot \bQ ~
\label{shiftQ}
\eeq
where $\calT$ and $\calN$ are {\mbox{$(32\times 32)$}}-dimensional matrices 
and where `$\cdot$' retains its Lorentzian signature for the index contraction that underlies matrix multiplication. 
The first constraint equation above then tell us that {\mbox{$\bQ^t \cdot \calT\cdot \bQ=0$}}, which implies that $\calT$ must be antisymmetric,
while the second constraint equation tells us that {\mbox{$\bQ^t \cdot ( \calN+ \calT^t\cdot \calT)\cdot \bQ =0$}}, which implies that
{\mbox{$\calN+ \calT^t \cdot \calT$}} must also be antisymmetric.    
It turns out that the precise value of {\mbox{$\calN + \calT^t \cdot \calT$}} will have no bearing on the Higgs mass.   We will therefore
set it to zero (which is indeed antisymmetric), implying that {\mbox{$\calN = - \calT^t \cdot \calT$}}.~   
Thus, while $\calT$ is antisymmetric, $\calN$ is symmetric.
Indeed, if we write our $\calT$-matrix in terms of left- and right-moving submatrices $\calT_{ij}$ in the form
\beq
       \calT ~=~ \begin{pmatrix}
               \calT_{11} & \calT_{12} \\
               \calT_{21} & \calT_{22} 
                   \end{pmatrix}~,
\eeq
then we must have {\mbox{$\calT_{11}^t = -\calT_{11}$}}, {\mbox{$\calT_{22}^t = -\calT_{22}$}}, and {\mbox{$\calT_{12}^t = -\calT_{21}$}}.
Likewise, we then find that
\beqn
        \calN_{11} ~&=&~  - \calT_{11}^t \calT_{11} + \calT_{21}^t \calT_{21}\nonumber\\ 
        \calN_{12} ~&=&~  - \calT_{11}^t \calT_{12} + \calT_{21}^t \calT_{22}\nonumber\\ 
        \calN_{21} ~&=&~  - \calT_{12}^t \calT_{11} + \calT_{22}^t \calT_{21}\nonumber\\ 
        \calN_{22} ~&=&~  - \calT_{12}^t \calT_{12} + \calT_{22}^t \calT_{22}~.
\label{NTrelations}
\eeqn

\subsection{Example:   The Standard-Model Higgs\label{sec:EWHiggs}}

In general, within any given string model, 
the deformation vectors 
$\bQ_a$, $\bQ_b$, $\tilde\bQ_a$, and $\tilde \bQ_b$ in Eq.~(\ref{Qdeform}) 
depend on the particular charge vector $(\bQ_R,\bQ_L)$ being deformed.
However the $\calT$- and $\calN$-matrices in Eq.~(\ref{shiftQ}) are universal for all charge vectors within the model.
It is therefore these matrices which carry all of the relevant information concerning the response of the theory 
to fluctuations of the particular Higgs field under study.
In general, these matrices depend on how the gauge groups and corresponding Higgs field are embedded
within the charge lattice.
Thus the precise forms of these matrices depend on the particular string model under study and the Higgs field
to which it gives rise.

To illustrate this point, it may be helpful to consider the special case of the Standard-Model (SM) Higgs.
For concreteness, we shall work within the framework of  heterotic  string models 
in which the Standard Model itself is realized at affine level {\mbox{$k=1$}} through a standard level-one $SO(10)$ embedding.
In the following we shall adhere to the conventions in Ref.~\cite{Dienes:1995sq}.
Since $SO(10)$ has rank $5$, this group can be minimally embedded within a 
five-dimensional sublattice $\lbrace Q_1,Q_2,Q_3,Q_4,Q_5\rbrace$ 
within the full 22-dimensional left-moving lattice $\lbrace \bQ_L\rbrace$.  
Within this sublattice, we shall take
the {\mbox{$\ell=1,2$}} directions as corresponding to 
the {\mbox{$U(2) = SU(2)\times U(1)$}} electroweak subgroup of $SO(10)$, 
while the {\mbox{$\ell =3,4,5$}} directions will correspond to the {\mbox{$U(3)= SU(3)\times U(1)$}} color subgroup.
By convention we will take the $SU(2)_L$ representations to lie along the line perpendicular to $(1,1,0,0,0)$
within the two-dimensional $U(2)$ sublattice,
and the $SU(3)_c$ representations to lie within the two-dimensional plane perpendicular to $(0,0,1,1,1)$
within the three-dimensional $U(3)$ sublattice.
It then follows that any state with charge vector $\bQ_L$ has $SU(2)$ quantum numbers determined by projecting $\bQ_L$ onto the 
$SU(2)$ line [thereby yielding an $SU(2)$ weight in the corresponding $SU(2)$ weight system]
and $SU(3)$ quantum numbers determined by projecting $\bQ_L$ onto the $SU(3)$ plane [thereby yielding an $SU(3)$ weight within the 
corresponding $SU(3)$ weight system].
Likewise, the $SO(10)$-normalized hypercharge $Y$ of any state with left-moving charge vector
$\bQ_L$ is given by {\mbox{$Y=\sum_{\ell=1}^5 a_{Y}^{(\ell)} Q_\ell$}} where  
\beq
                  {\bf a}_{Y}~=~ (\, \half,\, \half, ~ -\third,\, -\third,\, -\third\, )~
\eeq
(with all other components vanishing).
Thus, {\mbox{$Y\equiv  {\bf a}_{Y} \cdot \bQ_L$}}. 
Indeed we see that {\mbox{$k_Y\equiv  2 {\bf a}_{Y}\cdot {\bf a}_{Y} = 5/3$}}, as appropriate for the standard $SO(10)$ embedding
(as well as other non-standard $SO(10)$ embeddings~\cite{Dienes:1995sq}).
In a similar way, the electromagnetic charge $q_{\rm EM}$ of any state with charge vector $\bQ_L$ is given by 
{\mbox{$q_{\rm EM}= {\bf a}_{\rm EM} \cdot \bQ_L$}}, where
\beq
                  {\bf a}_{\rm EM}~=~ (\, 0, ~ 1, ~ -\third, \, -\third,\, -\third   )~
\eeq
(with all other components vanishing).
As a check we verify that {\mbox{$T_3 = {\bf a}_{T_3} \cdot \bQ_L$}},
where
{\mbox{${\bf a}_{T_3} = {\bf a}_{\rm EM} - {\bf a}_{Y} = (-\half, \half, 0,0,0) = \half \bQ_{T^+}$}}
where $\bQ_{T^+}$ is the charge vector (or root vector) associated with 
the $SU(2)$ gauge boson with positive $T_3$ charge.
 
Thus far, we have focused on the gauge structure of the theory.  As we have seen, the corresponding charge vectors
follow our usual group-theoretic expectations, just as they would in ordinary quantum field theory.
However, the charge vectors associated with the SM matter states 
in string theory are far more complex than would be expected in quantum field theory
and actually spill beyond the $SO(10)$ sublattice.

To see why this is so, it is perhaps easiest to 
consider the original $SO(10)$ theory {\it prior}\/ to electroweak breaking.
In this phase of the theory, the SM matter content consists of massless fermion and Higgs fields transforming 
in the ${\bf 16}$ and ${\bf 10}$ representations of $SO(10)$, respectively.
The former representations has charge vectors 
with $SO(10)$-sublattice components {\mbox{$Q^{(f)}_\ell =\pm 1/2$}} for each $\ell$ (with an odd net number of minus signs),
while the latter has
{\mbox{$Q^{(\phi)}_\ell =\pm \delta_{\ell k}$}} where {\mbox{$k=1,2,...,5$}}.
Thus, the ${\bf 16}$ and ${\bf 10}$ representations have 
conformal dimensions {\mbox{$h_{\bf 16}= 5/8$}} and {\mbox{$h_{\bf 10}=1/2$}}.
Indeed, according to the gauge embeddings discussed above,
the particular Higgs states which are electrically neutral have {\mbox{$Q_\ell =\pm \delta_{\ell 1}$}}. 
However, as a result of the non-trivial left-moving heterotic-string
vacuum energy {\mbox{$E_L= -1$}},
any  massless string state must correspond to worldsheet excitations contributing a total 
left-moving conformal dimension {\mbox{$h_L=1$}}.
Thus, even within the $SO(10)$ embedding specified above, string consistency constraints require 
that the SM fermion and Higgs states carry non-trivial 
charges not only {\it within}\/ the $SO(10)$ sublattice $\lbrace Q_1,Q_2,...,Q_5\rbrace$ {\it but also beyond it} --- \ie, 
elsewhere in the 17 remaining left-moving lattice directions {\mbox{$\bQ_{\rm int} \equiv \lbrace Q_6,...,Q_{\rm 22}\rbrace$}}
which {\it a priori}\/ correspond to gauge symmetries beyond those of the SM (such as those of potential hidden sectors).
Indeed, these additional excitations must contribute additional left-moving conformal dimensions $3/8$ and $1/2$ for
the SM matter and Higgs fields respectively, corresponding to
{\mbox{$[\bQ^{(f)}_{\rm int}]^2 = 3/4$}} for the fermions and {\mbox{$[\bQ^{(\phi)}_{\rm int}]^2 = 1$}} for the Higgs.

A similar phenomenon also occurs within the 10-dimensional {\it right-moving}\/ charge lattice, with 
components $\lbrace \tilde Q_1,..., \tilde Q_{10}\rbrace$.
The component associated with the non-zero component of the {\bf S}-vector discussed
below Eq.~(\ref{thetadef}) ---  henceforth chosen as $\tilde Q_1$  --- 
describes the spacetime spin-helicity of the state. 
As such, we must have {\mbox{$\tilde Q_1^{(f)}=\pm 1/2$}} for the SM fermions
and {\mbox{$\tilde Q_1^{(\phi)}=0$}} for the scalar Higgs. 
Of course, the right-moving side of the heterotic string has 
{\mbox{$E_R= -1/2$}}, requiring that all massless string states have total right-moving conformal dimensions {\mbox{$h_R=1/2$}}.   
We thus find that the SM fermion and Higgs fields 
must have additional nine-dimensional
charge vectors {\mbox{$\tilde \bQ_{\rm int}\equiv \lbrace\tilde Q_2,..., \tilde Q_{10}\rbrace$}}
(presumably corresponding to additional {\it right-moving}\/ gauge symmetries)
such that 
{\mbox{$[\tilde \bQ^{(f)}_{\rm int}]^2 = 3/4$}}
and 
{\mbox{$[\tilde \bQ^{(\phi)}_{\rm int}]^2 = 1$}}.

We see, then, that the electrically neutral Higgs field prior to electroweak symmetry breaking
must have a total $32$-dimensional charge vector of the form
\beqn
      \bQ_\phi ~&\equiv&~  
      (\bQ_L^{(\phi)}\,|\, \bQ_R^{(\phi)}) \nonumber\\ 
          ~&=&~  
      (1,0,0,0,0, \,  \bQ_{\rm int}^{(\phi)} \,|\,  0 , \, \tilde \bQ_{\rm int}^{(\phi)} ) 
\label{Higgsform}
\eeqn
where {\mbox{$[\bQ_{\rm int}^{(\phi)}]^2 = [\tilde \bQ_{\rm int}^{(\phi)}]^2 = 1$}}. 
In general, the specific forms of
$\bQ_{\rm int}^{(\phi)}$
and 
$\tilde \bQ_{\rm int}^{(\phi)}$
depend on the specific string model and the spectrum beyond the Standard Model.
However, those components which are specified within Eq.~(\ref{Higgsform}) are guaranteed 
by the underlying $SO(10)$ structure 
and by the requirement that the Higgs be electrically neutral.
Of course, the process of electroweak symmetry breaking can in principle alter the form of this vector.
However, we know that $U(1)_{\rm EM}$ necessarily remains unbroken.
Thus, even if the forms of the particular ``internal'' vectors $\bQ_{\rm int}^{(\phi)}$ and  
$\tilde \bQ_{\rm int}^{(\phi)}$ 
are shifted under electroweak symmetry breaking,
the zeros in the charge vector in Eq.~(\ref{Higgsform})
 ensure the electric neutrality of the Higgs field and must therefore be preserved.
This remains true not only for the physical Higgs field after electroweak symmetry breaking, but also
for its quantum fluctuations in the Higgsed phase.

This observation immediately allows us to constrain the form of the $\calT$-matrices which parametrize
the response of the charge lattice to small fluctuations of the Higgs field around its minimum.
Because the zeros in the charge vector in Eq.~(\ref{Higgsform}) must remain vanishing --- and
indeed because the electromagnetic charges and spin-statistics of {\it all}\/ string states must remain
unaltered 
under such fluctuations ---
we see that the (necessarily anti-symmetric) $\calT$-matrix can at most have the general form
\beq
\calT ~\sim~ \left(
   \begin{array}{cccccc|cc}
    ~ & ~ & ~ & ~ & ~                           &     \bt  &    0     & \tilde\bt    \\     
    ~ & ~ & ~ & ~ & ~                           &     \bz    &    0     &    \tilde\bz         \\     
    ~ & ~ & {\bf 0}_{5\times 5}  & ~ & ~        &     \bz  &    0     &    \tilde \bz         \\     
    ~ & ~ & ~ & ~ & ~                           &     \bz  &    0     &    \tilde \bz         \\     
    ~ & ~ & ~ & ~ & ~                           &     \bz  &    0     &    \tilde \bz         \\     
    -\bt^t & \bz^t  & \bz^t  & \bz^t  & \bz^t                      &      \bt_{11}   &    \bz^t    &     \bt_{12}         \\ 
   \hline     
      \rule{0pt}{2.5ex}                                                        0 & 0 & 0 & 0 & 0                           &     \bz  &    0   &      \tilde\bz         \\     
    -\tilde 
      \bt^t & \tilde \bz^t  & \tilde \bz^t  & \tilde\bz^t  & \tilde\bz^t  &     -\bt_{12}^t  &    \tilde \bz^t &        
                 \bt_{22}              
\end{array}
\right)
\label{Tmatrixform}
\eeq
where 
$\bt$ is an arbitrary 
 17-dimensional row vector;
where $\tilde \bt$ is an arbitrary 
9-dimensional row vector;
where $\bt_{11}$, $\bt_{12}$, and $\bt_{22}$
are arbitrary 
matrices of dimensionalities {\mbox{$17\times 17$}}, {\mbox{$9\times 17$}}, and {\mbox{$9\times 9$}}, respectively, with
$\bt_{11}$ and $\bt_{22}$ antisymmetric;
and where $\bz$ and $\tilde \bz$ are respectively $17$- and $9$-dimensional
zero row vectors.  
Indeed, as we have seen,
only this form of the $\calT$-matrix can preserve the electromagnetic charges and spin-statistics
of the string states  
under small shifts in the Higgs field around its new minimum, assuming a heterotic string model
with a standard level-one $SO(10)$ embedding.
The precise forms of $\bt$, $\tilde \bt$, 
$\bt_{11}$, $\bt_{12}$, and $\bt_{22}$ then depend on 
more model-specific details of how the Higgs is realized within the theory --- details which go beyond
the $SO(10)$ embedding.

As indicated above, this is only one particular example of the kinds of $\calT$-matrices that can occur.
However, all of the results of this paper will be completely general,
and will not rest on this particular example.

\subsection{Calculating the Higgs mass}

We can now use the general results in Sect.~\ref{latdeform}
to calculate the mass of $\phi$.
In general, this mass can be defined as
\beq
         m_\phi^2 ~\equiv~  {d^2 \Lambda(\phi)\over d \phi^2 } \biggl|_{\phi=0}
\label{higgsdef}
\eeq
where
\beq
             \Lambda(\phi) ~\equiv~ -{{\cal M}^4\over 2} \int_{\cal F} \dmu \, {\cal Z}(\tau,\taubar,\phi)~.
\label{Lambdaphi}
\eeq
Indeed, $\Lambda(\phi)$ is the vacuum energy that governs the dynamics of $\phi$.
Here
$ d^2\tau /\tau_2^2$ is the modular-invariant integration measure, 
$\calF$ is the fundamental domain of the modular group
\beq
            \calF ~\equiv~ \lbrace \tau :\,  -\half <\tau_1\leq \half, \,  \tau_2>0,\, |\tau|\geq 1 \rbrace~,
\label{Fdef}
\eeq
and {\mbox{$\calM\equiv M_s/(2\pi)$}} 
is the reduced string scale.
In this expression, following Eq.~(\ref{preZQdef}), the shifted partition function is given by
\beq
        {\cal Z}(\tau,\taubar,\phi) ~=~ \tau_2^{-1} {1\over \overline{\eta}^{12} \eta^{24}} \, \sum_{\bQ_L,\bQ_R} 
             (-1)^F \qbar^{\bQ_R^2/2} 
             q^{\bQ_L^2/2}~
\label{ZQdef}
\eeq
where the left- and right-moving charge vectors $\bQ_L$ and $\bQ_R$ are now deformed as in Eq.~(\ref{Qdeform})
and thus depend on $\phi$.

Given this definition, we begin by evaluating the leading contribution to the Higgs mass by taking 
partial derivatives of ${\cal Z}$, {\it i.e.}\/,
\beq
   {\partial^2\calZ \over \partial\phi^2} ~=~ 
        \tau_2^{-1} {1\over \overline{\eta}^{12} \eta^{24}} \, \sum_{\bQ_L,\bQ_R\in L} 
             (-1)^F \,X\, \,\qbar^{\bQ_R^2/2} q^{\bQ_L^2/2}~
\label{eq:Zexp}
\eeq
where the summand insertion $X$ is given by
\beq
      X ~\equiv~ \pi i 
     {\partial^2\over \partial\phi^2} (\tau \bQ_L^2 - \taubar \bQ_R^2 ) 
      -  \pi^2  \left\lbrack {\partial\over \partial\phi} (\tau \bQ_L^2 - \taubar \bQ_R^2)\right\rbrack^2~.
\label{stuff}
\eeq
Note that it is the {\it partial}\/ derivative $\partial^2/\partial\phi^2$ in Eq.~(\ref{eq:Zexp}) which provides the 
leading contribution to the Higgs mass;
we shall return to this point shortly.
Expanding $X$ in powers of $\tau_1$ and $\tau_2$ and then setting {\mbox{$\phi= 0$}} yields
\beq
  X\bigl|_{\phi=0} ~=~ 
      A \,\tau_1 + B \,\tau_2 + C \,\tau_1^2 + D \,\tau_2^2 + E \,\tau_1\tau_2~,
\eeq
where
\beqn 
   A~&=&~0~,\nonumber\\
   B~&=&~ -2\pi\alpha' (
            \bQ_a^2 +  \tilde \bQ_a^2 + \bQ_b^t \bQ_L +  \tilde \bQ_b^t \bQ_R )~,\nonumber\\   
   C~&=&~0~,\nonumber\\
   D~&=&~     4\pi^2 \alpha' \,(\bQ_a^t \bQ_L + \tilde \bQ_a^t \bQ_R)^2~ , \nonumber\\
   E~&=&~0~.
\label{intermed}
\eeqn 
Note that $A$, $C$, and $E$ each vanish as the result of the constraints in 
Eq.~(\ref{levelmatching}).  This is consistent, as these are the quantities which are proportional 
to powers of $\tau_1$, which multiplies $\bQ_L^2-\bQ_R^2$ within Eq.~(\ref{stuff}). 

Using Eqs.~(\ref{shiftQ}) and (\ref{NTrelations}),
we can now express the shift vectors within Eq.~(\ref{intermed}) directly in terms 
of $\bQ_L$ and $\bQ_R$.
For convenience we define
\beq
        \bQ_h \equiv \calT_{21} \bQ_L~,~~~~~
 \tilde \bQ_h \equiv \calT_{12} \bQ_R~,
\label{Qhdef}
\eeq
and likewise define 
\beq
        \bQ_j \equiv \calT_{11} \bQ_L~,~~~~~
 \tilde \bQ_j \equiv \calT_{22} \bQ_R~.
\label{Qjdef}
\eeq
We then find
\beqn
   B ~&=&~  -4\pi \alpha' (\bQ_h^2 + \tilde \bQ_h^2 - \tilde \bQ_j^t \bQ_h -  \bQ_j^t \tilde \bQ_h)~,\nonumber\\
   D ~&=&~  4\pi^2 \alpha' (\bQ_R^t \bQ_h - \bQ_L^t \tilde \bQ_h)^2~.
\label{eq:finalb}
\eeqn
Note the identity {\mbox{$\bQ_R^t \bQ_h = - \bQ_L^t \tilde \bQ_h$}}, as a result of which
our expression for $D$ can actually be collapsed into one term.
However, we have retained this form for $D$ in order to make manifest the symmetry between left- and right-moving contributions.
Our overall insertion into the partition function is then given by
{\mbox{$X|_{\phi=0} \equiv {\calX}/\calM^2$}}, where
\beqn
  \calX ~&=&~  
          \tau_2^2 \, (\bQ_R^t \bQ_h - \bQ_L^t \tilde \bQ_h)^2  \nonumber\\
   && ~~ - {\tau_2\over \pi } \left(\bQ_h^2 + \tilde \bQ_h^2 - \tilde \bQ_j^t \bQ_h -  \bQ_j^t \tilde \bQ_h\right)~.~~~~~~
\label{Xdef}
\eeqn

\subsection{Modular completion and additional Higgs-mass contributions \label{sec:completion}}

Thus far, we have calculated the leading contribution to the Higgs mass 
by evaluating $\partial^2 {\cal Z}/\partial \phi^2$.
However, the full contribution $d^2 {\cal Z}/d\phi^2$ (with full rather than 
partial $\phi$-derivatives) also includes
various additional effects on the partition function ${\cal Z}$ 
that come from fluctuations of the Higgs field.  For example, such fluctuations deform   
the background moduli fields (such as the metric that contracts compactified components of  $\bQ_L^2$ and $\bQ_R^2$ within the 
charge lattice).
Such effects produce additional contributions to the total Higgs mass.

It turns out that 
we can calculate all of these extra contributions in a completely general way 
through the requirement of modular invariance. 
Indeed, because modular invariance remains unbroken even when the theory is Higgsed,
the final expression for the total Higgs mass must not only be modular invariant but also arise
through a modular-covariant sequence of calculational operations.
As we shall demonstrate, the above expression for the insertion $\calX$ 
in Eq.~(\ref{Xdef}) does not have this property.    We shall therefore determine the additional contributions
to the Higgs mass by performing the ``modular completion'' of $\calX$ --- {\it i.e.}\/, by determining the
additional contribution to $\calX$ which will render 
this insertion consistent with modular invariance.
 
In general, prior to the insertion of $\calX$, the partition-function trace in Eq.~(\ref{preZQdef}) 
[or equivalently the trace in Eq.~(\ref{ZQdef}) evaluated at {\mbox{$\phi=0$}}]
is presumed to already be modular invariant, as required for the consistency of the underlying string.
In order to determine the modular completion of the quantity $\calX$ in Eq.~(\ref{Xdef}),
we therefore need to understand the modular-invariance effects 
that arise when $\calX$ is inserted into this partition-function trace.
Because $\calX$ involves various combinations of components of charge vectors,
let us begin by investigating the effect of inserting powers of a single charge vector 
component $Q_\ell$ (associated with the $\ell^{\rm th}$ lattice direction) 
into our partition-function trace.
Within the partition functions described in Eqs.~(\ref{partfZ}) and (\ref{partft}),
inserting $Q_\ell^n$ for any power $n$
is tantamount to replacing
\beq
     \vartheta_\ell ~\to~ \sum_{Q_\ell\in \mathbb{Z} +\alpha_\ell}  e^{2\pi i \beta_\ell Q_\ell} ~Q_\ell^n~ q^{Q_\ell^2/2}~.
\label{Qinsert}
\eeq
However, one useful way to proceed is to recognize that this latter sum
can be rewritten as
\beq
     {1\over (2\pi i)^n} \, \frac{\partial^n }{\partial z_\ell^n} \vartheta_\ell(z_\ell|\tau) \biggl|_{z_{\ell=0}}
\label{zderiv}
\eeq
where the generalized $\theta_\ell (z_\ell|\tau)$ 
function is defined as
\beq
        \vartheta_\ell(z_\ell|\tau) 
              ~\equiv~
              \sum_{Q_\ell\in \mathbb{Z} +\alpha_\ell} e^{2\pi i (\beta_\ell +z_\ell) Q_\ell} \,q^{Q_\ell^2/2}~.~~
\label{thetadefz}
\eeq
Indeed, we see that $\vartheta_\ell(\tau)$ is nothing but $\vartheta_\ell (z_\ell|\tau)$
evaluated at {\mbox{$z_\ell=0$}}.
However, for arbitrary $z$, these 
generalized $\vartheta(z|\tau)$ functions have the schematic modular-transformation properties
\beq 
      \vartheta_\ell\left(z\biggl|\frac{a\tau+b}{c\tau+d}\right) ~\sim~ 
              (c\tau+d)^{1/2} \, e^{ \pi i c(c\tau+d) z^2}\, 
      \vartheta_{\ell} ((c\tau+d) z | \tau)~.
\label{modtransthetaz}
\eeq
It then follows that 
\beq
   \vartheta\left(z\biggl|\frac{a\tau+b}{c\tau+d}\right) 
           \biggl|_{z=0} ~\sim~  (c\tau+d)^{1/2} \,
  \vartheta(z|\tau)\biggl|_{z=0}~,
\label{firstderiv}
\eeq
and likewise 
\beq
  \frac{\partial}{\partial z} 
   \vartheta\left(z\biggl|\frac{a\tau+b}{c\tau+d}\right) 
 \biggl|_{z=0} ~\sim~  (c\tau+d)^{3/2} \,\frac{\partial}{\partial z} 
  \vartheta(z|\tau)\biggl|_{z=0}~.
\label{secondderiv}
\eeq
This indicates that while the function {\mbox{$\vartheta(z|\tau)|_{z=0}$}} transforms covariantly with modular weight $1/2$,
its first derivative {\mbox{$[\partial\vartheta(z|\tau)/\partial z]|_{z=0}$}} transforms covariantly with modular weight $3/2$.

At first glance, one might expect this pattern to continue, with the second derivative
{\mbox{$[\partial^2 \vartheta(z|\tau)/dz^2]|_{z=0}$}} transforming
covariantly with modular weight $5/2$.   However, this is not what happens.   Instead, 
from Eq.~(\ref{modtransthetaz}) we find
\beqn
  \frac{\partial^2}{\partial z^2} 
   \vartheta\left(z\biggl|\frac{a\tau+b}{c\tau+d}\right) 
\biggl|_{z=0} &\sim&~   (c\tau+d)^{5/2} \,\frac{\partial}{\partial z} 
  \vartheta(z|\tau)\biggl|_{z=0} \nonumber\\
 && ~ + ~2\pi i \, c\, (c\tau+d)^{3/2}\, \vartheta(\tau)~.~~~~~~\nonumber\\
\eeqn
While the term on the first line is the expected result,
the term on the second line represents a {\it modular anomaly} which destroys the 
modular covariance of the second derivative.

Since modular covariance must be preserved, we must perform
a {\it modular completion}.  In this simple case, this means that
we must replace 
${\partial^2 / \partial z^2}$ with a {\it modular-covariant second derivative}\/ $D^2_z$
such that $D^2_z $ not only contains
$\partial^2 /\partial z^2$ but also has the property that {\mbox{$D^2_z \theta(z|\tau)|_{z=0}$}} transforms covariantly with weight $5/2$.
It is straightforward to show that the only such modular-covariant derivative is
\beq
      D^2_z ~\equiv~  \frac{\partial^2}{\partial z^2} + \frac{\pi}{\tau_2}~,
\label{modcovderiv}
\eeq
and with this definition one indeed finds
\beq
      \left\lbrace \left[ D^2_z \vartheta( z|\tau)\right]_{\tau\to \frac{a\tau+b}{c\tau+d}}  \right\rbrace\biggl|_{z=0} 
           ~\sim~ (c\tau+d)^{5/2} \, D^2_z \vartheta(z|\tau)\biggl|_{z=0}~,
\eeq
thereby continuing the pattern set by Eqs.~(\ref{firstderiv}) and (\ref{secondderiv}).
It turns out that this modular-covariant second $z$-derivative 
is equivalent to the
modular-covariant $\tau$-derivative  
\beq         D_\tau  ~\equiv ~ {\partial \over \partial \tau} ~-~ {ik\over 2\tau_2}~ 
\eeq
which preserves the modular covariance of any modular function of weight $k$.
Indeed, 
     our $\vartheta(z|\tau)$ functions have {\mbox{$k=1/2$}} and satisfy 
the heat equation 
$\partial^2 \vartheta(z|\tau) /\partial z^2 = 4\pi i \, \partial \vartheta(z|\tau) /\partial \tau$.
In this sense, the $z$-derivative serves as a ``square root'' of the $\tau$-derivative and gives us a precise
means of extracting the individual charge insertions (rather than their squares).
         In this connection, we emphasize that there is a tight correspondence between the Higgs field and the
            $z$-parameter.    Specifically, when we deform a theory through a continuous change in the value of the Higgs VEV, 
             its partition function deforms through a corresponding continuous change in the $z$-parameter.

In principle we could continue to examine higher $z$-derivatives 
(all of which will also suffer from modular anomalies),
but the results we have thus far will be sufficient for our purposes.
Recalling the equivalence between the expressions in Eqs.~(\ref{Qinsert}) and (\ref{zderiv}),
we thus see that the insertion of a single power of any given $Q_\ell$ does not disturb the
modular covariance of the corresponding holomorphic (or anti-holomorphic) factor in the 
partition-function trace, but the insertion of a quadratic term $Q_\ell^2$
along the $\ell^{\rm th}$ lattice direction
does {\it not}\/ lead to a modular-covariant result and must, according to Eq.~(\ref{modcovderiv}),
be replaced by the modular-covariant insertion 
$Q_\ell^2 - 1/(4\pi\tau_2)$.
Thus, our rules for modular completion 
through second-order in charge-vector components are given by
\beq
\begin{cases} 
   ~\bQ_\ell &\to~~ \bQ_\ell \\
   ~\bQ_\ell\bQ_{\ell'} &\to~~ \bQ_\ell \bQ_{\ell'}  - \frac{1}{4\pi \tau_2} \delta_{\ell,\ell'}~.
\end{cases}
\label{completionrules}
\eeq
These general results hold for all lattice directions $(\ell,\ell')$ regardless of whether 
they correspond to left- or right-moving lattice components.
Such modular completions have also arisen in other contexts, 
such as within string-theoretic threshold corrections~\cite{Kiritsis:1994ta,Kiritsis:1996dn,Kiritsis:1998en}. 

With these modular-completion rules in hand, we can now investigate the modular completion of the 
expression for $\calX$ in Eq.~(\ref{Xdef}).
It is simplest to begin by focusing on
the quartic terms, \ie, the terms in the top line of Eq.~(\ref{Xdef}).
Given the identity just below Eq.~(\ref{eq:finalb}),
these terms are proportional to $(\bQ_L^t \tilde \bQ_h)^2$.
With $Q_{L\ell}$ denoting the $\ell^{\rm th}$ component of $\bQ_L$, {\it etc.}\/, 
we find 
\beqn
 (\bQ_L^t \tilde \bQ_h)^2
      &=&  \left(  \sum_{\ell=1}^{22} \sum_{m=1}^{10}  Q_{L\ell} (\calT_{12})_{\ell m} Q_{Rm} \right)^2 \nonumber\\
       &=& \sum_{\ell,\ell' =1}^{22} \sum_{m,m'=1}^{10} 
          (\calT_{12})_{\ell m}  (\calT_{12})_{\ell' m'} \nonumber\\  
         && ~~~~~~~~~~~\times~ Q_{Rm} Q_{Rm'} Q_{L\ell} Q_{L\ell'}~.~~~~~~~~~~~~~~
\label{intterm}
\eeqn
Following the rules in Eq.~(\ref{completionrules}), we can readily 
obtain the modular completion of this expression by replacing the final line in Eq.~(\ref{intterm})
with
\beq
          \left( Q_{Rm} Q_{Rm'} - \frac{1}{4\pi \tau_2} \delta_{mm'} \right) 
          \left( Q_{L\ell} Q_{L\ell'} - \frac{1}{4\pi \tau_2} \delta_{\ell\ell'} \right)~. 
\label{substi}
\eeq
Substituting Eq.~(\ref{substi}) into Eq.~(\ref{intterm}) and recalling  
that {\mbox{$\calT_{12}^t = -\calT_{21}$}}, 
we thus find that the modular completion 
of the quartic term $(\bQ_L^t \tilde \bQ_h)^2$
within $\calX$
is given by
\beq
 (\bQ_L^t \tilde \bQ_h)^2 - \frac{1}{4\pi \tau_2} \left( \bQ_h^2 + \tilde \bQ_h^2\right) + \frac{\xi}{(4\pi\tau_2)^2}~~~ 
\label{quadr1}
\eeq
where 
\beqn
       \xi  ~&\equiv&~ 
         {\rm Tr} (\calT_{12}^t \calT_{12})
       ~=~ {\rm Tr} (\calT_{21}^t \calT_{21})~\nonumber\\
       && ~~~=~ - {\rm Tr} (\calT_{12} \calT_{21}) ~=~ - {\rm Tr} (\calT_{21} \calT_{12})~.~~~~~~~~
\eeqn

Remarkably, the quadratic terms $\bQ_h^2 + \tilde \bQ_h^2$ that are generated within Eq.~(\ref{quadr1}) 
already appear on the second line of Eq.~(\ref{Xdef}).
In other words, even if we had not already known of these quadratic terms, we could have deduced their existence
through the modular completion of our quartic terms!
Conversely, we could have generated the quartic terms through a modular completion of these quadratic terms --- {\it i.e.}\/,
each set of terms provides the modular completion of the other.
Thus, the only remaining terms within Eq.~(\ref{Xdef}) that might require modular completion are 
the final quadratic terms on the second line of Eq.~(\ref{Xdef}), namely
$\tilde \bQ_j^t \bQ_h +  \bQ_j^t \tilde \bQ_h$.
However, 
$\bQ_h$ and $\bQ_j$ involve only left-moving components of the lattice while
$\tilde \bQ_h$ and $\tilde \bQ_j$ involve only right-moving components.
Thus 
$\tilde \bQ_j^t \bQ_h +  \bQ_j^t \tilde \bQ_h$
is already modular complete.
Putting all the pieces together, we therefore find that the total expression for $\calX$ in Eq.~(\ref{Xdef}) has
a simple (and in fact universal) modular completion:
\beq
      \calX ~\to~ 
      \calX ~+~ \frac{\xi}{4\pi^2} ~.  
\label{Xmodcomplete}
\eeq
Indeed, this sole remaining extra term generated by the modular completion stems from 
the final term in Eq.~(\ref{quadr1}).
It is noteworthy that this extra term 
is entirely independent of the charge vectors.
This is consistent with our expectation that such additional terms
represent the contributions from the deformations of the moduli fields under Higgs fluctuations --- deformations
which act in a universal (and hence $Q$-independent) manner.

Some remarks are in order regarding
the uniqueness of the completion
in Eq.~(\ref{Xmodcomplete}).
In particular, at first glance one might wonder how the modular completion of 
the quadratic terms $\tilde \bQ_j^t \bQ_h +  \bQ_j^t \tilde \bQ_h$ could 
uniquely determine the quartic terms in $\calX$, given that the modular-completion rules 
within Eq.~(\ref{completionrules}) only seem to generate extra terms 
which are of lower powers in charge-vector components.
However, the important point is that the rules in Eq.~(\ref{completionrules}) only ensure
the modular covariance of the individual (anti-)holomorphic components
of the partition-function trace.  
In particular, these rules do not, in and of themselves, ensure that 
we continue to satisfy 
the additional constraint in Eq.~(\ref{genexp2}) 
that arises when stitching these holomorphic and anti-holomorphic components together
as in Eq.~(\ref{genexp}). 
However, given that $\bQ_h^2$ increases the modular weight of the 
holomorphic component by two units without increasing the modular weight of the
anti-holomorphic component,
and given that  
$\tilde \bQ_h^2$ does the opposite, the only way to properly modular-complete their sum
is by ``completing the square'' and realizing these terms as the off-diagonal terms that are generated
through a factorized modular completion as in Eq.~(\ref{substi}).
This then compels the introduction of the appropriate quartic diagonal terms, as seen above.

In this connection, it is also important to note that modular completion involves
more than simply demanding that our final result be modular invariant.
After all, we have seen in Eq.~(\ref{Xmodcomplete}) that the modular completion of $\calX$ 
involves the addition of a pure number, \ie, the addition of 
a quantity which is intrinsically modular-invariant on its own (or more precisely, 
a quantity whose
insertion into the partition-function summand automatically preserves
the modular invariance of the original partition function).   
However, 
as we have stated above, modular completion 
ensures more than the
mere modular invariance of our final result --- it also ensures that
this result is obtainable
through a modular-covariant sequence of calculational operations.
As we have seen, the extra additive constant that forms the modular completion
of $\calX$ in Eq.~(\ref{Xmodcomplete}) is crucial in allowing us to ``complete the square'' 
and thereby cast our results into the factorized form
of Eq.~(\ref{substi}) --- a form which itself emerged as a consequence of 
our underlying modular-covariant
$z$-derivatives $D^2_z$.
As such, the constant appearing in Eq.~(\ref{Xmodcomplete}) is an intrinsic  part 
of our resulting expression for $m_\phi^2$.

\subsection{Classical stability condition\label{stability}}

Thus far, we have focused on deriving an expression for the Higgs mass, as defined in Eq.~(\ref{higgsdef}). 
However, our results presuppose that we are discussing a classically stable 
particle.   In other words, while we are identifying the mass with the second $\phi$-derivative of the
classical potential, we are implicitly assuming that the first $\phi$-derivative vanishes so that we are sitting
at a minimum of the Higgs potential.
Thus, there is an extra condition that we need to impose, namely 
\beq
          {d \Lambda(\phi)\over d \phi } \biggl|_{\phi=0} ~=~ 0~.
\label{linearcond}
\eeq
This condition must be satisfied for the particular vacuum state 
within which our Higgs-mass calculation has been performed.

It is straightforward to determine the ramifications of this condition.
Proceeding exactly as above, we find in analogy with
Eq.~(\ref{stuff})
that {\mbox{$\partial\calZ/\partial \phi|_{\phi=0}$}} corresponds to an insertion given by
{\mbox{$Y|_{\phi=0} = {\calY/\calM}$}}, where
\beq
    {\calY} ~\sim~ \tau_2 \left( \bQ_R^t \bQ_h - \bQ_L^t \tilde \bQ_h\right) ~\sim~ \tau_2 \,(\bQ_R^t \calT_{21} \bQ_L)~. 
\label{Ydef}
\eeq
Given this result, there are {\it a priori}\/ three distinct ways in which 
the condition in Eq.~(\ref{linearcond}) can be satisfied 
within a given string vacuum.
First, $\calY$ might vanish for each state in the corresponding string spectrum.
Second, $\calY$ might not vanish for each state in the string spectrum 
but may vanish in the {\it sum}\/ over the string states
(most likely in a pairwise fashion between chiral and anti-chiral states with opposite charge vectors).
However, there is also a third possibility:
the entire partition-function trace may be non-zero, even with the $\calY$ insertion,
but nevertheless vanish when integrated over the fundamental domain of the modular group, as in Eq.~(\ref{Lambdaphi}). 
In general, very few mathematical examples are known of situations in which this latter
phenomenon occurs~{\mbox{\cite{Moore:1987ue,Dienes:1990qh,Dienes:1990ij}}},
although the fact that this would involve an integrand with vanishing modular weight offers
unique possibilities.

Two further comments regarding this condition are in order.
First, it is easy to verify that this condition respects modular invariance, as it must.
Indeed, the quantity $\calY$, as defined above, is already modular complete.
At first glance, this might seem surprising, given that the quartic terms within $\calX$ are nothing but the square of $\calY$,
and we have already seen that these quartic terms are not modular complete by themselves.
However, it is the squaring of $\calY$ 
that introduces the higher powers of charge-vector components
which in turn induce the modular anomaly. 
Second, if $\calY$ vanishes when summed over all of the string states,
then it might be tempting to hope that the quartic terms within $\calX$ 
also vanish when summed over the string states.
Unfortunately, this hope is not generally realized, since important sign information is lost when these quantities
are squared.
Of course, if $\calY$ vanishes for each individual state in the string spectrum,
then the quartic terms within $\calX$ will also evaluate to zero in any calculation of the corresponding Higgs mass.
This would then simplify the explicit evaluation of $\calX$ for such a string vacuum.

\subsection{A relation between the Higgs mass and the cosmological constant}

Let us now collect our results for the Higgs mass.
For notational simplicity we define 
\beq
 \langle A \rangle ~\equiv~ \int_{\cal F} \dmu 
        ~\frac{\tau_2^{-1}}{\overline{\eta}^{12} \eta^{24}} \, \sum_{\bQ_L,\bQ_R} 
             (-1)^F  A~ \qbar^{\bQ_R^2/2} q^{\bQ_L^2/2}~
\label{expvalue}
\eeq
where the charge vectors $\lbrace \bQ_L,\bQ_R\rbrace$ in the sum over states are henceforth understood as 
unperturbed (\ie, with {\mbox{$\phi=0$}}) and thus correspond directly to the charges that arise at the minimum of the Higgs potential.  
Our results then together imply that 
\beq 
       m_\phi^2 ~=~ -\frac{\calM^2}{2} \langle \calX\rangle  ~-~ \frac{\calM^2}{2} \frac{\xi}{4\pi^2} \langle {\bf 1} \rangle
\label{preresult}
\eeq
where $\calX$ is given in Eq.~(\ref{Xdef}).
As indicated above, these results implicitly assume that
{\mbox{$\langle \calY\rangle=0$}}, 
where $\calY$ is defined in Eq.~(\ref{Ydef}).
However, we immediately recognize that the quantity $\langle {\bf 1}\rangle$ within Eq.~(\ref{preresult}) is nothing other than 
the one-loop zero-point function (cosmological constant) $\Lambda$!~
More precisely, we may identify $\Lambda$  as {\mbox{$\Lambda(\phi)|_{\phi=0}$}} 
[where $\Lambda(\phi)$ is given in Eq.~(\ref{Lambdaphi})], or equivalently
\beq
       \Lambda ~=~ -\frac{\calM^4}{2} \,\langle {\bf 1}\rangle~.
\label{lambdadeff}
\eeq
We thus obtain the relation
\beq
           m_\phi^2 ~=~ \frac{\xi}{4\pi^2} \, \frac{\Lambda}{\calM^2}  ~-~ \frac{\calM^2}{2} \, \langle \calX \rangle~.
\label{relation1}
\eeq
Indeed, retracing our steps in arbitrary spacetime dimension $D$,
we obtain the analogous relation
\beq
           m_\phi^2 ~=~ \frac{\xi}{4\pi^2} \, \frac{\Lambda}{\calM^{D-2}}  ~-~ \frac{\calM^2}{2} \, \langle \calX \rangle~
\label{relation1b}
\eeq
where the cosmological constant $\Lambda$ now has mass dimension $D$. 

Remarkably, this is a general relation between the Higgs mass and the one-loop cosmological constant! 
Because this relation rests on nothing but modular invariance,
it holds generally for {\it any}\/ perturbative 
closed string in any arbitrary spacetime dimension $D$. 
The cosmological-constant term in Eq.~(\ref{relation1}) is universal, 
emerging as 
the result of a modular anomaly that required a modular completion, or equivalently
as the result of a universal shift in the moduli.
 By contrast, the second term depends on the particular charges that are inserted into the partition-function trace.

For weakly coupled heterotic strings,
we can push this relation one step further.
In such theories the string scale {\mbox{$M_s\equiv 2\pi \calM$}} and Planck scale $M_p$ are connected through the 
relation {\mbox{$M_s= g_s M_P$}} where
$g_s$ is the string coupling 
whose value is set by the vacuum expectation value of the dilaton.
Depending on the particular string model, $g_s$ in turn sets the values of the individual gauge couplings.
Likewise, the canonically normalized scalar field $\phi$ is 
{\mbox{$\widehat \phi \equiv  \phi/g_s$}}. 
We thus find that our relation in Eq.~(\ref{relation1}) equivalently takes the form
\beq
       m_{\widehat \phi}^2 ~=~ \frac{\xi}{M_P^2} \, \Lambda ~-~ \frac{g_s^2 \calM^2}{2} \, \langle \calX \rangle~.
\label{relation2}
\eeq

In quantum field theory, we would not expect to find a relation between a Higgs mass
and a cosmological constant.   Indeed, quantum field theories do not involve gravity and are thus
insensitive to the absolute zero of energy.
Even worse, in quantum field theory, the one-loop zero-point function is badly divergent. 
String theory, by contrast, not only unifies 
gauge theories with gravity but also yields a {\it finite}\/ $\Lambda$ (the latter
occurring as yet another byproduct of modular invariance).
Thus, it is only within a string context that such a relation could ever arise, and indeed
Eqs.~(\ref{relation1}) and (\ref{relation2}) are precisely the relations 
that arise for all weakly-coupled  four-dimensional heterotic strings. 
We expect that this is but the tip of the iceberg, and that other modular-invariant string constructions
lead to similar results.
It is intriguing that such relations join together precisely the two quantities ($m_\phi$ and $\Lambda$) whose
values lie at the heart of the two most pressing hierarchy problems in modern physics.

\section{Regulating the Higgs mass:~   From amplitudes to supertraces           \label{sec3}}

In Eq.~(\ref{relation1}) we obtained a result in which the Higgs mass, via the definition in Eq.~(\ref{expvalue}),
 is expressed in terms of certain one-loop string amplitudes
consisting of 
modular integrals of various traces over the entire string spectrum.
As discussed below Eq.~(\ref{eq:massRL}),
these traces include the contributions of not only {\it physical}\/ (\ie, level-matched) string states with {\mbox{$M_L^2=M_R^2$}},
but also {\it unphysical}\/ (\ie, non level-matched) string states with {\mbox{$M_L^2 \not= M_R^2$}}.
This distinction between physical and unphysical string states is important because only 
the physical string states can serve as {\it bona-fide}\/ in- and out-states.
By contrast, the unphysical states are intrinsically stringy and have no field-theoretic analogues.

We now wish to push our calculation several steps further.
In particular, there are three aspects to our result 
in Eq.~(\ref{relation1}) which we will need to understand 
in order to allow us to make contact with traditional quantum-field-theoretic expectations. 
The first concerns the fact that while the one-loop vacuum energy $\Lambda$ which appears in these
results is finite for all tachyon-free string models  --- even without spacetime supersymmetry --- the
remaining amplitude $\langle \calX\rangle$ which appears in these 
expressions is generically divergent.
Note that this is not in conflict with string-theoretic expectations;   
in particular, as we shall discuss in Sect.~\ref{UVIRequivalence},
string theory generally softens various field-theoretic divergences 
but need not remove them entirely.  
Thus, our expression for the Higgs mass is formally divergent and requires some sort of regulator 
in order to extract finite results.   
Second, while these results are expressed in terms of sums over the entire string spectrum,
we would like to be able to express
the Higgs mass directly in terms of supertraces over only the {\it physical}\/ string states --- \ie, the states
with direct field-theoretic analogues.
This will ultimately allow us to express the Higgs mass in a form that might be recognizable within ordinary quantum field
theory, and thereby extract an effective field theory (EFT) description of the Higgs mass
in which our Higgs mass experiences an effective renormalization-group ``running''. 
This will also allow us
to extract a stringy effective potential for the Higgs field. 
Finally, as a byproduct, we would also like to implicitly perform the stringy modular integrations 
inherent in Eq.~(\ref{expvalue}). 

As it turns out, these three issues are intimately related.
However, appreciating these connections requires 
a deeper understanding of the properties of the modular functions
on which which our Higgs-mass calculations rest.
In this section, we shall therefore outline the  
mathematical procedures which will enable us to address all three of our goals.
Many of these methods originated in the classic mathematics papers of Rankin~{\mbox{\cite{rankin1a,rankin1b}}} and Selberg~\cite{selberg1} 
from the late 1930s, and were later extended in an important way by Zagier~\cite{zag} in the early 1980s.
Some of the Rankin-Selberg results also later independently found their way 
into the string literature in various forms~{\mbox{\cite{McClain:1986id,OBrien:1987kzw,Kutasov:1990sv}}},
and have occasionally been studied and further developed (see, {\it e.g.}\/, Refs.~\cite{Dienes:1994np, Dienes:1995pm,
Dienes:2001se,
Angelantonj:2010ic,
Angelantonj:2011br,
Angelantonj:2012gw,
Angelantonj:2013eja,
Pioline:2014bra,
Florakis:2016boz}).
Our purpose in recounting these results here is not only to pull them all together  and explain their logical connections
in relatively non-technical terms, but also to extend them in certain directions which will be important for our work in Sect.~\ref{sec4}.~
This conceptual and mathematical groundwork
will thus form the underpinning for our further analysis of the Higgs mass in Sect.~\ref{sec4}.

\subsection{The Rankin-Selberg technique\label{sec:RStechnique}}

We are interested in 
modular integrals such as those in Eq.~(\ref{expvalue}) 
which generically take the form
\beq
                 I~\equiv ~\int_{\mathcal{F}}\dmu \,F(\tau,\taubar)~,
\label{eq:I}
\eeq
where $\calF$ is the modular-group fundamental domain given in Eq.~(\ref{Fdef}),
where $d\tau_1 d\tau_2/\tau_2^2$ is the modular-covariant integration measure
(with {\mbox{$\tau\equiv \tau_1+i\tau_2$}}, {\mbox{$\tau_i\in\mathbb{R}$}}), 
and where the integrand $F$ is modular invariant.
In general the integrands $F$ take the form
\beq
           F~\equiv~ \tau_2^k\, \sum_{m,n}  a_{mn} \qbar^m q^n
\label{integrand}
\eeq
where {\mbox{$q\equiv e^{2\pi i\tau}$}}
and where $k$
is the modular weight of the holomorphic and anti-holomorphic modular functions whose products contribute
to $F$.
Note that integrands of this form include those in Eq.~(\ref{Zform}):   we simply power-expand
the $\eta$-function denominators and absorb these powers into 
$m$ and $n$.
Thus, with string integrands written as in Eq.~(\ref{integrand}) we can now directly identify
{\mbox{$m= \alpha' M_R^2/4$}} and
{\mbox{$n= \alpha' M_L^2/4$}}.
The quantity $a_{mn}$ then tallies the number of bosonic minus fermionic string degrees of freedom
contributing to each $(M_R^2,M_L^2)$ term.

Invariance under {\mbox{$\tau\to \tau+1$}} guarantees that every term within $F$ has {\mbox{$m-n\in\mathbb{Z}$}}.
The {\mbox{$m=n$}} terms represent the contributions from physical string states with spacetime masses {\mbox{$\alpha' M^2 = 2(m+n)= 4n$}}, 
while the {\mbox{$m\not=n$}} terms represent the contributions from off-shell (\ie, unphysical) string states.
Within the {\mbox{$\tau_2\geq 1$}} integration subregion within $\calF$,
the {\mbox{$m\not=n$}} terms make no contribution to the integral $I$ 
because these contributions are eliminated when we perform the 
$\int_{-1/2}^{1/2} d\tau_1$ integral.
[Indeed, within this subregion of $\calF$
expressions such as Eq.~(\ref{eq:I}) come
with an implicit instruction 
that we are to perform the $\tau_1$ integration prior to performing the $\tau_2$ integration.]
However, the full integral $I$ does receive {\mbox{$m\not=n$}} contributions 
from the {\mbox{$\tau_2<1$}} subregion within $\calF$.
Thus, in general, both physical and unphysical string states contribute to amplitudes such as $I$.

Our goal is to express $I$ in terms of contributions from the physical string states alone.
Clearly this could be done if we could somehow transform the region of integration within $I$
from the fundamental domain $\calF$ to the positive half-strip 
\beq
            \calS ~\equiv~ \lbrace \tau :\,  -\half <\tau_1\leq \half, \,  \tau_2>0 \rbrace~,
\label{Sdef}
\eeq
for we would then have
\beq
          \int_\calS \dmu  \,F(\tau,\taubar) ~=~ \int_0^\infty {d\tau_2 \over \tau_2^2}  \, g(\tau_2)
\eeq
where $g(\tau_2)$ is our desired trace over only the physical string states:
\beq
               g(\tau_2) ~=~ \int_{-1/2 }^{1/2} d\tau_1 \,F(\tau,\taubar)~=~ \tau_2^k\, \sum_{n} a_{nn} \,e^{-4\pi\tau_2 n}~.~~~~
\label{gtrace}
\eeq  

Fortunately, there exists a well-known method for ``unfolding'' $\calF$ into $\calS$.
While $\calF$ is the fundamental domain of the modular group $\Gamma$ generated by both {\mbox{$\tau\to -1/\tau$}} and {\mbox{$\tau\to \tau+1$}},
the strip $\calS$ is the fundamental domain of the modular {\it subgroup}\/ {\mbox{$\Gamma_\infty$}} generated solely by {\mbox{$\tau\to \tau+1$}}.
(Indeed, this is the subgroup that preserves the cusp at {\mbox{$\tau=i\infty$}}.) 
Thus the strip $\calS$ can be realized as the sum of the images of $\calF$ transformed under all modular transformations $\gamma$ (including the identity) in 
the coset {\mbox{$\Gamma_\infty \backslash \Gamma$}}:
\beq
           \calS ~=~ \bigcup\limits_{\gamma\in \Gamma_\infty \backslash \Gamma} \gamma\cdot \calF ~.
\label{stripF}
\eeq
It then follows 
for any integrand $\widetilde F(\tau,\taubar)$ 
that
\beq
      \int_\calS \dmu \, \widetilde F(\tau,\taubar) ~=~ \int_\calF \dmu 
           \sum_{\gamma\in \Gamma_\infty \backslash \Gamma} \widetilde F_\gamma(\tau,\taubar)~,
\label{unfold}
\eeq
where $\widetilde F_\gamma(\tau,\taubar)$ is the $\gamma$-transform of $\widetilde F(\tau,\taubar)$.
Moreover, if $\widetilde F(\tau,\taubar)$ is invariant under {\mbox{$\tau\to\tau+1$}}, then the total integrand on
the right side of Eq.~(\ref{unfold}) is modular invariant.

At this stage, armed with the result in Eq.~(\ref{unfold}), 
we see that we are halfway towards our goal.
However, two fundamental problems remain.
First, while choosing $\widetilde F$ as our original integrand $F$ would allow us to express the left side of Eq.~(\ref{unfold}) 
directly in terms of the desired trace in Eq.~(\ref{gtrace}), our need to relate this to the original integral $I$ in 
Eq.~(\ref{eq:I}) would instead seem to require choosing $\tilde F$ such that 
{\mbox{$F= \sum_{\gamma\in \Gamma_\infty \backslash \Gamma} \widetilde F_\gamma$}}.
Second, the manipulations underlying Eq.~(\ref{unfold}), such as the exchanging of sums and regions of integration, implicitly 
assumed that the integrand on the right side of Eq.~(\ref{unfold}) converges sufficiently rapidly as {\mbox{$\tau_2\to\infty$}} 
[or equivalently that the integrand on the left side of Eq.~(\ref{unfold}) converges sufficiently rapidly as {\mbox{$\tau_2\to 0$}}]
so that all relevant integrals are absolutely convergent.
However this is generally not the case for the physical situations that will interest us.

It turns out that these problems together motivate a unique choice for $\widetilde F$.
Note that $g(\tau_2)$ generally has a form resembling that in Eq.~(\ref{gtrace}), consisting of an infinite sum multiplied by a power of $\tau_2$.   As {\mbox{$\tau_2\to 0$}}, 
the successive terms in this sum are less and less suppressed by the exponential factor $e^{-4\pi n\tau_2}$.   We therefore expect the infinite sum within $g(\tau_2)$ to experience an increased tendency to diverge as {\mbox{$\tau_2\to 0$}}.   Let us assume for the moment that the divergence of this infinite sum grows no faster than some inverse power of $\tau_2$ as {\mbox{$\tau_2\to 0$}}.    In this case, 
the divergence of the sum within $g(\tau_2)$ will cause 
$g(\tau_2)$ itself to diverge as {\mbox{$\tau_2\to 0$}} unless  $g(\tau_2)$ also includes a prefactor consisting of sufficiently many powers of $\tau_2$ to hold the divergence of the sum in check.  We can therefore {\it regulate}\/ our calculation by introducing sufficiently many 
extra powers of $\tau_2$ into $g(\tau_2)$.  In other words, in such cases we shall take
\beq
         \widetilde F(\tau,\taubar) ~=~ \tau_2^s \, F(\tau,\taubar)
\label{extratau2}
\eeq
where $s$ is chosen sufficiently large (typically requiring {\mbox{$s>1$}}) so as to guarantee convergence.
Indeed, since the number of powers of $\tau_2$ within $g(\tau_2)$ is generally correlated in string theory
with the number of uncompactified spacetime dimensions,
we may view this insertion of extra powers of $\tau_2$ as a stringy version of dimensional regularization,
taking {\mbox{$D\to D_{\rm eff} \equiv D-2s$}}.
However, since our original integrand $F(\tau,\taubar)$ is presumed modular invariant,
the choice in Eq.~(\ref{extratau2}) in turn implies that the integrand on the right side of Eq.~(\ref{unfold}) must
be taken as
\beq
       \sum_{\gamma \in \Gamma_\infty \backslash \Gamma}  ({\rm Im}\, \gamma\cdot \tau )^{s}  \,F_\gamma(\tau,\taubar)  ~=~
              E(\tau,\taubar,s) \,F(\tau,\taubar)
\eeq
where $E(\tau,\taubar,s)$ is the {\it non-holomorphic Eisenstein series}, often simply denoted $E(\tau,s)$ 
and  defined by 
\beq
          E(\tau,s) ~\equiv~ \sum_{\gamma\in \Gamma_\infty \backslash \Gamma} 
          [{\rm Im}\, (\gamma\cdot \tau) ]^{s} ~=~ 
             \half \, \sum_{(c,d)=1}  \frac{\tau_2^s}{|c\tau+d|^{2s}}~
\label{Eisenstein}
\eeq
with the second sum in Eq.~(\ref{Eisenstein}) restricted to integer, relatively prime values of $c,d$.
Thus, with these choices, we now have
\beq
           \int_\calF \dmu \, E(\tau,s) F(\tau,\taubar)~=~ \int_0^\infty d\tau_2 \,\tau_2^{s-2} g(\tau_2) ~ 
\label{stepone}
\eeq
where the expression on the right side depends on only the physical string states.

The Eisenstein series $E(\tau,s)$ 
has a number of important properties.  
It is convergent for all {\mbox{$s>1$}}, 
but can be analytically continued to all values of $s$.
It is not only modular invariant (consistent with $\calF$ as the corresponding region of 
integration), but its insertion on the left side of Eq.~(\ref{stepone}) relative to our original starting point in 
Eq.~(\ref{eq:I}) softens the divergence as {\mbox{$\tau_2\to\infty$}}, as required. 
Most importantly for our purposes,
however, this function has a simple pole at {\mbox{$s=1$}}, with a $\tau$-independent residue $3/\pi$.
The fact that this residue is $\tau$-independent means that we can formally extract 
our original integral $I$ in Eq.~(\ref{eq:I}) by taking the {\mbox{$s=1$}} residue of both sides of Eq.~(\ref{stepone}):
\beq
           I ~=~ \frac{\pi}{3}\, \oneRes\, \int_0^\infty d\tau_2 \,\tau_2^{s-2} \,g(\tau_2) ~. 
\label{RSresult}
\eeq
We have therefore succeeded in expressing our original modular integral $I$ in terms of only the contributions
from the physical states.
The result in Eq.~(\ref{RSresult}) was originally obtained by Rankin and Selberg in 1939 
(see, {\it e.g.}\/, Refs.~{\mbox{\cite{rankin1a,rankin1b,selberg1}}}), 
and has proven useful for a number of applications in both physics and pure mathematics.  

At this stage,
three important comments are in order.   
First, it may seem that the result in Eq.~(\ref{RSresult}) implies that the unphysical states 
ultimately make no contributions to the amplitude $I$.    However, this is untrue:  the result in Eq.~(\ref{RSresult}) was derived under
the supposition that our original integrand $F(\tau,\taubar)$ is modular invariant, and this modular invariance 
depends crucially on the existence of both physical and unphysical states in the full string spectrum.
For example, through the requirement of modular invariance,
the distribution of unphysical states in the string spectrum has a profound effect~{\mbox{\cite{Dienes:1994np,Dienes:1995pm}}} 
on the values of the physical-state degeneracies $\lbrace a_{nn}\rbrace$ 
which appear in Eq.~(\ref{gtrace}).

As our second comment, we point out that the above results can be reformulated in a manner which
eliminates the $\tau_2$ integration completely and which depends directly on the integrand $g(\tau_2)$.
To see this, we note if we define $I(s)$ as the term on the left
side of Eq.~(\ref{stepone}),
then the relation in 
Eq.~(\ref{stepone}) simply states that 
$I(s)$ is nothing but the
Mellin transform of $g(\tau_2)/\tau_2$. 
One can therefore use the inverse Mellin transform to write $g(\tau_2)/\tau_2$ directly in terms of $I(s)$.
While such an inverse relation is useful in many contexts, 
for our purposes it will be sufficient to note that such an inverse relation implies a direct connection
between the poles of $I(s)$ and the asymptotic behavior of $g(\tau_2)$ as {\mbox{$\tau_2\to 0$}}.
Specifically, one finds a correlation
\beqn
      && {\rm poles~of}~ I(s) ~{\rm at}~ s=s_n ~{\rm with~residues} ~ c_n \nonumber\\
      && ~~~~\Longrightarrow~~ g(\tau_2)\sim \sum_n  c_n \tau_2^{1-s_n} ~~{\rm as}~~\tau_2\to 0~.~~~~~~~ 
\label{correlations}
\eeqn
As we have seen, $I(s)$ has a single pole along the real axis at {\mbox{$s=1$}}, with residue $3I/\pi$. 
However, $I(s)$ also has an infinite number of poles at locations {\mbox{$s_n= \rho_n/2$}}, where
$\rho_n$ are the non-trivial zeros of the Riemann $\zeta$-function $\zeta(s)$.  
According to the Riemann hypothesis, these zeros all have the form {\mbox{$\rho_n = \half \pm i\gamma_n $}}
where {\mbox{$\gamma_n\in\mathbb{R}$}}.
The fact that {\mbox{${\rm Re}\/(s_n)<1$}} for all of these additional poles of $I(s)$  
then implies that the amplitude $I$ dominates the leading behavior of $g(\tau_2)$ as {\mbox{$\tau_2\to 0$}},
allowing us to write~{\mbox{\cite{zag,Kutasov:1990sv}}}
\beq
        I~=~ \frac{\pi}{3}\, \lim_{\tau_2\to 0} g(\tau_2)~.
\label{reformulation}
\eeq
Of course, from Eq.~(\ref{correlations}) we see that the {\mbox{$\tau_2\to 0$}} limit of $g(\tau_2)$ also contains
subleading oscillatory terms~\cite{zag} corresponding to the non-trivial zeros of the $\zeta$-function.
This suggests, through Eq.~(\ref{gtrace}), that the $a_{nn}$ coefficients tend to oscillate in sign
as {\mbox{$n\to \infty$}}.  This oscillating sign is in fact a consequence of the so-called 
``misaligned supersymmetry''~{\mbox{\cite{Dienes:1994np,Dienes:1995pm,Dienes:2001se}}}  
which is a generic property of all tachyon-free non-supersymmetric string models ---
a property whose existence is a direct consequence
of modular invariance in general situations where $I$ is finite and {\mbox{$F(\tau,\taubar)\not=0$}}.

Our final comment, however, is perhaps the most crucial.
As we have seen, the results in Eqs.~(\ref{RSresult}) and (\ref{reformulation}) were derived under the assumption,
as stated within the above derivation,
that the infinite sum within the definition of $g(\tau_2)$ in Eq.~(\ref{gtrace}) 
diverges no more rapidly than some inverse power of $\tau_2$ as {\mbox{$\tau_2\to 0$}}. This requirement was needed 
so that the introduction of sufficiently many $\tau_2$ prefactors could suppress this divergence and render a finite result.
Undoing the modular transformations involved in Eq.~(\ref{unfold}),
we see that this is equivalent to demanding that our original integrand $F(\tau,\taubar)$ either fall, remain constant,
or grow less rapidly than $\tau_2$ as {\mbox{$\tau_2\to \infty$}}.
Indeed, these are the conditions under which the Rankin-Selberg analysis is valid.
Not surprisingly, these are also the conditions under which any integrand $F$ lacking terms with {\mbox{$m=n<0$}} will produce a finite value for $I$.

\subsection{Regulating divergences\label{Regulators}}

The techniques discussed in Sect.~\ref{sec:RStechnique} are completely adequate
for situations in which the original amplitude $I$ is finite, with 
an integrand $F(\tau,\taubar)$ remaining finite or diverging less rapidly than $\tau_2$ as {\mbox{$\tau_2\to \infty$}}.
However, many physical situations of interest
(including those we shall ultimately need to consider in this paper) 
lead to integrands $F(\tau,\taubar)$ which diverge more rapidly than this as {\mbox{$\tau_2\to\infty$}}.
As a result, the corresponding integral $I$ formally diverges and must be regulated.

In this section we shall discuss three different methods of regulating such amplitudes.
These methods are appropriate for cases 
--- such as we shall ultimately face  ---
in which 
the integrand experiences a power-law divergence $\sim \tau_2^p$ with {\mbox{$p\geq 1$}} as {\mbox{$\tau_2\to\infty$}}.
As we shall see, these regulators each have different strengths and weaknesses,
and thus it will prove useful to have all three at our disposal.
In particular, two of these regulators will explicitly break modular invariance,
but are closer in spirit to those that are traditionally
employed in ordinary quantum field theory.
By contrast, the third regulator will be fully modular invariant.
By comparing the results we will then be able 
to discern the novel effects 
that emerge through a fully modular-invariant regularization procedure
and understand the reasons why such a regulator is greatly superior to the others.

All three of these regulators proceed from the same fundamental observation.
Let us suppose that $F(\tau,\taubar)$ diverges
at least as quickly as $\tau_2$ as {\mbox{$\tau_2\to\infty$}}.
Clearly, this behavior 
will cause the integral $I$ to diverge on the left side 
of Eq.~(\ref{RSresult}).
However, this behavior will also cause
$g(\tau_2)$ to diverge as {\mbox{$\tau_2\to \infty$}}, which means that the
right side of Eq.~(\ref{RSresult}) will also diverge.
Thus, in principle, a relation such as that in Eq.~(\ref{RSresult}) will be rendered meaningless.
However, if there were a consistent way of {\it subtracting}\/ or {\it regulating}\/ the appropriate divergence on each 
side of Eq.~(\ref{RSresult}),
we can imagine that we might then obtain an analogous relation between a  finite regulated 
integral $\widetilde I$
and a corresponding finite regulated physical-state trace $\widetilde g(\tau_2)$.
As we shall see, all three of the regulators we shall discuss have this property and 
lead to results which are analogous to the result in Eq.~(\ref{RSresult}) and relate regulated integrals to
regulated physical-state supertraces.

\subsubsection{Minimal regulator\label{sec:minimal}}

Perhaps the simplest and most minimal regulator that can be envisioned~\cite{zag} is one in which we
directly excise the divergence from the integral $I$ without disturbing the rest of the integral. 
Because the divergences on both sides of Eq.~(\ref{RSresult}) arise as {\mbox{$\tau_2\to \infty$}}, 
we can formally excise this region of integration
from $\calF$ by defining a truncated region $\calF_t$  to be the same as $\calF$ but with the additional
restriction that {\mbox{$\tau_2<t$}} for some truncation cutoff {\mbox{$t\geq 1$}}.   
We can then define our regulated integral $\widetilde I$ as
\beq
       \widetilde I ~\equiv~ \lim_{t\to \infty} \left\lbrack \int_{\calF_t} \dmu \, F - \Phi_I(t) \right\rbrack~
\label{tildeIdef}
\eeq
where the function $\Phi_I(t)$ describes the manner in which $\int_{\calF_t} \dmu \,F$ diverges as {\mbox{$t\to \infty$}}.
The explicit subtraction of $\Phi_I(t)$ within Eq.~(\ref{tildeIdef}) --- although not modular invariant --- 
thus renders $\widetilde I$ finite.

Likewise, let us imagine that $\Phi_g(\tau_2)$ describes the manner in which $g(\tau_2)$ diverges as {\mbox{$\tau_2\to\infty$}}.
We can then define 
\beq
                   \widetilde g(\tau_2) ~\equiv~  g(\tau_2) - \Phi_g(\tau_2) ~.
\label{tildegdef}
\eeq
Of course, given these definitions, one might then hope that the result in Eq.~(\ref{RSresult}) remains intact,
only with the replacements {\mbox{$I\to \widetilde I$}} and {\mbox{$g(\tau_2)\to \widetilde g(\tau_2)$}}.
Unfortunately, things are not this simple
because the {\mbox{$\tau_2<t$}} truncation of $\calF$ to $\calF_t$
greatly complicates the ``unfolding'' procedure that underlies 
the intermediate algebraic step in Eq.~(\ref{unfold}).     Starting from $\int_{\calF_t} \dmu\, F$, 
one must therefore follow the effects of this truncation
through all of the modular transformations involved in the unfolding.
This procedure is performed in Ref.~\cite{zag} and 
ultimately generates numerous extra terms beyond those appearing in Eq.~(\ref{RSresult}).   While many of these
extra terms give rise to the subtractions that appear within the definitions of $\widetilde I$ and $\widetilde g(\tau_2)$ 
in Eqs.~(\ref{tildeIdef}) and (\ref{tildegdef}),
some of these extra terms go beyond these subtractions and survive
the final {\mbox{$t\to\infty$}} limit.
Thus, with $\widetilde \Phi$ denoting these additional terms, the Rankin-Selberg result in Eq.~(\ref{RSresult})
generalizes to take the form~\cite{zag} 
\beq
           \widetilde I ~=~ \frac{\pi}{3}\, \oneRes \, \int_0^\infty d\tau_2 \,\tau_2^{s-2}\, \widetilde g(\tau_2)  ~+~ 
                \widetilde\Phi~
\label{Zresult}
\eeq
where the extra terms $\widetilde\Phi$ must also be included.
We thus see that with this regulator, our original integral $I$ continues to be
expressible in terms of a physical-state supertrace.

The crux of the matter, then, is to determine $\Phi_I(t)$, $\Phi_g(\tau_2)$, and $\widetilde \Phi$.
In general, let us suppose that 
{\mbox{$F(\tau,\taubar)\sim  \Phi(\tau_2)+...$}} as {\mbox{$\tau_2\to\infty$}}, where  $\Phi(\tau_2)$ is a function which
diverges at least as as quickly as $\tau_2$ as {\mbox{$\tau_2\to\infty$}}.
Upon performing the $\tau_1$ integration we then immediately see that
$g(\tau_2)$ diverges as {\mbox{$\tau_2\to\infty$}} in exactly the same manner as does $F(\tau,\taubar)$.
We can therefore identify $\Phi_g(\tau_2)$ with $\Phi(\tau_2)$ itself.
Likewise, it is also easy to verify that $\Phi_I(\tau_2)$ is nothing but the anti-derivative of $\Phi(\tau_2)/\tau_2^2$,
and of course we are free to disregard any terms within $\Phi_I(t)$ that vanish as {\mbox{$t\to\infty$}} since
such terms will not contribute within Eq.~(\ref{tildeIdef}).
The difficult part, of course, is to evaluate the extra constant term $\widetilde \Phi$. 
While this term often vanishes, such a cancellation is not guaranteed.
However, a general expression for this term is given in Ref.~\cite{zag} for any divergence function $\Phi(\tau_2)$.

For our later purposes we shall only need to consider one particular case,
namely that with 
\beq
           \Phi(\tau_2) ~=~ c_0 + c_1 \tau_2~.
\label{ourcase}
\eeq
For this divergence structure, it then follows that {\mbox{$\Phi_g=c_0+c_1\tau_2$}}.
Likewise,  we have {\mbox{$\Phi_I(t) = -c_0/t + c_1 \log t$}}, which we may equivalently take
to be simply {\mbox{$\Phi_I(t) = c_1 \log t$}} under the {\mbox{$t\to\infty$}} limit in Eq.~(\ref{tildeIdef}).
Finally, for this case it turns out~\cite{zag} that $\widetilde\Phi$ is given by
\beqn
   \widetilde \Phi ~&\equiv& ~ 2\, \oneRes \left\lbrack  
                   \frac{c_0\, \zeta^\ast(2s)}{s-1}  - \frac{c_1\, \zeta^\ast(2s-1)}{s-1}\right\rbrack \nonumber\\
                &=&~ \frac{\pi}{3} \, c_0 + \log\left( 4\pi e^{-\gamma}\right) c_1~
\eeqn
where {\mbox{$\zeta^\ast(s) \equiv \pi^{-s/2} \Gamma(s/2) \zeta(s)$}} is the so-called ``completed'' Riemann $\zeta$-function
and where {\mbox{$\gamma\approx 0.577$}} is the Euler-Mascheroni constant.
Thus, pulling the pieces together, we find that if {\mbox{$F(\tau,\taubar)\sim c_0 + c_1 \tau_2$}} as {\mbox{$\tau\to\infty$}}, 
then our regulated integral 
\beq
       \widetilde I ~\equiv~ \lim_{t\to \infty} \left(\int_{\calF_t} \dmu \, F ~ -~ c_1 \log\,t  \right)~
\label{Ifinite}
\eeq
can be expressed in terms of purely physical string states as
\beqn
           \widetilde I ~&=&~ \frac{\pi}{3}\, \oneRes \, \int_0^\infty d\tau_2 \,\tau_2^{s-2}\, 
             \biggl\lbrack g(\tau_2) -c_0-c_1 \tau_2 \biggr\rbrack \nonumber\\
          &&~ ~~~ + \frac{\pi}{3} \, c_0 + \log\left( 4\pi e^{-\gamma}\right) c_1~.~~~
\label{Zagierresult}
\eeqn

Note that the left side of Eq.~(\ref{Ifinite}) is independent of $c_0$.
Indeed, since the $c_0$ term within the asymptotic behavior of $F(\tau)$ does not actually
lead to a divergence, our regulator need not depend on $c_0$ in any way.
However, this implies that the right side of Eq.~(\ref{Zagierresult})
must also be independent of $c_0$.
We shall confirm this behavior explicitly in Sect.~\ref{sec4}.~ 
Of course, this does not imply that we can 
simply set {\mbox{$c_0=0$}} on the right side of Eq.~(\ref{Zagierresult});  
the presence of $c_0$ within the $\tau_2$ integral on the right side of 
Eq.~(\ref{Zagierresult}) ensures
that this integral diverges in precisely the correct way to
yield the correct residue at {\mbox{$s=1$}}.  Thus, in some sense, the appearance of $c_0$ within
the right side of Eq.~(\ref{Zagierresult}) 
acts precisely as a regulator should, adding a term in one place to help achieve
convergence and then subtracting it somewhere else in order to yield a $c_0$-independent result.   
Note that 
this $c_0$-independence of the right side of Eq.~(\ref{Zagierresult}) 
also allows the result in Eq.~(\ref{Zagierresult}) to reduce 
to the Rankin-Selberg result in Eq.~(\ref{RSresult}) when {\mbox{$c_1= 0$}} for {\it any}\/ value of $c_0$.
Indeed, we know that the result in Eq.~(\ref{Zagierresult}) 
must reduce in this way because with {\mbox{$c_1=0$}} we have no divergence at all, even when {\mbox{$c_0\not=0$}}.
The original Rankin-Selberg result in Eq.~(\ref{RSresult}) must therefore also apply 
when {\mbox{$c_1=0$}}, even when {\mbox{$c_0\not=0$}}.

\subsubsection{Non-minimal regulators\label{sec:nonminimal}}

The subtraction in Eq.~(\ref{tildeIdef}) is minimal, yielding a finite result
without introducing any new parameters related to the subtraction or altering any portion
of the integrand other than its divergence structure.
Even though a regulating parameter $t$ is introduced in Eq.~(\ref{tildeIdef}),
this quantity must be taken to infinity in order to 
encapsulate all relevant aspects of the original integral $I$.

However, it is also possible to define a similar regulator 
which encapsulates all parts of the original integral $I$ and yet
produces a finite result for {\it arbitrary}\/ finite values of {\mbox{$t\geq 1$}}.  
 Since the divergences within Eq.~(\ref{tildeIdef})
appear only in the {\mbox{$\tau_2 > t$}} region (specifically as {\mbox{$\tau_2\to\infty$}}),
we can 
``undo'' the {\mbox{$\tau_2\to \infty$}} limit and alternatively define
\beq
  \widetilde I(t) ~\equiv~ \int_{\calF_t} \dmu \, F + \int_{\calF-\calF_t} \dmu \, [ F- \Phi(\tau_2)]~
\label{Itdef}
\eeq 
where we shall continue to assume that {\mbox{$F(\tau,\taubar)\sim \Phi(\tau_2)+...$}} as {\mbox{$\tau_2\to\infty$}}.
  Note that $\widetilde I(t)$ is convergent for all finite $t$, as we desire.  Moreover,
because the second term in Eq.~(\ref{Itdef}) has an integrand which is convergent throughout 
the integration region $\calF-\calF_t$, taking the {\mbox{$t\to\infty$}} limit eliminates the second term and
$\widetilde I(t)$ reproduces our original unregulated integral $I$ in Eq.~(\ref{eq:I}).
Thus $\widetilde I(t)$ represents an alternative, $t$-dependent method of regulating our original
integral $I$, one which is distinct from the minimal regularization $\widetilde I$ 
in Eq.~(\ref{tildeIdef}).

These two regularizations are deeply connected, however 
 --- a fact which will also enable us to express $\widetilde I(t)$ in terms of supertraces,
just as we did for $\widetilde I$.
Note that the only $t$-dependence within $\widetilde I(t)$ arises from the 
integration of the subtraction term $\Phi(\tau_2)$
along the $t$-dependent lower boundary 
of the integration region $\calF-\calF_t$.
We thus see that the subtraction term  $\Phi(\tau_2)$
which regularizes $\widetilde I(t)$ 
introduces a non-trivial dependence on $t$ such that
\beq
           \widetilde I(t)~=~ \Phi_I(t) + {\cal C}
\label{tdependence}
\eeq
           where we recall that $\Phi_I(\tau_2)$ is the anti-derivative of $\Phi(\tau_2)/\tau_2^2$
and where ${\cal C}$ is an as-yet unknown $t$-independent quantity.
However, it is easy to solve for ${\cal C}$.
Given that {\mbox{${\cal C}= \widetilde I(t) - \Phi_I(t)$}}, we immediately see 
by taking the {\mbox{$t\to \infty$}} limit of both sides and comparing with Eq.~(\ref{tildeIdef}) that
{\mbox{$\lim_{t\to \infty}{\cal C} = \widetilde I$}}.
However, ${\cal C}$ is independent of $t$, which means that {\mbox{${\cal C}=\widetilde I$}}
for {\it any}\/ value of {\mbox{$t\geq 1$}}.
We thus obtain a general relation, valid for all {\mbox{$t\geq 1$}},
between our two regulators: 
\beq
             \widetilde I(t) ~=~ \widetilde I  +\Phi_I(t)~.
             \eeq
Our previous result for $\tilde I$ in 
Eq.~(\ref{Zresult}) then yields~\cite{zag} 
\beq
      \widetilde I(t) ~=~ \frac{\pi}{3}\, \oneRes \, \int_0^\infty d\tau_2 \,\tau_2^{s-2}\, \widetilde g(\tau_2)  ~+~ 
                \Phi_I(t)~+~ \widetilde\Phi~.
\eeq

Thus, just as with our minimal regulator, we find that our $t$-dependent regulator
produces a finite integral $\widetilde I(t)$ which continues to be expressible
in terms of a physical-state supertrace.
Indeed, for the divergence structure {\mbox{$\Phi(\tau_2)= c_0+c_1\tau_2$}},
we find that
our $t$-dependent regularized integral 
\beq
  \widetilde I(t) ~\equiv~ \int_{\calF_t} \dmu \, F + \int_{\calF-\calF_t} \dmu \, ( F- c_0 - c_1\tau_2)~
\label{I2}
\eeq 
is given by
\beqn
   \widetilde I(t) 
           ~&=&~ \frac{\pi}{3}\, \oneRes \, \int_0^\infty d\tau_2 \,\tau_2^{s-2}\, 
             \biggl\lbrack g(\tau_2) -c_0-c_1 \tau_2 \biggr\rbrack ~~~~\nonumber\\
          &&~ ~ + \left( \frac{\pi}{3} - {1\over t}\right) \, c_0 + 
              \log\left( 4\pi \,t\, e^{-\gamma}\right) c_1~.
\label{Zagierresult2}
\eeqn

Once again, $c_0$ plays a special role in this result
because the presence of the $c_0$ term within $\Phi(\tau_2)$ does not lead
to a divergence.
Indeed, given that the region $\calF-\calF_t$ has volume $1/t$ with respect to the $\dmu$ measure, 
we see that the subtraction of $c_0$ within Eq.~(\ref{I2}) simply removes
a finite quantity $c_0/t$ from the value of $\widetilde I(t)$.
For integrands having this divergence structure
we can therefore define a {\it modified}\/ (or {\it improved}\/) non-minimal regulator
\beqn
  \widehat I(t) ~&\equiv&~ \widetilde I(t) + c_0/t \nonumber\\
    ~&=&~ \int_{\calF_t} \dmu \, F + \int_{\calF-\calF_t} \dmu \, ( F - c_1\tau_2)~,~~~~
\label{I3}
\eeqn 
whereupon we find from Eq.~(\ref{Zagierresult2}) that
\beqn
   \widehat I(t) 
           ~&=&~ \frac{\pi}{3}\, \oneRes \, \int_0^\infty d\tau_2 \,\tau_2^{s-2}\, 
             \biggl\lbrack g(\tau_2) -c_0-c_1 \tau_2 \biggr\rbrack ~~~~~\nonumber\\
          &&~ ~ + \frac{\pi}{3} \, c_0 + \log\left( 4\pi \,t\, e^{-\gamma}\right) c_1~.
\label{Zagierresult3}
\eeqn
In other words, the $1/t$-dependence on the right side of Eq.~(\ref{Zagierresult2})
was in some sense spurious, reflecting a corresponding $1/t$-dependence that was
needlessly inserted into the regulator definition in Eq.~(\ref{I2}) and which has
now been removed from Eqs.~(\ref{I3}) and (\ref{Zagierresult3}).
The right side of Eq.~(\ref{Zagierresult3}) is then independent of $c_0$ in the manner discussed
below Eq.~(\ref{Zagierresult}) for the minimal regulator.

\subsubsection{Modular-invariant regulators \label{sec:modinvregs}}

Although our results in Eqs.~(\ref{Zagierresult}), (\ref{Zagierresult2}), and (\ref{Zagierresult3}) 
were each derived in a manner that remained true to the modular-invariant unfolding procedure,
neither side of these relations is modular invariant 
by itself.   
In other words, even though 
these relations correctly
allow us to 
express our regulated
integrals $\widetilde I$, $\widetilde I(t)$, and $\widehat I(t)$ in terms of a corresponding 
regulated physical-state supertrace $\widetilde g(\tau_2)$,
neither $\widetilde I$, $\widetilde I(t)$, nor $\widehat I(t)$ is itself a modular-invariant quantity. 
This is an important observation because 
these latter quantities will ultimately correspond to physical observables 
within the modular-invariant string context.
We must therefore additionally require that these observables themselves be modular invariant.

The issue, of course, is that 
neither $\widetilde I$, $\widetilde I(t)$, nor $\widehat I(t)$ 
incorporates a modular-invariant way of eliminating the associated divergences as {\mbox{$\tau_2\to\infty$}}. 
However, it is possible to design regulators in which such divergences are indeed eliminated in a 
fully modular-invariant way.
In this work we shall present a particular set of modular-invariant regulators which will have several useful properties for our purposes.

In order to define these regulators, let us first recall that the partition function of
a bosonic worldsheet field compactified on a circle of radius $R$ is given by
\beqn
     Z_{\rm circ}(a,\tau) &=&
     \sqrt{ \tau_2}\,
    \sum_{m,n\in\mathbb{Z}} \,
      \overline{q}^{(ma-n/a)^2/4}  \,q^{(ma+n/a)^2/4}\nonumber\\
     &=& \sqrt{ \tau_2}\,
    \sum_{m,n\in\mathbb{Z}} \,
        e^{-\pi \tau_2 (m^2 a^2 + n^2 /a^2)} \, e^{2\pi i mn \tau_1}~~~
\nonumber\\
\label{Zcircdef}
\eeqn
where we have defined the dimensionless inverse radius
{\mbox{$a\equiv \sqrt{\alpha'}/R$}}.
Here the sum over $m$ and $n$ represents  
the sum over all possible KK momentum and winding modes, respectively.
Note that {\mbox{$Z_{\rm circ}\to 1/a$}} as {\mbox{$a\to 0$}},
while {\mbox{$Z_{\rm circ}\to a$}} as {\mbox{$a\to \infty$}}.
As expected, $Z_{\rm circ}(a,\tau)$ is modular invariant for any $a$.
Using $Z_{\rm circ}(a,\tau)$, we shall then 
regulate any divergent integral of the form in Eq.~(\ref{eq:I}) 
by defining
a corresponding series of regulated integrals $\widetilde I_\rho (a)$:
\beq
       \widetilde I_\rho (a) ~\equiv~ \int_{\calF} \dmu \, F(\tau) \, {\cal G}_\rho(a,\tau)
\label{Iadef}
\eeq
where our regulator functions  
$\calG_\rho(a,\tau)$ are defined for any {\mbox{$\rho\in \mathbb{R}^+$}}, {\mbox{$\rho\not=1$}}, as
\beq
    \calG_\rho(a,\tau) ~\equiv~ 
    A_\rho\,  a^2 \frac{\partial}{\partial a} \biggl\lbrack Z_{\rm circ}( \rho a,\tau) - Z_{\rm circ}(a,\tau)\biggr\rbrack~ 
\label{regG}
\eeq 
where {\mbox{$A_\rho\equiv \rho/(\rho-1)$}} is an overall normalization factor.
Note that $\calG_\rho(a,\tau)$ inherits its modular invariance from $Z_{\rm circ}$, thereby rendering the regulated
integral $\widetilde I_\rho(a)$ in Eq.~(\ref{Iadef}) fully modular invariant for any $a$ and $\rho$. 
We further note that $\calG_\rho(a,\tau)$ satisfies the identity
\beq
           \calG_\rho(a,\tau) ~=~ \calG_{1/\rho}(\rho a,\tau)~.
\label{rhoflipidentity}
\eeq
We can therefore take {\mbox{$\rho>1$}} without loss of generality.

These functions $\calG_\rho(a,\tau)$ have two important properties which 
make them suitable as regulators
when {\mbox{$a\ll 1$}}.
First, as {\mbox{$a\to 0$}}, we find that {\mbox{$\calG_\rho(a,\tau)\to 1$}} for all $\tau$.
Thus the {\mbox{$a\to 0$}} limit restores our original unregulated theory.
Second, for any {\mbox{$a>0$}}, we find that {\mbox{$\calG_\rho(a,\tau)\to 0$}} {\it exponentially rapidly}\/ as {\mbox{$\tau_2\to\infty$}}.
Thus the insertion of $\calG_\rho (a,\tau)$ into the integrand of Eq.~(\ref{Iadef})
successfully eliminates whatever power-law divergence  
might have otherwise arisen from the original integrand $F(\tau)$. 
Indeed, we see that this now happens in a smooth, fully modular-invariant way rather than through
a sharp, discrete subtraction.
Motivated by these two properties, we shall therefore focus on situations in which {\mbox{$a\ll 1$}}, 
as these are the situations in which our regulator preserves as much of the original theory as possible (as we expect
of a good regulator) while simultaneously eliminating all power-law divergences as {\mbox{$\tau_2\to\infty$}}.
In fact, for the special case {\mbox{$\rho=2$}} and for the specific values 
{\mbox{$a= 1/\sqrt{k+2}$}} 
 the insertion of this regulator  even has a direct physical interpretation, arising through a procedure in which 
the various fields in the background string geometry are turned on in such 
a way that the CFT associated with the flat four-dimensional spacetime
is replaced by that associated with a $SU(2)_k$ WZW model~\cite{Kiritsis:1994ta,Kiritsis:1996dn,Kiritsis:1998en}.

That said, there is one further property of these regulator functions $\calG_\rho(a,\tau)$ 
which will prove useful for our purposes.
When {\mbox{$a\ll 1$}}, the contributions from all non-zero winding modes 
within $Z_{\rm circ}$ (and ultimately within $\calG$)
are exponentially suppressed relative to those of the KK momentum modes.
In other words, when {\mbox{$a\ll 1$}} we can effectively restrict our summation in Eq.~(\ref{Zcircdef})
to cases with {\mbox{$n=0$}}.
We then find that $\calG_\rho(a,\tau)$ loses its dependence on $\tau_1$,
rendering $\calG_\rho(a,\tau)$ a function of $\tau_2$ alone.
In such cases we shall simply denote our regulator function as $\calG_\rho(a,\tau_2)$.

\begin{figure*}[t!]
\centering
\includegraphics[keepaspectratio, width=0.51\textwidth]{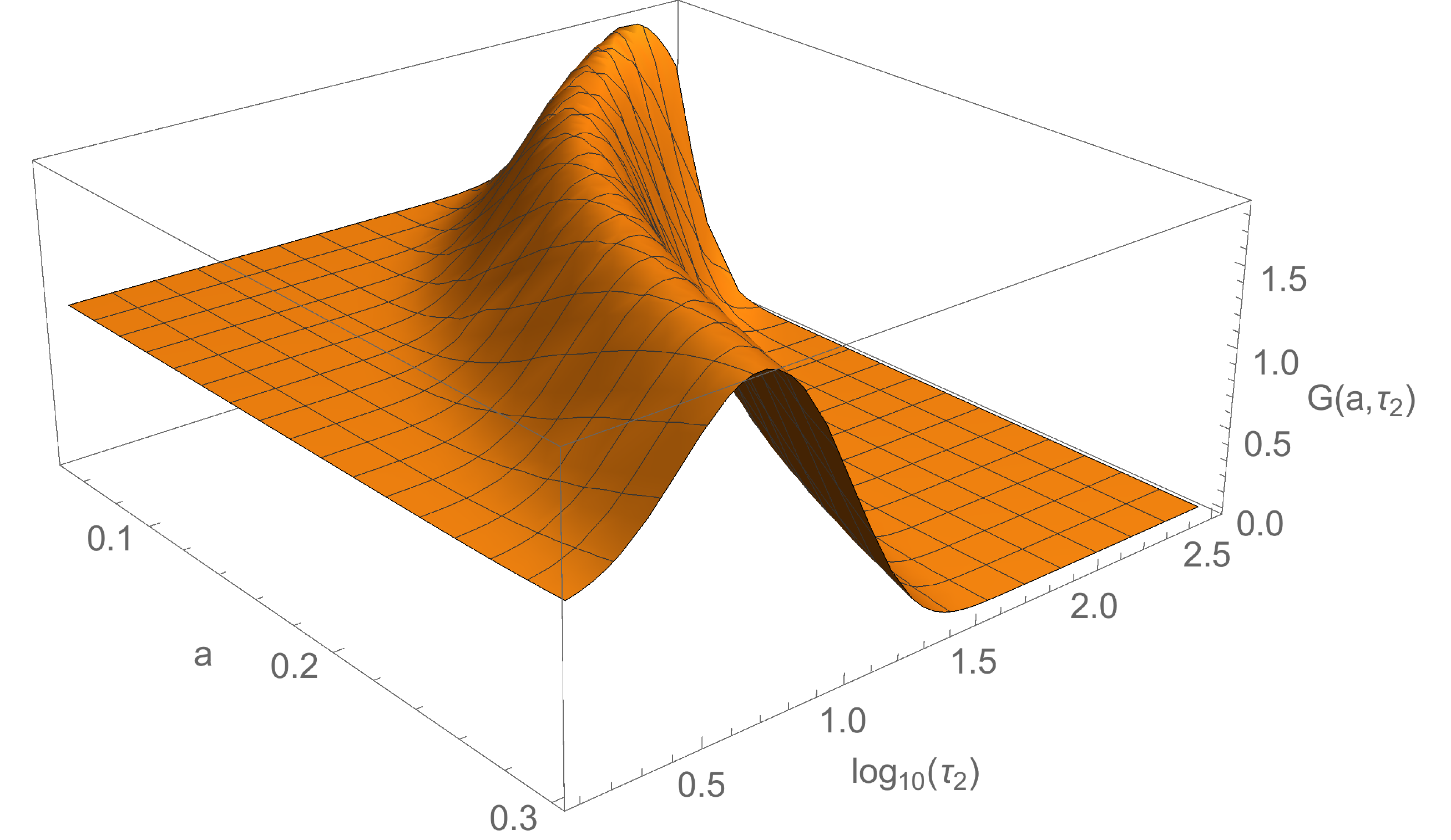}
\hskip 0.12 truein
\includegraphics[keepaspectratio, width=0.45\textwidth]{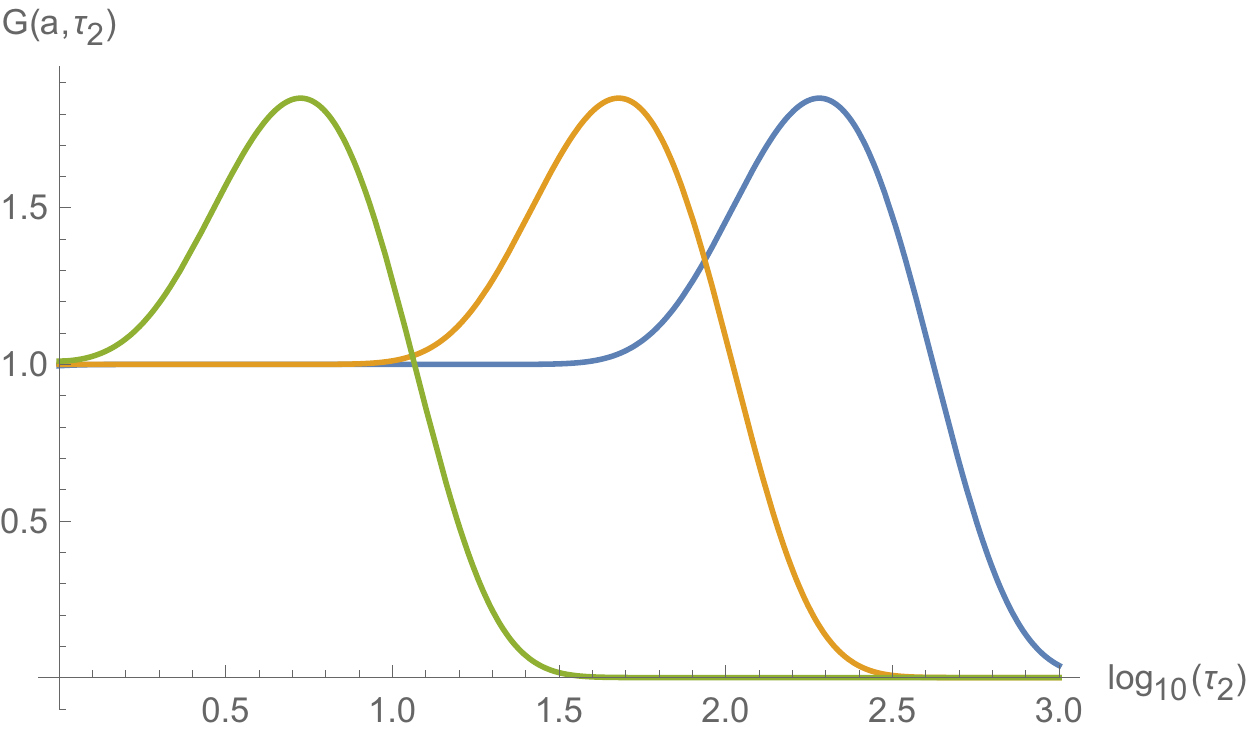}
\caption{
{\it Left panel}\/:  The modular-invariant regulator function $\calG_2(a,\tau_2)$, 
plotted within the $(a,\tau_2)$ plane for {\mbox{$a\ll 1$}} and {\mbox{$\tau_2\geq 1$}}.
{\it Right panel}\/:  The modular-invariant regulator $\calG_2(a,\tau_2)$, 
plotted as a function of $\tau_2$ for
{\mbox{$a=0.05$}} (blue), {\mbox{$a=0.1$}} (orange), and {\mbox{$a=0.3$}} (green).
In all cases we see that {\mbox{$\calG_2(a,\tau_2) \to 0$}} for all {\mbox{$a>0$}} as {\mbox{$\tau_2\to\infty$}},
while {\mbox{$\calG_2(a,\tau_2)\to 1$}} for all {\mbox{$\tau_2\geq 1$}} as {\mbox{$a\to 0$}}.
Indeed, for {\mbox{$a=0.05$}}, we see that {\mbox{$\calG_2(a,\tau_2)\approx 1$}} for all {\mbox{$\tau_2\lsim 100$}}.
Thus for small non-zero $a$ this regulator succeeds in suppressing the divergences
that might otherwise arise as {\mbox{$\tau_2\to\infty$}} while nevertheless 
having little effect throughout the rest of
the fundamental domain.}
\label{regulator_figure}
\end{figure*}

In Fig.~\ref{regulator_figure}, we have plotted 
the regulator function $\calG_2(a,\tau_2)$ within the $(a,\tau_2)$ plane for {\mbox{$\tau_2\geq 1$}} (left panel)
and as a function of {\mbox{$\tau_2\geq 1$}} for various discrete values of {\mbox{$a\ll 1$}} (right panel).
We see, as promised, that
{\mbox{$\calG_2(a,\tau_2) \to 0$}} for all {\mbox{$a>0$}} as {\mbox{$\tau_2\to\infty$}},
while {\mbox{$\calG_2(a,\tau_2)\to 1$}} for all {\mbox{$\tau_2\geq 1$}} as {\mbox{$a\to 0$}}.
We also note that this suppression for large $\tau_2$ is quite pronounced,
even for {\mbox{$a\ll 1$}}.

For any $a$ and $\rho$, we see 
from Fig.~\ref{regulator_figure}
that there is a corresponding value $\tau_2^\ast$ 
which can be taken as characterizing the approximate $\tau_2$-location of the transition between
the unregulated region with {\mbox{$\calG_\rho(a,\tau_2)\approx 1$}} and 
the regulated region with {\mbox{$\calG_\rho(a,\tau_2)\approx 0$}}.
For example, we might define $\tau_2^\ast$ as the critical 
value corresponding to the top of the ``ridge'' in the left panel 
of Fig.~\ref{regulator_figure} (or equivalently the maximum in the right panel of Fig.~\ref{regulator_figure}).
Alternatively, given the shapes of the functions in the right panel of Fig.~\ref{regulator_figure},
we might define $\tau_2^\ast$ as the location 
at which $\calG_\rho(a,\tau_2)$ experiences an inflection from being concave-down to concave-up.
Finally, a third possibility might be to define $\tau_2^\ast$ as
the value of $\tau_2$ at which {\mbox{$\calG_\rho(a,\tau_2) =1/2$}},
representing the ``midpoint'' between {\mbox{$\calG=1$}} and {\mbox{$\calG=0$}}.
For the {\mbox{$\rho=2$}} case shown in Fig.~\ref{regulator_figure},
we then find for {\mbox{$a\ll 1$}} that each of these has a rather straightforward  
scaling behavior with $a^{-2}$:
\beqn
   \hbox{ridge top:}~~~           && ~~ \tau_2^\ast ~\approx~  \frac{3}{2\pi a^2} ~\approx~ \frac {0.477}{a^2}~\nonumber\\
   \hbox{inflection:}~~~          && ~~ \tau_2^\ast ~\approx~  \frac{3+\sqrt{6}}{2\pi a^2} ~\approx~ \frac {0.867}{a^2}~
                                          ~~~~~~\nonumber\\
   \hbox{{\mbox{$\calG=1/2$}}:}~~~         && ~~ \tau_2^\ast ~\approx~ \frac{1.411}{a^2}~.
\label{tau2ast}
\eeqn
Indeed, each of these results becomes increasingly accurate as {\mbox{$a\to 0$}}.
Moreover, although the 
numerical coefficient in the third case depends significantly on $\rho$,
the numerical coefficients in the first two cases are actually independent of $\rho$.
In all cases, however, 
we see that $\calG_\rho(a,\tau_2)$ suppresses the contributions from regions of the fundamental domain
with {\mbox{$\tau_2\gg \tau_2^\ast$}} while preserving the contributions from regions with {\mbox{$1< \tau_2\ll \tau_2^\ast$}}.
Indeed, this property holds regardless of our precise definition for $\tau_2^\ast$. 

Armed with these regulator functions $\calG_\rho(a,\tau_2)$, 
we now wish to express the integral in Eq.~(\ref{Iadef})
in terms of an appropriately regulated supertrace over physical string states.
However, given that $\widetilde I_\rho(a)$ is fully modular invariant and convergent as {\mbox{$\tau_2\to\infty$}},
we can simply use the original Rankin-Selberg result in Eq.~(\ref{RSresult}).
We thus have
\beq
      \widetilde I_\rho (a) ~=~ \frac{\pi}{3}\, \oneRes \, \int_0^\infty d\tau_2 \,\tau_2^{s-2} \,\widetilde g_\rho(a,\tau_2) ~ 
\label{Irhoa}
\eeq
where, in analogy with Eq.~(\ref{gtrace}), we have 
\beq
           \widetilde g_\rho(a,\tau_2) ~\equiv~ \int_{-1/2}^{1/2} d\tau_1  \, F(\tau) \, {\calG}_\rho(a,\tau)~.
\label{gFGdef}
\eeq

In general, for arbitrary $a$,
the regulator
$\calG_\rho(a,\tau)$ will have a traditional $(q,\qbar)$ power-expansion
of the form {\mbox{$\calG\sim \sum_{r,s} b_{rs} \qbar^r q^s$}}, just as  we have
{\mbox{$F\sim \sum_{m,n} a_{mn}\qbar^m q^n$}} in  Eq.~(\ref{integrand}). 
Given this, 
we find that the $\tau_1$ integral in Eq.~(\ref{gFGdef}) projects onto those states for which {\mbox{$n-m= r-s$}}.
However, $\calG_\rho(a,\tau)$ generally  receives contributions from states with many different values of $r-s$.
As a result, $\widetilde g_\rho(a,\tau_2)$ will generally receive contributions from not only the {\it physical}\/ 
{\mbox{$m=n$}} states within $F(\tau)$  but also 
some of the {\it unphysical}\/ {\mbox{$m\not=n$}} states.
In other words, for general $a$, our regulator function $\calG_\rho(a,\tau)$ becomes entangled
with the physical-state trace in a way that allows unphysical states to contribute.

As we have seen, it is useful for practical purposes
that $\calG_\rho(a,\tau)$ loses its dependence on $\tau_1$ when {\mbox{$a\ll 1$}}.
In other words, 
for {\mbox{$a\ll 1$}}
we find that the contributions from terms with {\mbox{$r\not=s$}} within $\calG$ are suppressed.
The $\tau_1$ integral in Eq.~(\ref{gFGdef}) then projects onto only the {\mbox{$m=n$}} physical states, as desired,
and to a good approximation our expression for $\widetilde g_\rho(a,\tau_2)$ in Eq.~(\ref{gFGdef}) 
simplifies to
\beq
           \widetilde g_\rho(a,\tau_2) ~\approx~ g(\tau_2) \, {\calG}_\rho(a,\tau_2)~
\label{gFGdef2}
\eeq
where $g(\tau_2)$ is our original unregulated physical-state trace in Eq.~(\ref{gtrace}).
Thus, for {\mbox{$a\ll 1$}}, the same regulator function $\calG_\rho(a,\tau_2)$ which smoothly softens the 
{\mbox{$\tau_2\to\infty$}} divergence
in the integrand $\widetilde I_\rho(a)$ also smoothly softens the 
{\mbox{$\tau_2\to\infty$}} divergence
in the physical-state trace $\widetilde g_\rho(a,\tau_2)$ --- all without introducing contributions from unphysical states.
However, we shall later demonstrate that the integral 
in Eq.~(\ref{Iadef}) can actually be performed exactly, yielding
an expression in terms of purely physical states for all values of $a$.

While these regulator functions $\calG_\rho(a,\tau_2)$ are suitable for many applications,
it turns out that we can use these functions in order to 
construct additional modular-invariant regulators
whose symmetry properties transcend even those of $\calG_\rho(a,\tau_2)$.
To do this, we first observe from the modular invariance of $\calG_\rho(a,\tau_2)$ that 
\beq
           \calG_\rho(a, 1/\tau_2) ~=~ \calG_\rho(a, \tau_2) ~
\label{StransG}
\eeq
for any $\rho$, $a$, and $\tau_2$.  Indeed, invariance under {\mbox{$\tau_2\to 1/\tau_2$}} follows directly from
invariance under the modular transformation {\mbox{$\tau\to -1/\tau$}} for {\mbox{$\tau_1=0$}}.
Second, the identity in Eq.~(\ref{rhoflipidentity}) tells us that the parameters $(\rho,a)$
which define our $\calG$-functions
have a certain redundancy, such that  the $\calG$-function with $(\rho,a)$ is the same as
the $\calG$-function with $(1/\rho, \rho a)$. 
Indeed, only the combination {\mbox{$a'\equiv \sqrt{\rho} a$}} is invariant under this redundancy. 
        Thus, while Eq.~(\ref{StransG}) provides a symmetry under reciprocal flips in $\tau_2$,
Eq.~(\ref{rhoflipidentity}) provides a symmetry under reciprocal flips in $\rho$.

Given these two symmetries, it is natural to wonder
whether $\calG_\rho(a,\tau)$ also exhibits a reciprocal flip symmetry in
the one remaining variable {\mbox{$a'\equiv \sqrt{\rho} a$}}.
This would thus be a symmetry under {\mbox{$a\to 1/\rho a$}}, or  equivalently under {\mbox{$\rho a^2\to 1/(\rho a^2)$}}. 
Indeed, we shall find in Sect.~\ref{sec:alignment} 
that such an additional symmetry will be very useful and 
render the modular symmetry manifest in certain cases where it would otherwise have been obscure.  
Unfortunately, $\calG_\rho(a,\tau)$ does not exhibit such a symmetry.
One might nevertheless wonder whether it is possible to modify this regulator
function in such a way that it might exhibit this additional symmetry as well.

It turns out that this enhanced symmetry structure is relatively easy to arrange.
First, we observe that $Z_{\rm circ}(a,\tau)$ is itself invariant under
{\mbox{$a\to 1/a$}} for any $\tau$;  indeed, this is the symmetry underlying T-duality for closed strings.
Given this, it is then straightforward to verify that 
the functions
\beq
      \widehat \calG_\rho(a,\tau) ~\equiv~ \frac{1}{1+ \rho a^2} \, \calG_\rho(a,\tau)~
\label{hatGdef}
\eeq
not only inherit all of the regulator properties and symmetries 
discussed above for $\calG_\rho(a,\tau)$ when {\mbox{$a\ll 1$}},
but are also  manifestly invariant under {\mbox{$a'\to 1/a'$}}, or equivalently {\mbox{$a\to 1/(\rho a)$}}, for any $\tau$: 
\beq
     \widehat \calG_\rho(a,\tau) ~=~ \widehat\calG_\rho ( 1/\rho a, \tau)~.
\label{newest}
\eeq
   We shall therefore take these $\widehat \calG$-functions as defining our enhanced modular-invariant regulators.
We shall likewise define
corresponding enhanced regularized 
integrals $\widehat I_\rho(a)$ as in Eq.~(\ref{Iadef}), but with $\calG_\rho(a,\tau)$
replaced by $\widehat \calG_\rho(a,\tau)$.
We then find that we can express $\widehat I_\rho(a)$ in terms of corresponding
physical-state traces $\widehat g_\rho(a,\tau_2)$ as in
Eqs.~(\ref{Irhoa}) through (\ref{gFGdef2}),
except with $\calG_\rho(a,\tau_2)$ replaced by $\widehat \calG_\rho(a,\tau_2)$
throughout.  

The enhanced regulators in Eq.~(\ref{hatGdef})
can also be understood in a completely different way, through analogy with what we
have already observed for our non-minimal regulators in Sect.~\ref{sec:nonminimal}.~
As discussed after Eq.~(\ref{Zagierresult3}),
the quantity $\widetilde I(t)$ defined through our non-minimal regulator
ultimately contained 
a spurious $t$-dependence that could be removed without disturbing
the suitability of the regulator itself.
It is for this reason that 
we were able to transition from our original
non-minimal regulator in Eq.~(\ref{I2}) to 
our improved non-minimal regulator in Eq.~(\ref{I3}) in which
such spurious terms were eliminated.

It turns out that a similar situation arises for our original modular-invariant 
regulators $\calG_\rho(a,\tau_2)$.
Indeed, as we shall find  in Sect.~\ref{sec4}, use of these regulators would have led to results with analogously spurious terms --- \ie,
terms which obscure the underlying symmetries of the theory.
However, just as with Eq.~(\ref{I3}), it is possible to define
improved modular-invariant regulators 
in which such spurious effects are eliminated.
Indeed, these improved modular-invariant regulators 
are nothing but the regulators $\widehat \calG_\rho(a,\tau)$
introduced in Eq.~(\ref{hatGdef}).
Additional reasons for adopting the $\widehat \calG_\rho(a,\tau)$ regulators
will be discussed in Sect.~\ref{sec:Conclusions}.~
These improved regulators will therefore be our main interest in this paper.

\subsubsection{Aligning the non-minimal and modular-invariant regulators\label{sec:alignment}}

Needless to say, the most important feature of
our modular-invariant regulators is precisely that they are modular invariant.
Use of these regulators therefore provides a way of controlling the divergences that might
appear in string amplitudes while simultaneously preserving the modular invariance that rests at the heart
of all that we are doing in this paper.

This becomes especially apparent upon comparing these modular-invariant regulators with
the non-minimal regulators of Sect.~\ref{sec:nonminimal}.   Recall that the non-minimal regulators operate by isolating those terms
within the partition function $F(\tau)$ which would have led to a divergence as {\mbox{$\tau_2\to \infty$}}, and then performing
a brute-force subtraction of those terms over the entire region of the fundamental domain $\calF$
with {\mbox{$\tau_2\geq t$}}.
In so doing, modular invariance is broken twice:   first, in artificially separating those terms 
within the partition function which would have led to a divergence from those 
which do not;  and second, in then selecting a particular sharp location {\mbox{$\tau_2=t$}} at which 
to perform the subtraction of these divergence-inducing terms, essentially multiplying these terms 
by $\Theta(t-\tau_2)$ where $\Theta$ is the Heaviside function.
By contrast, our modular-invariant regulator keeps the entire partition function $F$ intact
and then multiplies $F$
by a single modular-invariant regulator function $\widehat \calG_\rho(a,\tau)$.
As such it does not induce a sharp Heaviside-like subtraction at any particular location within the fundamental domain,
but rather (as illustrated in the right panel of Fig.~\ref{regulator_figure})
induces a smooth damping which operates most strongly for  {\mbox{$\tau_2\gg \tau_2^\ast$}} and which can be removed (or pushed off
towards greater and greater values of $\tau_2^\ast$) as {\mbox{$a\to 0$}}. 
All of these crucial differences are induced by the modular invariance of the regulator
and render our modular-invariant regulators 
wholly different from the non-minimal regulator of Sect.~\ref{sec:nonminimal}.

These two regulators do share one common feature, however:   they both introduce suppressions 
into the integrands of our string amplitudes.
Within the non-minimal regulator this takes the form of a 
sharp subtraction that occurs at {\mbox{$\tau_2=t$}},
while the modular-invariant regulator
gives rise to a smoother suppression, a transition from
{\mbox{$\widehat\calG\approx 1$}} to {\mbox{$\widehat\calG\approx 0$}} 
 that occurs near {\mbox{$\tau_2\approx \tau_2^\ast$}}.
To the extent that these two regulators share this single common feature, it is therefore possible 
to ``align''
them by choosing a particular definition for $\tau_2^\ast$
within the modular-invariant regulator and then identifying
\beq
              t ~=~ \tau_2^\ast~.
\label{prealignment}
\eeq 
In general, we have seen in Eq.~(\ref{tau2ast})
that $\tau_2^\ast$ takes the general form
\beq
               \tau_2^\ast ~=~ \frac{\xi}{a^2} ~=~ \frac{\xi \rho}{\rho a^2}~,
\label{tau2asta}
\eeq
where $\xi$ is a numerical coefficient which depends on the particular definition of $\tau_2^\ast$ that is chosen.
Thus, for any value of $t$,
we can correspondingly tune our choices for {\mbox{$\rho>1$}} and $\rho a^2$ in order to enforce Eq.~(\ref{prealignment}) 
and in this sense bring our regulators into alignment.

This alone is sufficient to align our regulators.
However, in keeping with the spirit of symmetry-enhancement that motivated our transition
from $\calG$ to $\widehat \calG$,
we can push this one step further.
We have already seen that our $\widehat\calG$ regulator has a symmetry
under {\mbox{$\tau_2\to 1/\tau_2$}} (and thus under {\mbox{$\tau_2^\ast \to 1/\tau_2^\ast$}})
as well as a symmetry under {\mbox{$a\to 1/\rho a$}} [or equivalently under {\mbox{$\rho a^2 \to 1/(\rho a^2)$}}].
Although these are independent symmetries,
the fact that $\tau_2^\ast$ and $\rho a^2$ are related through Eq.~(\ref{tau2asta}) suggests
that we can align these   
two symmetries as well by further demanding that {\mbox{$\xi \rho=1$}}.

For either of the first two $\tau_2^\ast$ definitions in Eq.~(\ref{tau2ast}), this 
is a very easy condition to enforce:   we simply take {\mbox{$\rho = 1/\xi$}}.    
This is possible because the ``ridge-top'' and ``inflection'' definitions 
lead to values of $\xi$ which are independent of $\rho$.
By contrast, for the {\mbox{$\calG = 1/2$}} condition (where {\mbox{$\calG\approx \widehat\calG$}} in the {\mbox{$\rho a^2 \ll 1$}} limit),
the value of $\xi$ is itself highly $\rho$-dependent and it turns out that the constraint {\mbox{$\rho \xi(\rho)=1$}} has
no solution for $\rho$.

It is easy to understand why these different $\tau_2^\ast$ definitions lead to such different outcomes for $\rho$.
For {\mbox{$a\ll 1$}} and {\mbox{$\rho>1$}}, the contributions to $\widehat \calG$ from $Z_{\rm circ}(\rho a,\tau)$ are hugely 
suppressed compared with those from $Z_{\rm circ}(a,\tau)$.
As a result,
any defining condition for $\tau_2^\ast$ which depends on the 
actual values of $\widehat \calG$ will carry a sensitivity to $\rho$ only through
the $\widehat \calG$-prefactor {\mbox{$A_\rho = \rho/(\rho-1)$}}.    
By contrast, any defining condition for $\tau_2^\ast$ which depends on the 
vanishing of {\it derivatives}\/ of $\widehat \calG$ becomes insensitive to the overall
scale factor $A_\rho$ 
and thus independent of $\rho$.
Indeed, the ``ridge-top'' and ``inflection'' definitions depend on the vanishing of
the first and second  $\widehat \calG$-derivatives respectively. 
Such conditions therefore lead to a vastly simpler algebraic structure for $\tau_2^\ast$ as
a function of $\rho$.

Thus, pulling the pieces together, we see that we can align our modular-invariant regulator with
our non-minimal regulator by adopting a particular definition for $\tau_2^\ast$
and then choosing the values of the $(\rho,a)$ parameters  
within our modular-invariant regulator such that
\beq
               t ~=~  \frac{\xi \rho}{\rho a^2}~.
\label{identification}
\eeq
Moreover, we can further enhance the symmetries underlying
this identification by restricting our attention to $(\rho,a)$ choices 
for which {\mbox{$\xi \rho=1$}}.
However, because modular invariance essentially smoothes out the
sharp transition at {\mbox{$\tau_2=t$}}
that otherwise existed within the non-minimal regulator, 
we face an  inevitable uncertainty in how we define $\tau_2^\ast$.
In the following, we shall therefore adopt 
\beq
              t ~=~ \frac{1}{\rho a^2}~
\label{alignment}
\eeq
as our alignment condition.
Directly enforcing this condition 
enables us to sidestep
the issues associated with choosing a particular value of $\xi$ or a 
particular definition for $\tau_2^\ast$.
However, in enforcing this condition we should remain mindful of our regulator condition 
that {\mbox{$a\ll 1$}}.
Likewise, whenever needed, our choices for $\rho$ should lie within
a range that is sensibly close to the approximate values of $\xi^{-1}$ that characterize
the transition from {\mbox{$\widehat \calG\approx 1$}} to {\mbox{$\widehat\calG\approx 0$}}.
For example, when needed for the purposes of illustration, we shall 
choose the fiducial value {\mbox{$\rho=2$}}.
Indeed, such a value is very close to the value that would be required for the ``ridge-top''
definition, yielding {\mbox{$\xi \rho = 0.954$}}.
However, by enforcing Eq.~(\ref{alignment}) directly, we will be able to maintain alignment
without needing to identify a particular definition 
for $\tau_2^\ast$.  Moreover, as already noted, the combination $\rho a^2$ is
invariant under the symmetry in Eq.~(\ref{rhoflipidentity}). 
This combination will therefore appear naturally in many of our future calculations,
thereby largely freeing us from the need to specify $\rho$ and $a$ individually.

Of course, we see from Eq.~(\ref{alignment}) that choosing $\rho$ within this range and taking {\mbox{$a\ll 1$}} will be possible
only if {\mbox{$t\gg 1$}}.   Thus, although the choice of $t$ is completely arbitrary
within the non-minimal regulator, only those non-minimal regulators with {\mbox{$t\gg 1$}} can
be aligned with our modular-invariant regulators in a meaningful way.

\section{Towards a field-theoretic interpretation: 
 The Higgs mass as a supertrace over physical string states
          \label{sec4}}

Equipped with the mathematical machinery from Sect.~\ref{sec3}, we now 
seek to express
our result for the Higgs mass given in Eq.~(\ref{relation1}) 
in terms of the supertraces over only the physical string states.
In so doing we will be 
developing an 
understanding of our results from a  field-theory perspective
--- indeed, as a string-derived effective field theory (EFT) valid at low energies.
All of these results will be crucial for allowing us to understand how
the Higgs mass ``runs'' within such an EFT, and
ultimately allowing us to extract 
a corresponding ``stringy'' effective Higgs potential in Sect.~\ref{sec5}.

\subsection{Modular invariance, UV/IR equivalence, and the passage to an EFT \label{UVIRequivalence}}

Our first task is to understand the manner through which
one may extract an EFT description
of a theory with modular invariance.
This is a subtle issue because such theories, as we shall see, possess a certain
UV/IR equivalence.  However, understanding this issue is ultimately crucial for the physical interpretations that we will be providing
for our results in the rest of this section, especially as they relate to the effects
of the mathematical regulators
we have presented in Sect.~\ref{sec3}.

In this paper, our interest has thus far focused on performing a fully string-theoretic calculation
of the Higgs mass.  Given that modular invariance is a fundamental symmetry of perturbative closed strings,
we have taken great care to preserve modular invariance at every step of our calculations (or to note the extent
to which this symmetry has occasionally been violated, such as for two of the three possible regulators discussed in
Sect.~\ref{sec3}).~ 
However, modular transformations
mix the contributions of individual string states into each other in 
highly non-trivial ways across the entire string spectrum.
Indeed, we shall see that modular invariance even leads to a fundamental 
equivalence between ultraviolet (UV) and infrared (IR) divergences.
Thus a theory such as string theory can be modular invariant
only if all of its states across the {\it entire}\/ string spectrum 
are carefully balanced against each other~\cite{Dienes:1994np} and treated similarly, as a coherent whole.
EFTs, by contrast, are predicated on an approach that treats UV physics and IR physics
in fundamentally different ways, retaining the dynamical degrees of freedom associated with the IR physics 
while simultaneously ``integrating out'' the degrees of freedom associated with the UV physics.
As a result, any attempt to develop
a true EFT description of a modular-invariant 
theory such as string theory inherently breaks modular invariance.

It is straightforward to see that modular invariance leads to an equivalence between UV and IR divergences.
In general, one-loop closed-string amplitudes are typically 
expressed in terms of modular-invariant integrands $F(\tau)$ which
are then integrated over the fundamental domain $\calF$ of the modular group.
If such an amplitude diverges, this divergence will arise from the {\mbox{$\tau_2\to \infty$}}
region within $\calF$.
Given that the contributions from the heavy string states 
within the integrand are naturally suppressed as {\mbox{$\tau_2\to\infty$}},
it would be natural to interpret this divergence
as an IR divergence involving low-energy physics.

However, such an interpretation would be inconsistent
within a modular-invariant theory.
In any modular-invariant theory with a modular-invariant integrand $F(\tau)$, 
we can always rewrite
our amplitude through the identity
\beq 
         \int_\calF  \dmu ~ F(\tau) =  
         \int_\calF  \dmu ~ F(\gamma \cdot \tau) =  
         \int_{\gamma\cdot \calF}  \dmu ~ F(\tau)~~~
\label{anychoice}
\eeq
which holds for any modular transformation $\gamma$.
From Eq.~(\ref{anychoice}) we see that 
choosing $\calF$ as our region of integration
is mathematically equivalent to choosing
any of its images {\mbox{$\gamma\cdot \calF$}} under any modular transformation $\gamma$.
One of these equivalent choices is {\mbox{$\calF'\equiv \gamma_S\cdot \calF$}} 
where $\gamma_S$ is the {\mbox{$\tau\to -1/\tau$}} modular transformation.
This region is explicitly given as
\beqn
            \calF' ~&\equiv&~ \lbrace \tau :\,    \tau_2>0,\, |\tau|\leq 1,  \,
              (\tau_1 +1)^2 + \tau_2^2 \geq 1, ~~~~~~~~\nonumber\\
             && ~~~~~~~~~~~~~~ (\tau_1 -1)^2 + \tau_2^2 \geq 1\, \rbrace~,~~~~~
\eeqn
and as such includes the {\mbox{$\tau_2\to 0$}} region but no longer includes the {\mbox{$\tau_2\to\infty$}} region.
Indeed, via the identity in Eq.~(\ref{anychoice}) we see that the
divergence of our amplitude now appears as {\mbox{$\tau_2\to 0$}}.
However, 
there is no suppression of the contributions from the heavy string states 
within the integrand
as {\mbox{$\tau_2\to 0$}}.
Instead, any divergence as {\mbox{$\tau_2\to 0$}} arises through the accumulating contributions of the heavy
string states and would therefore naturally be interpreted as a UV divergence. 
Thus, by trading $\calF$ for $\calF'$ through Eq.~(\ref{anychoice}), we see that we can 
always mathematically recast what would naively appear 
to be an IR divergence as {\mbox{$\tau_2\to \infty$}} 
into what would naively appear to be a UV divergence as {\mbox{$\tau_2\to 0$}} --- all without
disturbing the integrand of our amplitude in any way.
A similar conclusion holds for the many other $\calF''$ domains that could equivalently have been chosen for
other choices of the modular transformation $\gamma$.

This is a fundamental observation.
When we calculate an amplitude in string theory, 
we are equipped with an integrand which reflects
the spectrum of string states but  we must
choose an appropriate fundamental domain of the modular group.
This choice is not something dictated within the theory itself, but instead
amounts to a {\it convention}\/ which is adopted for the sake of performing a calculation.
It is possible, of course, that the amplitude in question diverges.
As we have seen, if we choose the fundamental domain $\calF$ as defined in Eq.~(\ref{Fdef}) 
then this divergence will manifest itself
as an IR divergence.
However, if we choose $\calF'$ as our fundamental domain, this same divergence of
the amplitude will manifest itself as a UV divergence.
Both interpretations are equally valid 
because the divergence 
of a one-loop modular-invariant string amplitude
is neither intrinsically UV nor intrinsically IR.~
Indeed, such a divergence 
is a property {\it of the amplitude itself}\/ and is not intrinsically tied
to any particular value of $\tau$.
Such a divergence is then merely {\it represented}\/ as a UV or IR divergence depending on our choice of
a region of integration.

This observation can also be understood through a comparison with our expectations from quantum field theory.
As we have seen in Eq.~(\ref{unfold}), there is a tight relation between
the fundamental domain $\calF$ and the strip $\calS$ defined in Eq.~(\ref{Sdef}): 
essentially $\calF$ is a ``folded'' version of $\calS$.   Likewise, the modular-invariant integrand $F(\tau)$  
that is integrated over $\calF$ is nothing but the sum of the images of the {\it non}\/-invariant integrand which 
would be integrated over $\calS$.
Thus, through the unfolding procedure in Eq.~(\ref{unfold}),  we have two equivalent representations
for the same physics.  These are often called the $\calF$- and $\calS$-representations.

It is through the $\calS$-representation that we can most directly
make contact with the results that would come from a quantum field theory based on point particles.  
Within the $\calS$-representation, we can identify $\tau_2$ as the Schwinger proper-time
parameter, with {\mbox{$\tau_2\to\infty$}} corresponding to the field-theoretic IR limit
and with {\mbox{$\tau_2\to 0$}} corresponding to the field-theoretic UV limit.
Indeed, within field theory these limits are physically distinct, just as they are geometrically
distinct within the strip.
However, 
upon folding the strip $\calS$ 
into the fundamental domain $\calF$, we see that {\it both}\/ the UV {\it and}\/ IR field-theoretic regions
within $\calS$ are together mapped onto the {\mbox{$\tau_2\to\infty$}} region within $\calF$.
Indeed, the distinct UV and IR regions of the strip $\calS$ are now ``folded'' so as 
to lie directly on top of each other within $\calF$.
Thus, within the $\calF$-representation, the {\mbox{$\tau_2\to\infty$}} limit in some sense
represents {\it both}\/ the UV and IR field-theory limits simultaneously --- limits
which would have been viewed as distinct within field theory but which are now related to each other in
string theory through modular invariance.   
An identical argument also holds for the {\mbox{$\tau_2\to 0$}} region within $\calF'$.

We can therefore summarize the situation as follows.
For a modular-invariant string-theoretic amplitude there is only one kind of divergence.   
It can be represented as either a UV divergence or an IR divergence depending on our choice
of fundamental domain (region of integration).   However, in either case, this single 
string-theoretic divergence can be mapped back to what can 
be considered a modular-invariant {\it combination}\/ of UV and IR field-theoretic divergences in 
field theory (\ie, on the strip $\calS$).  Indeed, we may schematically write
\beq
     \underbrace{ 
   {\rm IR}_{\calF} \,=\, {\rm UV}_{\calF'} }_{\hbox{string theory}}
        ~\Longleftrightarrow~
     \underbrace{  {\rm IR}_{\calS} \,\oplus\, {\rm UV}_{\calS}  }_{\hbox{field theory}}~~
\label{UVIR}
\eeq
where `$\oplus$' signifies a modular-invariant combination.
We shall obtain an explicit example of such a combination below.
It is ultimately in this way, through Eq.~(\ref{UVIR}), that  
our modular-invariant string theory loses its ability to distinguish between UV and IR physics.
We will discuss these issues further in Sect.~\ref{sec:Conclusions}.

Our discussion in this paper has thus far been formulated with $\calF$ chosen
as our fundamental domain.  In this way we have been implicitly casting 
our string-theoretic divergences as infrared.    In the following, we shall therefore continue along this line and attach corresponding 
physical interpretations to our mathematical results as far as possible. 
However, we shall also 
occasionally
indicate how our results might alternatively 
appear within the $\calF'$-representation, or within the 
unfolded $\calS$-representation of ordinary quantum field theory. 
This will ultimately be important for extracting an EFT for the Higgs mass,
for understanding how our Higgs mass ``runs'' within such an EFT,
and for eventually interpreting our results in terms of a stringy effective potential.

One further comment regarding the nature of these divergences is in order.
The above discussion has focused on the manner in which modular invariance mixes
UV and IR divergences
when passing from field theory to string theory.
However, it is also important to remember that modular invariance likewise affects
the {\it strengths}\/ of these divergences.
To understand this, we recall from Eq.~(\ref{stripF}) that the strip $\calS$, which serves as the field-theoretic region of integration,
is nothing but the sum of the images of $\calF$, a string-theoretic region of integration,
under each of the modular transformations $\gamma$ in the coset {\mbox{$\Gamma_\infty\backslash \Gamma$}}.
However, there are an infinite number of such modular transformations within this coset.
This means, in essence, that our string-theoretic divergences (if any) are added together
an infinite number of times when $\calF$ is unfolded into $\calS$,
implying that the resulting field-theoretic divergences are far more severe than those of the string.
Phrased somewhat differently, we see that modular invariance softens a given
field-theoretic divergence by allowing us to reinterpret part of this divergence as resulting from an infinity
of identical copies of a weaker (modular-invariant) string divergence, whereupon we are authorized to
select only one such copy. 

This observation is completely analogous to what happens within field theory in the presence of a gauge symmetry.
If we were to disregard the gauge symmetry when calculating a field-theoretic amplitude, 
we would integrate over an infinite number of gauge slices when performing our path integrals. 
This would result in divergences which are spuriously severe.
However, modular invariance is similar to gauge symmetry in the sense that both represent redundancies of 
description.  
(In the case of modular invariance, the redundancy arises from the fact that all values of {\mbox{$\gamma \cdot \tau$}}
for {\mbox{$\gamma\in\Gamma$}} correspond to the same worldsheet torus.)
In a modular-invariant theory, we therefore divide out by the (infinite) volume of the 
redundancy coset {\mbox{$\Gamma_\infty\backslash \Gamma$}} and consider only one modular-invariant ``slice''. 
Indeed, this is precisely what is happening when we pass from $\calS$ to $\calF$ (or any of its images)
as the appropriate region of integration in a modular-invariant theory, where  the particular choice of image
is nothing but the particular choice of slice.
This passage from $\calS$ to a particular modular-invariant slice
then softens our field-theoretic divergences and in some cases even eliminates them entirely.

We have already seen one example of this phenomenon:  the one-loop vacuum energy (cosmological
constant) $\Lambda$ is badly divergent in quantum field theory, yet finite in any tachyon-free closed string theory.
Indeed, it is modular invariance which is alone responsible for this phenomenon.
As we shall see, a similar softening of divergences also occurs for the Higgs mass.

We conclude this discussion with one additional comment.
It is a common assertion that string theory lacks UV divergences.
The rationale usually provided for this is that string theory intrinsically has a minimum length 
scale, namely $M_s^{-1}$, and that this provides a ``cutoff'' that eliminates  
all physics from arbitrarily short length scales and thereby eliminates the associated UV divergences.  
However, this argument fails to acknowledge that IR divergences may still remain,
and of course in a modular-invariant theory the UV and IR divergences are mixed.
Indeed, as we have explained, there is no modular-invariant way of disentangling these
two kinds of divergences.   Thus string theory is not free of divergences.
These divergences are simply softer than they would have been in field theory.

\subsection{The divergence structure of the Higgs mass}

With the above comments in mind, we now consider the divergence structure of the Higgs mass.
We begin by recalling from Eq.~(\ref{relation1}) that 
the Higgs mass $m_\phi$ 
within any four-dimensional heterotic string 
is given by  
\beq
    m_\phi^2 ~=~ -\frac{\calM^2}{2} \biggl(
            \langle \calX_{1a}\rangle 
            + \langle \calX_{1b}\rangle 
          + \langle \calX_2\rangle  \biggr)
          + \frac{\xi}{4\pi^2} \frac{\Lambda}{\calM^2}~~~
\label{Higgsmass}
\eeq
where
\beqn
  \calX_{1a} ~&\equiv &~
        \frac{\tau_2}{\pi} 
              \left( \tilde \bQ_j^t \bQ_h +  \bQ_j^t \tilde \bQ_h \right) \nonumber\\ 
  \calX_{1b} ~&\equiv &~
        - \frac{\tau_2}{\pi} 
              \left(  \bQ_h^2 +  \tilde \bQ_h^2 \right) \nonumber\\ 
  \calX_2 ~&\equiv &~  
          \tau_2^2 \, \left(\bQ_R^t \bQ_h - \bQ_L^t \tilde \bQ_h \right)^2 \nonumber\\
         && ~~~ = ~    
          4 \tau_2^2\, (\bQ_R^t \bQ_h)^2  ~=~ 4 \tau_2^2 \, (\bQ_L^t \tilde \bQ_h )^2 ~~~~~~
\label{Xidef}
\eeqn
and where $\Lambda$ is the one-loop cosmological constant.
Note that we have explicitly separated those terms $\calX_{1a}$ and $\calX_{1b}$
which are quadratic in charge
insertions 
from those terms $\calX_2$ which are quartic, as these will shortly play very different roles.
Moreover, within the quadratic terms, we have further distinguished 
those insertions $\calX_{1a}$ within which each term consists of a paired contribution 
of a left-moving charge with a right-moving charge
from those insertions $\calX_{1b}$
in which each term consists of two charges which are 
both either left- or right-moving.
Indeed, we recall from Sect.~\ref{sec2} that only 
$\langle \calX_{1a}\rangle$ 
and the sum 
$\langle \calX_{1b}\rangle + \langle \calX_2\rangle$ 
are modular invariant;
in particular, 
$\langle \calX_{1b}\rangle$
and 
  $\langle \calX_2\rangle$ 
are the modular completions of each other
and thus  
neither 
is modular invariant by itself.
That said, it will prove convenient in this section to  
simply define
\beqn
  \calX_{1} ~&\equiv &~ \calX_{1a} + \calX_{1b}\nonumber\\
     &= & ~
 \frac{\tau_2}{\pi} 
              \left( \tilde \bQ_j^t \bQ_h +  \bQ_j^t \tilde \bQ_h 
              -  \bQ_h^2 -  \tilde \bQ_h^2 \right) ~,
\label{Xidef2}
\eeqn
so long as we remember that only the 
full combination $\langle \calX_1\rangle +\langle \calX_2\rangle$ is modular invariant.

As discussed in Sect.~\ref{sec2}, these results are completely general and apply to any
scalar $\phi$ whose VEV determines the vacuum structure of the theory.
Indeed, the various charge insertions 
$\bQ_h$, $ \tilde \bQ_h$, $ \bQ_j$, and $\tilde \bQ_j$
in Eq.~(\ref{Xidef})
are defined in Eqs.~(\ref{Qhdef}) and (\ref{Qjdef}) in terms of the
$\calT$-matrices which encapsulate the relevant information concerning
specific scalar under study.

Unlike the other terms in Eq.~(\ref{Higgsmass}), the final term $\Lambda$ 
emerges as the result of a universal shift in the background moduli.
As such, this quantity is wholly independent of the specific $\calT$-matrices,
and merely   provides a uniform shift to the masses of all scalars 
in the theory regardless of the specific roles these scalars might play in breaking gauge symmetries 
or otherwise affecting the vacuum state of the theory.
In other words, $\Lambda$ provides what is essentially a mere ``background'' contribution
to our scalar masses.  Moreover, as the one-loop cosmological constant of the theory,
$\Lambda$ is an independent physical observable  unto itself.
For this reason, we shall defer our discussion of $\Lambda$ to Sect.~\ref{Lambdasect} 
and focus instead on the effects coming from the $\calX_i$ insertions
in Eq.~(\ref{Higgsmass}).

In order to make use of the machinery in Sect.~\ref{sec3}, we must
first understand the divergence structure that can arise from each of these $\calX_i$
insertions as {\mbox{$\tau_2\to \infty$}}.
For any string model in four spacetime dimensions,
the original partition function prior to any $\calX_i$ insertions has
the form indicated in Eq.~(\ref{Zform}),
with an overall factor of $\tau_2^{-1}$. 
Thus, the insertion of 
$\calX_1$  leads to 
integrands without a leading factor of $\tau_2$,
while the insertion of
$\calX_2$ leads to integrands with a 
leading factor of $\tau_2^{+1}$. 

Determining the possible divergences 
as {\mbox{$\tau_2\to\infty$}}  requires that we
also understand the spectrum of low-lying states 
that contribute to these integrands.
We shall, of course, assume that our string model
is free of of physical (on-shell) tachyons.
Thus, expanding the partition function $\calZ$ of our string model as 
in Eq.~(\ref{integrand}) with {\mbox{$k= -1$}}, 
we necessarily have {\mbox{$a_{nn}=0$}} for all {\mbox{$n< 0$}} in Eq.~(\ref{integrand}).

There is, however,
an {\it off-shell}\/ tachyonic state which must always appear within the spectrum of any self-consistent heterotic string model:
this is the so-called {\it proto-graviton}~\cite{Dienes:1990ij} with {\mbox{$(m,n)=(0,-1)$}}, and no possible GSO projection can eliminate this state 
from the spectrum.
Although this state is necessarily a singlet under all of the gauge symmetries of the model,
it transforms as a vector under the spacetime Lorentz group.   Consequently 
the degrees of freedom that compose this state have non-vanishing charge vectors of the form
\beq
                 {\bf Q}_{\hbox{\scriptsize proto-graviton}} ~=~               ({\bf 0}_{22} \, | \pm 1, {\bf 0}_9 )
\label{protocharge}
\eeq
where we have written this charge vector in the same basis as used in Eq.~(\ref{Tmatrixform}),
with the non-zero charge component in Eq.~(\ref{protocharge}) lying along the spacetime-statistics direction
discussed in Sect.~\ref{sec:EWHiggs}.

Because of this non-zero charge component, the proto-graviton state has the possibility of contributing to one-loop string
amplitudes even when certain charge insertions occur.
However, we have seen in Eq.~(\ref{Tmatrixform})
that the $\calT$-matrices appropriate for shifts in the Higgs VEV do not disturb the spin-statistics of the states
in the spectrum, and thus necessarily have zeros along the corresponding columns and rows. 
Indeed, these zeros are a general feature which would apply to all such $\calT$-matrices
regardless of the specific Higgs scalar under study or its particular gauge embedding.
As a result, the would-be contributions from the proto-graviton state
do not survive either of the $\calX_i$ insertions in Eq.~(\ref{Xidef}).
Indeed, similar arguments also apply to potential contributions from the proto-gravitino states (such as would appear in the
spectra of string models exhibiting spacetime supersymmetry).

In general, a heterotic string model can also contain other off-shell tachyonic $(m,n)$ states with {\mbox{$m\not=n$}} but {\mbox{$m+n<0$}}.   Unlike the proto-graviton state with {\mbox{$(m,n)=(0,-1)$}}, such states would generally have {\mbox{$(m,n)= (k+1,k)$}} where {\mbox{$-1<k<-1/2$}}.     However, even if a given string model were to contain such states, these states --- like the proto-graviton state --- would also likely not have non-zero charges in the appropriate Higgs directions.    Indeed, this closely mirrors the situation that emerges for the analogous calculation of string threshold corrections in a variety of semi-realistic string models, as discussed in Ref.~\cite{Dienes:1995bx}, where it was explicitly demonstrated that none of the potential off-shell tachyonic states that appeared in such models carried the sorts of non-trivial gauge charges that were relevant for the corresponding gauge threshold calculations.   
We shall therefore make the same assumption here.  

The net result, then, is that the lightest states that can contribute in the presence of 
the $\calX_i$ insertions are the massless on-shell string states.   These states contribute to the {\mbox{$(m,n)=(0,0)$}} term 
in the integrand, and thus their contributions cause the integrand to scale 
as $\tau_2^k$ in the {\mbox{$\tau_2\to \infty$}} limit, 
where {\mbox{$k=0$}} for the $\calX_1$ insertion   and {\mbox{$k=1$}} for the $\calX_2$ insertion. Moreover, all heavier states make contributions which are exponentially suppressed as {\mbox{$\tau_2\to \infty$}},
regardless of the $\tau_2^k$ prefactor, and thus do not lead to divergences in this limit.
We thus see that the $\calX_1$ and $\calX_2$ insertions  
together produce an integrand exhibiting
the {\mbox{$\tau_2\to \infty$}} divergence structure given in Eq.~(\ref{ourcase}), where we now identify
\beqn
               c_0 ~&=&~  -\half \, \calM^2 \,  \, \zStr  ~ \mathbbX_1 \nonumber\\    
               c_1 ~&=&~  -\half \, \calM^2  \, \, \zStr  ~ \mathbbX_2~.  
\label{c0c1}
\eeqn
Here the $\mathbbX_i$ are the same as the charge insertions $\calX_i$ but without their leading $\tau_2$ prefactors:
\beq
     \mathbbX_1 \equiv \tau_2^{-1} \calX_1~,~~~~
     \mathbbX_2 \equiv \tau_2^{-2} \calX_2~.
\label{bbXdefs}
\eeq
Likewise, in writing the expressions in Eq.~(\ref{c0c1}) we are introducing the {\it supertrace}\/ notation 
\beq
              \Str \, A ~\equiv~   \sum_{{\rm physical}~ i}  (-1)^{F_i} \, A_i \, \label{supertracedef}
\eeq
where the sum is over the {\it physical}\/ states $i$ in the string spectrum,
where $F_i$ is the spacetime fermion number of state $i$, 
and where $A_i$ is the value of $A$ for that state. 
Note that the off-shell string states with {\mbox{$m\not=n$}} are not propagating string degrees of freedom
and thus our definition for the supertrace `Str' in Eq.~(\ref{supertracedef}) does not include them.
The supertrace `Str' therefore includes the contributions from only those 
string states which have direct field-theory analogues. 
In Eq.~(\ref{c0c1}) the `{\mbox{$M=0$}}' subscripts on `Str' indicate that these supertraces
are further restricted to include only those states which are massless.
We thus see that the divergence in the Higgs mass arises, 
not unexpectedly, from the contributions of {\it massless}\/ string states
(specifically those massless states which are 
charged under the different $\mathbbX_i$).
This is exactly as we expect for an infrared divergence.   However, as discussed in Sect.~\ref{UVIRequivalence},
the interpretation 
of this divergence as being infrared in nature and arising from massless states
depends crucially on our choice to work in the $\calF$-representation.

Given the divergence structure in Eq.~(\ref{ourcase}) with coefficients given in Eq.~(\ref{c0c1}), we see
that our expression for the Higgs mass in Eq.~(\ref{Higgsmass}) is generically logarithmically divergent as {\mbox{$\tau_2\to \infty$}}
(and finite only if {\mbox{$c_1\sim \, \zStr \mathbbX_2$}} 
happens to vanish within a particular string model).
We thus see that
\begin{itemize}
\item   Just as the one-loop vacuum energy in any tachyon-free closed string theory
            is {\it finite}\/ as a result of modular invariance, the corresponding 
             Higgs mass is at most {\it logarithmically divergent}\/.
\end{itemize}
Modular invariance has thus induced a significant softening of the Higgs divergence,
reducing what would have been a {\it quadratic}\/ UV Higgs divergence in field theory
into a logarithmic Higgs divergence in string theory.
However, even though the Higgs divergence has been softened 
within Eq.~(\ref{Higgsmass}),
we must still regulate the logarithmic divergence that remains. 
In the following we shall do this using the regulators discussed in Sect.~\ref{sec3}
and interpret the resulting expressions in terms of an effective field theory (EFT).~
Because each of these regulators has different strengths and weaknesses,
we shall apply each of these regulators in turn.
Comparing the corresponding results will ultimately enable us to understand 
the full power of the modular-invariant regulator.

\subsection{Results using the minimal regulator\label{higgsmin}}

Using the minimal regulator discussed in Sect.~\ref{sec:minimal},
we can regulate the Higgs mass in Eq.~(\ref{Higgsmass}) to take the form
\beqn
    \widetilde m_\phi^2 \, &=&\,  -\half \, \calM^2  \lim_{t\to\infty}
           \biggl[ 
            \langle \calX_1\rangle_t 
          + \langle \calX_2\rangle_t \nonumber\\
            && ~~~ - (\, \zStr \,\mathbbX_2)  \log\,t \biggr] 
     + \frac{\xi}{4\pi^2} \frac{\Lambda}{\calM^2}~~~~~~
\label{regHiggsmass1}
\eeqn
where, in analogy with Eq.~(\ref{expvalue}), we have now defined
\beq
 \langle A \rangle_t ~\equiv~ \int_{\calF_t} \dmu 
        ~\frac{\tau_2^{-1}}{\overline{\eta}^{12} \eta^{24}} \, \sum_{\bQ_L,\bQ_R} 
             (-1)^F  A~ \qbar^{\bQ_R^2/2} q^{\bQ_L^2/2}
\label{expvaluet}
\eeq
where $\calF_t$ is the truncated domain of integration discussed above
Eq.~(\ref{tildeIdef}).
Note that unlike $m_\phi^2$ in Eq.~(\ref{Higgsmass}), our expression for the regulated $\widetilde m_\phi^2$ is
manifestly finite, as the logarithmic divergence arising due to the massless string states charged under $\calX_2$
has been explicitly excised.
Indeed, this explicit subtraction of the contributions from massless states
is similar to what occurs in the 
calculation of string-theoretic gauge threshold corrections in Ref.~\cite{Kaplunovsky:1987rp} 
and in the calculation of string-theoretic kinetic mixing in Ref.~\cite{Dienes:1996zr}.

While the expectation values $\langle \calX_i \rangle$ in Eq.~(\ref{regHiggsmass1}) receive
contributions from both physical and unphysical string states, our results from Sect.~\ref{sec:minimal} 
allow us to express $\widetilde m_\phi^2$ purely in terms of
physical string states.
Indeed, disregarding the contributions from the finite $\Lambda$-term in
Eq.~(\ref{Higgsmass}), 
we find that our unregulated physical-state trace $g(\tau_2)$ 
is given by
\beq
             g(\tau_2) \,=\, -\half \calM^2 \left[
      \Str\, (\mathbbX_1 + \tau_2 \mathbbX_2) \,e^{-\pi \alpha' M^2 \tau_2} \right]~,~
\label{gdefn}
\eeq
consistent with the divergence structure {\mbox{$\Phi_g(\tau_2)= c_0+c_1\tau_2$}} discussed above.
Our regulated physical-state trace $\widetilde g(\tau_2)$ 
in Eq.~(\ref{tildegdef}) is then given by
\beqn
      \widetilde g(\tau_2) \, &\equiv& \, 
       -\half \calM^2 \biggl\lbrace
            \Str\, \left[( \mathbbX_1 + \tau_2 \mathbbX_2  ) \, 
               e^{-\pi \alpha' M^2 \tau_2}\right] \nonumber\\
       && ~~~~~~ - \, \zStr \, \mathbbX_1  - (\, \zStr \, \mathbbX_2)\,\tau_2 \biggr\rbrace\nonumber\\
       &=& 
       \, -\half \calM^2~
            \, \pStr \, \left[( \mathbbX_1 +  \tau_2 \mathbbX_2 ) 
              \, e^{-\pi \alpha' M^2 \tau_2}\right]~,
      \nonumber\\ 
\label{phystrac}
\eeqn
and thus we see that only the {\it massive}\/ string states
contribute to $\widetilde g(\tau_2)$.
However, integrating our result for $\widetilde g(\tau_2)$ in Eq.~(\ref{phystrac}) over $\tau_2$ and taking the residue at {\mbox{$s=1$}} as in Eq.~(\ref{Zagierresult}), we find
\beqn
&& \oneRes \int_0^\infty d\tau_2 \, \tau_2^{s-2}\, \widetilde g(\tau_2) \nonumber\\
  &&  ~~\,=\,  -\half \calM^2 \, \oneRes \, \biggl\lbrace 
           \Gamma(s-1) \,\, \pStr \left\lbrack   \mathbbX_1 (\pi \alpha' M^2)^{1-s} \right\rbrack
\nonumber\\
&&     ~~~~~~~~~~~~~~~~~~+ 
           \Gamma(s) \,\, \pStr \left\lbrack   \mathbbX_2 (\pi \alpha' M^2)^{-s} \right\rbrack 
               \biggr\rbrace\nonumber\\
&&  ~~\,=\,  -\half \calM^2  \, \pStr \mathbbX_1 ~.
\label{integrateg}
\eeqn 
Thus taking the residue has the effect of further projecting out from the $\widetilde g(\tau_2)$ term  
the contributions that emerge from the $\mathbbX_2$ insertion.
Following Eq.~(\ref{Zagierresult}) we then obtain our final result for the regulated Higgs mass: 
\beqn
    \widetilde m_\phi^2 \, &=&\,       -\half\calM^2 \biggl[
          \frac{\pi}{3} \,\, \pStr \mathbbX_1 \nonumber\\
&&~~~~+ \frac{\pi}{3} \, \, \zStr \mathbbX_1 +
                         (\, \zStr \,\mathbbX_2) \log 4\pi e^{-\gamma} \biggr]\nonumber\\
        &=& \, -\half\calM^2 \biggl[
          \frac{\pi}{3} \, \Str\, \mathbbX_1 +
                         (\, \zStr \,\mathbbX_2) \log 4\pi e^{-\gamma} \biggr]\nonumber\\
\label{firstresult}
\eeqn
where we have not displayed the additional universal $\Lambda$-term that appears in Eq.~(\ref{regHiggsmass1}).
Note, in particular, that the contributions from the massless states 
within the $\mathbbX_1$ insertion, after having been subtracted
in Eq.~(\ref{phystrac}),
 have been {\it restored}\/ in Eq.~(\ref{firstresult}).
This is precisely in accord with our expectations regarding the role of the $c_0$ parameter, 
as discussed below Eq.~(\ref{Zagierresult}).

One important comment is in order.   In our derivation in Eq.~(\ref{integrateg}), we 
implicitly assumed that the residue of the supertrace sum is the same as the 
supertrace sum of the individual residues.    This allowed us to exchange the order of the supertrace-summation 
and residue-extraction procedures.
This kind of calculation will occur repeatedly throughout this 
paper, and we shall do this in each instance.   This exchange of ordering is justified because we 
are working within a regulated theory in which there are no additional divergences that might 
arise from the supertrace sum beyond those which were already encapsulated within our 
original assertion that the Higgs mass has at most a logarithmic divergence when 
{\mbox{$\zStr \mathbbX_2\not=0$}}, or equivalently that $g(\tau_2)$ diverges no more rapidly than 
{\mbox{$c_0+c_1\tau_2$}} as $\tau_2\to\infty$.   This will be discussed further in Sect.~\ref{sec:Conclusions}.

We see, then, that use of the minimal regulator discussed in Sect.~\ref{sec:minimal} leads
to a final parameter-independent expression for the Higgs mass purely in terms of the contributions
of physical string states.
Moreover, this expression 
eliminates the explicit contributions from the massive states 
within the quartic insertion $\mathbbX_2$.
This initially surprising observation makes sense if we think of the logarithmic divergence of $m_\phi^2$ 
as an ultraviolet one,
given that the contributions of the massive states 
to the quartic terms are the contributions which are expected to grow the most rapidly 
as we proceed upwards through string spectrum. 

Despite the form of Eq.~(\ref{firstresult}),
we note that $\widetilde m_\phi^2$ is
not actually insensitive to the massive spectrum resulting from the $\calX_2$ insertion, just as
$\widetilde m_\phi^2$ 
is not insensitive to all of the {\it unphysical}\/ string states 
whose contributions also originally appeared within Eq.~(\ref{regHiggsmass1}). 
Indeed {\it all}\/ of these states ---  along with the full spectrum of physical 
states from the $\calX_1$ insertion ---
are part of what makes our original expression for $m_\phi^2$ 
invariant under the fundamental modular symmetry 
which lies at the core of the procedure we have followed in 
casting $\widetilde m_\phi^2$ as a supertrace over only physical string states.
Indeed, as we have already seen in Sect.~\ref{sec:completion}, the insertion
$\calX_1$ is intrinsically part
of the modular completion of $\calX_2$, and {\it vice versa}\/.
We should therefore simply interpret the result in Eq.~(\ref{firstresult}) as telling us
that modular invariance is sufficiently powerful a symmetry that the 
spectra resulting from the $\calX_1$ and $\calX_2$ insertions are no longer 
independently adjustable, but rather are so locked together that
it is no longer necessary to explicitly sum over all of them
independently
when expressing our regulated Higgs mass $\widetilde m_\phi^2$ in this manner.

That said, it is significant that
our regulated result for the Higgs mass
in Eq.~(\ref{firstresult})
treats $\mathbbX_1$ and $\mathbbX_2$ in fundamentally different ways.
As noted below Eq.~(\ref{Xidef}), only the strict combination $\calX_1+\calX_2$ preserves
modular invariance.
Indeed, $\calX_1$ and $\calX_2$ appeared symmetrically 
in our original {\it unregulated}\/ expression in Eq.~(\ref{Higgsmass}).
We therefore see that 
although the string spectrum itself is strictly modular invariant, as discussed above,
our regulated result 
for the Higgs mass in Eq.~(\ref{firstresult}) is not. 
Of course, this outcome is completely expected because
our regulator in this case is built upon a method for subtracting divergences which 
is not modular invariant.

\subsection{Results using the non-minimal regulator\label{higgsnonmin}}

Let us now investigate how these results change if we instead regulate our Higgs mass
using the modified {\it non}\/-minimal regulator of Eqs.~(\ref{I3}) and (\ref{Zagierresult3}).
As discussed in Sect.~\ref{sec3},
this regulator --- like the minimal regulator --- is also
not modular invariant.
However, it will lead to a richer structure than that obtained
from the minimal regulator --- a structure which will enable us to make a comparison 
with field-theoretic expectations.

Using the modified non-minimal regulator, we follow Eq.~(\ref{I3})
in defining the regulated form
of our logarithmically divergent Higgs mass $m_\phi^2$ as
\beqn
    \widehat m_\phi^2(t) && ~ \equiv\,   -\half\, \calM^2 \, \biggl[ 
             \langle \calX_1\rangle + \langle \calX_2\rangle  \nonumber\\  
             && \, - \int_{\calF-\calF_t} \dmu ~ \tau_2\, (\, \zStr \,\mathbbX_2) \biggr]  
     +  \frac{\xi}{4\pi^2} \frac{\Lambda}{\calM^2}~.~~~~~~~~~
\label{regHiggsmass2}
\eeqn
Unlike our expression for 
$\widetilde m_\phi^2$ in Eq.~(\ref{regHiggsmass1}) which employed the minimal regulator,
we see that this new regulated Higgs mass $\widehat m_\phi^2(t)$ is a function of an arbitrary dimensionless
parameter {\mbox{$1\leq t<\infty$}}.
Given the discussion in Sect.~\ref{sec:nonminimal}, along with our result for the minimal regulator in 
Eq.~(\ref{firstresult}), it then follows that $\widehat m_\phi^2(t)$ can
be expressed purely in terms of physical string states as
\beq
   \widehat m_\phi^2(t) \,=\,  
        -\half \calM^2 \, \biggl[
          \frac{\pi}{3} \, \Str\, \mathbbX_1 
    +  (\, \zStr \,\mathbbX_2) \log 4\pi t e^{-\gamma} \biggr]~\\
\label{nonminresult}
\eeq
where we have again refrained from displaying the additional universal $\Lambda$-term.
Indeed, the extra crucial factor within Eq.~(\ref{nonminresult}) 
relative to the result in Eq.~(\ref{firstresult}) is a new
logarithmic dependence on the regulator parameter $t$.

The expression for $\widehat m_\phi^2 (t)$ in 
Eq.~(\ref{nonminresult}) is the exact string-theoretic result arising from our non-minimal 
regulator.  As such, this result is complete unto itself for any {\mbox{$1\leq t<\infty$}} and requires no further manipulations.
That said, we would nevertheless like to broaden the discussion to 
make contact with a potential field-theoretic interpretation.
Towards this end, we shall now make two further adaptations.

Normally, in an effective field theory (EFT), we divide our states
into two categories, ``light'' and ``heavy'', with respect to some physical scale $\mu$.
         Within the EFT, 
those states which are designated light then play a different role relative
to those which are designated heavy.  
However, because the boundary between light and heavy is set in relation to the scale $\mu$, 
any calculation which distinguishes between these two categories 
inevitably results in physical quantities (such as Higgs masses) which 
depend on (or ``run'' with) the scale $\mu$.  In other words, we would expect
to obtain a regulated Higgs mass $\widehat m_\phi^2(\mu)$
which is a {\it function}\/ of the scale $\mu$.
Moreover, given that
our unregulated Higgs mass $m_\phi^2$ is only logarithmically divergent prior to regularization,
we would expect the regulated quantity $\widehat m_\phi^2(\mu)$ to have at most a logarithmic dependence
on $\mu$.  Indeed, such a relation would be nothing but a renormalization-group equation (RGE) for the Higgs mass.

Our string-theoretic result in Eq.~(\ref{nonminresult}) already resembles such an RGE.~
Indeed, as stated above, there are only two adaptations that we must make in order to turn this into an actual RGE.~
First, we must somehow identify the regulator parameter $t$ which appears in Eq.~(\ref{nonminresult}) as corresponding to
a physical scale $\mu$.
Second, we wish to generalize our notion of {\it massless}\/ states --- states which play a special role  in Eq.~(\ref{nonminresult})
--- to states which are simply light with respect to $\mu$.

These two issues are intertwined and have a common resolution.
Recall that our string-theoretic calculation, as outlined above, isolated the
strictly {\it massless}\/  states as special based on the behavior of
their potentially divergent contributions as {\mbox{$\tau_2\to\infty$}}.
Indeed, as we have seen, the contributions from states with masses {\mbox{$M>0$}} have a built-in Boltzmann-like 
suppression $\sim e^{-\pi \alpha' M^2 \tau_2}$ as {\mbox{$\tau_2\to\infty$}}, while massless states
do not.   
Thus massless states are unprotected by the Boltzmann suppression factor as {\mbox{$\tau_2\to\infty$}},
which is why their contributions are subtracted as part of the regularization procedure.

Within the {\it non}\/-minimal regulator, however, we distinguish between
two different ranges for $\tau_2$:   one range with {\mbox{$1\leq \tau_2 \leq t$}}, and a
second range with {\mbox{$t\leq \tau_2 < \infty$}}.
Only within the second range do we subtract the contributions from the massless states;
indeed, massless states are considered ``safe'' within the first range.
But for any finite $t$, it is possible that there are many light states which do not have
appreciable Boltzmann suppression factors at {\mbox{$\tau_2=t$}}.
Such light (or ``effectively massless'') states are therefore essentially indistinguishable from 
truly massless states as far
as their Boltzmann suppression factors are concerned.
Indeed, it is only as {\mbox{$\tau_2\to\infty$}} that we can distinguish the truly massless
states relative to all the others. 

This suggests that for any finite value of $t$, we can assess whether a given state 
of mass $M$ is effectively light or heavy
according to the magnitude of its corresponding Boltzmann suppression factor at {\mbox{$\tau_2=t$}} within the 
partition function.
Recalling that the contribution from a physical string state of 
mass $M$ to the string partition function scales as $e^{-\pi \alpha' M^2 \tau_2}$,
we can establish an arbitrary criterion for the magnitude of the Boltzmann suppression 
of a state with mass {\mbox{$M=\mu$}} at the cutoff $t$:
\beq
                 e^{-\pi \alpha' \mu^2 t} ~\sim~ e^{-\varepsilon}
\label{massscale}
\eeq
where {\mbox{$\varepsilon\geq 0$}} 
is an arbitrarily chosen dimensionless parameter.
According to this criterion, states whose Boltzmann factors at {\mbox{$\tau_2= t$}} 
exceed $e^{-\varepsilon}$ have not experienced significant Boltzmann suppression and 
can then be considered light relative to that choice of $t$, 
while all others can be considered heavy.
We thus find that our division between light and heavy states can be demarcated by
a running mass scale $\mu (t)$ defined as
\beq
           \mu^2(t) ~\equiv~ \frac{\varepsilon}{\pi \alpha' t}~.
\label{mudefeps}
\eeq
Note, as expected, that {\mbox{$\mu(t)\to 0$}} as {\mbox{$t\to\infty$}}.  Thus, as expected, 
the only states that can be considered light as {\mbox{$t\to\infty$}} are those which are exactly massless. 

Ultimately, the choice of $\varepsilon$ 
determines an overall scale for the mapping between $t$ and $\mu$ and is thus a matter of convention.
For the sake of simplicity within Eq.~(\ref{mudefeps}) and our subsequent expressions, we shall henceforth 
choose {\mbox{$\varepsilon = \pi$}}, whereupon Eq.~(\ref{mudefeps}) reduces to
\beq
           \mu^2(t) ~=~ \frac{1}{\alpha' t}~.
\label{mut}
\eeq
   
With these adaptations, our result for the regulated Higgs mass $\widehat m_\phi^2 (t)$ in Eq.~(\ref{nonminresult}) 
can be rewritten as
\beqn
   \widehat m_\phi^2(\mu) \,&=&\,  \frac{\xi}{4\pi^2} \frac{\Lambda}{\calM^2} 
        +\half \calM^2 \, \biggl[
          - \frac{\pi}{3} \, \Str\, \mathbbX_1 \nonumber\\
    && ~~~~~~ +  (\,\newzStr \,\mathbbX_2) \log \left(\frac{\mu^2}{\calM_\ast^2}\right)  \biggr]~~~~~~~~~~
\label{nonminresultFT}
\eeqn
where we have defined 
\beq
            \calM_\ast^2 ~\equiv~  4\pi \,e^{-\gamma} \,M_s^2 ~=~ 16\pi^3\, e^{-\gamma}\, \calM^2~
\eeq
and where we have restored the additional universal $\Lambda$-term in Eq.~(\ref{nonminresultFT}).
We thus see that while the first two terms in Eq.~(\ref{nonminresultFT}) are independent of $\mu$ and 
together constitute what may be considered an overall threshold term, 
the logarithmic $\mu$-dependence within $\widehat m_\phi^2(\mu)$ for any $\mu$ arises from those
physical string states which are charged under $\mathbbX_2$ with masses {\mbox{$M\leq \mu$}}.

As we have seen, the enhanced non-minimal regulator we are using here 
operates by explicitly subtracting the contributions 
of the $\mathbbX_2$-charged massless states from all regions {\mbox{$\tau_2\geq  t$}}.
This is a sharp cutoff, and it is natural to wonder 
how such a cutoff actually maps back onto the strip under the unfolding process.
Indeed, answering this question will give us some idea about how this sort of cutoff might be interpreted 
in field-theoretic language.
As expected, imposing a sharp cutoff {\mbox{$\tau_2\leq t$}} within the $\calF$ representation
produces both an IR cutoff as well as a UV cutoff on the strip.
The IR cutoff is inherited directly from the string-theory cutoff and takes the same form {\mbox{$\tau_2\leq t$}}, thereby excising
all parts of the strip with $\tau_2$ exceeding $t$, independent of $\tau_1$.
However, the corresponding UV cutoff is highly non-trivial and is actually sensitive to $\tau_1$ as well ---
a degree of freedom that does not have a direct interpretation in the field theory.
Mathematically, this UV cutoff excises from the strip 
that portion of the region~\cite{zag}
\beq
       \bigcup_{(a,c)=1} \, S_{a/c}
\label{circles}
\eeq
which lies within the range {\mbox{$-1/2\leq \tau_1 \leq 1/2$}},
where $S_{a/c}$ denotes the disc
of radius $(2c^2 t)^{-1}$  
which is tangent to the {\mbox{$\tau_2=0$}} axis at
{\mbox{$\tau_1= a/c$}} and where the union in Eq.~(\ref{circles}) includes all such disks
for all relatively prime integers $(a,c)$.
Thus, as one approaches the {\mbox{$\tau_2=0$}} axis of the strip from above, the excised region 
consists of an infinite series of smaller and smaller discs which are all tangential to this axis
in an almost fractal-like pattern.
Clearly, 
all points which actually lie along the {\mbox{$\tau_2=0$}} axis with {\mbox{$\tau_1\in \mathbb{Q}$}} are excised
for any finite $t$  
(and strictly speaking the other points along the {\mbox{$\tau_2=0$}} axis with {\mbox{$\tau_1\not\in \mathbb{Q}$}} are 
not even part of the strip).
Thus, through this highly unusual UV regulator, all UV divergences on the strip are indeed eliminated 
for any finite $t$.
Of course, this excised UV region is nothing but the image of the IR-excised region {\mbox{$\tau_2\geq t$}} under
all of the modular transformations (namely those within the coset {\mbox{$\Gamma_\infty\backslash \Gamma$}}) 
that play a role in building the strip from $\calF$.
However, in field-theoretic language this amounts to a highly unusual UV regulator indeed!

\subsection{Results using the modular-invariant regulator}

Finally, we turn to the results for the Higgs mass 
that are obtained using the fully modular-invariant regulator 
$\widehat \calG_\rho(a,\tau_2)$ in Eq.~(\ref{hatGdef}).
As we have stressed, only such results can be viewed as faithful to the modular  symmetry
that underlies closed string theory, and therefore only such results can be viewed
as truly emerging from closed string theories.

We have seen in Eq.~(\ref{Higgsmass}) that the string-theoretic Higgs mass 
$m_\phi^2$ has two contributions:   one of these stems from the $\calX_i$
insertions and requires regularization, while the other --- namely the
cosmological-constant term --- is finite within any tachyon-free modular-invariant
theory and hence does not.
When discussing the possible regularizations of the Higgs mass using
the minimal and non-minimal regulators in Sects.~\ref{higgsmin} and \ref{higgsnonmin},
we simply carried the cosmological-constant  
term along within our calculations and focused
on applying our regulators to the contributions with $\calX_i$ insertions.
This was adequate for the minimal and non-minimal regulators because these regulators
involve the explicit subtraction of divergences 
and thus have no effect on quantities which are already finite and therefore lack
divergences to be subtracted.
Our modular-invariant regulator, by contrast, operates by deforming the theory.
Indeed, this deformation has the effect of
multiplying the partition function of the theory
with a new factor $\widehat \calG_\rho(a,\tau)$.
As such, this regularization procedure 
can be expected to have an effect even when acting on finite quantities such as $\Lambda$.
When regularizing the Higgs mass in this manner,
we must therefore consider how this regulator affects both classes of Higgs-mass contributions --- 
those involving non-trivial $\calX_i$ insertions, and those coming from the 
cosmological constant.
Indeed, with
    $\widehat m_\phi^2(\rho,a)\bigl|_{\calX,\Lambda}$
respectively denoting these two classes of contributions 
to the $\widehat \calG$-regulated version 
    $\widehat m_\phi^2(\rho,a)$
of the otherwise-divergent string-theoretic Higgs mass in Eq.~(\ref{Higgsmass}),
we can write 
\beqn
    \widehat m_\phi^2(\rho,a) 
    ~&\equiv&~ \widehat m_\phi^2(\rho,a)\Bigl|_{\cal X}  + ~ \widehat m_\phi^2(\rho,a)\Bigl|_\Lambda~ \nonumber\\ 
    ~&\equiv&~ \widehat m_\phi^2(\rho,a)\Bigl|_{\cal X}  + ~ \frac{\xi}{4\pi^2 \calM^2} \,\widehat \Lambda(\rho,a)~.~~~~~~  
\label{twocontributions}
\eeqn
We shall now consider each of these contributions in turn.

\subsubsection{Contributions from terms with charge insertions\label{chargeinsertions}}

Our first contribution in Eq.~(\ref{twocontributions})
is given by 
\beq
    \widehat m_\phi^2(\rho,a)\Bigl|_{\cal X}  ~\equiv~ -\frac{\calM^2}{2}  
            \Bigl\langle \calX_1+ \calX_2\Bigr\rangle_\calG  
           \label{Higgsmass1}
\eeq
 where
\beq
 \langle A \rangle_\calG ~\equiv~ \int_{\cal F} \dmu
      \left\lbrace 
       \left\lbrack 
        \tau_2^{-1} \sum_{m,n} (-1)^F  A~ \qbar^{m} q^{n} 
       \right\rbrack
     \widehat \calG_\rho(a,\tau)\right\rbrace ~
\label{AG}
\eeq
with {\mbox{$m\equiv  \alpha' M_R^2/4$}}, {\mbox{$n\equiv \alpha' M_L^2/4$}}. 
Indeed, the insertion of $\widehat \calG_\rho(a,\tau)$ into the integrand of 
Eq.~(\ref{AG}) is what tames the logarithmic divergence.
Following the result in Eq.~(\ref{Irhoa}) we then find that 
$\widehat m_\phi^2 (\rho,a)$ can be expressed as
\beq
      \widetilde m_\phi^2(\rho,a)\Bigl|_{\calX}  ~=~ 
          \frac{\pi}{3}\, \oneRes \, \int_0^\infty d\tau_2 \,\tau_2^{s-2} \,\widehat g_\rho(a,\tau_2) ~ 
\label{Irhoa2}
\eeq
where 
\beqn
           && \widehat g_\rho(a,\tau_2) \,\equiv\, 
         -\frac{\calM^2}{2}
  \int_{-1/2}^{1/2} d\tau_1 \nonumber\\   
       && ~~~~~~~~~ 
          \times\, \left\lbrace 
\left\lbrack 
   \sum_{m,n} (-1)^F  \,(\mathbbX_1 + \tau_2 \mathbbX_2)\,  \qbar^{m} q^{n} \right\rbrack 
       \,\widehat \calG_\rho(a,\tau)\right\rbrace ~~~\nonumber\\
&& ~~~~~ \approx ~
          -\frac{\calM^2}{2}
           \left\lbrack \Str \,(\mathbbX_1 + \tau_2 \mathbbX_2)\, e^{-\pi \alpha' M^2 \tau_2} \right\rbrack 
           \widehat \calG_\rho(a,\tau_2) ~.~~~ \nonumber\\
\label{gFGdef3}
\eeqn
Note that in passing to the approximate factorized form in the final expression of Eq.~(\ref{gFGdef3}), we 
have followed the result in Eq.~(\ref{gFGdef2}) 
and explicitly restricted our attention to those cases
with {\mbox{$a\ll 1$}}, as appropriate for the 
regulator function $\widehat \calG_\rho(a,\tau)$.
Indeed, the term within square brackets in the second line of 
Eq.~(\ref{gFGdef3}) is our desired supertrace over physical string states,
while the regulator function $\widehat\calG_\rho(a,\tau_2)$ --- an example of which 
is plotted in the right panel of Fig.~\ref{regulator_figure} --- generally eliminates the
divergence that would otherwise have arisen as {\mbox{$\tau_2\to \infty$}} for any {\mbox{$a>0$}}.
Moreover, we learn that
as a consequence of the
identity in Eq.~(\ref{StransG})
--- an identity which holds
for $\widehat \calG$ as well
as for $\calG$ itself ---
the behavior shown in
the right panel of Fig.~\ref{regulator_figure} 
can be symmetrically ``reflected'' through {\mbox{$\tau_2=1$}}, resulting in the same 
suppression behavior as {\mbox{$\tau_2\to 0$}}.

The next step is to substitute Eq.~(\ref{gFGdef3}) back into Eq.~(\ref{Irhoa2})
and evaluate the residue at {\mbox{$s=1$}}.   
In general, the presence of the regulator function $\widehat\calG_\rho(a,\tau_2)$ within 
Eq.~(\ref{gFGdef3}) renders this calculation somewhat intricate.    However, we 
know that {\mbox{$\widehat \calG_\rho(a,\tau_2)\to 1$}} as {\mbox{$a\to 0$}}.
Indeed, having already exploited our regulator in 
allowing us to pass from Eq.~(\ref{Higgsmass1}) to Eq.~(\ref{Irhoa2}),
we see that taking {\mbox{$a\to 0$}} corresponds to the limit in which we subsequently remove our regulator.
Let us first
focus on the contributions from massive states.
In the {\mbox{$a\to 0$}} limit, we then obtain
\beqn
&& \oneRes \int_0^\infty d\tau_2 \, \tau_2^{s-2}\, \widehat g_\rho(a,\tau_2) \nonumber\\
&&  ~~\,=\,  -\half \calM^2 \, \oneRes \, \bigl\lbrack
                          \Gamma(s-1) \, \pStr \mathbbX_1 \,(\pi \alpha' M^2)^{1-s} ~~~~\nonumber\\  
          && ~~~~~~~~~~~~~~~~~~~~~~+ \Gamma(s) \, \pStr \mathbbX_2 \,(\pi \alpha' M^2)^{-s}\bigr\rbrack \nonumber\\  
&& ~~\,=\,  -\half \calM^2  \, \pStr \mathbbX_1~,
\label{cheat}
\eeqn
whereupon we find that the contribution from massive states yields
\beq
  M>0:~~~\lim_{a\to 0} \widehat m_\phi^2 (\rho,a)\Bigl|_\calX \,=\, 
                      - \frac{\pi}{6} \calM^2 \, \pStr \mathbbX_1~.
\label{prelimitingcase}
\eeq
 This result is independent of $\rho$.
Moreover, as expected for massive states, this contribution is finite.
Of course, there will also be contributions from massless states.
In general, these contributions are more subtle to evaluate, and we know
that as {\mbox{$a\to 0$}} the effective removal of 
the regulator will lead to divergences coming from 
potentially non-zero values of $\zStr \mathbbX_2$ (since
it is the massless states which are charged under $\mathbbX_2$ which
cause the Higgs mass to diverge).
However, massless states charged under $\mathbbX_1$ --- like the massive
states --- do not lead to divergences.  We might therefore imagine restricting our
attention to cases with {\mbox{$\zStr \mathbbX_2=0$}}, and deforming
our theory slightly so that these massless $\mathbbX_1$-charged states accrue small non-zero masses.
In that case, the calculation in Eq.~(\ref{cheat}) continues to apply.
We can imagine removing this deformation without encountering any divergences.
This suggests that the full result for the regulated Higgs mass in the 
{\mbox{$a\to 0$}} limit should be the same as in Eq.~(\ref{prelimitingcase}), but
with massless $\mathbbX_1$-charged states also included.
We therefore expect
\beq
  \lim_{a\to 0} \,\widehat m_\phi^2 (\rho,a)\Bigl|_\calX  ~=~
            - \frac{\pi}{6} \calM^2 \, \Str\, \mathbbX_1~
\label{limitingcase}
\eeq
in cases for which {\mbox{$\zStr \mathbbX_2=0$}}.
We shall rigorously confirm this result below.

As discussed in Sect.~\ref{sec:modinvregs},
the two quantities $(\rho, a)$ that parametrize our modular-invariant regulator
are analogous to the quantity $t$ that parametrized our non-minimal regulator.
Indeed, these quantities effectively specify the value of the ``cutoff'' imposed by these regulators,
and as such we can view these quantities as corresponding to a floating physical mass scale $\mu$.
This scale $\mu$ is defined in terms of $t$ for the non-minimal regulator
in Eq.~(\ref{mut}), 
and we have already seen that
maintaining alignment between this regulator and our modular-invariant regulator requires
that we enforce the condition in Eq.~(\ref{alignment}).
We shall therefore identify a physical scale $\mu$ for our modular-invariant regulator as 
\beq
        \mu^2(\rho,a) ~\equiv ~ \frac{\rho a^2}{\alpha'}~.
\label{mudef}
\eeq
Since {\mbox{$\rho \sim {\cal O}(1)$}}, 
the {\mbox{$a\ll 1$}} region 
for our regulator corresponds to the restricted region {\mbox{$\mu \ll M_s$}}.

The identification in Eq.~(\ref{mudef})
enables us to rewrite
our result in Eq.~(\ref{limitingcase}) in the more suggestive form
\beq
 \lim_{\mu \to 0} \,\widehat m_\phi^2 (\mu) \Bigl|_\calX  ~=~
                      - \frac{\pi}{6} \calM^2 \,\Str\,\mathbbX_1~
\label{limitingcasemu}
\eeq
in cases for which {\mbox{$\zStr\mathbbX_2=0$}}.
In EFT language, we can therefore regard this result as holding in the deep infrared.

The natural question that arises, then, is to determine how our regulated Higgs 
mass $\widehat m_\phi^2(\mu)$
{\it runs}\/ as a function of the scale $\mu$.
In order to do this,
we need to evaluate 
$\widehat m_\phi^2 (\rho,a)$ 
as a function of $a$ for small {\mbox{$a\ll 1$}} {\it without}\/ taking the full {\mbox{$a\to 0$}} limit.

As indicated above, this calculation is somewhat intricate and is presented in Appendix~\ref{higgsappendix}.~
The end result, given in Eq.~(\ref{finalhiggsmassa}),
is an expression for $\widehat m_\phi^2(\rho,a)$ 
which is both {\it exact}\/ and valid for all $a$.
Using the identification in Eq.~(\ref{mudef}) and henceforth taking the benchmark value {\mbox{$\rho=2$}},
the result in Eq.~(\ref{finalhiggsmassa}) can then be expressed in terms of the scale $\mu$,
yielding
\beqn
 && \widehat m_\phi^2(\mu)\Bigl|_\calX  \,=\, \frac{\calM^2}{1+\mu^2/M_s^2} \Biggl\lbrace \nonumber\\ 
     && ~~\phantom{+} 
          \, \zStr \mathbbX_1 \left\lbrack - \frac{\pi}{6}\left(1+\mu^2/M_s^2\right)               \right\rbrack \nonumber\\
     && ~+ \, \zStr \mathbbX_2 \left\lbrack  
      \log\left( \frac{  \mu}{2\sqrt{2} e M_s}\right) 
               \right\rbrack \nonumber\\
     && ~+ \, \pStr \mathbbX_1 \, \Biggl\lbrace - \frac{\pi}{6}  
     - \frac{1}{2\pi} \left(\frac{M}{\calM}\right)^2  \times \nonumber\\
     && ~~~~~~~~~~~~  
      \times \left\lbrack    
        \calK_0^{(0,1)}\!\left( \frac{2\sqrt{2}\pi M}{\mu} \right) + 
        \calK_2^{(0,1)}\!\left( \frac{2\sqrt{2}\pi M}{\mu} \right)  
      \right\rbrack \Biggr\rbrace \nonumber\\
     && ~+ \, \pStr \mathbbX_2 \, 
     \Biggl \lbrack 
        2\calK_0^{(0,1)}\!\left( \frac{2\sqrt{2}\pi M}{\mu} \right)  
             -                   \calK_1^{(1,2)}\!\left( \frac{2\sqrt{2}\pi M}{\mu}\right) \Biggr\rbrack  
              \Biggr\rbrace
\nonumber\\
                            \label{finalhiggsmassmu}
\eeqn
where we have defined the Bessel-function combinations
\beq
     \calK_\nu^{(n,p)} (z) ~\equiv~ \sum_{r=1}^\infty ~ (rz)^{n} \Bigl\lbrack 
       K_\nu(rz/\rho) - \rho^p K_\nu(rz)  \Bigr\rbrack~, 
\label{Besselcombos}
\eeq
with $K_\nu(z)$ denoting
the modified Bessel function of the second kind.
We see, then, that 
the contributions to 
the running of 
$\widehat m_\phi^2(\mu)\bigl|_\calX$ 
from the different
states in our theory 
depend rather non-trivially on their masses and on their various $\mathbbX_1$ 
and $\mathbbX_2$ charges, with
the contributions
from each string state with non-zero mass $M$
governed by various combinations of Bessel functions $K_\nu(z)$ with
arguments {\mbox{$z\sim M/\mu$}}.

There is a plethora of physics wrapped within Eq.~(\ref{finalhiggsmassmu}), and we shall
unpack this result in several stages.
First, it is straightforward to take the {\mbox{$\mu\to 0$}} limit of Eq.~(\ref{finalhiggsmassmu}) 
in order to verify our expectation in Eq.~(\ref{limitingcase}).
Indeed, in the {\mbox{$\mu\to 0$}} limit, we have {\mbox{$z\to \infty$}} for all {\mbox{$M>0$}}.
Since
\beq
          \calK_\nu^{(n,p)}(z) ~\sim~ \sqrt{\frac{\pi \rho}{2}} \,z^{n-1/2} \,e^{-z/\rho} ~~~~{\rm as}~ z\to \infty~,
\label{asymptoticform}
\eeq
it then follows that
all of the terms 
involving Bessel functions in Eq.~(\ref{finalhiggsmassmu})
vanish exponentially in the {\mbox{$\mu\to 0$}} limit.
For cases in which {\mbox{$\zStr \mathbbX_2 =0$}} [\ie, cases in which the original Higgs mass $m_\phi^2$
is finite, with no massless states charged under $\mathbbX_2$],  
we thus reproduce the result in Eq.~(\ref{limitingcase}).

Using the result in Eq.~(\ref{finalhiggsmassmu}),
we can also study the running of $\widehat m_\phi^2(\mu)$ as a function of {\mbox{$\mu>0$}}.
Of course, given that our $\widehat \calG$-function acts as a regulator only for {\mbox{$a\ll 1$}},
our analysis is restricted to the {\mbox{$\mu\ll M_s$}} region.
Let us first concentrate on the contributions from the terms within 
Eq.~(\ref{finalhiggsmassmu}) that do not involve Bessel functions.
These contributions are given by
               \beq
  {\calM^2} \,\biggl\lbrace 
      - \frac{\pi}{6}\, \Str\, \mathbbX_1 
      + \, \zStr \mathbbX_2 \log\left( \frac{  \mu}{2\sqrt{2} e M_s}\right)\biggr\rbrace~.
\label{finalhiggsmasslimit}
\eeq
From this we see 
that our deep-infrared 
contribution to $\widehat m_\phi^2$ in Eq.~(\ref{limitingcase}) 
actually persists as an essentially constant contribution for all scales {\mbox{$\mu\ll M_s$}}. 
We also see from Eq.~(\ref{finalhiggsmasslimit}) that each massless string state  
also contributes an additional logarithmic running 
which is proportional to its $\mathbbX_2$ charge and which
persists all the way into the deep infrared.
Given that massless $\mathbbX_2$-charged states are
precisely the states that led to the original logarithmic 
divergence in the {\it unregulated}\/ Higgs mass $m_\phi^2$,
this logarithmic running is completely expected.
Indeed, it formally leads to a divergence in our regulated
Higgs mass $\widehat m_\phi^2(\mu)$ in the full {\mbox{$\mu\to 0$}} limit
(at which our regulator is effectively removed),
but otherwise produces a finite contribution for all other {\mbox{$\mu>0$}}.
The issues connected with this logarithm are actually no different from those
that arise in an ordinary field-theoretic calculation. 
We shall discuss these issues in more detail in Sect.~\ref{sec:Conclusions}
but in the meantime this term will not concern us further.

The remaining contributions are those arising from the terms
within Eq.~(\ref{finalhiggsmassmu}) involving supertraces over Bessel functions.
Although our analysis is restricted to the {\mbox{$\mu\ll M_s$}} region,
our supertraces receive contributions from the entire string spectrum.
This necessarily includes states with masses {\mbox{$M\gsim M_s$}}, but may also include
potentially light states with non-zero masses far below $M_s$.
The existence of such light states depends on our string construction and
on the specific string model in question.  
Indeed, such states are particularly relevant for the kinds of string models that motivate
our analysis, namely (non-supersymmetric) string models 
in which the Standard Model is realized directly within the low-energy spectrum.

The Bessel functions 
corresponding to states with masses {\mbox{$M\gsim M_s$}}
have arguments {\mbox{$z\sim M/\mu \gg 1$}} when {\mbox{$\mu\ll M_s$}}.
As a result,
in accordance with Eqs.~(\ref{finalhiggsmassmu}) and (\ref{asymptoticform}), 
the contributions from these states to the running of $\widehat m_\phi^2(\mu)\bigl|_\calX$ 
are exponentially suppressed.
It then follows that the dominant contributions to
the Bessel-function running of $\widehat m_\phi(\mu)$ 
within the {\mbox{$\mu\ll M_s$}} region 
come from the correspondingly light states, \ie, states with masses {\mbox{$M\ll M_s$}}.
However, for states with masses {\mbox{$M\ll M_s$}},
we see from Eq.~(\ref{finalhiggsmassmu}) 
that the corresponding Bessel-function contributions which are proportional to their $\mathbbX_1$ charges
are all suppressed by a factor $(M/\calM)^2$.
We thus conclude that
the contribution from a state of non-zero mass {\mbox{$M\ll M_s$}} within the string spectrum 
is sizable only when this state carries a non-zero $\mathbbX_2$ charge.
Indeed, we see from Eq.~(\ref{finalhiggsmassmu}) that this contribution
for each bosonic degree of freedom of mass $M$ is given by
\beq
        2\calK_0^{(0,1)}\!\left(z\right)  -  \calK_1^{(1,2)}\!\left(z\right) 
\label{lightcontribution}
\eeq
per unit of $\mathbbX_2$ charge, 
where {\mbox{$z\equiv 2\sqrt{2}\pi M/\mu$}}.

\begin{figure}[t!]
\centering
\includegraphics[keepaspectratio, width=0.48\textwidth]{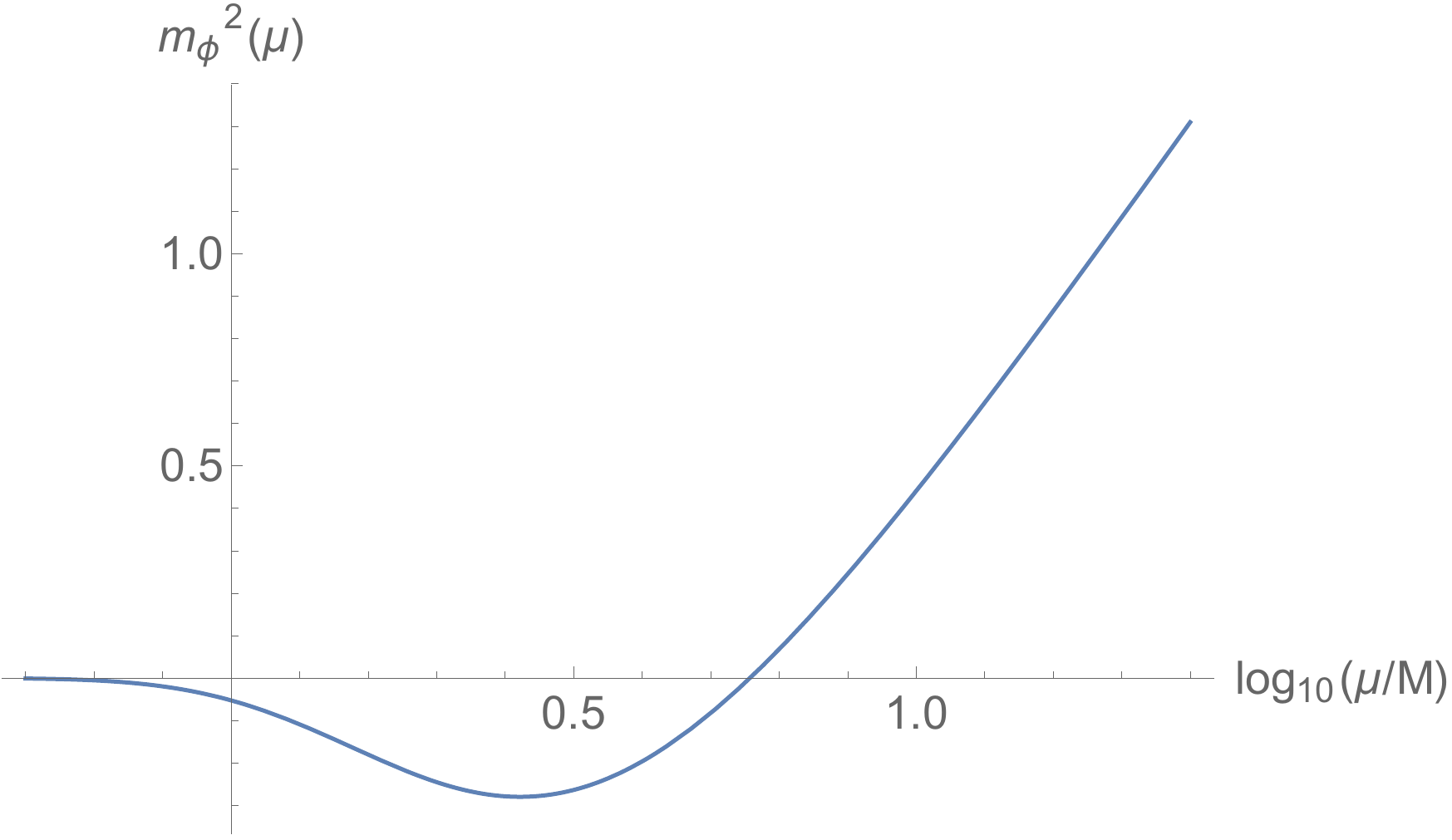}
\caption{
The expression in Eq.~(\ref{lightcontribution}), plotted as a function of $\mu/M$.
This quantity is the Bessel-function contribution per unit $\mathbbX_2$ charge 
to the running of the regulated Higgs mass $\widehat m_\phi^2(\mu)\bigl|_{\calX}/\calM^2$
from a bosonic state of non-zero mass $M$.
This contribution is universal for all $\mu/M$ and assumes only that {\mbox{$\mu \ll M_s$}}.
When {\mbox{$\mu\gg M$}}, the state is fully dynamical and produces a running which is effectively logarithmic.
By contrast, when {\mbox{$\mu \ll M$}}, the state is heavier than the scale $\mu$ and is effectively
integrated out, thereby suppressing any contributions to the running.  Finally, within the
intermediate {\mbox{$\mu\approx M$}} region, 
the Bessel-function expression in Eq.~(\ref{lightcontribution}) 
provides a smooth connection between 
these two asymptotic behaviors and even gives rise to a transient ``dip''
in the overall running.
Note that for a fixed scale $\mu$, adjusting the mass $M$ of the relevant state
upwards or downwards simply corresponds to shifting this curve
rigidly to the right or left, respectively.
In this way one can imagine summing over all such contributions to the running
as one takes the supertrace over the entire $\mathbbX_2$-charged string spectrum.}   
\label{transientfigure}
\end{figure}

In Fig.~\ref{transientfigure},  we plot this contribution
as a function of $\mu/M$.
As expected, we see that states with {\mbox{$M\gg \mu$}} produce no running and can be ignored --- essentially
they have been ``integrated out'' of our theory at the scale $\mu$ and leave behind only an exponential tail.
By contrast, states with {\mbox{$M\lsim \mu$}} are still dynamical at the scale $\mu$. 
We see from Fig.~\ref{transientfigure} that their effective contributions are then effectively {\it logarithmic}\/.
Indeed, as {\mbox{$z\to 0$}}, one can show that~\cite{Paris}
\beqn
          \calK_0^{(0,1)}(z)~&\sim &~ - \half \log\,z  + \half\left[ \log\,(2\pi) - \gamma\right]  \nonumber\\           \calK_1^{(1,2)}(z)~&\sim &~ 1~
\label{Kasymp}
\eeqn
where $\gamma$ is the Euler-Mascheroni constant.
This leads to an 
asymptotic logarithmic running of the form 
\beq
            \log\left[ \frac{1}{\sqrt{2}}\,e^{-(\gamma+1)} \frac{\mu}{M}\right]
\label{loglimit}
\eeq
for {\mbox{$\mu\gg M$}} in Fig.~\ref{transientfigure}.
Finally, between these two behaviors, we see that the expression in Eq.~(\ref{lightcontribution})
interpolates smoothly and even gives rise to a transient ``dip''.
This is a uniquely string-theoretic behavior resulting from the 
specific combination of Bessel functions in Eq.~(\ref{lightcontribution}).
Of course, the statistics factor $(-1)^F$ within the supertrace
flips the sign of this contribution for degrees of freedom which are fermionic.

Thus far we have focused on the 
Higgs-mass running, as shown in Fig.~\ref{transientfigure},
 from a single massive string degree of freedom of mass $M$.
However, the contribution from another string state with a different mass $M'$ can be
simply obtained by rigidly sliding this curve towards the left or right (respectively corresponding to 
cases with {\mbox{$M'<M$}} and {\mbox{$M'>M$}}, respectively).
The complete supertrace contribution in Eq.~(\ref{finalhiggsmassmu})
is then obtained by summing over all of these curves, each with its appropriate horizontal displacement and each weighted by 
the corresponding net (bosonic minus fermionic) number of degrees of freedom.
The resulting net running from the final term within 
Eq.~(\ref{finalhiggsmassmu}) is therefore highly sensitive to the properties of the 
massive $\mathbbX_2$-charged part of the string spectrum.  
This will be discussed further in Sect.~\ref{seehow}.~
Of course, as discussed above, the contributions from states with {\mbox{$M'\gg \mu$}} are
exponentially suppressed.   Thus, for any $\mu$, the only states which contribute meaningfully to
this Bessel-function running of the Higgs mass are those with {\mbox{$M\lsim \mu$}}.

Thus, combining these Bessel-function contributions with those from Eq.~(\ref{finalhiggsmasslimit})
and keeping only those (leading) terms which dominate when {\mbox{$M\ll \mu\ll M_s$}},
we see that we can approximate the exact result in Eq.~(\ref{finalhiggsmassmu}) as
\beqn
&& \widehat m_\phi^2(\mu)\Bigl|_\calX  \,\approx\,
      - \frac{\pi}{6}\, \calM^2\, \Str\, \mathbbX_1 
      + \calM^2 \, \zStr \mathbbX_2 \,\log\left( \frac{  \mu}{2\sqrt{2} e M_s}\right)\nonumber\\
  && ~~~~~~~ + \calM^2 \,\effStr 
  \mathbbX_2 \,\log\left[ \frac{1}{\sqrt{2}}\,e^{-(\gamma+1)} \frac{\mu}{M}\right]~.~~~~~~~~
\label{approxhiggsmassmu}
\eeqn
Interestingly, we see that to leading order,
the $\mathbbX_1$ charges of the string states
only contribute to an overall constant term in Eq.~(\ref{approxhiggsmassmu}),
and they do this for all states regardless of their masses.
By contrast, it is the $\mathbbX_2$ charges 
of the states which induce a corresponding running,
and this only occurs for those
states within the EFT at the scale $\mu$ --- \ie, those light 
states with masses {\mbox{$M\lsim \mu$}}.

The net running produced by the final term in Eq.~(\ref{approxhiggsmassmu})
can exhibit a variety of behaviors.
To understand this, let us consider the behavior of this term
as we increase $\mu$ from the deep infrared.
Of course, this term does not produce any running at all 
until we reach {\mbox{$\mu \sim M_{\rm lightest}$}},
where $M_{\rm lightest}$ is the mass of the lightest massive string state
carrying a non-zero $\mathbbX_2$ charge.
This state then contributes a logarithmic
running which persists for all higher $\mu$.
However, as $\mu$ increases still further,
additional $\mathbbX_2$-charged string states 
enter the EFT and contribute their own individual logarithmic contributions. 
Of course, if these additional states 
have masses {\mbox{$M\gg M_{\rm lightest}$}},
the logarithmic nature of the running shown in Fig.~\ref{transientfigure}
from the state with mass $M_{\rm lightest}$
will survive intact until {\mbox{$\mu \sim M$}}.
However, if the spectrum of states is relatively dense 
beyond $M_{\rm lightest}$, the logarithmic contributions from each of these states
must be added together, leading to a far richer behavior.
 
One important set of string models exhibiting the latter property
are those involving a relatively large compactification radius $R$.
In such cases, we can identify {\mbox{$M_{\rm lightest}\sim 1/R$}},
whereupon we expect an entire  tower of corresponding Kaluza-Klein (KK) states
of masses {\mbox{$M_k\sim k/R$}}, {\mbox{$k\in\mathbb{Z}^+$}}, each sharing a common charge $\mathbbX_2$ and
a common degeneracy of states $g$.
For any scale $\mu$, the final term in Eq.~(\ref{approxhiggsmassmu}) 
then takes the form
\beqn
         && \calM^2 \,g \,\mathbbX_2 \,\sum_{k=1}^{\mu R}
  \,\log\left[ \frac{1}{\sqrt{2}}\,e^{-(\gamma+1)} \, \frac{\mu R}{k}\right]~\nonumber\\
         && ~~=~ \calM^2 \,g \,\mathbbX_2 \,\left\lbrace 
       \mu R \,\log\left[  \frac{1}{\sqrt{2}}\,e^{-(\gamma+1)} \, \mu R\right] - \log \,(\mu R)!\right\rbrace\nonumber\\ 
         && ~~=~ \calM^2 \,g \,\mathbbX_2 \,
       \,  \left\lbrace 
  \log\left[  \frac{1}{\sqrt{2}}\,e^{-(\gamma+1)} \right] +1 \right\rbrace \, \mu R
\label{newpower}
\eeqn
where in passing to the third line we have used Stirling's approximation
{\mbox{$\log N! \approx N \log N - N$}}.
We thus see that in such cases our sum over logarithms actually produces a {\it power-law}\/ 
running!  In this case the running is linear, 
but in general the KK states associated with $d$ large toroidally-compactified dimensions
collectively yield a regulated Higgs mass whose running scales as $\mu^d$.

This phenomenon whereby a sum over KK states deforms a running from logarithmic to power-law
is well known from phenomenological studies of theories with large extra dimensions,
where it often plays a crucial role 
(see, \eg, Refs.~\cite{Dienes:1998vh, Dienes:1998vg, Dienes:1998qh}).
This phenomenon can ultimately be understood from 
the observation that a large compactification radius
effectively increases the overall spacetime dimensionality of the theory,
thereby shifting the mass dimensions of quantities such 
as gauge couplings and Higgs masses and simultaneously shifting their corresponding runnings.
Indeed, as discussed in detail in Appendices~A and B 
of Ref.~\cite{Dienes:1998vg}
(and as illustrated in Fig.~11 therein),
the emergence of power-law running from logarithmic running is surprisingly robust. 

Of course, it may happen that 
the spectrum of light states not only has a lightest mass $M_{\rm lightest}$
but also a heaviest mass $M_{\rm heaviest}$, with a significant mass gap beyond this
before reaching even heavier scales. 
If such a situation were to arise (but clearly does not within the large extra-dimension
scenario described above),
then the corresponding running of $\widehat m_\phi^2(\mu)\bigl|_\calX$
would only be power-law within the range {\mbox{$M_{\rm lightest}\lsim \mu\lsim M_{\rm heaviest}$}}.
For {\mbox{$\mu > M_{\rm heaviest}$}}, by contrast, the running would then revert back to logarithmic. 

In summary, we see that while the first term within
Eq.~(\ref{approxhiggsmassmu}) 
represents an overall constant contribution arising from the entire spectrum of $\mathbbX_1$-charged states,
the second term represents an overall logarithmic contribution from precisely the massless $\mathbbX_2$-charged states
which were the source of the original divergence of the unregulated Higgs mass $m_\phi^2$.
By contrast, the final term
in Eq.~(\ref{approxhiggsmassmu}) 
represents the non-trivial contribution to the running from
the massive $\mathbbX_2$-charged states.
As we have seen,
this latter contribution can exhibit a variety of behaviors, ranging from logarithmic (in cases with
relatively large mass splittings between the lightest massive $\mathbbX_2$-charged
states) to power-law (in cases with relatively small uniform mass splittings between
such states).
Of course, 
depending on the details of the underlying string spectrum,
mixtures between these different behaviors are also possible.

\subsubsection{Contribution from the cosmological constant \label{Lambdasect}} 

Let us now turn to the 
second term in Eq.~(\ref{twocontributions}).
This contribution lacks $\calX_i$ insertions
and arises from the cosmological-constant term in Eq.~(\ref{Higgsmass}).
Although this contribution is the result of a universal shift in the background moduli
and is thus independent of the specific $\calT$-matrices,
we shall now demonstrate that it too can be expressed as a supertrace over the physical string spectrum.
It also develops a scale dependence when subjected to our modular-invariant regulator.

Within the definition of $\Lambda$ in Eq.~(\ref{lambdadeff}),
the integrand function $\calF(\tau,\taubar)$ is simply $(-\calM^4/2) \calZ(\tau,\taubar)$
where $\calZ(\tau,\taubar)$ is the partition function of the string in the Higgsed phase.
Of course, if this theory exhibits unbroken spacetime supersymmetry, 
the contributions from the bosonic states in the spectrum cancel level-by-level against
those from their fermionic superpartners.  In such cases we then have {\mbox{$\calZ=0$}}, implying {\mbox{$\Lambda=0$}}.
Otherwise, for heterotic strings, we necessarily have {\mbox{$\calZ\not =0$}}.
Indeed, it is a theorem (first introduced in Ref.~\cite{Dienes:1990ij} and discussed more recently, {\it e.g.}\/, in 
Ref.~\cite{Abel:2015oxa})
that any non-supersymmetric heterotic string model in $D$ spacetime dimensions must contain an off-shell
tachyonic {\it proto-graviton}\/ state
whose contribution to the partition function remains uncancelled.  This then results in a string partition function
whose power-series expansion has the leading behavior {\mbox{$Z=(D-2)/q + ...$}} 

In principle this proto-graviton contribution would appear to introduce 
an exponential divergence as {\mbox{$\tau_2\to\infty$}}, thereby taking us beyond
the realm of validity for the mathematical techniques presented in Sect.~\ref{sec:RStechnique}.~
However, this tachyonic state is off-shell and thus does not appear in the actual physical string spectrum.
Indeed, as long as there are no additional {\it on-shell}\/ tachyons present in the theory,
the corresponding integral $\Lambda$ is fully convergent
because the integral over the fundamental domain $\calF$ comes
with an explicit instruction that we are  to integrate across $\tau_1$  in the {\mbox{$\tau_2>1$}} region of $\calF$
{\it before}\/ integrating over $\tau_2$.
This integration therefore prevents the proto-graviton state from contributing to $\Lambda$ 
within the {\mbox{$\tau_2>1$}} region of integration, and likewise prevents this state from contributing to $g(\tau_2)$.

Assuming, therefore, that we can disregard the proto-graviton contribution to $\calZ$ as {\mbox{$\tau_2\to\infty$}},
we find that {\mbox{$\calZ\sim \tau_2^{-1}$}} as {\mbox{$\tau_2\to\infty$}}.
Thus $\calZ$ is effectively of rapid decay and we can use the 
original Rankin-Selberg results in Eq.~(\ref{RSresult}).
In this connection, we note that this assumption regarding the proto-graviton contribution
finds additional independent  
support through the arguments presented in Ref.~\cite{Kutasov:1990sv} which 
demonstrate that any contributions from the proto-graviton beyond those in Eq.~(\ref{RSresult}) 
are suppressed by an infinite volume factor in all spacetime dimensions {\mbox{$D>2$}}.
A similar result is also true in string models
with exponentially suppressed cosmological constants~\cite{Abel:2015oxa}.
 
With $\calZ$ taking the form in Eq.~(\ref{integrand})
and with the mass $M$ of each physical string state 
identified via {\mbox{$\alpha' M^2=2(m+n)=4m$}},
we then have
\beq
         g(\tau_2) ~=~   -\frac{\calM^4}{2} \, \tau_2^{-1}\, \Str\, e^{-\pi \alpha' M^2 \tau_2}~.
\label{glambda}
\eeq
Inserting this result into Eq.~(\ref{RSresult}) and performing the $\tau_2$ integral
then yields~\cite{Dienes:1995pm}
\beqn
     \Lambda ~&=&~ - \frac{\calM^4}{2}\, \frac{\pi}{3} \,\oneRes
                \biggl[\pi^{2-s} \,\Gamma(s-2) \,\Str\, (\alpha' M^2)^{2-s}\biggr\rbrack\nonumber\\
             ~&=&~  \frac{\calM^4}{2} \frac{\pi^2}{3} \Str\, (\alpha' M^2) \nonumber\\
             ~&=&~ \frac{1}{24} \calM^2 \, \Str\, M^2.
\label{eq:lamlam}
\eeqn 
We thus see that $\Lambda$ is given as a universal supertrace 
over {\it all}\/ physical string states,
and not only those with specific charges relative to the Higgs field.

As evident from the form of the final supertrace in Eq.~(\ref{eq:lamlam}), 
massless states do not ultimately contribute within this expression for $\Lambda$.
Strictly speaking, our derivation in Eq.~(\ref{eq:lamlam}) already implicitly assumed
this, given that the intermediate steps in Eq.~(\ref{eq:lamlam}) are valid only for {\mbox{$M>0$}}.
However, it is easy to see that the contributions
from massless states lead to a $\tau_2$-integral whose divergence has no residue at {\mbox{$s=1$}}.
Thus, massless states make no contribution to this expression, 
and the result in Eq.~(\ref{eq:lamlam}) stands.

This does {\it not}\/ mean that massless states do not contribute to $\Lambda$, however.
Rather, this just means that the constraints from modular invariance 
so tightly connect 
the contributions to $\Lambda$ from the massless states to those from the 
massive states  
(and also those from the unphysical string states of any mass)
that an expression for $\Lambda$ as in Eq.~(\ref{eq:lamlam}) becomes possible.

For further insight into this issue,
it is instructive to obtain this same result through Eq.~(\ref{reformulation}). 
We then have
\beq
     \Lambda~=~ -\frac{\pi}{3} \frac{\calM^4}{2} \lim_{\tau_2\to 0} 
      \biggl\lbrack \tau_2^{-1} \,\Str\, \exp\left( -\pi \alpha' M^2 \tau_2\right)\biggr\rbrack~.
\eeq
Expanding the exponential {\mbox{$e^{-x}\approx 1 -x + ...$}} and taking the {\mbox{$\tau_2\to 0$}} limit of each term separately, 
we find that the linear term leads directly to the result in Eq.~(\ref{eq:lamlam}) while the contributions from all
of the higher terms vanish.
Interestingly, the constant term would {\it a priori}\/ appear to lead to a divergence for $\Lambda$.
The fact that $\Lambda$ is finite in such theories then additionally tells us that~\cite{Dienes:1995pm}
\beq 
      {\rm Str}\, {\bf 1} ~=~0~.
\label{eq:lamlam0}
\eeq
As apparent from our derivation, this constraint must hold for any 
tachyon-free modular-invariant theory 
(\ie, any modular-invariant theory in which $\Lambda$ is finite).  
Indeed, this is one of the additional constraints from modular invariance which
relates the contributions of the physical string states
which are massless to the contributions of those which are massive.
Thus, we may regard the result in Eq.~(\ref{eq:lamlam}) --- like all of the results of this paper --- 
as holding within a modular-invariant context 
in which other constraints such as that in Eq.~(\ref{eq:lamlam0}) are also simultaneously 
satisfied.
We also see from this analysis that 
our supertrace definition in Eq.~(\ref{supertracedef})
may be more formally defined as~\cite{Dienes:1995pm}
\beq
              \Str \, A ~\equiv~   
       \lim_{y\to 0} \, \sum_{{\rm physical}~ i}  (-1)^{F_i} \, A_i \,  e^{- y \alpha' M_i^2}~.
\label{supertracedef2}
\eeq

The supertrace results in Eqs.~(\ref{eq:lamlam}) and (\ref{eq:lamlam0}) were first derived in 
Ref.~\cite{Dienes:1995pm}.
As discussed in Refs.~{\mbox{\cite{Dienes:1995pm,Dienes:2001se}}}, 
these results hold for all tachyon-free heterotic strings in four dimensions,
and in fact similar results hold in all spacetime dimensions {\mbox{$D>2$}}.
For theories exhibiting spacetime supersymmetry,
these relations are satisfied rather trivially.
However, even if the spacetime supersymmetry is broken
--- and even if the scale of supersymmetry breaking is relatively large or at the Planck scale ---
these results nevertheless continue to hold.
In such cases, these supertrace relations do not arise as the results of pairwise cancellations
between the contributions of bosonic and fermionic string states.
Rather, these relations emerge as the results of conspiracies that occur across the {\it entire}\/
string spectrum,  with the bosonic and fermionic string states  
always carefully arranging themselves 
at all string mass levels 
so as to exhibit a so-called ``misaligned supersymmetry''~{\mbox{\cite{Dienes:1994np,Dienes:2001se}}}.
No pairing of bosonic and fermionic states 
occurs within misaligned supersymmetry,
yet misaligned supersymmetry ensures that these supertrace relations are always satisfied.
These results therefore constrain the extent to which supersymmetry can be broken in tachyon-free string theories
while remaining consistent with modular invariance.

The results that we have obtained thus far pertain to the cosmological constant $\Lambda$.
As such, they would be sufficient if we were aiming to understand this quantity unto itself,
since $\Lambda$ is finite in any tachyon-free modular-invariant
theory and hence requires no regulator.
However, in this paper our interest in this quantity stems from the fact that $\Lambda$ is
an intrinsic contributor to the total Higgs mass in Eq.~(\ref{relation1}),
and we already have seen that the Higgs mass requires regularization.
At first glance, one might imagine regulating the terms with non-zero $\calX_i$ insertions
while leaving the $\Lambda$-term alone.
However, it is ultimately inappropriate to regularize only a subset of terms that contribute
to the Higgs mass --- for consistency we must apply the same regulator to the entire expression
at once.
Indeed, we recall from Sect.~\ref{sec2} that the entire Higgs-mass expression including $\Lambda$ forms
a modular-invariant unit, with $\Lambda$ emerging from the modular completion
of some of the terms with non-trivial $\calX_i$ insertions.
For this reason,
we shall now study the analogously regulated 
cosmological constant
\beq
             \widehat \Lambda (\rho,a) ~\equiv~ \int_\calF \dmu~ {\cal Z}(\tau) \, \widehat\calG_\rho(a,\tau)~
\label{Lambdahatdef}
\eeq
and determine the extent to which this regularized cosmological constant
can also be expressed in terms of supertraces over the physical string states.

Our discussion proceeds precisely as for the terms involving the $\calX_i$ insertions.
Following the result in Eq.~(\ref{Irhoa}) we find that 
$\widehat \Lambda (\rho,a)$ can be expressed as
\beq
      \widehat \Lambda(\rho,a)  ~=~ 
\frac{\pi}{3}\, \oneRes \, \int_0^\infty d\tau_2 \,\tau_2^{s-3} \,\widehat g_\rho(a,\tau_2) ~ 
\label{Irhoa3}
\eeq
where 
\beqn
           && \widehat g_\rho(a,\tau_2) \,\equiv\, 
         -\frac{\calM^4}{2}
  \int_{-1/2}^{1/2} d\tau_1 \nonumber\\   
       && ~~~~~~~~~ 
          \times\, \left\lbrace 
\left\lbrack 
   \sum_{m,n} (-1)^F  \,\qbar^{m} q^{n} \right\rbrack 
       \,\widehat \calG_\rho(a,\tau)\right\rbrace ~~~\nonumber\\
&& ~~~~~ = ~
          -\frac{\calM^4}{2}
           \left\lbrack \Str \, e^{-\pi \alpha' M^2 \tau_2} \right\rbrack 
           \widehat \calG_\rho(a,\tau_2) ~.~~~~~ 
\label{gFGdef4}
\eeqn
In the second line the sum over $(m,n)$ indicates a sum over the entire spectrum 
of the theory, while
in passing to the factorized form in the third line of Eq.~(\ref{gFGdef4}) we 
have again followed the result in Eq.~(\ref{gFGdef2}) 
and explicitly restricted our attention to those cases
with {\mbox{$a\ll 1$}}, as appropriate for the 
regulator function $\widehat \calG_\rho(a,\tau)$.
Indeed, the term within square brackets in the second line of 
Eq.~(\ref{gFGdef3}) is our desired supertrace over physical string states,
while the regulator function $\widehat\calG_\rho(a,\tau_2)$ 
provides a non-trivial $\tau_2$-dependent weighting to the different
terms within $g_\rho(a,\tau_2)$.

Once again, 
the next step is to substitute Eq.~(\ref{gFGdef4}) back into Eq.~(\ref{Irhoa3})
and evaluate the residue at {\mbox{$s=1$}}.   
In general, the presence of the regulator function $\widehat\calG_\rho(a,\tau_2)$ within 
Eq.~(\ref{gFGdef3}) renders this calculation somewhat intricate.    However, 
just as for the terms with non-trivial $\calX_i$ insertions,
we know that {\mbox{$\widehat \calG_\rho(a,\tau_2)\to 1$}} as {\mbox{$a\to 0$}}.
In this limit, we therefore expect to obtain our original (finite) unregulated $\Lambda$:
\beq
  \lim_{a\to 0} \,\widehat \Lambda(\rho,a) ~=~ \Lambda ~=~ 
             \frac{1}{24} \calM^2 \, \Str\, M^2~
\label{limitingcase2}
\eeq
where in the final equality we have utilized the result in Eq.~(\ref{eq:lamlam}).
Equivalently, upon identifying the physical scale $\mu$ as in Eq.~(\ref{mut}),
we thus expect
\beq
    \lim_{\mu\to 0} \,\widehat \Lambda (\mu)   ~=~ \Lambda~.
\label{lambdamulimit}
\eeq

Let us now determine how $\widehat\Lambda(\mu)$ runs
as a function of the scale $\mu$.
To do this, we need to evaluate $\widehat \Lambda(\rho,a)$ explicitly as a function of $\rho$ and $a$.
This question is tackled in Appendix~\ref{lambdaappendix}, yielding the exact result
in Eq.~(\ref{lambdaresult}).   Written in terms of the physical scale $\mu$ in Eq.~(\ref{mudef}) 
this result then takes the form
\beqn
   \widehat\Lambda (\mu ) ~&=&~ \frac{1}{1+\mu^2/M_s^2} \, \Biggl\lbrace
    \frac{\calM^2}{24} \, \Str\,M^2 \nonumber\\
  && ~~   - \frac{7}{960 \pi^2} \,(n_B-n_F)\,  \mu^4  
~\nonumber\\
  && ~~ -  \frac{1}{2\pi^2}\,
    \pStr  M^4 \,\Biggl[ \calK_1^{(-1,0)}\!\left( \frac{2\sqrt{2} \pi M}{\mu} \right) ~~~~~~\nonumber\\
  && ~~~~~~~~~ +      4\,  \calK_2^{(-2,-1)}\!\left( \frac{2\sqrt{2} \pi M}{\mu}\right)\nonumber\\
  && ~~~~~~~~~ +       \calK_3^{(-1,0)}\!\left( \frac{2\sqrt{2} \pi M}{\mu}   \right)
     \Biggr] ~  \Biggr\rbrace~
\label{lambdamuresult}
\eeqn
where we have again taken {\mbox{$\rho=2$}} as our benchmark value, 
where $\calK_\nu^{(n,p)}(z)$ are the Bessel-function combinations defined in Eq.~(\ref{Besselcombos}),
and where $n_{B}$ and $n_F$ are the numbers of massless bosonic and fermionic degrees of freedom in the theory
respectively (so that {\mbox{$\zStr {\bf 1} = n_B-n_F$}}).

It is straightforward to verify that 
this result is consistent 
with the result in Eq.~(\ref{lambdamulimit}) in the {\mbox{$\mu\to 0$}} limit.
Because all of the Bessel-function combinations within Eq.~(\ref{lambdamuresult}) vanish
exponentially rapidly as their arguments grow to infinity, 
only the first term in Eq.~(\ref{lambdamuresult}) survives in this limit.
We therefore find that the {\mbox{$\mu\to 0$}} limit of Eq.~(\ref{lambdamuresult}) 
yields the anticipated result in Eq.~(\ref{lambdamulimit}). 

From Eq.~(\ref{lambdamuresult}) 
we can also understand the manner in which $\widehat \Lambda(\mu)$ runs as a function of $\mu$ for all {\mbox{$0<\mu\ll M_s$}}.
Let us first focus on the Bessel-function terms within the square brackets in Eq.~(\ref{lambdamuresult}).
By themselves, these terms behave in much the same way as shown in Fig.~\ref{transientfigure}, except without the 
transient dip and with the asymptotic behavior for {\mbox{$\mu\gsim M$}} scaling as a power (rather than logarithm) of $\mu$.
More specifically, to leading order in $\mu/M$ and for {\mbox{$\mu\gsim M$}}, we find using the techniques developed
in Ref.~\cite{Paris} that
\beqn
        && \calK_1^{(-1,0)}\!\left( z \right) 
   +      4 \,\calK_2^{(-2,-1)}\!\left( z\right) +   \calK_3^{(-1,0)}\!\left( z \right) ~~~~~~~\nonumber\\ 
        && ~~~~~~~~~~~~~~~~~~~~~~~\sim~ \frac{7}{480} \left( \frac{\mu}{M}\right)^4~
\eeqn
where {\mbox{$z\equiv 2\sqrt{2}\pi M/\mu$}}.
By contrast, for {\mbox{$\mu \lsim M$}}, this quantity is exponentially suppressed.
Thus, 
recalling 
the result in Eq.~(\ref{eq:lamlam})
for our original unregulated (but nevertheless finite) cosmological constant $\Lambda$ 
and once again keeping only those (leading) running terms which dominate for {\mbox{$M\ll \mu \ll M_s$}},
we find that Eq.~(\ref{lambdamuresult}) simplifies to take the approximate form
                                                             \beqn
   \widehat\Lambda (\mu ) ~&\approx&~ \Lambda   - \frac{7}{960 \pi^2} \left[ \left(\zStr    {\bf 1} \right) +
                                  \left( \effStr {\bf 1} \right) \right]\! \mu^4 ~\nonumber\\
       &\approx&~   \Lambda - \frac{7}{960 \pi^2 } \left( \zeffStr {\bf 1} \right) \mu^4 ~.
\label{lambdamuresult2}
\eeqn
We once again emphasize that we have retained the second term (scaling as $\mu^4$) as this is the 
leading $\mu$-dependent term when {\mbox{$M\ll \mu \ll M_s$}}. 
Just as for $\widehat m_\phi^2(\mu)\bigl|_\calX$, there also generally 
exist additional running terms  
which scale as $\mu^2$ and $\log\,\mu$, but these terms are subleading
 relative to the above $\mu^4$ term when {\mbox{$M\ll \mu \ll M_s$}}. 
We shall discuss these subleading terms further in Sect.~\ref{sec5}.~
Moreover, just as we saw for $\widehat m_\phi^2(\mu)\bigl|_\calX$, the $\mu^4$ scaling
behavior can be enhanced to an even greater effective power $\mu^n$ with {\mbox{$n>4$}} if the spectrum of light states
is sufficiently dense when taking the supertrace over string states.
However, even this leading  $\mu^n$ scaling is generally subleading compared with the constant term $\Lambda$.
Thus the regulated quantity $\widehat \Lambda(\mu)$ --- unlike $\widehat m_\phi^2(\mu)\bigl|_\calX$ 
in Eq.~(\ref{approxhiggsmassmu}) ---
is dominated by a constant term and exhibits at most a highly suppressed running relative to this constant.

\FloatBarrier
\subsubsection{The Higgs mass in string theory:  See how it runs!\label{seehow}}

We now finally combine both contributions 
$\widehat m_\phi^2(\mu)\bigl|_{\calX,\Lambda}$
as in Eq.~(\ref{twocontributions})
in order to obtain our final result for the 
total modular-invariant regulated Higgs mass $\widehat m_\phi^2(\mu)$.
The exact result, of course, is given by the sum of Eqs.~(\ref{finalhiggsmassmu})
and (\ref{lambdamuresult}), with the  
latter first multiplied by $\xi/(4\pi^2 \calM^2)$.
However, once again taking the corresponding approximate forms in
Eqs.~(\ref{approxhiggsmassmu}) and (\ref{lambdamuresult2})
which are valid for {\mbox{$M\ll \mu\ll M_s$}},
we see that the $\mu^4$ running within 
Eq.~(\ref{lambdamuresult2})
is no longer the dominant running for 
$\widehat m_\phi^2(\mu)$ as a whole, 
as it is extremely suppressed compared with the running coming from
Eq.~(\ref{approxhiggsmassmu}).
We thus find that to leading order,
the net effect of adding 
Eqs.~(\ref{approxhiggsmassmu}) and (\ref{lambdamuresult2})
is simply to add the overall constant $\xi \Lambda/(4\pi^2 \calM^2)$ to 
the result in Eq.~(\ref{approxhiggsmassmu}).
We therefore find that the total regulated Higgs mass has the leading
running behavior
\beqn
  && \widehat m_\phi^2(\mu)  ~\approx~
       \frac{\xi}{4\pi^2} \frac{\Lambda}{\calM^2}
      - \frac{\pi}{6}\, \calM^2\, \Str\, \mathbbX_1\nonumber\\
  && ~~~~~~~~~~ 
      + \calM^2 \, \zStr \mathbbX_2 \,\log\left( \frac{  \mu}{2\sqrt{2} e M_s}\right)\nonumber\\
  && ~~~~~~~~~~ 
      + \calM^2 \,\effStr 
  \mathbbX_2 \,\log\left[ \frac{1}{\sqrt{2}}\,e^{-(\gamma+1)} \frac{\mu}{M}\right]~~~~~~~~~
\label{totalhiggsrunning}
\eeqn
where we have retained only the terms 
that are leading for {\mbox{$M\ll \mu\ll M_s$}}.
Once again, just as for $\widehat m_\phi^2(\mu)\bigl|_\calX$,
we see that to this order the $\mathbbX_2$ charges of the string states
lead to non-trivial running while their  $\mathbbX_1$ charges only contribute
to an overall additive constant. 
Indeed, in the {\mbox{$\mu\to 0$}} limit, we find
\beq
  \lim_{\mu\to 0}  \widehat m_\phi^2(\mu)  ~=~
       \frac{\xi}{4\pi^2} \frac{\Lambda}{\calM^2}
      - \frac{\pi}{6}\, \calM^2\, \Str\, \mathbbX_1
\label{asymplimit}
\eeq
when {\mbox{$\zStr \mathbbX_2=0$}}.
Of course, when {\mbox{$\zStr \mathbbX_2\not=0$}}, the {\mbox{$\mu\to 0$}} limit diverges, as expected from the 
fact that the massless $\mathbbX_2$-charged states are precisely the states that led
to a divergence in the original unregulated Higgs mass $m_\phi^2$. 
As discussed in Sect.~\ref{chargeinsertions},
we nevertheless continue to obtain a finite result for the regulated Higgs mass $\widehat m_\phi^2(\mu)$ 
for all {\mbox{$\mu>0$}} even when
{\mbox{$\zStr \mathbbX_2\not=0$}}.

In order to understand what the running in Eq.~(\ref{totalhiggsrunning}) looks like for {\mbox{$0<\mu\ll M_s$}}, 
let us begin by considering the contribution from 
a single $\mathbbX_2$-charged string state with a given mass {\mbox{$0<M\ll M_s$}}.
In this case, we have {\mbox{$\zStr \mathbbX_2=0$}}.  It then follows that
the approximate form in 
Eq.~(\ref{totalhiggsrunning}) reduces to
\beqn
  && \widehat m_\phi^2(\mu)\Bigl|_\calX  ~\approx~
       \frac{\xi}{4\pi^2} \frac{\Lambda}{\calM^2}
      - \frac{\pi}{6}\, \calM^2\, \Str\, \mathbbX_1\nonumber\\
          && ~~~~~~~~~~ 
      + \calM^2 \,\effStr 
  \mathbbX_2 \,\log\left[ \frac{1}{\sqrt{2}}\,e^{-(\gamma+1)} \frac{\mu}{M}\right]~~~~~~~~~
\label{totalhiggsrunning2}
\eeqn
where $\Lambda$ now represents the contribution to the total cosmological constant 
from this single state and where the supertraces $\Str\, \mathbbX_{1,2}$ now simply reduce to the (statistics-weighted) 
$\mathbbX_{1,2}$ charges of that state. 
For {\mbox{$\mu\gg M$}}, the final term in
Eq.~(\ref{totalhiggsrunning2}) produces a logarithmic
running.   Of course, the approximate result in Eq.~(\ref{totalhiggsrunning2}) is valid only for {\mbox{$\mu\gg M$}}.
For {\mbox{$\mu \ll M$}}, we instead know that our running asymptotically approaches
the constant
in Eq.~(\ref{asymplimit}).
Likewise, for {\mbox{$\mu \sim M$}}, we know that the running interpolates between
these two behaviors via the transient ``dip'' shown in Fig.~\ref{transientfigure}.

\begin{figure*}[t!]
\centering
\includegraphics[keepaspectratio, width=0.99\textwidth]{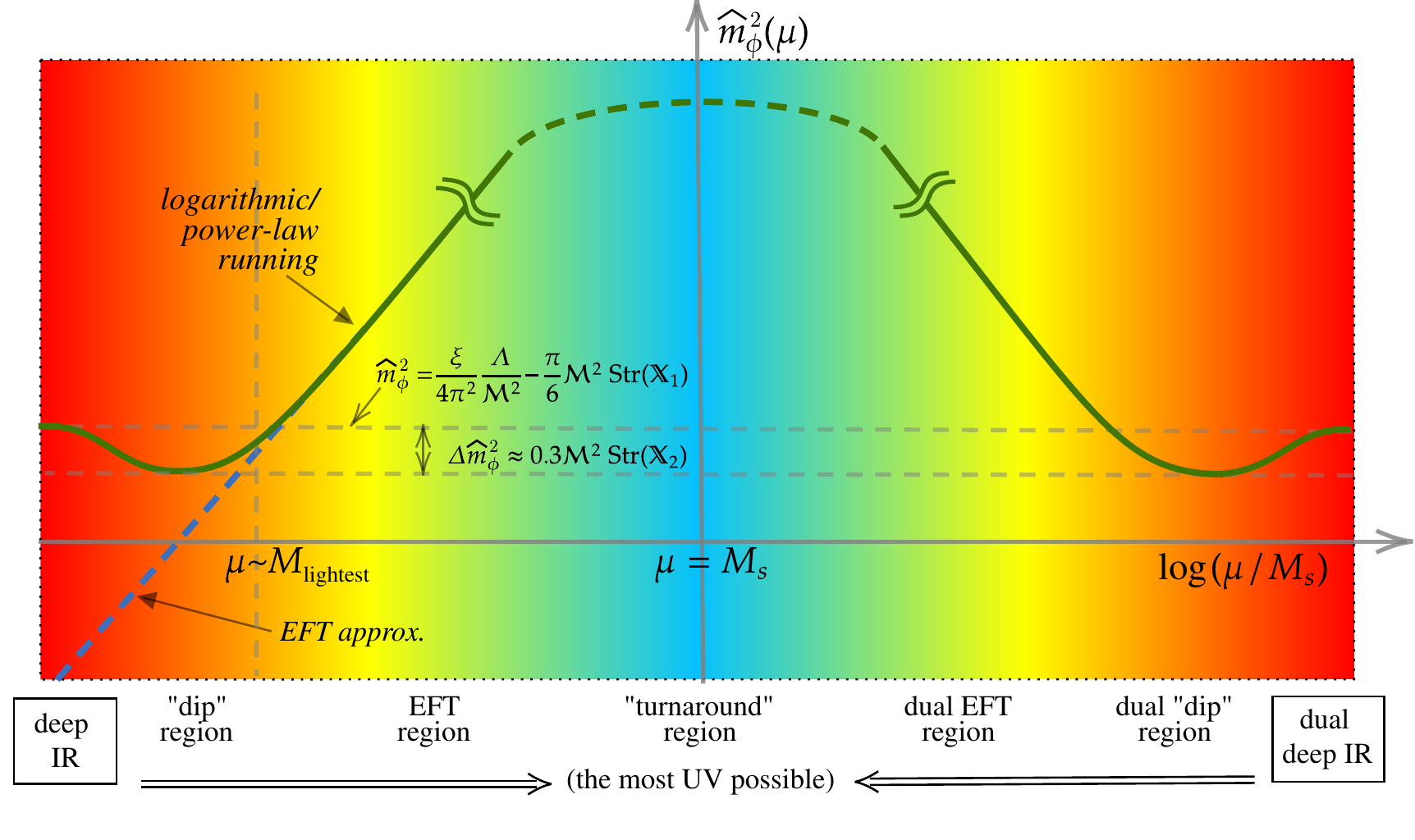}
\caption{The ``running'' of the regulated Higgs mass $\widehat m_\phi^2(\mu)$, as calculated
from first principles in a fully modular-invariant string framework. 
As discussed in the text, this running (green curve) exhibits a rather complicated anatomy.
In the deep infrared (as {\mbox{$\mu\to 0$}}), the Higgs mass approaches an asymptotic
value  which depends on the cosmological constant of the theory as well as on the supertrace
over the $\mathbbX_1$ charges of all of the string states.
Moving towards higher values of $\mu$, a non-trivial scale-dependence does not emerge until {\mbox{$\mu\sim M_{\rm lightest}$}}, 
where $M_{\rm lightest}$ collectively represents the masses of the lightest massive $\mathbbX_2$-charged states.
The ``dip'' that is observed in this region is a string-theoretic transient effect
which smoothly connects the asymptotic deep-IR region ({\mbox{$\mu\ll M_{\rm lightest}$}}) to
an EFT-like region ({\mbox{$M_{\rm lightest}\lsim \mu \ll M_s$}}).
Moving beyond this dip region, the theory then enters the EFT-like region
in which    the Higgs mass experiences a running which is either logarithmic or power-law,
depending on the density of string states with masses in the range {\mbox{$M_{\rm lightest} \lsim M \ll M_s$}}.
Note that this logarithmic/power-law running behavior will actually persist all the way into the deep infrared 
(dashed blue curve) if $M_{\rm lightest}$ is exceedingly small. 
Finally, as $\mu$ approaches $M_s$, the regulator we have employed is no longer valid and thus we cannot
explicitly calculate the running.  One possible running is sketched (dashed green curve).
{\it However, as a general principle, modular invariance requires that the running of the Higgs mass (and indeed the
running of {\it any}\/ physical quantity) exhibit an invariance under {\mbox{$\mu \to M_s^2/\mu$}}.}
Thus, as $\mu$ increases beyond $M_s$, we inevitably begin to re-enter an IR-like regime
which we may associate with a ``dual'' EFT.~ 
The background colors of this sketch indicate the transition from the deep IR (red) to the UV (blue) and
then back to IR (red).
This symmetry under  {\mbox{$\mu\to M_s^2/\mu$}} 
implies that the self-dual scale {\mbox{$\mu\sim M_s$}} exhibits the ``maximum possible'' UV behavior, in the sense
that further increases in $\mu$ only serve to push the theory back towards IR behavior.
Of course, as discussed in Sect.~\protect\ref{UVIRequivalence},
a fully modular-invariant string theory does not distinguish between the UV and the IR.~
Indeed, the UV and IR labels in this figure only arise  
upon attempting to extract a field-theoretic description of our theory, as we are implicitly doing 
when discussing the ``running'' of the Higgs mass.  }
\label{anatomy}
\end{figure*}

With all of these pieces combined,
the net running contributed from a single bosonic $\mathbbX_2$-charged state of mass $M$ has the behavior sketched as 
the green curve within the {\mbox{$\mu\ll M_s$}} portion of Fig.~\ref{anatomy}, where we interpret {\mbox{$M_{\rm lightest}\sim M$}}.   
Indeed, this curve is essentially the same as that shown in Fig.~\ref{transientfigure}, but with the 
addition of the asymptotic constant in Eq.~(\ref{asymplimit}).

Given this result from a single state of mass $M$, we now must take the supertrace over the entire spectrum of states in the theory.
However, as discussed in Sect.~\ref{chargeinsertions},
increasing (respectively decreasing) the mass of the contributing state simply shifts the corresponding contribution 
rigidly to the right (respectively left).
Taking the supertrace then simply amounts to adding these different shifted contributions together, weighted
by their corresponding $\mathbbX_2$ charges and statistics factors. 
Of course, as discussed in Sect.~\ref{chargeinsertions},
heavy states whose masses exceed $\mu$ are effectively integrated out of the theory:  they contribute
to the overall asymptotic constant in Eq.~(\ref{asymplimit}) but  
produce no effective running beyond this.
For any value of $\mu$ we therefore need only sum over the contributions from those light states whose 
masses lie below $\mu$.

The net result of this summation over string states is as follows.
As explained in the discussion surrounding Eq.~(\ref{newpower}),
this summation has the potential to turn the logarithmic
running into a power-law running for scales $\mu$ which lie within the spectrum of masses of the light $\mathbbX_2$-charged
states --- \ie, for scales {\mbox{$\mu> M_{\rm lightest}$}}, where
$M_{\rm lightest}$ denotes the mass 
of the lightest $\mathbbX_2$-charged states.
Indeed, as discussed previously, whether an effective power-law running emerges depends on the density of states in the theory
with masses {\mbox{$M\gsim M_{\rm lightest}$}}. 
It is for this reason that we have indicated in Fig.~\ref{anatomy} 
that the net running within the {\mbox{$\mu>  M_{\rm lightest}$}} region can be either logarithmic
or power-law.
However, as we progress to lower scales {\mbox{$\mu\sim M_{\rm lightest}$}}, 
we enter the ``dip region''
where this logarithmic/power-law running shuts off.
Finally, for {\mbox{$\mu<M_{\rm lightest}$}}, all running ceases as we enter the deep infrared region.
It is here that we recover the asymptotic constant value predicted in Eq.~(\ref{asymplimit}),
which now represents the sum over the individual asymptotic contributions from all of the string states.
Note that this implies that the unregulated integral represents the IR value of the Higgs mass.

Depending on the relative values of $\Lambda$, $\Str\,\mathbbX_1$, and
$\pStr\, \mathbbX_2$,
the Higgs may actually become tachyonic within the ``dip'' region.
Possible phenomenological implications of this will be briefly discussed in Sect.~\ref{sec:Conclusions}.~
Of course, if $M_{\rm lightest}$ is exceedingly small (or similarly if {\mbox{$\zStr\mathbbX_2\not=0$}}), we
never hit the dip region or the asymptotic-constant region.   
In this case, our EFT-like logarithmic/power-law running persists all 
the way into the deep infrared (as indicated through the dashed blue line in Fig.~\ref{anatomy}).

All of these results are valid for the same ``infrared'' region in which our regulator itself is valid, namely the region
with {\mbox{$a\ll 1$}} or equivalently {\mbox{$\mu \ll M_s$}}.
However, it turns out that we also have information about what happens in the opposite region, namely that with
{\mbox{$\mu \gg M_s$}}:  as sketched in Fig.~\ref{anatomy}, {\it we simply enter a ``dual infrared'' 
region in which this same infrared behavior again emerges, but in reverse.}
This is a direct consequence of the modular invariance which we have been careful to maintain throughout our calculations.
Indeed, modular invariance ensures that this entire picture is symmetric under
the {\it scale-inversion}\/ duality transformation
\beq
                \mu ~\to~ \frac{M_s^2}{\mu}~.
\label{muduality}
\eeq
As a result, when plotted as a function of $\log(\mu/M_s)$, 
the behavior of $\widehat m_\phi^2 (\mu)$ for {\mbox{$\mu\ll M_s$}} is reflected symmetrically
through the self-dual point {\mbox{$\mu_\ast=M_s$}} to yield the reverse behavior as {\mbox{$\mu\gg M_s$}}.
Of course, this tells us nothing about the behavior 
of $\widehat m_\phi^2(\mu)$ 
near the self-dual region with {\mbox{$\mu\sim M_s$}}, 
except that any running at the self-dual point {\mbox{$\mu_\ast=M_s$}} must ultimately become exactly flat.
We have sketched one possible shape for this running with a dashed (rather than solid) curve
within the self-dual region
in Fig.~\ref{anatomy}. 

The origins of the scale-duality symmetry in Eq.~(\ref{muduality}) are easily understood.
We have seen in Eq.~(\ref{newest}) that our regulator functions $\widehat G_\rho(a,\tau)$  ---
and hence our regulated Higgs masses $\widehat m_\phi^2(\rho,a)$ ---  have an invariance under
{\mbox{$\rho a^2 \to 1/(\rho a^2)$}}.
Likewise, we have seen in Eq.~(\ref{mudef}) that the running scale $\mu^2$ is given by $\rho a^2 M_s^2$.
These two relations then directly imply that $\widehat m_\phi^2(\mu)$ is invariant under
the scale-duality transformation in Eq.~(\ref{muduality}).
However, the origins of this scale-duality symmetry actually run deeper 
than any particular modular-invariant regulator we might choose,
and are directly connected
to the underlying modular invariance of theory. 
To see this, we recall from the discussion surrounding Eqs.~(\ref{massscale}) and (\ref{mudefeps}) that
the contributions of string states of mass $M$ to the one-loop partition function
experience Boltzmann suppressions scaling as $e^{-\pi \alpha' M^2 \tau_2}$.
Thus, for any particular benchmark value {\mbox{$\tau_2= t$}}, 
we can separate our string spectrum into two groups:   ``heavy'' states (whose Boltzmann suppressions
at {\mbox{$\tau_2=t$}} are significant according to some convention, and whose contributions therefore do not require regularization) 
and ``light'' states (whose Boltzmann suppressions are not significant, and whose contributions
therefore require regularization).  
Indeed, taking {\mbox{$t\to \infty$}} ensures that the only states whose contributions to the partition
function remain unsuppressed are those which are strictly massless.
On this basis, with an eye towards extracting an EFT with a running scale $\mu$,
we are directly led to identify $\mu^2$ inversely with $t$, as in Eq.~(\ref{mudefeps}). 
However, modular invariance tells us that any physical quantities which
depend on $\tau$ must be invariant under {\mbox{$\tau\to -1/\tau$}}.
Along the {\mbox{$\tau_1=0$}} axis, this becomes an invariance under {\mbox{$\tau_2\to 1/\tau_2$}}.   This then immediately
implies an invariance 
under {\mbox{$t \to 1/t$}}, or equivalently 
under {\mbox{$\mu \to \mu_\ast^2 /\mu$}} where $\mu_\ast$ is an arbitrary self-dual mass scale.
Of course, the choice of normalization for $\mu$ in relation to $t$ is purely a matter of convention,
and for convenience in this paper we have chosen our normalization for $\mu$ such that
{\mbox{$\mu_\ast = M_s$}}.
We thus see that while the particular choice of self-dual scale $\mu^\ast$ is a matter of 
convention, the existence of a scale-inversion duality symmetry of the form {\mbox{$\mu \to \mu_\ast^2/\mu$}} 
is inevitable, emerging directly from the underlying modular invariance of the theory.
This issue will be discussed further in Sect.~\ref{sec:Conclusions}.

Although this scale-duality symmetry follows directly from modular invariance, 
its implications are profound.
{\it Ultimately, the existence of such a symmetry signals the existence of an ultimate limit 
on the extent to which 
our EFT way of thinking can possibly remain valid in string theory.}
Indeed, as discussed in the Introduction, string theory is rife with duality symmetries which defy EFT notions:  an immediate example of this
is T-duality, under which the physics associated with a closed string 
propagating on a spacetime with a compactified dimension of radius $R$
is indistinguishable from the physics associated with a closed string propagating
on a spacetime with a compactified dimension of radius {\mbox{$R'\equiv \alpha'/R$}}.  This is true as
an exact symmetry not only for the string spectrum but also for all interactions.
Thus such strings cannot distinguish between large and small compactification geometries,
thereby preventing us from establishing a linear EFT-like ordering of length scales from large to small, 
or equivalently from IR to UV.~
What we are seeing now is that a similar phenomenon is guaranteed by modular invariance.
Although we can identify {\mbox{$\mu\ll M_s$}} as the deep infrared region of our EFT,
and although we may legitimately identify the passage towards larger scales $\mu$ as a passage towards an increasingly UV region of this EFT, we see that the validity of this identification has
a fundamental limit.
Indeed, pushing $\mu$ beyond  $M_s$ only serves to reintroduce the original IR behavior of our theory ---
a behavior which we may now associate with the {\it dual}\/ energy scale 
{\mbox{$\mu'\equiv M_s^2/\mu$}} associated with a ``dual'' EFT.~
In this sense, the energy scales near $M_s$ exhibit the ``most possible UV'' behavior
that can be realized.
This is indicated through the background colors of Fig.~\ref{anatomy}, with red indicating the IR
regions of our theory and blue indicating the UV.

As we have emphasized, all of this is the inevitable consequence of modular invariance. 
At first glance, it may not seem that our results for $\widehat m_\phi^2(\mu)$ are modular invariant, much less
symmetric under the scale-duality symmetry in Eq.~(\ref{muduality}).    Indeed, this symmetry is hardly manifest
within our exact expressions for 
$\widehat m_\phi^2(\mu)\bigl|_\calX$ and $\widehat \Lambda(\mu)$ 
in Eqs.~(\ref{finalhiggsmassmu}) and (\ref{lambdamuresult}), respectively.
However, this symmetry is ultimately ensured through
intricate relations satisfied by the various supertraces involved.
Indeed, such relations are themselves manifestations of the underlying modular invariance of the theory.
In this paper we have not focused on the identities satisfied by these supertraces. 
However, we have already seen two such identities, namely the simple expressions for $\Str \,{\bf 1}$ and
$\Str \,M^2$ in Eqs.~(\ref{eq:lamlam0}) and (\ref{eq:lamlam}), respectively.
These identities, which were originally derived in Ref.~\cite{Dienes:1995pm},
hold in any modular-invariant string theory, arising as a consequence of 
the so-called ``misaligned supersymmetry''~\cite{Dienes:1994np}
that tightly constrains the distributions of bosonic and fermionic states across any modular-invariant 
string spectrum. 
In a similar vein, there will also exist 
identities satisfied
by {\it all}\/ of the supertraces that appear within Eqs.~(\ref{finalhiggsmassmu})
and (\ref{lambdamuresult}), especially when $\mathbbX_i$ insertions are involved.
These identities are likely to be more intricate and interwoven than those for $\Str\, {\bf 1}$ and $\Str\, M^2$,
but together they act to ensure the modular invariance of our results for $\widehat m_\phi^2(\mu)$.

We conclude, then, that the duality symmetry in Eq.~(\ref{muduality}) is a fundamental property
of the running of any physical quantity in a modular-invariant theory.
As such, there is a maximum degree to which our theory can approach the UV:   
once the energy scale $\mu$ passes the self-dual 
point {\mbox{$\mu\sim M_s$}}, further increases in $\mu$ only push us towards increasingly IR behavior.
Of course, as we explained in Sect.~\ref{UVIRequivalence},
the full modular-invariant string theory does not distinguish between the UV and IR --- this distinction only 
has meaning when we attempt to extract an EFT description, as we are doing when discussing the 
``running'' of the Higgs mass.
Such, then, are the unique properties of UV/IR-mixed theories such as those exhibiting modular invariance.

\FloatBarrier
\section{Transcending the charge lattice \label{sec5}}

Thus far in this paper, our guide has been modular invariance --- an exact symmetry of closed strings.
However, right from the beginning of Sect.~\ref{sec2}, we have further assumed that our closed-string models have
associated charge lattices.
This has been the language of our analysis,
and as such this has given our calculations a certain concreteness, enabling us to obtain and express our main results 
in a rather direct and understandable fashion as the supertrace of physical string states weighted by their eigenvalues with respect
to certain combinations of worldsheet charge operators. 
This language also allowed us to understand the origins of the modular anomaly that ultimately connected the Higgs mass and the
cosmological constant.

It is certainly the case that many classes of closed-string 
models can be described in terms of their associated charge lattices.
For example, charge lattices appear for a wide variety of geometric compactifications
and thus play an essential role in many corresponding 
free-field constructions (such as those based on free worldsheet bosons and complex fermions).
However, not all string theories can be described in this manner.

Fortunately, in hindsight, it is not difficult to demonstrate that our results
are actually general and do not ultimately rely on the existence of such charge lattices.
Indeed, as we shall now demonstrate, 
most of our results follow from modular invariance alone, and can be expressed in a more general language that
makes no specific reference to charge operators.
Thus, phrasing our results in this more general language 
demonstrates that our results actually transcend their charge-lattice roots
and have a more encompassing generality.
This will also help us discern the existence of 
a ``stringy'' effective potential for the Higgs.

\subsection{$\calX$'s without $\bQ$'s:  A reformulation of the partition-function insertions}

In Sect.~\ref{sec2}, we established a framework for calculating the Higgs mass 
in which the charge lattice played a central role.
By extracting the $\eta$-functions as explicit prefactors in Eq.~(\ref{Zform}),
we implicitly separated out the contributions from the oscillator modes
and thereby implicitly cast the partition function of our theory into a form
in which the $(m,n)$ exponents in Eq.~(\ref{Zform}) were
directly related to the lengths of the charge vectors in an underlying
charge lattice, as in Eq.~(\ref{mnQ}).
By contrast, the {\it spacetime masses}\/ $(M_L,M_R)$ of the corresponding string states
receive contributions from not only the charges (\ie, the compactification momentum modes) 
but also the oscillator modes within the $\eta$'s.
Indeed, these contributions are ultimately added together, as in Eq.~(\ref{eq:massRL}).

Taken together, this means that we can always rewrite our partition
function $\calZ$ in a general form 
which is closer to what appears in Eq.~(\ref{integrand}), namely
\beq
           \calZ(\tau) ~=~ \tau_2^{-1} \sum_{\rm states} (-1)^F  \, \qbar^{\alpha' M_R^2/4} \, q^{\alpha' M_L^2/4}~
\label{Zmasses}
\eeq
where we are summing over the states in the string spectrum with left- and right-moving 
spacetime mass contributions $(M_L,M_R)$.
Although Eq.~(\ref{Zmasses}) reduces to Eq.~(\ref{Zform}) for theories with a charge lattice,
the expression in Eq.~(\ref{Zmasses}) is more general and applies to {\it any}\/ closed string.
Just as in Sect.~\ref{sec2}, this partition function is assumed to describe the theory
in its Higgsed phase.

Following the arguments in Sect.~\ref{sec2}, we seek to evaluate the Higgs mass
by exploring the response of the theory to small fluctuations $\phi$ of the Higgs
field around its minimum $\langle \phi \rangle$.   Previously we described
the response of the system in terms of a deformation of the charge lattice, as in Eq.~(\ref{Qdeform}).
However, we may more generally simply describe the response of our theory in terms of the corresponding deformations
to our left- and right-moving masses, so that $M_L$ and $M_R$ for a given string state $s$ 
now become functions of $\phi$:
\beqn 
                  M_L^{(s)} &\to& ~M_L^{(s)} + \delta M_L^{(s)}(\phi) ~\equiv~ M_L^{(s)}(\phi)~\nonumber\\
         M_R^{(s)} &\to& ~M_R^{(s)} + \delta M_R^{(s)}(\phi) ~\equiv~ M_R^{(s)}(\phi)~.
\label{Mshifts}
\eeqn
Of course, for a given Higgs field,
not all states in the string spectrum will have their masses shifted.   
Indeed, mass shifts will arise for only those states which couple to the fluctuations 
parametrized by $\phi$;   the masses of the other states will remain independent of $\phi$.
It is for this reason that we have explicitly attached a state index $s$ to the masses in Eq.~(\ref{Mshifts}) --- namely
to clarify that the mass shifts depend on the particular state $s$ and not merely on the unperturbed masses $(M_L,M_R)$. 
This is completely analogous to the fact that only certain charge vectors in Eq.~(\ref{Qdeform}) 
will be deformed.
In the following, for the sake of parallelism with Eq.~(\ref{Qdeform}), we shall
suppress the state index $s$ in Eq.~(\ref{Mshifts}) with the understanding that
whether certain $M_{L,R}(\phi)$ are truly $\phi$-dependent depends not on $M_{L,R}$
but rather on the identity of the state $s$ from which these contributions emerge.
In this context, we remark that changing the value of $\phi$ generally does more than shift the masses of certain states ---   
it will also typically {\it mix}\/ these states, thereby changing the corresponding mass eigenstates.
We should thus understand the index $s$ as continuously following a given mass eigenstate as $\phi$ is changed.

Just as in the charge-vector formalism, 
the choices of $\delta M_{L,R}(\phi)$ are not arbitrary;   modular invariance [in this case,
the invariance of $\calZ(\phi)$ under {\mbox{$\tau\to \tau+1$}}] must still be maintained.
This implies that
\beq
           \delta M_L(\phi) ~=~ \delta M_R(\phi)~,
\label{TMs}
\eeq
which is the generalization of the constraints in Eq.~(\ref{levelmatching}).

Given these observations, we can then calculate the Higgs mass precisely as in 
Eqs.~(\ref{higgsdef}) and (\ref{Lambdaphi}), except with $\calZ$ expressed
as in Eq.~(\ref{Zmasses}) with the masses $M_L$ and $M_R$ 
replaced by their $\phi$-dependent versions $M_L(\phi)$ and $M_R(\phi)$.
We then have
\beq
   {\partial^2\calZ \over \partial\phi^2} ~=~
        \tau_2^{-1}  \sum_{\rm states}
             (-1)^F \,X\, \,\qbar^{\alpha'M_R^2/4} \, q^{\alpha' M_L^2/4}~
\label{eq:Zexp2}
\eeq
where the summand insertion $X$ is precisely the same quantity as in Eq.~(\ref{stuff})
but now expressed as
\beqn
      X ~&\equiv&~ \frac{\pi i\alpha'}{2}\,
     \partial_\phi^2 (\tau M_L^2 - \taubar M_R^2 )\nonumber\\
   &&~~~~~
      -  \left(\frac{\pi \alpha'}{2}\right)^2  
       \left\lbrack {\partial_\phi } (\tau M_L^2 - \taubar M_R^2)\right\rbrack^2~.~~
\label{stuff2}
\eeqn
However, each term within Eq.~(\ref{stuff2}) with a non-zero power of $\tau_1$
also contains equally many powers of $M_L^2-M_R^2$, and we see from Eqs.~(\ref{Mshifts}) and
(\ref{TMs})   that
\beq
           \partial_\phi   (M_L^2-M_R^2) ~=~
           \partial_\phi  (\delta M_L^2 - \delta M_R^2) ~=~ 0~.
\eeq
Thus each factor of $\tau M_L^2 - \taubar M_R^2$ within
Eq.~(\ref{stuff2}) can be replaced with {\mbox{$i\tau_2 (M_L^2 + M_R^2) = 2 i\tau_2 M^2 $}} where
$M$ is the ($\phi$-dependent) shifted spacetime mass of the corresponding state. 
This then yields
\beq
      X ~=~ - \pi \alpha'\tau_2\, 
     \partial_\phi^2   M^2 
      +  \left(\pi \alpha' \tau_2 \right)^2 
       \left( \partial_\phi  M^2 \right)^2~.
\label{stuff3}
\eeq
Recalling 
{\mbox{$  \calX \equiv \calM^2 X|_{\phi=0}$}}
and separating out the different powers of $\tau_2$ in Eq.~(\ref{stuff3}) then immediately enables us to identify
\beqn
               \mathbbX_1 ~&=&~ - \frac{1}{4\pi}     \partial_\phi^2 M^2 \Bigl|_{\phi=0}~\nonumber\\
               \mathbbX_2 ~&=&~ \phantom{-} \frac{1}{16\pi^2 \calM^2} (\partial_\phi M^2)^2 \Bigl|_{\phi=0}~.
\label{newXi}
\eeqn

These, then, are the general forms of the $\mathbbX_i$ insertions --- forms which do not rely on the existence
of a charge lattice.   
As indicated above, only for those string states which couple to the Higgs will the corresponding
spacetime masses $M$ be affected by fluctuations of the Higgs field and thereby accrue a $\phi$-dependence.   
Thus, given the $\phi$-derivatives in Eq.~(\ref{newXi}), we see that
only these states will contribute to $\mathbbX_1$ and $\mathbbX_2$.
Of course, when the string model in question has an underlying charge lattice,
the masses of our string states can be written in terms of the lengths
of the vectors in that lattice, whereupon the results in Eq.~(\ref{newXi}) can be
evaluated to have the more specific forms presented in earlier sections.

Given the general results in Eq.~(\ref{newXi}), the rest of our analysis proceeds precisely as before.
The $\phi$-derivatives generally induce modular anomalies that require modular completions and lead us to add the
universal $\Lambda$-term, as before.
Our final result for the unregulated Higgs mass from Eq.~(\ref{relation1})
then takes the form
\beqn
   m_\phi^2 ~&=&~  \frac{\xi}{4\pi^2} \frac{\Lambda}{\calM^2} 
     - \frac{\calM^2}{2} \left\langle \tau_2 \,\mathbbX_1 + \tau_2^2 \,\mathbbX_2 \right\rangle \nonumber\\
    &=&~  \frac{\xi}{4\pi^2} \frac{\Lambda}{\calM^2} 
                + \frac{\calM^2}{8\pi} \, \left\langle \tau_2 \,\partial_\phi^2 M^2 \Bigl|_{\phi=0} \right\rangle \nonumber\\
    && ~~~~~~~~~~~ - \frac{1}{32\pi^2} \left\langle \tau_2^2 \,(\partial_\phi M^2)^2 \Bigl|_{\phi=0} \right\rangle~~~~
\eeqn  
where the bracket notation $\langle ...\rangle$ is defined in Eq.~(\ref{expvalue}).
This result can then be regulated, as discussed in Sect.~\ref{sec4},
leading to a regulated Higgs mass $\widehat m_\phi^2(\rho,a)$.
Indeed, none of the results presented after Sect.~\ref{sec2} depended
on the specific forms of $\mathbbX_1$ and $\mathbbX_2$ in any way.

\subsection{A stringy effective potential for the Higgs}

As we shall now demonstrate,
the results in Eq.~(\ref{newXi}) allow us to reach 
some powerful conclusions about the regulated Higgs mass $\widehat m_\phi^2(\rho,a)$.
They will also allow us to extract a stringy effective potential for the Higgs.
We begin by noting that $\widehat m_\phi^2 (\rho,a)$ 
has two contributions, 
$\widehat m_\phi^2(\rho,a)\bigl|_\calX$ and
$\widehat m_\phi^2(\rho,a)\bigl|_\Lambda$,
as indicated in Eq.~(\ref{twocontributions}).   
The first of these contributions
involves $\mathbbX_i$ insertions, while the second comes from the cosmological constant and lacks such insertions.
Of course, both contributions depend on the spectrum of masses of the states in our theory ---  masses which we are
now taking to be general functions of $\phi$ before eventually truncating to {\mbox{$\phi=0$}}.
However, we see from Eq.~(\ref{newXi}) that the $\mathbbX_i$ insertions depend on the {\it derivatives}\/
of these masses with respect to $\phi$, whereas no such derivatives appear within $\widehat m_\phi^2(\rho,a)\bigl|_\Lambda$.
Thus suggests that $\widehat m_\phi^2(\rho,a)\bigl|_\calX$ might be related to $\phi$-derivatives of 
$\widehat m_\phi^2(\rho,a)\bigl|_\Lambda$ prior to the truncation to {\mbox{$\phi=0$}}.
If so, the two Higgs-mass contributions 
$\widehat m_\phi^2(\rho,a)\bigl|_\calX$ 
and 
$\widehat m_\phi^2(\rho,a)\bigl|_\Lambda$ 
would actually be deeply connected to each other.
 
To investigate this possibility, we note
from Appendices~\ref{higgsappendix} and \ref{lambdaappendix} that
these two contributions to the Higgs mass can be identically expressed in terms of simpler 
quantities, $P_\calX$ and $P_\Lambda$, via the relations 
\beqn
  \widehat m_\phi^2(\rho,a)\Bigl|_\calX ~&=&~ \frac{1}{1+\rho a^2} \,A_\rho\,
          a^2 \frac{\partial}{\partial a} \,
          \biggl[   P_\calX(\rho a) - P_\calX(a) \biggr]~ \nonumber\\
  \widehat \Lambda(\rho,a)         ~&=&~ \frac{1}{1+\rho a^2} \,A_\rho\,
          a^2 \frac{\partial}{\partial a} \,
          \biggl[   P_\Lambda(\rho a) - P_\Lambda(a) \biggr]~\nonumber\\ 
\label{mtoP}
\eeqn
where \hbox{{\mbox{$A_\rho\equiv \rho/(\rho-1)$}}},
where we recall {\mbox{$\widehat m_\phi^2(\rho,a)\bigl|_\Lambda \equiv \xi \widehat \Lambda(\rho,a)/(4\pi^2\calM^2)$}},
and where 
$P_\calX(a)$ and 
$P_\Lambda(a)$ are
given in Eqs.~(\ref{finalPa})
and (\ref{Pacos}) respectively.
Since $\phi$-derivatives commute with the operators in Eq.~(\ref{mtoP}),
the question of whether $\widehat m_\phi^2(\rho,a)\bigl|_\calX$
might be related to $\phi$-derivatives of 
$\widehat m_\phi^2(\rho,a)\bigl|_\Lambda$
then boils down to whether
$P_\calX(a)$ might be related to $\phi$-derivatives of $P_\Lambda(a)$,
where each mass $M$ is now to be regarded
as a function of $\phi$ prior to truncation.  

Our procedure, then, will be 
to evaluate $\phi$-derivatives of $P_\Lambda(a)$ in Eq.~(\ref{Pacos}) and
determine the extent to which we reproduce $P_\calX(a)$ in Eq.~(\ref{finalPa}). 
To do this, we shall now treat each squared mass $M^2$ 
within $P_\Lambda(a)$ in Eq.~(\ref{Pacos}) 
as representing the {\mbox{$\phi\to 0$}} limit of a function $M^2(\phi)$ which,
for {\mbox{$\phi/\calM \ll 1$}}, we can imagine Taylor-expanding to take the generic form\footnote{
    It is significant that we are Taylor-expanding $M^2$ rather than $M$.
    In closed string theories, $M^2$ is the fundamental quantity, with worldsheet excitations
    making contributions directly to $\alpha' M^2$ as in Eq.~(\ref{eq:massRL}).
        As we shall shortly see when discussing the stability of these theories, 
    it is also significant that our Taylor expansion for $M^2$ generically includes a term linear in $\phi$.
    This follows directly from Eq.~(\ref{Qdeform}) in conjunction with Eqs.~(\ref{masssum}) and
    (\ref{eq:massRL}).} 
\beq
          M^2(\phi) ~=~ \calM^2\left\lbrack \beta_0 + \beta_1 \left( \frac{\phi}{\calM}\right) + \frac{1}{2}\, \beta_2 
              \left( \frac{\phi}{\calM}\right)^2 + ... \right\rbrack
\label{TaylorM}
\eeq
where {\mbox{$\beta_0\geq 0$}}.
The original physical squared mass of the state then corresponds to {\mbox{$M^2(\phi)\bigl|_{\phi=0}=\beta_0 \calM^2$}}, so that
massive states correspond to functions with {\mbox{$\beta_0>0$}} while
massless states have {\mbox{$\beta_0=0$}}.  However, we see that even massless 
states introduce a $\phi$-dependence prior to the truncation to {\mbox{$\phi=0$}}.

Focusing initially on the first term within $P_\Lambda(a)$ in Eq.~(\ref{Pacos}), we immediately see that 
\beqn
   && \partial_\phi^2 \left. \left[   \frac{\calM^2}{24 a} \, \Str\, M^2 \right] \right|_{\phi=0} 
   = ~ \left. \frac{\calM^2}{24a} \, \Str\, (\partial_\phi^2 M^2) \right|_{\phi=0} \nonumber\\
   && ~~~~~~~~~~~~~~~ = ~  - \frac{\calM^2}{2} \,\left[ \frac{\pi}{3a} \,\Str\, \mathbbX_1\right]~
\eeqn
where we have used the result in Eq.~(\ref{newXi}) in passing to the final expression.
This successfully reproduces the initial terms within $P_\calX(a)$ in Eq.~(\ref{finalPa}).
    Next, we evaluate the second $\phi$-derivative of
the Bessel-function terms within $P_\Lambda(a)$ in Eq.~(\ref{Pacos}).
To do this, we note the mathematical identity 
\beqn
  &&  \partial_\phi^2 \left[ M^2 K_2\arg\right] ~=~ 
          - \frac{r M}{2a\calM} \left( \partial_\phi^2 M^2\right) K_1\arg \nonumber\\
  &&~~~~~~~~~~~~~~~~~ +  \frac{r^2}{4 a^2 \calM^2} \left(\partial_\phi M^2\right)^2 K_0\arg~
\eeqn 
which follows from standard results for Bessel-function derivatives along with a judicious 
repackaging of terms.
Given this, and given the relations in Eq.~(\ref{newXi}), we then find that  
\beqn
&& \partial_\phi^2 \left. \left\lbrace  
   \frac{a}{\pi^2 } 
       \, \bpStr  \!\left\lbrack M^2 \sum_{r=1}^\infty
                \frac{1}{r^2} K_2\left(\frac{  r M}{ a \calM}\right)\right\rbrack\right\rbrace \right|_{\phi=0} \nonumber\\
&& ~~~~~~~~~ =~  
  \frac{2}{\pi}\, \pStr \mathbbX_1  \,\left\lbrack \sum_{r=1}^\infty \left(\frac{M}{r\calM}\right)
          \,K_1\left( \frac{r  M}{a\calM} \right)\right\rbrack \nonumber\\
&& ~~~~~~~~~ \phantom{=}~ +  
      \frac{4}{a} \,\pStr \mathbbX_2 \left\lbrack \sum_{r=1}^\infty  
            K_0\left(  \frac{ r  M}{a\calM} \right) \right\rbrack~.~~~~~~~~~~~~~~ 
\label{besselderivs}
\eeqn
where the supertrace in the first line is over all states whose mass functions $M^2(\phi)$ have {\mbox{$\beta_0>0$}}. 
We thus see that the result in Eq.~(\ref{besselderivs}) likewise successfully reproduces 
the Bessel-function terms within $P_\calX(a)$ in Eq.~(\ref{finalPa}).

Our final task is to evaluate $\partial_\phi^2$ acting on the second term in Eq.~(\ref{Pacos}).
At first glance, it would appear that this term does not yield any contribution since 
it is wholly independent of the mass $M$ and would thus not lead to any $\phi$-dependence.
Indeed, as evident from Eq.~(\ref{P2cosa}), this term represents a contribution to $P_\Lambda(a)$ 
from purely massless states, and as such the identification {\mbox{$M=0$}} has already been implemented
within this term.   This is why no factors of the mass $M$ remain within this term.
However, as discussed above, for the purposes of the present calculation we are to regard the masses $M$ 
as functions of $\phi$ before taking the $\phi$-derivatives.
          Thus, when attempting to take $\phi$-derivatives of the second term in Eq.~(\ref{Pacos}),
we should properly go back one step to the original derivation of this term that appears in Eq.~(\ref{P2cosa}) 
and reinsert a non-trivial mass function $M^2(\phi)$ with {\mbox{$\beta_0=0$}} into the derivation.
The remaining derivation of this term then algebraically mirrors the derivation of the {\it massive}\/ 
Bessel-function term in Eq.~(\ref{P2cosb}), only with $M^2$ now replaced by $M^2(\phi)$ with {\mbox{$\beta_0=0$}}.
In other words, for the purposes of our current calculation, we should formally identify
\beqn
&& \frac {\calM^4}{2}\, \frac{\pi^2}{45}\,(n_B-n_F)   \,{a^3} \nonumber\\
&& ~~~=~ 
   \frac{\calM^2}{2}\left. \frac{a}{\pi^2 } 
       \, \bzStr \left\lbrack M^2 \sum_{r=1}^\infty
                \frac{1}{r^2} K_2\left(\frac{  r M}{ a \calM}\right) \right\rbrack \right|_{\phi=0}~~~~~~~
\label{substitution}
\eeqn
and then evaluate the $\phi$-derivatives before 
truncating to {\mbox{$\phi=0$}}.
Aside from the overall factor of $- \calM^2/2$, acting with $\partial^2_\phi$ and then truncating to {\mbox{$\phi=0$}} yields 
the same result as on the right side of Eq.~(\ref{besselderivs}),
except with each supertrace over massive states 
replaced with a supertrace over massless states.
We thus need to evaluate these Bessel-function expressions at zero argument.
However, for small arguments {\mbox{$z\ll 1$}}, the Bessel functions have the leading asymptotic behaviors
\beq
        K_\nu(z) ~\sim~ \begin{cases}
          ~-\log (z/2) - \gamma +...& {\rm for}~~\nu=0 \\
          ~\phantom{-}\half \,\Gamma(\nu) \,(z/2)^\nu + ... & {\rm for}~~\nu>0 
          \end{cases}
\label{Bessellimits}
\eeq
where $\gamma$ is the Euler-Mascheroni constant.
Analyzing the $\Str \,\mathbbX_1$ term,
we thus see that
\beqn
  &&    \frac{2}{\pi}\, \lim_{M\to 0} \,\pStr \mathbbX_1   \left\lbrack \sum_{r=1}^\infty \left(\frac{M}{r\calM}\right)
          \,K_1\left( \frac{r  M}{a\calM} \right)\right\rbrack \nonumber\\
  && ~~~~~~~=~ \frac{2}{\pi}\, \zStr \mathbbX_1  \,\lim_{M\to 0} \left\lbrack \sum_{r=1}^\infty \left(\frac{M}{r\calM}\right)
          \left(\frac{a  \calM}{r M} \right)\right\rbrack \nonumber\\
  && ~~~~~~~=~ \frac{2a}{\pi} \,\sum_{r=1}^\infty \,\frac{1}{r^2} ~= ~     \frac{\pi}{3}\, a~,
\label{hidden}
\eeqn 
thereby successfully reproducing the corresponding term which appears in $P_\calX(a)$.
Indeed, we see that the {\mbox{$M\to 0$}} limit in Eq.~(\ref{hidden}) is convergent and continuous with the exact {\mbox{$M=0$}} result.

For theories in which {\mbox{$\zStr \mathbbX_2=0$}}, there are no further terms to consider.
The results of this analysis are then clear:
within such theories, we have found that
\beq
              P_\calX(a) ~=~ \partial_\phi^2 \, P_\Lambda(a,\phi) \bigl|_{\phi=0}~.
\eeq
Through Eq.~(\ref{mtoP}), this then implies that 
\beq
              \widehat m_\phi^2(\rho,a)\bigl|_\calX ~=~ 
                  \partial_\phi^2 \, \widehat\Lambda( \rho,a,\phi) \bigl|_{\phi=0}~,
\eeq
whereupon use of Eq.~(\ref{twocontributions}) tells us that
\beq
              \widehat m_\phi^2(\rho,a) ~=~ \left.\left( \partial_\phi^2 +  \frac{\xi}{4\pi^2 \calM^2} \right)  
               \, \widehat\Lambda( \rho,a,\phi) \right|_{\phi=0}~,
\eeq
or equivalently
\beqn
          \widehat m_\phi^2(\mu) ~&=&~  
             \left.\left( \partial_\phi^2 +  \frac{\xi}{4\pi^2 \calM^2} \right)  
               \, \widehat\Lambda( \mu,\phi) \right|_{\phi=0} ~~~~~\nonumber\\
             &=&~  \left. D_\phi^2   ~\widehat\Lambda( \mu,\phi) \right|_{\phi=0}~,
\label{finalCWresult}
\eeqn
where we have defined the modular-covariant derivative 
\beq
            D_\phi^2   ~\equiv~ \partial_\phi^2 +  \frac{\xi}{4\pi^2 \calM^2}~.
\label{Dphi2}
\eeq
Of course, for theories with {\mbox{$\zStr \mathbbX_2=0$}}, our original unregulated Higgs mass 
was already finite and {\it a priori}\/ there was no  
need for a regulator.   However, even within such theories, it is the use of our modular-invariant
regulator for both $\Lambda$ and $m_\phi^2$ which enabled us to extract EFT descriptions   
of these quantities and to analyze their runnings as functions of an effective scale $\mu$.

The result in Eq.~(\ref{finalCWresult}) is both simple and profound.
Indeed, comparing this result with our starting point in Eq.~(\ref{higgsdef})
and recalling the subsequent required modular completion in Eq.~(\ref{Xmodcomplete}),
we see that we have in some sense come full circle.
However, as stressed above, we have now demonstrated this result using only the general 
expressions for $\mathbbX_1$ and $\mathbbX_2$ in Eq.~(\ref{newXi}) and thus 
{\it entirely  without the assumption of a charge lattice}\/.
This result therefore holds for {\it any}\/ modular-invariant string theory with {\mbox{$\zStr\mathbbX_2=0$}}. 
Indeed, as indicated above, we can view $D_\phi^2$ as a modular-covariant
derivative, in complete analogy with the lattice-derived
covariant derivative $D_z^2$ in Eq.~(\ref{modcovderiv}).

But more importantly, we see from Eq.~(\ref{finalCWresult}) that 
within such theories
we can now identify $\widehat \Lambda(\mu,\phi)$ as {\it an effective potential for the Higgs}\/.  Strictly speaking, this is not the entire effective potential --- it does
not, for example, allow us to survey different minima  
as a function of $\phi$  in order to select the global and local minima, as would be needed in order
to determine the ground states of the theory in different possible phases (with unbroken and/or broken symmetries).
However, we see that $\widehat \Lambda(\mu,\phi)$ does provide a {\it piece}\/ of the full potential, namely 
the portion of the potential in the immediate vicinity of the assumed minimum (around which 
$\phi$ parametrizes the fluctuations, as always).
With this understanding, we shall nevertheless simply refer to $\widehat \Lambda(\mu,\phi)$ as the Higgs effective potential.
Indeed, as expected, we see from Eq.~(\ref{finalCWresult}) that 
the Higgs mass is related to the curvature of this potential around this minimum.   
One can even potentially imagine repeating the calculations in this paper {\it without}\/ implicitly assuming 
the stability condition in Eq.~(\ref{linearcond}), thereby dropping the implicit assumption that we are sitting
at a stable vacuum of the theory.   In that case, the first and second $\phi$-derivatives of $\widehat \Lambda(\mu,\phi)$
would describe the slope and curvature of the potential for arbitrary values of $\phi$, whereupon the methods in this paper
could provide a method of ``tracing out'' the 
shape of the full potential.   However, at best this would appear to be a challenging undertaking.

As remarked above, the form of Eq.~(\ref{finalCWresult}) makes sense from the perspective 
of Eq.~(\ref{higgsdef}), in conjunction with the subsequent modular completion.
At first glance, it may seem surprising that such a result would continue to survive 
even after imposing our modular-invariant regulator 
in order to generate our regulated expressions
for $\widehat m_\phi^2(\mu)$ and $\widehat \Lambda(\mu,\phi)$,
and perhaps even more surprising after 
the Rankin-Selberg techniques and their generalizations in Sect.~\ref{sec3}
are employed in order to express these regulated quantities in terms of supertraces over purely physical (level-matched)
string states.
Ultimately, however, 
the result in Eq.~(\ref{finalCWresult})
concerns the $\phi$-structure of the theory and the response of the theory to fluctuations in the Higgs field.
In theories with {\mbox{$\zStr\mathbbX_2=0$}},
these properties are essentially ``orthogonal'' to the manipulations that occurred in Sects.~\ref{sec3} and \ref{sec4},
which ultimately concern the regulators and the resulting behavior of these quantities as functions of $\mu$.
In other words, in such theories
the process of $\phi$-differentiation in some sense ``commutes''
with all of these other manipulations.
Thus the relation in Eq.~(\ref{finalCWresult}) holds not only for our original unregulated Higgs mass
and cosmological constant, but also for their regulated counterparts as well as for the running which describes
their dependence on the variables defining the regulator.

It is also intriguing that we are able to identify a modular-covariant 
derivative $D_\phi^2$ within the results in Eq.~(\ref{finalCWresult}).
Of course, this is the {\it second}\/ $\phi$-derivative.
By contrast, the {\it first}\/ $\phi$-derivative does not require modular completion.
We have already seen this in Sect.~\ref{stability}, where we found
that $\partial_\phi$ acting on the partition function $\calZ$ corresponds to insertion
of the factor $\calY$, which was already modular invariant.
In this sense, $\phi$-derivatives are similar to the $z$-derivatives 
discussed in Sect.~\ref{sec:completion}.~

The result in Eq.~(\ref{finalCWresult}) holds only for theories in which {\mbox{$\zStr\mathbbX_2=0$}}.
However, when {\mbox{$\zStr\mathbbX_2\not=0$}}, there is an additional term  
to consider within $P_\Lambda$.
Taking the {\mbox{$M\to 0$}} limit of the $\pStr\,\mathbbX_2$ result in 
Eq.~(\ref{besselderivs}) in conjunction with the limiting behavior in Eq.~(\ref{Bessellimits}),
we formally obtain
\beqn
      && \frac{4}{a} \,\lim_{M\to 0}\, \pStr \mathbbX_2 \left\lbrack \sum_{r=1}^\infty  
            K_0\left(  \frac{ r  M}{a\calM} \right) \right\rbrack\nonumber\\
      && ~~~~=~ \frac{4}{a} \,\zStr \mathbbX_2 \sum_{r=1}^\infty \left[  -\log\left( \frac{rM}{2a\calM}\right) -\gamma \right]~.~~~~~
\label{badstuff}
\eeqn
Unfortunately, this infinite $r$-summation is not convergent. 
It also does not correspond to what is presumably the 
exact {\mbox{$M=0$}} result within $P_\calX$.
We stress that these complications arise only when {\mbox{$\zStr \mathbbX_2\not=0$}}, 
which is precisely the condition under which the original unregulated Higgs mass is divergent.

In order to better understand this phenomenon,
we can perform a more sophisticated analysis 
by analytically performing the $r$-summation 
in complete generality before taking the {\mbox{$M\to 0$}} limit.
We begin by defining the Bessel-function combinations
\beq
   \mathbbK_\nu(z) ~\equiv~ 2\, \sum_{r=1}^\infty \, (r z)^{-\nu} K_\nu (r z)~.
\label{bbK}
\eeq
These Bessel-function combinations are relevant for both $P_\calX$ and $P_\Lambda$
in the same way that the combinations $\calK^{(n,p)}_\nu(z)$  
in Eq.~(\ref{Besselcombos})
were relevant for $\widehat m_\phi^2\bigl|_\calX$ and $\widehat \Lambda$, and indeed
\beq
     \calK^{(-\nu,p)}_\nu (z) ~=~ \half \bigl[
              \rho^{-\nu} \,\mathbbK_\nu(z/\rho) - \rho^p \,\mathbbK_\nu(z) \bigr]~.
\eeq 
Using the techniques in Ref.~\cite{Paris}, it is then straightforward (but exceedingly tedious) 
to demonstrate that $\mathbbK_\nu(z)$ for {\mbox{$z\ll 1$}} has a 
Maclaurin-Laurent series representation given by
\begin{widetext}
\beqn
\mathbbK_{\nu}(z) ~&=&~ 
        \sum_{p=1}^{\nu}\, 
                 2^{-\nu}\pi^{p}
             \frac{(-1)^{\nu-p}}{\left(\nu-p\right)!}
             \zeta^{\ast}(2p)
               \left(\frac{z}{2}\right)^{-2p} 
    ~+~ 2^{-\nu} \sqrt{\pi}  \,\Gamma\left( \half-\nu\right)\, \frac{1}{z} \nonumber\\
&&~~~~+~ 
       \frac{(-2)^{-\nu}}{\nu!}\left[\gamma-\frac{H_\nu}{2}+\log\left(z/4\pi\right)\right]
    ~+~ \sum_{p=1}^{\infty} \,2^{-\nu}\pi^{-p}
                \frac{(-1)^{\nu+p}}{\left(\nu+p\right)!}
             \zeta^{\ast}(2p+1)
                \left(\frac{z}{2}\right)^{2p}
\label{lyon}
\eeqn
                                                                                                                                             where {\mbox{$H_n\equiv \sum_{k=1}^{n}1/k$}} is the $n^{\rm th}$ harmonic number 
and where {\mbox{$\zeta^\ast(s) \equiv \pi^{-s/2} \Gamma(s/2) \zeta(s) = \zeta^\ast(1-s)$}} 
is the ``completed'' Riemann $\zeta$-function. 
The representation in Eq.~(\ref{lyon})  
is particularly useful for {\mbox{$z\ll 1$}}, allowing us to extract
the leading behaviors 
\beqn
\mathbbK_{0}(z) ~&=&~\frac{\pi}{z}+\left[\gamma+\log\left(\frac{z}{4\pi}\right)\right]
            -\frac{\zeta(3)z^2}{8\pi^{2}}+\frac{3\zeta(5)z^4}{128\pi^{4}}+\ldots\nonumber \\
\mathbbK_{1}(z) ~&=&~ \frac{\pi^{2}}{3z^{2}}-\frac{\pi}{z}
          -\frac{1}{2}\left[\gamma-\frac{1}{2} +\log\left(\frac{z}{4\pi}\right)\right]
             +\frac{\zeta(3)z^2}{32\pi^{2}} -\frac{\zeta(5)z^4}{256\pi^{4}}+\ldots\nonumber \\
\mathbbK_{2}(z) ~&=&~  \frac{2\pi^{4}}{45z^{4}}-\frac{\pi^{2}}{6z^{2}}+\frac{\pi}{3z}
            +\frac{1}{8}\left[\gamma-\frac{3}{4}+\log\left(\frac{z}{4\pi}\right)\right]
             -\frac{\zeta(3)z^2}{192\pi^{2}}+\frac{\zeta(5)z^4}{2048\pi^{4}}+\ldots
\label{Kseries}
\eeqn
\end{widetext}
Indeed, use of the expression for $\mathbbK_1(z)$ confirms our result in Eq.~(\ref{hidden}).

Armed with the expression for $\mathbbK_2(z)$ in Eq.~(\ref{Kseries}), we can now rigorously evaluate the leading terms within
$P_\Lambda(a)$ ---
and by extension within $\widehat \Lambda(\rho,a)$ --- in complete generality, even when massless states
are included.
Starting from Eq.~(\ref{Pacos}) in conjunction with the replacement in Eq.~(\ref{substitution}),
we now have
\beqn
  P_\Lambda(a) ~&=&~ \frac{\calM^2}{24a} \,\Str\, M^2 
                 - \frac{1}{4\pi^2 a} \,\Str\, M^4 \,\mathbbK_2\!\left( \frac{M}{a\calM}\right)~\nonumber\\
      ~&\approx&~ \frac{\calM^2}{24a} \,\Str\, M^2 
                 - \frac{1}{4\pi^2 a} ~\aMeffStr\, M^4 \,\mathbbK_2\!\left( \frac{M}{a\calM}\right)~\nonumber\\
\label{Plam}
\eeqn
where the final supertrace on the first line is over {\it all}\/ states in the theory, including those that are massless,
and where in passing to the second line we have recognized that $\mathbbK_2(z)$ is exponentially suppressed unless {\mbox{$z\ll 1$}}.
The fact that $\mathbbK_2(z)$ is now explicitly restricted to the {\mbox{$z\ll 1$}} regime implies that it is legitimate
to  insert the series expansion for $\mathbbK_2(z)$ from Eq.~(\ref{Kseries}) within Eq.~(\ref{mtoP}).
Identifying the physical scale $\mu$ as in Eq.~(\ref{mudef}) 
and retaining only the leading terms for {\mbox{$\mu\ll M_s$}},
we then obtain
\beqn
 \widehat \Lambda(\mu,\phi) \,&=&~  \frac{1}{1+\mu^2/M_s^2} \Biggl\lbrace \nonumber\\
       && \phantom{-}\frac{1}{24}\calM^2 \,\Str\, M^2  +  \zeffStr \left( \frac{M^2 \mu^2}{96\pi^2}  
            - \frac{7\mu^4}{960\pi^2}\right) \nonumber\\ 
   && - \frac{1}{32\pi^2} \,\zeffStr M^4 \log\left( \sqrt{2}\, e^{\gamma+1/4} \frac{M}{\mu}\right)+...\Biggr\rbrace ~\nonumber\\
 ~&=&~  \frac{1}{24}\calM^2 \,\Str\, M^2    
           -\Str\, \frac{M^2 \mu^2}{96\pi^2} \nonumber\\
     &&  + \zeffStr \left( \frac{M^2 \mu^2}{96\pi^2}  
            - \frac{7\mu^4}{960\pi^2} \right) \nonumber\\ 
   && - \frac{1}{32\pi^2} \,\zeffStr M^4 \log\left( \sqrt{2}\, e^{\gamma+1/4} \frac{M}{\mu}\right)+... \nonumber\\
\label{intermed2}
\eeqn
where we have continued to adopt our benchmark value {\mbox{$\rho=2$}}
and where we recall that each factor of $M$ carries a $\phi$-dependence through Eq.~(\ref{TaylorM}).
Note that in passing to the final expression in Eq.~(\ref{intermed2}) we have Taylor-expanded the overall 
prefactor and kept only those terms of the same order as those already shown.
However, we now see that the $\mu^2$ term from expanding the prefactor cancels the corresponding $\mu^2$
term from $\mathbbK_2$, leaving behind a net $\mu^2$ term which scales as the $M^2$ supertrace 
of only those states whose masses {\it exceed}\/ $\mu$.
We thus obtain our final result
\beqn
 \widehat \Lambda(\mu,\phi) \,&=&\, 
  \frac{1}{24}\calM^2 \,\Str\, M^2    
           -\antieffStr \frac{M^2 \mu^2}{96\pi^2}  
            - \zeffStr \frac{7\mu^4}{960\pi^2} \nonumber\\ 
   && - \frac{1}{32\pi^2} \,\zeffStr M^4 \log\left( \sqrt{2}\, e^{\gamma+1/4} \frac{M}{\mu}\right) \,+\, ... \nonumber\\
\label{Lambdafull}
\eeqn
 Indeed, this result provides the leading approximation to the exact expression
in Eq.~(\ref{lambdamuresult}).

The first and third terms in this result are consistent with 
those in Eq.~(\ref{lambdamuresult2}), and indeed the $\mu^4$ term is the contribution
from the massless states within the second supertrace in Eq.~(\ref{Plam}). 
However, to this order, we now see that there 
are two additional terms.
The first is a term scaling as $\mu^2$ which depends on the spectrum of states with masses {\mbox{$M\gsim \mu$}}.
This contribution lies outside the range {\mbox{$M\ll \mu$}} studied in Eq.~(\ref{lambdamuresult2})
but nevertheless generally appears for {\mbox{$M\gsim \mu$}}.
The second is a logarithmic term. 
This term is subleading when compared to the other terms shown, and 
massless states make no contribution to this term (divergent or otherwise) when evaluated at {\mbox{$\phi=0$}} because
of its $M^4$ prefactor.

This logarithmic term is nevertheless of critical importance 
when we consider the corresponding Higgs mass. 
As we have seen in Eq.~(\ref{finalCWresult}), the Higgs mass $\widehat m_\phi^2(\mu)$
receives a contribution which scales as $\partial_\phi^2 \widehat \Lambda(\mu)$.
Of course, all of the dependence on $\phi$ is carried within the masses $M$ which
appear in Eq.~(\ref{Lambdafull}), and as expected $\widehat\Lambda(\mu)$ depends not
on these masses directly but on their squares.
However, for any function $f(M^2)$ we have the algebraic identity
\beq
  \partial_\phi^2 f(M^2) 
               ~=~  (\partial_\phi^2 M^2)  \frac{\partial f}{\partial M^2} +
               (\partial_\phi M^2)^2 
            \frac{\partial^2 f}{(\partial M^2)^2}~.~~
\eeq 
Thus, identifying {\mbox{$f\sim \widehat\Lambda(\mu)$}} 
and recalling Eq.~(\ref{newXi}),  we obtain
\beqn
       \partial_\phi^2 \,\widehat \Lambda(\mu) \Bigl|_{\phi=0}  &=& 
        \left. -4\pi \,\Str \,\mathbbX_1 \,\frac{\partial \widehat \Lambda(\mu)}{\partial M^2}\right|_{\phi=0} \nonumber\\
       && ~~~+ \left. 16\pi^2 \calM^2 \,\Str \,\mathbbX_2 \,\frac{\partial^2 \widehat \Lambda(\mu)}{(\partial M^2)^2}\right|_{\phi=0}~
         ~~~~~~~~
\label{derivs}
\eeqn
where we have implicitly used the fact that only non-negative powers of $\phi$ appear within $\widehat\Lambda(\mu)$,
thereby ensuring that our truncation to {\mbox{$\phi=0$}} factorizes within each term.

In principle, both supertraces in Eq.~(\ref{derivs}) include massless states.
Moreover, we see that the $\Str \, \mathbbX_1$ term is proportional to the single 
$M^2$-derivative of $\widehat\Lambda(\mu)$, and when acting on the logarithm term within Eq.~(\ref{Lambdafull})
we find that massless states continue to be harmless, yielding no contribution (and therefore no divergences).
By contrast, we see that the $\Str \, \mathbbX_2$ term is proportional 
to the {\it second}\/ $M^2$-derivative of $\widehat \Lambda(\mu)$.
This derivative therefore leaves behind a logarithm with no leading $M^2$ factors remaining.
Thus, for {\mbox{$M= 0$}}, we obtain a logarithmic divergence for the Higgs mass --- as expected ---
so long as {\mbox{$\zStr \,\mathbbX_2\not=0$}}.
Indeed, all of this information is now directly encoded within the effective potential
$\widehat\Lambda(\mu)$ for this theory, as given in Eq.~(\ref{Lambdafull}).  

This situation is analogous to the behavior 
of the traditional Coleman-Weinberg potential $V(\varphi_c)$ as originally given in Refs.~{\mbox{\cite{Coleman:1973jx,Weinbergpost}}}.
In that case, it was shown that $V(\varphi_c)$ contains a term
scaling as 
\beq
          V(\varphi_c) ~\sim~ \varphi_c^4 \,\log \varphi_c^2
\eeq
where $\varphi_c$ are the fluctuations of the classical Higgs field around its VEV
and where one has assumed a $U(1)$-charged scalar field subject to a $\lambda \phi^4$ interaction.
The Higgs mass (which goes as the second derivative $\partial^2 V/\partial \varphi_c^2$)
therefore remains finite even as {\mbox{$\varphi_c\to 0$}}, whereas the {\it fourth}\/ derivative $\partial^4 V/\partial \varphi_c^4$
actually has a logarithmic singularity as {\mbox{$\varphi_c\to 0$}}.
Indeed, this fourth derivative describes the behavior of the coupling $\lambda$.
The cure for this disease, as suggested in Refs.~{\mbox{\cite{Coleman:1973jx,Weinbergpost}}}, is to
move away from the {\mbox{$\varphi_c=0$}} origin, and instead define the coupling $\lambda$ at this shifted point.

Of course, in our more general string context, we see that our potential scales like $M^4 \log M$.   Moreover, within the $\mathbbX_2$ term,
it is not the fourth derivative with respect to $M$ which leads to difficulties --- rather, it is the {\it second}\/ derivative
with respect to $M^2$.   As a consequence, this logarithmic divergence shows up in the Higgs mass rather than in a four-point coupling.
That said, it is possible that the cure for this disease may be similar to that discussed in 
Refs.~{\mbox{\cite{Coleman:1973jx,Weinbergpost}}}.
In particular, this suggests  that in string theories for which {\mbox{$\zStr \mathbbX_2\not=0$}}, a cure for our logarithmically divergent
Higgs mass and the fact that
radiative potential is not twice-differentiable there 
may be similarly found by avoiding the sharp {\mbox{$\phi=0$}} truncation that originally appears in Eq.~(\ref{higgsdef}),
and by instead deforming our theory away from the {\mbox{$\phi=0$}} origin in $\phi$-space.

Finally, given that we are now equipped with our effective Higgs potential $\widehat \Lambda(\mu,\phi)$, 
we can revisit our classical stability condition, as originally discussed in Sect.~\ref{stability}.~
In general, our theory will be sitting at an extremum of the potential as long as 
\beq
         \partial_\phi  \widehat\Lambda(\mu,\phi)\Bigl|_{\phi=0}  ~=~0~.
\label{stabcond}
\eeq
This, then, is a supplementary condition that we have implicitly assuming to be satisfied within our analysis.
Note that 
\beqn
         \partial_\phi \widehat\Lambda(\mu,\phi)\Bigl|_{\phi=0} 
         ~&=&~  \left. (\partial_\phi M^2) \frac{\partial \widehat \Lambda(\mu)}{\partial M^2} \right|_{\phi=0}\nonumber\\ 
         ~&=&~ 4\pi \calM \,\Str \,\mathbbY    \left.\frac{\partial \widehat \Lambda(\mu)}{\partial M^2} \right|_{\phi=0} ~
\label{connecttoY}
\eeqn
where
\beq
           \mathbbY ~\equiv~ \frac{1}{ 4\pi\calM} \,(\partial_\phi M^2)\Bigl|_{\phi=0}~.
\label{Ydef2}
\eeq
Indeed, for the case of theories with an underlying charge lattice, we have {\mbox{$\mathbbY = \tau_2^{-1} \calY$}}
where $\calY$ is given in Eq.~(\ref{Ydef}).
Thus the stability of the theory (and the possible existence of a destabilizing $\phi$-tadpole) is closely 
related to the values of $\mathbbY$ across the string spectrum, as already anticipated in Sect.~\ref{stability}.~
Substituting the exact expression in Eq.~(\ref{lambdamuresult})
into Eq.~(\ref{connecttoY}),
we find
\beqn
       &&  \partial_\phi \widehat\Lambda(\mu,\phi)\Bigl|_{\phi=0} 
         \,=~ \frac{\calM^3}{1+\mu^2/M_s^2} 
     \,\Str \,\mathbbY \,\Biggl\lbrace   
      \frac{\pi}{6}
         + \frac{1}{2\pi} \left(\frac{M}{\calM}\right)^2 \times \nonumber\\  
     && ~~~~~~~~\times 
      \left\lbrack    
        \calK_0^{(0,1)}\!\left( \frac{2\sqrt{2}\pi M}{\mu} \right) + 
        \calK_2^{(0,1)}\!\left( \frac{2\sqrt{2}\pi M}{\mu} \right)  
      \right\rbrack \Biggr\rbrace \nonumber\\ 
  && ~~~=~   \frac{\calM^3}{1+\mu^2/M_s^2} ~\Biggl\lbrace
      \zStr \mathbbY \left\lbrack  \frac{\pi}{6}\left(1+\mu^2/M_s^2\right)   \right\rbrack \nonumber\\
  && ~~~~~~~~+ \pStr \mathbbY \,\Biggl\lbrace   
      \frac{\pi}{6}
         + \frac{1}{2\pi} \left(\frac{M}{\calM}\right)^2 \times \nonumber\\  
     && ~~~~~~~~\times 
      \left\lbrack    
        \calK_0^{(0,1)}\!\left( \frac{2\sqrt{2}\pi M}{\mu} \right) + 
        \calK_2^{(0,1)}\!\left( \frac{2\sqrt{2}\pi M}{\mu} \right)  
      \right\rbrack \Biggr\rbrace \Biggr\rbrace~ \nonumber\\ 
\label{explicitBessel}
\eeqn
where in passing to the second expression we have explicitly separated the
contributions from the massless and massive string states, 
and where 
$\calK_\nu^{(n,p)} (z)$ continue to denote
the combinations of Bessel functions in Eq.~(\ref{Besselcombos}).

Interestingly (but not unexpectedly), the terms multiplying $\Str\,\mathbbY$ 
in Eq.~(\ref{explicitBessel}) are the same as the terms multiplying $\Str\,\mathbbX_1$ 
in Eq.~(\ref{finalhiggsmassmu}).
Equivalently, we can view the quantity in Eq.~(\ref{explicitBessel}) as the coefficient
of the tadpole term (linear in $\phi$) within the effective potential
$\Lambda(\mu,\phi)$ in Eq.~(\ref{lambdamuresult}) when the masses 
are Taylor-expanded as in Eq.~(\ref{TaylorM}).

There are several ways in which the expressions in Eq.~(\ref{explicitBessel})
might vanish for all $\mu$, as required for a stable vacuum.
In principle, for a given value of $\mu$, there might exist a spectrum of states
with particular masses $M$ such that the contributions from the Bessel and
non-Bessel terms together happen to cancel when tallied across the spectrum.   
Any continuous change in the 
value of $\mu$ might then induce a corresponding continuous change in the 
spectrum such that this cancellation is maintained.   This is not unlike what
happens in the traditional field-theoretic Coleman-Weinberg potential, where changing the
scale $\mu$ can change the vacuum state and the spectrum of excitations built upon it.
Of course in the present case we are 
working within the context of string theory rather than field theory.
As such, we are dealing with an infinite tower of 
string states and simultaneously maintaining modular invariance as the
spectrum is deformed.

Another possibility is to simply demand stability in the deep infrared region,
as {\mbox{$\mu\to 0$}}.   From Eq.~(\ref{explicitBessel}) 
we see that this would then require simply that 
\beq
       \Str \,\mathbbY ~=~0~
\label{fullsutrace}
\eeq
where the supertrace is over all string states, both massless and massive.

A final possibility is to guarantee stability for every value of $\mu$ by demanding
the somewhat stronger condition
\beq
       ~~~~~\Str \,\mathbbY \,=\,0~ ~~{\rm for~each~mass~level~individually}.~~
\label{bestcondition}
\eeq
Of course, Eq.~(\ref{bestcondition}) implies Eq.~(\ref{fullsutrace}), but 
the fact that $\Str\,\mathbbY$ vanishes for each mass level {\it individually}\/ ensures
that stability no longer rests on any $\mu$-dependent cancellations involving the
Bessel functions.

Comparing Eq.~(\ref{Ydef2}) with  Eq.~(\ref{newXi}), we see that {\mbox{$\mathbbX_2 = \mathbbY^2$}}.
However,
as discussed below Eq.~(\ref{Ydef}),
constraints on $\mathbbY$ do not necessarily become constraints on $\mathbbX_2$, even if the values of
$\mathbbY$ happen to cancel pairwise amongst degenerate states across the string spectrum
[which would guarantee Eq.~(\ref{bestcondition})]. 
Thus the requirement of stability does not necessarily lead 
to any immediate constraints on the supertraces of $\mathbbX_2$.

In summary, then, we have shown that for theories with {\mbox{$\zStr\,\mathbbX_2=0$}} there exists an effective Higgs potential
$\widehat\Lambda(\mu,\phi)$ from which the Higgs mass can be obtained through the modular-covariant double-derivative $D^2_\phi$,
as in Eq.~(\ref{finalCWresult}).
This effective potential is given exactly in Eq.~(\ref{lambdamuresult}), with
the leading terms given in Eq.~(\ref{Lambdafull}).
By contrast, for theories with {\mbox{$\zStr \,\mathbbX_2\not=0$}} we have found
that the effective potential $\widehat \Lambda(\mu,\phi)$ 
picks up an additional contribution whose second  
derivative is discontinuous at {\mbox{$\phi=0$}}.
In this sense, the Higgs mass is not well defined at {\mbox{$\phi=0$}}.
Of course, one option is to retain the expression obtained in Eq.~(\ref{finalhiggsmassmu});
this expression is not the second derivative of $\widehat\Lambda(\mu,\phi)$ when {\mbox{$\zStr\,\mathbbX_2\not =0$}},
but it is indeed finite except as {\mbox{$\mu\to 0$}}.
An alternative option is to define our Higgs mass away from the {\mbox{$\phi=0$}} origin.
Either way, these features exactly mirror those seen within the traditional
Coleman-Weinberg potential.

\section{Pulling it all together:  Discussion,  top-down perspectives, and future directions \label{sec:Conclusions}}

A central question when analyzing any string theory is to understand the properties of  
its ubiquitous scalars ---  its Higgs fields, its moduli fields, its axions, and so forth. 
To a great extent the behavior of a scalar is dominated by its 
mass, and in this paper we have developed a completely general 
framework for understanding the masses of such scalars at one-loop order in
closed string theories.
Our framework can be applied at all energy scales, is independent of any supersymmetry, and 
maintains worldsheet modular invariance and hence finiteness at all times. 
Moreover, our framework is entirely string-based and does not rely on establishing any 
particular low-energy effective field theory.  Indeed the notion of an effective 
field theory at a given energy scale ends up being an {\it output}\/ of our analysis,
and we have outlined the specific conditions and approximations under which such an EFT emerges
from an otherwise completely string-theoretic calculation.

Beyond the crucial role played by the scalar mass, another 
motivation for studying this quantity is its special status as 
the ``canary in the coal mine'' for UV completion.  
The scalar mass term is virtually the only operator that 
is both highly UV-sensitive and also IR-divergent when coupled to massless states. 
Thus, once we understand this operator, we understand much of the entire structure of the theory.  

We can appreciate the special status of this operator if we think about a typical EFT.~
Within such an EFT, the familiar result for the one-loop contributions to the Higgs mass 
takes the general form 
\begin{equation}
m_{\phi}^{2}~=~\frac{M_{\text{UV}}^{2}}{32\pi^{2}}\,\eftStr\,\partial_{\phi}^{2}M^{2}\,-\eftStr\,\partial_{\phi}^{2}\left[\frac{M^{4}}{64\pi^{2}}\log\left(c\frac{M^{2}}{M_{\text{UV}}^{2}}\right)\right]~
\label{eq:CW}
\end{equation}
where $M_{\rm UV}$ is an ultraviolet cutoff, where $c$ is a constant,
and where $\eftStr$ denotes a supertrace over the states in the effective theory.
This expression has both
a quadratic UV-divergence which we would normally subtract by a counter-term as well as a logarithmic
cutoff dependence which would normally be indicative of RG running.  Thus any UV-completion such as string theory
has to resolve two issues within this 
expression at once:  not only must it make the quadratic term finite, but it must also
be able to give us specific information about the running.    In particular, 
to what value does the Higgs mass actually run in the IR?~
Such information is critical in order to nail down the logarithmic running, anchoring 
it firmly as a function of scale.  

Prior to our work, such questions remained unanswered.
In retrospect, one clue could already be found in the earlier work of Ref.~\cite{Dienes:1995pm}, which in turn
 rested on previous results in Ref.~\cite{Kutasov:1990sv}.
In Ref.~\cite{Dienes:1995pm}, it was shown that
the one-loop cosmological constant $\Lambda$ 
for any non-tachyonic closed string 
can be expressed as a supertrace over
the entire infinite spectrum of level-matched physical string states:
\beq
 \Lambda~=~ \frac{1}{24}\,\calM^2\, \Str\, M^2~.
\label{eq:lam-rep}
\eeq
This result, which we have rederived in Eq.~(\ref{eq:lamlam}),
immediately suggests two things.
The first is that it might be possible to derive an analogous spectral supertrace formula
for the one-loop Higgs mass within such strings which, like that in Eq.~(\ref{eq:lam-rep}), depends on only the physical states in the theory.
The second, stemming from a comparison between the result in Eq.~(\ref{eq:lam-rep}) and 
the first term in Eq.~(\ref{eq:CW}), is that there might exist a possible
derivative-based connection between the one-loop Higgs mass and the one-loop cosmological constant.

In this study, we have addressed all of these issues.
Indeed, one of the central results of our study is an
equivalent spectral supertrace formula for the one-loop Higgs mass.
Like the calculation of the cosmological constant,
our calculation for the Higgs mass relies on nothing more than worldsheet modular invariance ---
an exact symmetry which maintains string finiteness and is preserved, even today.
Another of our central results is a deep connection between 
the Higgs mass and the cosmological constant.
However, we also found that unlike the cosmological constant, the Higgs mass may
actually have a leading logarithmic divergence.
Indeed, this issue depends on the particular string model under study,
and in particular the presence of massless states carrying specific charges.
As a result of this possible divergence,
and as a result of the extreme sensitivity of the Higgs mass to physics at all scales, arriving at a fully consistent treatment of the Higgs mass 
required us to broach several delicate issues.  
These encompassed varied aspects of regularization and renormalization and touched  
on the very legitimacy of extracting an effective field theory from a UV/IR-mixed theory. 
The scope of our study was therefore quite broad, with a number of 
important insights and techniques developed along the way.

Our first step was to understand how the Higgs and similar scalars
reside within a typical modular-invariant string theory. 
In particular, for closed string theories with charge lattices, 
we began by examining the manner in which fluctuations
of the Higgs field deform these charge lattices, all the while bearing in mind that these
deformations must preserve modular invariance.  
We were then able to express the contributions to the Higgs mass in terms of 
one-loop modular integrals with specific charge insertions $\calX_i$ incorporated into the
string partition-function traces.
However, we found that these insertions have an immediate consequence, producing
a modular anomaly which then requires us to 
perform a ``modular completion'' of the theory.
 This inevitably introduces an additional term into the Higgs mass, one which 
is directly related to the one-loop cosmological constant.
Our derivation of this term rested solely on considerations of modular invariance and thereby
endows this result with a generality 
that holds across the space of perturbative closed-string models.
In this way we arrived at one of the central conclusions of our work, namely the existence
of a universal relation between scalar masses and the cosmological constant in any tachyon-free closed string theory.
This relation is given in Eq.~(\ref{relation1}) for four-dimensional theories,
and in Eq.~(\ref{relation1b}) for theories in arbitrary spacetime dimensions $D$. 
Stemming only from modular invariance, this result is exact and holds regardless of 
other dynamics that the theory may experience.

Having established the generic structure of one-loop contributions to the Higgs mass,
we then pushed our calculation one step further with the aim of expressing our 
result for the Higgs mass as a supertrace over the purely physical level-matched spectrum of the theory. 
Indeed, we demonstrated that the requirements of modular covariance so deeply constrain 
the contributions to the Higgs mass from the unphysical states that these latter
contributions can be expressed in terms of contributions from the physical states
alone.  However, part of this calculation required dealing with the 
logarithmic divergences which can arise.
This in turn required that we somehow {\it regularize}\/ the Higgs mass.  

For this reason, we devoted a large portion of our study to establishing
a general formalism for regulating quantities such as the Higgs mass that
emerge in string theory. We initially considered two forms of 
what could be called ``standard'' regulators.
The ``minimal'' regulator is essentially a subtraction of the contributions
of the massless states.   We referred to this as a minimal regulator because
it does not introduce any additional parameters into the theory.   Thus, for
any divergent quantity, there is a single corresponding regulated quantity.
We also discussed what we referred to as a ``non-minimal'' regulator,
based on a mathematical regularization originally introduced in the mathematics
literature~\cite{zag}.   
This regulator introduces a new dimensionless parameter $t$, so that for any divergent quantity
there exist a set of corresponding regularized quantities parametrized by $t$, with the limit
{\mbox{$t\to \infty$}} corresponding to the removal of the regulator and the restoration of the original unregulated quantity.
This regulator is essentially the one used in Ref.~\cite{Kaplunovsky:1987rp} and later in Ref.~\cite{Dienes:1996zr}.

As we have explained in Sect.~\ref{sec3}, both of these regulators yield finite quantities which can be expressed in terms of supertraces over only those string states which are physical (\ie, level-matched).    Indeed, in each of these cases,
the relation between the regulated quantities and the corresponding supertraces respects modular invariance.  Thus, the regulated quantity and the corresponding supertrace in each case transform identically under modular transformations.   However, for both the minimal and non-minimal regulators, neither the regulated quantity nor the corresponding supertrace expression is modular invariant on its own.   While this additional criterion 
was not important for the purposes that led to the original development of these regulators in the mathematics literature,
this criterion is critical for us because we now wish these regulated quantities to correspond to physical observables (such as our regulated Higgs mass).   Each of these regulated quantities must therefore be independently modular invariant on its own. 

We therefore presented a third set of regulators --- those based on the functions $\widehat\calG_\rho(a,\tau)$.
These are our modular-invariant regulators, and they depend on two free parameters $(\rho,a)$.
Unlike the minimal and non-minimal regulators, 
these regulators do not operate by performing a sharp, brute-force subtraction of particular contributions 
within the integrals associated with one-loop string amplitudes.
Instead, we simply multiply the integrand of any one-loop string amplitude by the regulator function $\widehat \calG_\rho(a,\tau)$.  These functions have two important properties which 
make them suitable as regulators when {\mbox{$a\ll 1$}}.
First, as {\mbox{$a\to 0$}}, we find that {\mbox{$\widehat\calG_\rho(a,\tau)\to 1$}} for all $\tau$.
Thus the {\mbox{$a\to 0$}} limit restores our original unregulated theory.
Second,  {\mbox{$\widehat\calG_\rho(a,\tau)\to 0$}} exponentially quickly as {\mbox{$\tau\to i\infty$}} for all {\mbox{$a>0$}}.   
These functions thereby suppress all relevant divergences
which might appear in this limit.
But most importantly for our purposes, $\widehat \calG_\rho(a,\tau)$ is completely modular invariant.
In particular, this function is completely smooth, with no sharp transitions in its behavior.
As a result, multiplying the integrand of any one-loop string amplitude by $\widehat\calG_\rho(a,\tau)$ 
does not simply excise certain problematic contributions within the corresponding string amplitude, 
but rather provides a smooth, modular-invariant way of deforming (and thereby regulating) the entire theory.
This function even has a physical interpretation in the {\mbox{$\rho=2$}} special case, arising as 
the result of the geometric deformations discussed in  Refs.~\cite{Kiritsis:1994ta, Kiritsis:1996dn, Kiritsis:1998en}.

Armed with this regulator, we then demonstrated that 
our regulated Higgs mass can be expressed as the 
supertrace over only the physical string states.
Our result for $\widehat m_\phi^2(\rho,a)$ is given in Eq.~(\ref{twocontributions}),
where $\widehat m_\phi^2(\rho,a)\bigl|_\calX$ is given in 
Eq.~(\ref{finalhiggsmassa})
and where $\widehat \Lambda(\rho,a)$ is given in  
Eq.~(\ref{lambdaresult}).
We stress that this is the exact string-theoretic result for the
regulated Higgs mass, expressed as a function of the regulator parameters $(\rho,a)$.
Moreover $\widehat \Lambda(\rho,a)$ by itself is the corresponding regulated
cosmological constant.   As discussed in the text, 
the one-loop cosmological constant $\Lambda$ requires regularization in this context
even though it is already finite in all tachyon-free closed string theories.

We originally derived these results under the assumption that our underlying
string theory could be formulated with an associated charge lattice.
This assumption gave our calculations a certain concreteness,
allowing us to see exactly which states 
with which kinds of charges ultimately contribute to the Higgs mass.
However, we then proceeded to demonstrate that many of our results are actually more
general than this, and do not require a charge lattice at all.
This lattice-free reformulation also had an added benefit, allowing us to demonstrate a second
deep connection between the Higgs mass and the cosmological constant beyond that in Eq.~(\ref{relation1}).   
In particular, we were able to demonstrate that each of these quantities can be expressed
in terms of a common underlying quantity $\widehat \Lambda(\rho,a,\phi)$ via
relations of the form
\beq
   \begin{cases}
      ~\widehat \Lambda(\rho,a) &=~ \widehat \Lambda(\rho,a,\phi)\bigl|_{\phi=0} \\
      ~\widehat m_\phi^2 (\rho,a) &=~ D_\phi^2 \,\widehat \Lambda(\rho,a,\phi)\bigl|_{\phi=0}
   \end{cases}
\label{effpotl}
\eeq
where $D_\phi^2$ is the modular-covariant second $\phi$-derivative given in Eq.~(\ref{Dphi2}). 
These relations allow us to interpret
$\widehat\Lambda(\rho,a,\phi)$ as a stringy effective potential for the Higgs.
Indeed, these relations are ultimately the fulfillment of our original suspicion
that the Higgs mass might be related to the cosmological constant through a double $\phi$-derivative,
as discussed below Eq.~(\ref{eq:lam-rep}).
However, we now see from Eq.~(\ref{Dphi2}) that this is not just an ordinary $\phi$-derivative $\partial_\phi^2$, but
rather a {\it modular-covariant}\/ derivative, complete with anomaly term.   
The second relation in Eq.~(\ref{effpotl}) thereby {\it subsumes}\/
our original relation between the Higgs mass and the cosmological constant,
as expressed in Eq.~(\ref{relation1}).   Moreover, we see that 
the regulated
cosmological constant $\widehat\Lambda(\rho,a)$ is nothing but the {\mbox{$\phi=0$}} truncation of 
the same effective potential $\widehat\Lambda(\rho,a,\phi)$.
In this way, $\widehat \Lambda(\rho,a,\phi)$ emerges as the central object 
from which our other relevant quantities can be obtained.

At no step in the derivation of these results was modular invariance broken.
Thus all of these results are completely consistent with modular invariance, as required.
Moreover, expressed as functions of the worldsheet regulator parameters $(\rho,a)$, all of our
quantities are purely string-theoretic and there are no ambiguities in their definitions.

Our next goal was to interpret these regulated quantities in terms of a physical cutoff scale $\mu$.
Of course, if we had been working within a field-theoretic context, 
all of our regulator parameters would have had direct spacetime interpretations in terms of a spacetime scale $\mu$.
As a result, varying the values of these regulator parameters would have 
led us directly to a renormalization-group flow with an associated RGE.~
String theories, by contrast, are formulated 
not in spacetime but on the worldsheet --- for such strings, spacetime is nothing but a derived quantity.
As a result, although we were able to express our regulated quantities as functions of the two regulator parameters $(\rho,a)$,
the only way to extract an EFT description from these otherwise complete string-theoretic
expressions was to develop
a mapping between the worldsheet parameters $(\rho,a)$ and a physical spacetime scale $\mu$.

As we have seen, this issue of connecting $(\rho,a)$ to $\mu$ is surprisingly subtle,
and {\it it is at this step that we must make certain choices that break modular invariance.}\/
We already discussed some the issues surrounding IR/UV equivalence in Sect.~\ref{UVIRequivalence} ---
indeed, these issues already suggested that the passage to an EFT would be highly non-trivial
and involve the breaking of modular invariance.
But now, with our complete results in hand, we can take a bird's-eye view and finally map out the full structure of the problem.

\begin{figure*}[t!]
\centering
\includegraphics[keepaspectratio, width=0.95\textwidth]{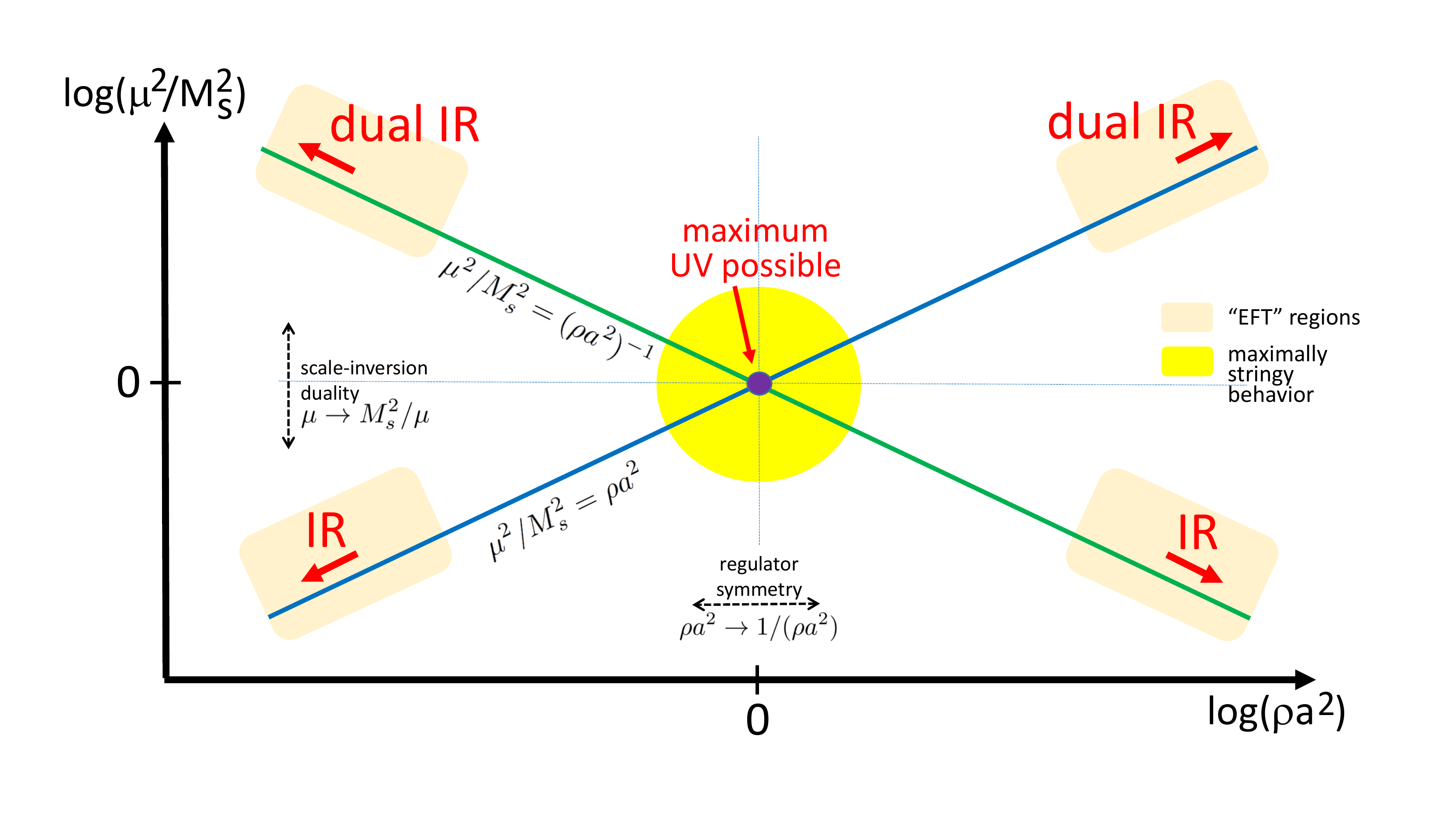}
\caption{
The scale structure of physical quantities in a modular-invariant string theory.
For the modular-invariant regulator function 
$\widehat \calG_\rho(a,\tau_2)$ discussed in this paper, 
the mapping between the regulator parameter $\rho a^2$ and the physical spacetime scale $\mu$
has two distinct branches.
The traditional branch (shown in blue) identifies {\mbox{$\mu^2/M_s^2 = \rho a^2$}}, 
but modular invariance implies 
the existence of an invariance under the scale-inversion duality symmetry 
{\mbox{$\mu\to M_s^2/\mu$}}.  This in turn implies 
the existence of a second branch 
(shown in green) on which we can alternatively identify
{\mbox{$\mu^2/M_s^2= 1/(\rho a^2)$}}.
Although $\widehat \calG_\rho(a,\tau_2)$ functions as a regulator for {\mbox{$\rho a^2<1$}},
its symmetry under {\mbox{$\rho a^2 \to 1/(\rho a^2)$}} 
implies that this function also acts as a regulator when 
extended into the {\mbox{$\rho a^2>1$}} region.
This then allows us to see the full four-fold modular structure of the theory.
The Higgs-mass plot shown in Fig.~\ref{anatomy} can now be understood as
following 
the {\mbox{$\mu^2/M_s^2 = \rho a^2$}} branch from the lower-left corner of this figure 
inward toward the central location at which {\mbox{$\mu=M_s$}},
after which it then follows the 
{\mbox{$\mu^2/M_s^2 = 1/(\rho a^2)$}} branch outward towards the upper-left corner.
However, in a modular-invariant theory, all four quadrants of this figure are equivalent and describe the same physics.
Likewise, in such theories there is no distinction between IR and UV.~
Thus one can exchange ``IR'' $\leftrightarrow$  ``UV'' within all labels of this sketch,
and  we have simply chosen to show 
those labels
that have the most natural interpretations
within the lower-left quadrant.  
Finally, 
regions with beige shading indicate locations where EFT descriptions exist, 
while stringy effects dominate in the yellow central region.
As a result, focusing on any one of the four EFT regions by itself
necessarily breaks modular invariance
because the choice of EFT region is tantamount to picking a preferred direction for the flow of the scale $\mu$ relative to 
the underlying string-theoretic regulator parameter $\rho a^2$. 
However, even within the EFT regions, string states {\it at all mass scales}\/ contribute non-trivially.
Thus even these EFT regions differ from what might be expected within quantum field theory.} 
\label{mappings_figure}
\end{figure*}

Our understanding of this issue is summarized in Fig.~\ref{mappings_figure}. 
Although the specific situation sketched in Fig.~\ref{mappings_figure} corresponds 
to our modular-invariant regulator function $\widehat\calG_\rho(a,\tau)$,
the structure of this diagram is general. 
Ultimately, the connection between worldsheet physics and spacetime physics
follows from the one-loop partition function, 
which for physical string states of spacetime masses $M_i$ takes the general form 
\beq
         {\cal Z}\/ ~\sim~  \tau_2^{-1} \, \sum_i \,e^{- \pi \alpha' M_i^2 \tau_2}~.
\label{pfZ}
\eeq
These $M_i$ are precisely the masses 
which ultimately appear in our physical supertrace formulas.
However, our regulator function $\widehat \calG_\rho(a,\tau)$ 
cannot regulate the divergences that might arise from light and/or massless states 
as {\mbox{$\tau\to i\infty$}} unless it suppresses contributions
to the partition function within the region {\mbox{$\tau_2\gsim \tau_2^\ast$}} for some $\tau_2^\ast$.  
We thus see that whether a given state contributes significantly to one-loop amplitudes
in the presence of the regulator
depends on the magnitude of $\alpha' M_i^2 \tau_2^\ast$.
This immediately leads us to identify a corresponding spacetime physical RG scale 
{\mbox{$\mu^2\sim 1/(\alpha' \tau^\ast_2)$}}.
Indeed, this was precisely the logic that originally led us to Eq.~(\ref{mut}).
Moreover, for our specific regulator function $\widehat\calG_\rho(a,\tau)$, we have 
{\mbox{$\tau_2^\ast \sim 1/(\rho a^2)$}},
thus leading to the natural identification {\mbox{$\mu^2/M_s^2=\rho a^2$}}.

However, modular invariance does not permit us to 
identify just one special
point {\mbox{$\tau_2= \tau_2^\ast$}} along the {\mbox{$\tau_1=0$}} axis within the fundamental domain. 
Indeed, for every such special point, the corresponding point with {\mbox{$\tau_2= 1/\tau_2^\ast$}} is equally special,
since these two points along the {\mbox{$\tau_1=0$}} axis are related by the {\mbox{$\tau\to -1/\tau$}} modular transformation.
In other words, although our $\widehat\calG_\rho(a,\tau)$ regulator function suppresses contributions
from the {\mbox{$\tau_2\gsim \tau_2^\ast$}} region, the modular invariance of this function requires that it also
simultaneously suppress contributions from the region with {\mbox{$\tau_2 \lsim 1/\tau_2^\ast$}}.
(In general a modular-invariant regulator function equally suppresses the contributions  
from the regions that approach {\it any}\/ of the modular cusps, but for the purposes of mapping to a physical
spacetime scale $\mu$  our concern is limited to the cusps along the {\mbox{$\tau_1=0$}} axis.)
We thus see that for any value of the spacetime scale $\mu$ that we attempt to identify as corresponding
to $\rho a^2$, there always exists a second scale $M_s^2/\mu$ which we might equally validly identify as
corresponding to $\rho a^2$.
This is the implication of the {\mbox{$\mu\to M_s^2/\mu$}} scale-inversion symmetry discussed in Sect.~\ref{sec4}.

The upshot of this discussion is that the mapping from $\rho a^2$ to $\mu$
in any modular-invariant theory actually
has {\it two branches}\/, as shown in Fig.~\ref{mappings_figure}.
Along the first branch we identify {\mbox{$\mu^2/M_s^2=1/\tau_2^\ast= \rho a^2$}}, but along the second
branch we identify  
{\mbox{$\mu^2/M_s^2=\tau_2^\ast= 1/(\rho a^2)$}}. 
These branches contain the same physics, but the choice of either branch breaks modular invariance.
In this respect modular invariance is much like another description-redundancy symmetry, namely gauge invariance:
all physical quantities must be gauge invariant, but the choice of 
a particular gauge slice (which is tantamount to the choice of a particular branch) 
necessarily breaks the underlying symmetry.

In most of our discussions in this paper,
we focused on the behavior of our 
regulator functions within the {\mbox{$a\ll 1$}} regime.
However, as we have seen in Eq.~(\ref{newest}),
these functions exhibit a symmetry under {\mbox{$\rho a^2 \to 1/(\rho a^2)$}}.
The logical necessity for this extra symmetry will be discussed below,
but this symmetry implies that $\widehat \calG_\rho(a,\tau)$ also acts as a regulator when 
extended into the {\mbox{$a\gg 1$}} region.
This then allows us to see the full four-fold modular structure of the theory,
as shown in Fig.~\ref{mappings_figure}.
Given this structure,
we can also revisit the Higgs-mass plot shown in Fig.~\ref{anatomy}.   We now see that
we can interpret this plot as following 
the {\mbox{$\mu^2/M_s^2 = \rho a^2$}} branch from the lower-left corner of Fig.~\ref{mappings_figure}
toward the central location at which {\mbox{$\mu=M_s$}},
and then 
following the 
{\mbox{$\mu^2/M_s^2 = 1/(\rho a^2)$}} branch outward towards the upper-left corner.

Given the sketch in Fig.~\ref{mappings_figure},
we can also understand more precisely how the passage from string theory 
to an EFT breaks modular invariance.  Within this sketch,
regions with beige shading indicate locations where EFT descriptions exist
(and where our regulators are designed to function most effectively, with {\mbox{$a\ll 1$}} or {\mbox{$a\gg 1$}}). 
By contrast, stringy effects dominate in the yellow central region, which is
the only region that locally exhibits the full modular symmetry, lying on both branches simultaneously.
As a result, we necessarily break modular invariance by choosing to focus on any one of the four 
EFT regions alone.
Indeed, each EFT region intrinsically exhibits a certain direction for the flow of the scale $\mu$
relative to the flow of the underlying worldsheet parameter $\rho a^2$.
However, the relative direction of this flow is not modular invariant,
as evidenced from the fact that this flow is reversed in switching from one branch
to the other.

At first glance, the fact that the EFT regions appear only at the extreme ends of each branch 
in Fig.~\ref{mappings_figure} might lead 
one to believe that only extremely light states contribute within the EFT
and that the infinite towers of heavy string states can be ignored within such regions.
However, as we have repeatedly stressed throughout this paper, even this seemingly mild 
assertion would be incorrect.   For example, even within the {\mbox{$\mu\to 0$}} limit,
we have seen in Eq.~(\ref{asymplimit}) that
the Higgs mass receives contributions from $\mathbbX_1$-charged states of all masses across
the entire string spectrum.  Likewise, $\Lambda$ receives contributions from {\it all}\/ string states, 
regardless of their mass. 
We have also seen that the Higgs mass accrues a $\mu$-dependence which transcends
our field-theoretic expectations, even for {\mbox{$\mu\ll M_s$}}.   A particularly 
vivid example of this is the unexpected ``dip'' region shown in Fig.~\ref{anatomy} --- an
effect which is the direct consequence of the stringy Bessel functions whose form is dictated by
modular invariance.
Thus modular invariance continues to govern the behavior of the Higgs mass at all scales,
even within the EFT regions.

Likewise, within such theories there is no distinction between IR and UV.~
We can already see this within Fig.~\ref{mappings_figure},
where the points near the upper end of the figure (\ie, with large $\mu$) are 
designated not as ``UV'' but as ``dual IR'', since they are the images of the IR regions
with small $\mu$ under the duality-inversion symmetry.
But even this labeling is not truly consistent with 
modular invariance, since there is no reason to adopt the 
language of the small-$\mu$ 
region in asserting that the bottom part of the figure corresponds to the IR.~
Thanks to the equivalence under {\mbox{$\mu\to M_s^2/\mu$}}, we might as well have decided to label the upper
portion of the figure as ``UV'' and the lower portion of the figure as ``dual UV''.
In that case, the center of the figure would represent the most IR-behavior that is possible,
rather than the most UV.~
The upshot is that the mere distinction between ``IR'' and ``UV'' 
itself breaks modular invariance.
In a modular-invariant theory, what we would normally call a UV divergence 
is not distinct from an IR divergence ---  they are one and the same.  
Indeed, we have seen that the quadratic divergences normally associated with the Higgs mass in field theory 
are softened to mere logarithmic divergences --- such is the power of modular invariance ---
but in string theory there is no deeper physical interpretation
to this remaining divergence as either UV or IR in nature until we decide to introduce one.

In this connection, we note that it 
might have seemed tempting to look at the EFT expression in Eq.~(\ref{eq:CW}) and suppose that
in a UV-complete theory one could have set about the calculation in a piecemeal
manner, dividing the contributions into a UV contribution
and a much less lethal logarithmically divergent IR contribution
and then evaluating each one separately.
This is certainly the kind of reasoning that is suggested by the notion
of softly broken symmetries, for example.   
However, because there is no intrinsic notion of
UV and IR in a modular-invariant theory, no such separation can exist.
Instead, all we have in string theory 
are amplitudes which may be divergent,
and the question as to whether such divergences are most naturally interpreted as UV or IR in nature
ultimately boils down to a {\it convention}\/ as to which modular-group fundamental domain  
is selected as our region of integration.
Although these arguments are expressed in terms of one-loop amplitudes, 
similar arguments extend to higher loops as well. 
Of course, most standard textbook recipes for evaluating one-loop modular
integrals in string theory adopt the 
fundamental domain which includes the cusp at {\mbox{$\tau\to i\infty$}}.
This choice then leads to an IR interpretation for the divergence.
But when we derived
our supertrace expressions involving only the physical string states,
our calculations required that we sum over an infinite number of such fundamental domains which 
are all related to each other under modular transformations, as in Eq.~(\ref{stripF}).
It is only in this way that we were able to transition from the fundamental domain to the strip
and thereby obtain supertraces involving only the physical string states.
Thus UV and IR physics are inextricably mixed within such supertrace expressions.

Given this bird's-eye view, we can now also understand in a deeper way why it was necessary 
for us to switch from our original modular-invariant regulator functions $\calG_\rho(a,\tau)$ 
in Eq.~(\ref{regG}) to our enhanced modular-invariant functions $\widehat \calG_\rho(a,\tau)$ in Eq.~(\ref{hatGdef})
which exhibited the additional symmetry under {\mbox{$a\to 1/(\rho a)$}}.
At the level of the string worldsheet,
our original functions $\calG_\rho(a,\tau)$ would have been suitable, since they already 
satisfied the two critical criteria
which made them suitable as regulators:
\begin{itemize}
\item   {\mbox{$\calG_\rho(a,\tau)\to 1$}} for all $\tau$ as {\mbox{$a\to 0$}}, so that the {\mbox{$a\to 0$}} 
           limit restores our original unregulated theory;  and
\item   {\mbox{$ \calG_\rho(a,\tau)\to 0$}} sufficiently rapidly for any {\mbox{$a>0$}} as
       $\tau_2$ approaches the appropriate cusps ({\mbox{$\tau\to i\infty$}}, or equivalently  {\mbox{$\tau\to 0$}}), so that
     $f$ is capable of regulating our otherwise-divergent integrands for all {\mbox{$a>0$}}.
\end{itemize}
Indeed, for any divergent string-theoretic quantity $I$, these functions would have led to
a corresponding set of finite quantities $\widetilde I_\rho(a)$ for each value of $(\rho,a)$.
We further saw that these $\calG$-functions had a redundancy under {\mbox{$(\rho,a)\to (1/\rho,\rho a)$}},
so that the only the combination $\rho a^2$ was invariant.

However, while such functions would have been suitable at the level of the string worldsheet, there is one additional
condition that must also be satisfied if we want to be able to interpret our results 
in {\it spacetime}\/, with the invariant combination $\rho a^2$ identified as a running spacetime scale $\mu^2/M_s^2$.
As we have argued below Eq.~(\ref{pfZ}),
modular-invariant string theories necessarily exhibit an invariance
under {\mbox{$\mu\to M_s^2/\mu$}};  indeed,
this scale-duality symmetry rests on very solid foundations.
However, given this scale-inversion symmetry, we see that we 
would not have been able to consistently identify $\rho a^2$ with the spacetime scale
$\mu^2/M_s^2$ unless our regulator function itself also exhibited such an inversion symmetry, with an invariance under
{\mbox{$\rho a^2 \to 1/(\rho a^2)$}} [or equivalently under {\mbox{$a\to 1/(\rho a)$}}].
This was the ultimately the reason we transitioned from the $\calG$-functions to the $\widehat \calG$-functions,
as in Eq.~(\ref{hatGdef}).   This not only preserved the first two properties itemized above, 
but also ensured a third:
\begin{itemize}
\item {\mbox{$\widehat \calG_\rho(a,\tau) = \widehat \calG_\rho(1/\rho a,\tau)$}} for all $(\rho,a)$.
\end{itemize}
In other words, while our first two conditions ensured proper behavior for our regulator functions on the string worldsheet,
it was the third condition which allowed us to endow our regulated string theory with an interpretation
in terms of a renormalization flow with a spacetime mass scale $\mu$. 
Indeed, we see from Fig.~\ref{mappings_figure} that in some sense this extra symmetry was forced on 
us the moment we identified {\mbox{$\mu^2/M_s^2 = \rho a^2$}} and recognized the existence of the scale-duality
symmetry under {\mbox{$\mu\to M_s^2/\mu$}}.
A similar symmetry structure would also need to hold for any alternative regulator functions that might be chosen.

Given these insights, we then proceeded to derive expressions for
our regulated Higgs mass $\widehat m_\phi^2(\mu)$ and regulated cosmological constant
(effective potential) $\widehat \Lambda(\mu)$  
as functions of $\mu$.
The exact results for these quantities are given in Eqs.~(\ref{finalhiggsmassmu})  
and (\ref{lambdamuresult}), respectively.    
Once again, we stress that these results are fully modular invariant except for the fact
that we have implicitly chosen to work within the lower-left branch of Fig.~\ref{mappings_figure}.
For {\mbox{$\mu\ll M_s$}},
we were then able to derive the corresponding approximate EFT running 
for these quantities in Eqs.~(\ref{approxhiggsmassmu})  
and (\ref{lambdamuresult2}).
Indeed, as we have seen in Eq.~(\ref{Lambdafull}),
 our final result for the running 
effective potential $\widehat \Lambda(\mu,\phi)$ takes 
the general form
\beqn
 \widehat \Lambda(\mu,\phi) \,&=&\, 
  \frac{1}{24}\calM^2 \,\Str\, M^2    
           -c' \antieffStr M^2 \mu^2 \nonumber\\  
   && - \zeffStr \left[ \frac{M^4}{64\pi^2} \log\left( c \frac{M^2}{\mu^2}\right) 
            + c''\mu^4\right] ~~~~~~~
\label{eq:lambdaconclusions} 
\eeqn
where 
{\mbox{$c= 2e^{2\gamma+1/2}$}},
{\mbox{$c'=1/(96\pi^2)$}}, and
{\mbox{$c''= 7 c'/10$}},
and where of course we regard the masses $M^2$ as a functions of $\phi$ as in Eq.~(\ref{TaylorM}).
These specific values of $\lbrace c,c',c''\rbrace$
were of course calculated with our regulator function taken as $\widehat \calG_\rho(a,\tau)$
assuming the benchmark value {\mbox{$\rho=2$}},
and with $\mu$ defined along the lower-left branch in Fig.~\ref{mappings_figure}.
However, in general these constants depend on the precise profile of our regulator function.
Finally, given our effective potential, we also discussed the general conditions under which 
our theory is indeed sitting at a stable minimum as a function of $\phi$.

With the results in Eq.~(\ref{eq:lambdaconclusions}) in conjunction with
the relations in Eq.~(\ref{effpotl}),
we have now obtained an understanding of the Higgs mass 
as emerging from $\phi$-derivatives 
of an infinite spectral supertrace of regulated effective potentials.
We can now also perceive the critical similarities and differences relative to the 
EFT expectations in Eq.~(\ref{eq:CW}) and thereby address the questions posed
at the beginning of this section.
For example, 
from the first term within Eq.~(\ref{eq:lambdaconclusions})
we see that the Higgs mass within the full modular-invariant theory 
contains a term of the form $\frac{1}{24} \calM^2 \Str \partial_\phi M^2$.
Comparing this term 
with first term within Eq.~(\ref{eq:CW}),  
we might be tempted to identify
{\mbox{$M_{\rm UV}= \sqrt{3/2} \pi \calM$}}.
However, despite the superficial resemblance between these terms,
we see that the full string-theoretic term is very different
because the relevant supertrace is over the {\it entire}\/ spectrum of states in the theory
and not just the light states in the EFT.~

It is also possible to compare the logarithmic terms within
Eqs.~(\ref{eq:CW})
and (\ref{eq:lambdaconclusions}).
Of course, as in the standard treatment, the logarithmic term in Eq.~(\ref{eq:CW}) can be regulated by subtracting 
a term of the form $\log(M_{\rm UV}/\mu)$, thereby obtaining an effective running.
We then see that both logarithmic terms actually agree.
While it is satisfying to see this agreement, it is nevertheless 
remarkable that we have obtained such a logarithmic EFT-like running 
from our string-theoretic result.
As we have seen, our full string results in 
Eqs.~(\ref{finalhiggsmassmu})  and (\ref{lambdamuresult}) did
not contain logarithms --- they contained Bessel functions.
Moreover, unlike the term discussed above, their contributions were {\it not}\/ truncated
to only the light states with {\mbox{$M\lsim \mu$}} --- they involved supertraces over {\it all}\/
of the states in the string spectrum, as expected for a modular-invariant theory.
However, the behavior of the Bessel functions themselves 
smoothly and automatically suppressed the contributions from states with {\mbox{$M\gsim \mu$}}.
Thus, we did not need to {\it impose}\/
the {\mbox{$M\lsim \mu$}} restriction on the supertrace of the logarithm
term in Eq.~(\ref{eq:lambdaconclusions})
based on a prior EFT-based expectation, as in Eq.~(\ref{eq:CW});
this restriction, and thus an EFT-like interpretation, 
emerged naturally from the Bessel functions themselves.
It is, of course, possible to verify the appearance of such a term directly within the
context of a given compactification  through a direct calculation of the two-point function of the Higgs field
(and indeed we verified this explicitly for various compactification choices),
but of course the expression in 
Eq.~(\ref{eq:lambdaconclusions}) 
is completely general
and thus holds regardless of the specific compactification.

We can also now answer the final question posed at the beginning of this section:
 to what value does the Higgs mass actually run as {\mbox{$\mu\to 0$}}?
Assuming {\mbox{$\zStr \mathbbX_2=0$}}, the answer is clear from Eq.~(\ref{asymplimit}):
\beqn
  \lim_{\mu\to 0}  \widehat m_\phi^2(\mu)  \, &=& \,
       \frac{\xi}{4\pi^2} \,\frac{\Lambda}{\calM^2}
      - \frac{\pi}{6} \,\calM^2\, \Str\, \mathbbX_1 \nonumber\\
   &=&\, 
       \frac{\xi}{96\pi^2} \,\Str\, M^2 
      + \frac{1}{24}\, \calM^2\, \Str\, \partial_\phi^2 M^2 \Bigl|_{\phi=0} \nonumber\\
           &=&\, \frac{\calM^2 }{24}\,D_\phi^2 \,\Str\, M^2\Bigl|_{\phi=0}~. 
\label{asymplimit2}
\eeqn
From a field-theory perspective, this is a remarkable result:   all running actually
stops as {\mbox{$\mu\to0$}}, and the Higgs mass approaches a constant whose value is set by
a supertrace over {\it all}\/ of the states in the string spectrum.
This behavior is clearly not EFT-like.
However, the underlying reason for this has to do with UV/IR equivalence and the scale-inversion
symmetry under {\mbox{$\mu\to M_s^2/\mu$}}.
Regulating our Higgs mass ensures that our theory no longer diverges as {\mbox{$\mu\to \infty$}};  rather,
the Higgs mass essentially ``freezes'' to a constant in this limit.
It is of course natural that in this limit the relevant constant includes contributions from
all of the string states.
The scale-inversion symmetry then implies that the Higgs mass must also ``freeze'' 
to exactly the same value as {\mbox{$\mu\to 0$}}.
We thus see that although a {\it portion}\/ of the running of the Higgs mass is EFT-like
when {\mbox{$\mu\ll M_s$}}, this EFT-like behavior does not persist all the way to {\mbox{$\mu=0$}} because
the scale-inversion symmetry forces 
the behavior as {\mbox{$\mu\to 0$}} to mirror
the behavior as {\mbox{$\mu\to \infty$}}.
Indeed, the ``dip'' region is nothing but the stringy transition between these two
regimes.

Given the results in Eq.~(\ref{asymplimit2})
we also observe that we can now write
\beq
    m_\phi^2 ~=~ \left. \frac{1}{24} \calM^2 \,\Str \left[ D_\phi^2  M^2(\phi)\right]\,\right|_{\phi=0}~~~
\eeq
This result is thus the Higgs-mass analogue 
of the $\Lambda$-result in Eq.~(\ref{eq:lam-rep}).
We can also take the $a\to 0$ (or equivalently $\mu\to 0$) limit of Eq.~(\ref{effpotl}),
yielding the simple relations
\beq
   \begin{cases}
      ~ \Lambda  &=~ \Lambda(\phi)\bigl|_{\phi=0} \\
      ~ m_\phi^2 &=~ D_\phi^2 \,\Lambda(\phi)\bigl|_{\phi=0}~.
   \end{cases}
\label{effpotl2}
\eeq
Indeed, for theories with {\mbox{$\zStr \mathbbX_2=0$}}, 
these are exact relations amongst finite quantities.

The final results of our analysis 
are encapsulated within Fig.~\ref{anatomy}.
Indeed, this figure graphically illustrates many of the most important conclusions of this paper.
In Fig.~\ref{anatomy}, we have dissected the anatomy of the Higgs-mass running,
illustrating how this running
passes through different distinct stages as $\mu$ increases. 
Starting from  the ``deep IR/UV''
region near {\mbox{$\mu\approx 0$}}, the Higgs mass passes through the ``dip'' region and the ``EFT'' region
before ultimately reaching the ``turnaround'' region. 
Beyond this, the theory enters the 
``dual EFT'' region, followed by the ``dual dip'' region
and ultimately the ``dual deep IR/UV'' region.
Above all else, this figure clearly illustrates
how in a modular-invariant theory our normal understanding
of ``running'' is turned on its head.  The Higgs mass does not somehow
get ``born'' in the UV and then run to some possibly undesirable
value in the IR.~ 
Instead, we may more properly consider the Higgs mass to be ``born'' at {\mbox{$\mu=M_s$}}.
It then runs symmetrically towards both lesser and greater values of $\mu$
until it eventually asymptotes to a constant as {\mbox{$\mu\to 0$}} and as {\mbox{$\mu\to \infty$}}.

We conclude this discussion with two comments regarding technical points.
First, as discussed in Sect.~\ref{sec4}, we have freely assumed throughout this paper 
that the residue of a supertrace sum is equivalent to the supertrace sum of the individual residues.    In other words, as discussed below Eq.~(\ref{integrateg}), we have assumed that the supertrace sum does not introduce any additional divergences beyond those already encapsulated within our assertion that the four-dimensional Higgs mass is at most logarithmically divergent, or equivalently that the level-matched integrand has a divergence structure
{\mbox{$g(\tau)\sim c_0+c_1\tau_2$}} as {\mbox{$\tau_2\to\infty$}}.   Indeed, this assumption is justified because we are working within the presence of a regulator which is sufficiently powerful to render our modular integrals finite, given this divergence structure.  Moreover, the divergence structure of our original unregulated Higgs mass is completely general for theories in four spacetime dimensions, since only a change in spacetime dimension can alter the numbers of $\tau_2$ prefactors which emerge.   Of course, four-dimensional string models generically contain many moduli, and some of these moduli may correspond to the radii associated with possible geometric compactifications from our original underlying 10- and/or 26-dimensional worldsheet theories.   If those moduli are extremely large or small, one approaches a decompactification limit in which our theory becomes effectively higher-dimensional.   For any finite or non-zero value of these moduli, our results still hold as before.   However, in the full limit as these moduli become infinite or zero, new divergences may appear which are related to the fact that the effective dimensionality of the theory has changed.   Indeed, extra spacetime dimensions generally correspond to extra factors of $\tau_2$, thereby increasing the strengths of the potential divergences.    Although all of our results in Sects.~\ref{sec2} and \ref{sec3} are completely general for all spacetime dimensions, our results in Sect.~\ref{sec4} are focused on the case of four-dimensional string models 
for which {\mbox{$g(\tau_2)\sim c_0+c_1\tau_2$}} as {\mbox{$\tau_2\to\infty$}}.      
As a result,  the supertrace-summation and residue-extraction procedures will not commute in the decompactification limit, and additional divergences can arise.   However, this does not pose a problem for us --- we simply use the same regulators we have already outlined in Sect.~\ref{sec3}, but instead work directly in a higher-dimensional framework in which $g(\tau_2)$ as 
{\mbox{$\tau_2\to\infty$}}    
takes a form appropriate for the new effective spacetime dimensionality. 
Once this is done, we are once again free to exchange the orders of residue-extraction and supertrace-summation, knowing that our results must once again be finite.

Our second technical point relates to the concern that has occasionally been expressed in the prior literature
about the role played by the off-shell tachyons which necessarily appear 
within the spectra of all heterotic strings, and the exponential one-loop divergences
they might seem to induce in the absence of supersymmetry
as {\mbox{$\tau\to i\infty$}}.    
In this paper, we discussed this issue briefly in the paragraph surrounding
Eq.~(\ref{protocharge}). 
Ultimately, however, we believe that this concern is spurious.
First, as discussed below Eq.~(\ref{protocharge}), such states typically lack the non-zero charges 
needed in order to contribute to the relevant one-loop string amplitudes. 
Second, 
within such one-loop amplitudes,
our modular integrations 
come with an implicit instruction 
that within the {\mbox{$\tau_2>1$}} region of the fundamental domain
we are to perform the $\tau_1$ integration prior to performing the $\tau_2$ integration.
This then eliminates the contributions from the off-shell tachyons in the {\mbox{$\tau\to i\infty$}} limit.
This integration-ordering prescription 
is tantamount to replacing the divergence as {\mbox{$\tau\to i\infty$}} with its average along the line segment {\mbox{$-1/2\leq \tau_1\leq 1/2$}},
which makes sense in the {\mbox{$\tau_2\to\infty$}} limit as this line segment moves infinitely far up the fundamental domain.
Another way to understand this is to realize that under a modular transformation no information can be lost, yet this entire
line segment as {\mbox{$\tau_2\to \infty$}} is mapped to the single point with {\mbox{$\tau_1=\tau_2=0$}} under the modular transformation {\mbox{$\tau\to -1/\tau$}}.
Finally, through the compactification/decompactification argument in presented in Ref.~\cite{Kutasov:1990sv},    
one can see directly that this off-shell tachyon makes no contribution in all spacetime dimensions {\mbox{$D>2$}}. 
Thus no exponential divergence arises.
However, we note that even if 
an exponential divergence were to survive, it would also be automatically regulated through 
our modular-invariant regulator $\widehat G_\rho(a,\tau)$ --- or sufficiently many higher powers thereof --- given
that  $\widehat G_\rho(a,\tau)$ itself exhibits an exponential suppression as {\mbox{$\tau\to i\infty$}}.

The results in this paper have touched on many different topics.  Accordingly, there are 
several directions that future work may take.

First, although we have focused in this paper on the mass of the
Higgs, it is clear that this UV/IR-mixed picture of running provides a general paradigm
for how one should think about the behavior of a modular-invariant theory as a whole. 
  For example, one question that naturally arises from 
our discussion concerns the renormalization of the dimensionless couplings. 
This was the subject of the seminal work in  Ref.~\cite{Kaplunovsky:1987rp}. 
Even though a regulator was chosen in Ref.~\cite{Kaplunovsky:1987rp} which 
was not consistent with modular invariance,
this was one of the first calculations in which the contributions from the full 
infinite towers of string states were incorporated within a calculation of gauge couplings and their behavior.
It would therefore be interesting to revisit these issues and analyze the running and beta functions
of the dimensionless gauge couplings that would emerge in the presence of a fully modular-invariant regulator. 
The first steps in this direction have already been taken in Refs.~\cite{Kiritsis:1994ta, Kiritsis:1996dn, Kiritsis:1998en}.
However, using the techniques we have developed in this paper, it is now possible to
extend these results to obtain full scale-dependent RG
flows for the gauge couplings
 as functions of $\mu$, and in a continuous way that simultaneously incorporates both UV and IR physics and which does not artificially separate the results into a field-theoretic running with a string-theoretic threshold correction.
Moreover, due to the {\mbox{$\mu\to M_s^2/\mu$}} symmetry
we expect that the coefficients of {\it all}\/ operators in the theory 
should experience symmetric runnings with vanishing gradients at {\mbox{$\mu=M_s$}}.
For operators with zero engineering dimension, this then translates to a vanishing
beta function at {\mbox{$\mu=M_s$}}, suggesting the existence of an unexpected (and ultimately unstable) ``UV'' 
fixed point at that location.

In the same vein, it would also be interesting to study the behavior of scattering amplitudes
within a full modular-invariant context.
We once again expect significant deviations from our field-theoretic expectations at all scales ---
including those at energies relatively far below the string scale ---  but it would be interesting
to obtain precise information about how this occurs and what shape the deviations take.

Given our results thus far,
perhaps the most important and compelling avenue to explore concerns the gauge hierarchy problem.
As discussed in the Introduction,
it remains our continuing hope that modular symmetries might provide a new perspective on this problem,
one that transcends our typical field-theoretic expectations.
Some ideas in this direction were already sketched in Ref.~\cite{Dienes:2001se},
along with suggestions 
that the gauge hierarchy problem
might be connected with the cosmological-constant problem,
and that these both might be closely connected with the question of vacuum stability.
It was also advocated in the Conclusions section of Ref.~\cite{Dienes:2001se} 
that these insights might be better understood through calculational frameworks that did not involve
discarding the contributions of the infinite towers of string states, but which instead
incorporated all of these contributions in order 
to preserve modular invariance and the string finiteness that follows.

The results of this paper 
enable us to begin the process of 
fulfilling these ambitions.
In particular, the effective potential
in Eq.~(\ref{eq:lambdaconclusions}) is a powerful first step because
this result 
provides a ``UV-complete'' effective potential 
which yields the raw expressions
for radiative corrections written in terms of the spectrum of whatever theory
one may be interested in studying. Moreover it is an expression that
is applicable at all energy scales, including the scales associated with the 
cosmological constant and the electroweak  physics 
where such results are critical.

Given our results, we can develop a string-based reformulation of both 
of these hierarchy problems.
Our expression for the cosmological constant in Eq.~(\ref{eq:lam-rep})
[or equivalently taking {\mbox{$\lim_{\mu \to 0} \widehat \Lambda(\mu)$}}]
implicitly furnishes us with a constraint 
of the form {\mbox{$\Str\,M^2 \sim 24 M_\Lambda^4/\calM^2$}}
where {\mbox{$M_\Lambda \sim \Lambda^{1/4}\approx 2.3\times 10^{-3}\,$}}{\rm eV} is the
mass scale associated with the cosmological constant.
Likewise, we see that
{\mbox{$\Lambda\ll 4\pi^{2}M_{\rm EW}^{2}\mathcal{M}^{2}$}} where
{\mbox{$M_{\rm EW}\sim {\cal O}(100)~{\rm GeV}$}} denotes the electroweak scale.
Thus, 
with $\phi$ representing the Standard-Model Higgs 
and roughly identifying the physical Higgs mass as
{\mbox{$\lim_{\mu \to 0} \widehat m_\phi^2 (\mu) \sim M_{\rm EW}^2$}},
we see from Eq.~(\ref{asymplimit2}) that we can obtain a second constraint
of the form
{\mbox{$\partial_\phi^2 \Str\, M^2\bigl|_{\phi=0}\sim 24 M_{\rm EW}^2/\calM^2$}}.
We therefore see that our two hierarchy conditions now respectively take the forms
\beq
   \begin{cases}
         ~\phantom{\partial_\phi^2\,} \Str\, M^2 \Big|_{\phi=0} \!\! &\sim~   24\,  M_\Lambda^4/\calM^2      \\
         ~\partial_\phi^2\,           \Str\, M^2 \Big|_{\phi=0} \!\! &\sim~   24\,  M_{\rm EW}^2/\calM^2 
   \end{cases}
\label{stringVeltman}
\eeq
where we continue to regard our masses $M^2$ as functions of the 
Higgs fluctuations $\phi$, as in Eq.~(\ref{TaylorM}).
To one-loop order, these are the hierarchy conditions that must be satisfied by the
the spectrum of any modular-invariant string theory. 
Indeed, substituting the masses in Eq.~(\ref{TaylorM}), these two conditions reduce to
the forms
\beq
\begin{cases}
     ~\Str\, \beta_0   \!\! &\sim~   24\,  M_\Lambda^4/\calM^4 \\
     ~\Str\, \beta_2   \!\! &\sim~   24\,  M_{\rm EW}^2/\calM^2 ~. 
\end{cases}
\label{stringVeltman2}
\eeq
Although every massive string state has a non-zero $\beta_0$ and therefore
contributes to the first constraint, only those string states 
which couple to the Higgs field
have a non-zero $\beta_2$ and thereby contribute to the second.
Of course, given the form of Eq.~(\ref{TaylorM}), 
the non-zero $\beta_i$'s for each state are still expected to 
be $\sim {\cal O}(1)$, which is precisely why these constraints are so
difficult to satisfy. 
Moreover, as we know in the case of string models exhibiting charge lattices,
these $\beta_i$-coefficients are related to the charges of the individual string states
and therefore can be discrete in nature.

    Given the constraints in Eq.~(\ref{stringVeltman2}), 
    it is natural to wonder why there is no hierarchy condition 
    corresponding to $\Str\,\beta_1$.
    Actually, such a condition exists, although this is not normally treated as a hierarchy 
    constraint.   This is nothing but our stability condition 
    {\mbox{$\partial_\phi \widehat \Lambda(\mu,\phi)\bigl|_{\phi=0}=0$}}  in
    Eq.~(\ref{stabcond2}), which can be considered on the same footing as
    the other two relations in Eq.~(\ref{effpotl}).  As we have seen,
    this leads directly to the relations {\mbox{$\Str\,\mathbbY=0$}} or equivalently
    {\mbox{$\partial_\phi \Str\,M^2\bigl|_{\phi=0}=0$}}, which can be considered
    alongside the relations in Eq.~(\ref{stringVeltman}).
    This then leads to the constraint {\mbox{$\Str \,\beta_1=0$}}.
    Of course, it is always possible that there exists a 
    non-zero Higgs tadpole, as long as this tadpole is sufficiently small as to have
    remained unobserved (\eg, at colliders, or cosmologically), 
    leading to string models which are not truly stable but only metastable.  
    Such models would be analogous to non-supersymmetric
    string models in which the {\it dilaton}\/ tadpole is non-vanishing
    but exponentially suppressed to a sufficient degree that the theory 
    is essentially stable on cosmological timescales~\cite{Abel:2015oxa}.
    In such cases involving a non-zero Higgs potential, 
    we can define an associated mass scale $M_{\rm stab}$ 
    which characterizes the maximum possible Higgs instability we can tolerate
    experimentally and/or observationally.
    Our corresponding ``hierarchy'' condition would then take the form
\beq
\Str\, \beta_1 ~\lsim~ {M_{\rm stab}}/{\calM}~.
\label{stabcond2}
\eeq
    Of course, this condition differs from
    the others in that it does not describe a phenomenological constraint on a particular
    vacuum but rather helps to determine whether that vacuum even exists.
    All conditions nevertheless determine whether a given value of $\langle \phi\rangle$ (in this case
    defined as {\mbox{$\langle \phi\rangle =0$}}) is viable. 
       In general, such ``hierarchies'' exist for each scalar $\phi$ in the theory.

Despite their fundamentally different natures, these two types of hierarchies can actually
be connected to each other.
In the case that $\phi$ represents the Standard-Model Higgs,  
this connection will then allow us to relate $M_{\rm stab}$ to $M_{\rm EW}$.
The fundamental reason for this connection is that 
a tadpole corresponds to a linear term in an effective potential for the Higgs.
This is in addition to the quadratic mass term.
However, we can eliminate the linear term by completing the square, which
of course simply shifts the corresponding Higgs VEV.~
The maximum size of this tadpole diagram is therefore  
also bounded by $M_{\rm EW}$.   More precisely,
we find for the Standard-Model Higgs that
\beq
       M_{\rm stab} ~\sim~ 24\, M_{\rm EW}^3/\calM^2~,
\eeq
whereupon Eq.~(\ref{stabcond2}) takes the form
\beq
\Str\, \beta_1 ~\lsim~  24\, M_{\rm EW}^3/\calM^3~.
\label{stabcond2alt}
\eeq
Indeed, in this form Eq.~(\ref{stabcond2alt}) 
more closely resembles the relations in 
Eq.~(\ref{stringVeltman2}).

It is remarkable that in string theory
the constraints from the cosmological-constant problem
and the gauge hierarchy problem in Eq.~(\ref{stringVeltman2}) take such similar algebraic forms.
Indeed in some sense $\beta_0$ and $\beta_2$ measure the responses of our 
individual string states to mass (or gravity) and to 
fluctuations of the Higgs field, respectively,
with $\beta_2$ related to the {\it charges}\/ of these states
with respect to Higgs couplings.
It is also noteworthy that these conditions 
each resemble the so-called ``Veltman condition''~\cite{Veltman:1980mj}
of field theory.
Recall that the Veltman condition for addressing the gauge hierarchy
in an effective field theory such as the Standard Model
calls for cancelling the quadratic divergence of the Higgs mass
by requiring the vanishing of the (mass)$^2$ supertrace $\Str \,M^2$ 
when summed over all light EFT states which couple to the Higgs.
However, we now see that in string theory 
the primary difference is that the supertraces $\Str\,M^2$ in Eq.~(\ref{stringVeltman2}) 
are evaluated over  
the {\it entire}\/ spectrum of string states
and not merely the light states within the EFT.~ 
This is an important difference because the vanishing of this supertrace
when restricted to the EFT generally tells us  nothing
about its vanishing in the full theory, or vice versa.
These are truly independent conditions, and we see that
string theory requires the latter, not the former.

One of the virtues of modular invariance --- and indeed an indication of
its overall power as a robust, unbroken symmetry --- is that the string naturalness
conditions in Eqs.~(\ref{stringVeltman}) and (\ref{stringVeltman2}) 
necessarily include the effects of {\it all}\/ physics occurring 
at intermediate scales.  This includes, for example,
the effects of a possible GUT phase transition.
As discussed earlier in this paper,
this is true because modular invariance is an exact symmetry governing
not only all of the states in the string spectrum but also their interactions.
Thus all intermediate-scale physics --- even including phase transitions ---
must preserve 
modular invariance.  This in turn implies that as the masses and degrees of
freedom within the theory evolve,
they all evolve together in a carefully balanced way such that modular invariance is preserved. 
Thus, given that relations such as that in Eq.~(\ref{asymplimit2}) are general and rest solely on 
modular invariance, they too will remain intact. 
Relations such as those in Eqs.~(\ref{stringVeltman}) and (\ref{stringVeltman2}) 
then remain valid.

Thus far we have reformulated the constraints associated
with the cosmological-constant and gauge hierarchy problems,
providing what may be viewed as essentially ``stringy'' versions of the traditional Veltman condition.
However our results also suggest new stringy mechanisms by which such constraints might actually be satisfied ---
mechanisms by which such hierarchies might
actually emerge within a given theory. 
Given the general running behavior of the Higgs mass 
in Fig.~\ref{anatomy},
we observe two interesting features that may be
relevant for hierarchy problems.
First, let us imagine that we apply our formalism for the running of the Higgs mass in the 
original {\it unbroken}\/ phase of the theory.
We will then continue to obtain a result for the Higgs running 
with the same shape as that shown in Fig.~\ref{anatomy}, 
only with the relevant quantities $\Lambda$, $\mathbbX_1$, and $\mathbbX_2$ evaluated
in the unbroken phase.
Concentrating on the region with {\mbox{$\mu\leq M_s$}},
we see that there is a relatively slow (logarithmic) running which stretches all the way from
the string scale $M_s$ down to the energy scales associated with the lightest massive string states,
followed by  a transient ``dip'' region within which the Higgs mass experiences
a sudden local minimum.
This therefore provides a natural scenario in which electroweak symmetry breaking
might be triggered at an energy scale hierarchically below the fundamental high energy scales
in the theory.
Note that the dip region indeed produces a {\it minimum}\/ for the Higgs mass only if {\mbox{$\Str \,\mathbbX_2>0$}}; 
otherwise the logarithmic running changes 
sign and the Higgs mass would already be tachyonic 
at high energy scales near the string scale,
signifying (contrary to assumptions) that our theory was not sitting at a stable minimum in $\phi$-space
at high energies.  (We also note that 
even though {\mbox{$\mathbbX_2\geq 0$}},
the supertrace $\Str\,\mathbbX_2$ 
can have either sign 
depending on how these $\mathbbX_2$-charges are distributed between
bosonic and fermionic states.) 
However, with {\mbox{$\Str\,\mathbbX_2>0$}},
this transient minimum in Fig.~\ref{anatomy} will cause the Higgs to become tachyonic as long as
\beq
       \frac{\pi}{6}\, \Str\,\mathbbX_1 + 
       \frac{3}{10}\, \Str\,\mathbbX_2
      ~\gsim~ \frac{\xi}{4\pi^2} \frac{\Lambda}{\calM^4}~
\eeq
where the factor of $3/10$ represents the
approximate value $\approx 0.3$ parametrizing the ``dip depth'' from Fig.~\ref{anatomy}.~
It is remarkable that this condition 
links the scale of electroweak symmetry breaking with the value of the one-loop
cosmological constant.
Just as with our other conditions, this condition can be also expressed as a constraint 
on the values of our $\beta_i$ coefficients:
\beq
       \frac{9}{5} \,  \Str\,\beta_1^2
       -4\pi^2\, \Str\,\beta_2   
      ~\gsim~ \xi\, \Str\,\beta_0~.
\eeq
This is then our condition for triggering electroweak symmetry breaking at small scales
hierarchically below $\calM$.
Of course,  after this breaking occurs, we would need to work in the broken phase
wherein $\phi$ returns to representing the Higgs fluctuations relative to the new broken-phase vacuum.

The second feature illustrated within Fig.~\ref{anatomy} that may be relevant for the
hierarchy problems concerns the scale-duality symmetry {\mbox{$\mu\to M_s^2/\mu$}}.
As we have discussed at numerous points throughout this paper,
this symmetry implies an equivalence between UV physics and IR physics --- an observation
which already heralds a major disruption of our understanding of the relationship between
high and low energy
scales compared with field-theoretic expectations.    
Given that hierarchy problems not only emerge 
within the context of low-energy EFTs but also assume traditional field-theoretic relationships between UV and IR physics,
it is possible to speculate that such hierarchy problems are not fundamental
and do not survive in string theory in the manner we normally assume.
Furthermore, we have already seen that modular invariance not only leads to this UV/IR mixing but
also softens divergences so dramatically that certain otherwise-divergent amplitudes
(such as the cosmological constant)
are rendered finite.
Taken together, these observations suggest that modular invariance may hold the key to an entirely new way of thinking 
about hierarchy problems --- a point originally made in Ref.~\cite{Dienes:2001se}
and which we will develop further in upcoming work~\cite{SAAKRDinprep}.

The results of this paper also prompt a number of additional lines of research.
For example, although most of our results are completely general and hold across all
modular-invariant string theories, much of our analysis in this paper has been restricted to one-loop order.
It would therefore be interesting to understand what occurs at higher loops.
In this connection, we note that it is often asserted in the string literature 
that modular invariance is only a one-loop symmetry, seeming to imply that it should no longer apply at higher loops.
However, this is incorrect:   modular invariance is an {\it exact}\/ worldsheet symmetry of (perturbative) closed strings,
and thus holds at all orders.  This symmetry is merely {\it motivated}\/ 
by the need to render one-loop string amplitudes consistent with the underlying conformal invariance of the string worldsheet.
Once imposed, however, this symmetry affects the entire string model --- all masses and interactions, to any order.
Likewise, one might wonder whether there are {\it multi-loop}\/ versions of modular invariance 
which could also be imposed, similarly motivated by considerations of higher-loop amplitudes.
However, it has been shown~\cite{Kawai:1987ew}
that within certain closed string theories, amplitude factorization and 
physically sensible state projections
together ensure that one-loop modular invariance automatically implies multi-loop modular invariance.
Thus one-loop modular invariance is sufficient, and no additional 
symmetries of this sort are needed.

Because modular invariance is an {\it exact}\/ worldsheet symmetry,
we expect that certain features we have discussed 
in this paper (such as the existence of the scale-duality symmetry under {\mbox{$\mu\to M_s^2/\mu$}})
will remain valid to all orders.
We believe that the same is true of other consequences of modular invariance,
such as our supertrace relations and the ``misaligned supersymmetry''~{\mbox{\cite{Dienes:1994np,Dienes:1995pm,Dienes:2001se}}}
from which they emerge.

That said, modular invariance is a symmetry of closed strings.
For this reason, we do not expect modular invariance to hold for Type~I strings, which contain
both closed-string and open-string sectors.
However, within Type~I strings there are tight relations between the closed-string and open-string
sectors, and certain remnants of modular invariance survive even into the open-string sectors.
For example, certain kinds of misaligned supersymmetry have been found 
to persist even within open-string sectors~\cite{Cribiori:2020sct}.
It will therefore be interesting to determine the extent to which the results and techniques of this paper
might extend to open strings.

The results described in this paper have clearly covered a lot of territory, stretching 
from the development of new techniques for calculating Higgs masses to the development 
of  modular-invariant methods 
of regulating divergences.   
We have also tackled critical questions concerning UV/IR mixing and the extent to which one can extract
effective field theories from modular-invariant string theories, complete with Higgs masses and a cosmological
constant that run as functions of a spacetime mass scale.
We have demonstrated that there are unexpected relations between the Higgs mass and the one-loop cosmological
constant in any modular-invariant string model, and that it is possible to extract 
an entirely string-based effective potential for the Higgs. 
Moreover, as indicated in the Introduction,
our results apply to {\it all}\/ scalars in the theory --- even beyond the Standard-Model Higgs ---  and apply
whether or not spacetime supersymmetry is present. 
As such, we anticipate that there exist numerous areas of exploration that may be prompted by these developments.
But perhaps most importantly for phenomenological purposes, we believe that 
the results of this paper can ultimately serve as the launching point for a rigorous investigation of 
the gauge hierarchy problem in string theory.
Much work therefore remains to be done.

\begin{acknowledgments}

We are happy to thank Carlo Angelantonj, Athanasios Bouganis, and Jens Funke for insightful discussions. 
The research activities of SAA were supported by the STFC grant 
ST/P001246/1 and partly by a CERN Associateship and 
Royal-Society/CNRS International Cost Share Award IE160590.
The research activities of KRD were supported in part by the U.S.\ Department of Energy
under Grant DE-FG02-13ER41976 / DE-SC0009913, and also 
by the U.S.\ National Science Foundation through its employee IR/D program.
The opinions and conclusions
expressed herein are those of the authors, and do not represent any funding agencies.

\end{acknowledgments}

\appendix

\section{Evaluating the Higgs mass with the modular-invariant regulator:~
Explicit calculation}\label{higgsappendix} 

Our goal in Appendices~\ref{higgsappendix} and \ref{lambdaappendix} is to provide an explicit calculation of the
regulated Higgs mass $\widehat m_\phi^2(\rho,a)$ given in Eq.~(\ref{twocontributions}), 
and to express the result directly in terms
of supertraces over the physical string states.
In this Appendix we shall focus on the contribution $\widehat m_\phi^2(\rho,a)\bigl|_\calX$
which comes from the terms 
with non-trivial $\calX_i$ insertions.
The contribution $\widehat m_\phi^2(\rho,a)\bigl|_\Lambda$ 
from the $\Lambda$-term will be discussed in Appendix~\ref{lambdaappendix}.

Because our regulator function $\widehat \calG_\rho(a,\tau)$ 
is built upon the circle partition function $Z_{\rm circ}(a,\tau)$
defined in Eq.~(\ref{Zcircdef}), the core of our calculation 
of $\widehat m_\phi^2(\rho,a)\bigl|_\calX$
ultimately rests
on evaluating the integral
\beq
       P(a) ~\equiv~ \int_\calF \dmu~ F(\tau)\,Z_{\rm circ}(a,\tau)~
\label{mainint}
\eeq
where $F(\tau)$ is the modular-invariant string partition function 
with the $\calX_i$ insertions indicated in Eq.~(\ref{Higgsmass1}). 
We shall therefore begin by focusing on this integral.
Note that the {\mbox{$a\to 1/a$}} symmetry of $Z_{\rm circ}(a)$ ensures that
{\mbox{$P(a)= P(1/a)$}}.
Once we have evaluated $P(a)$, we can then easily evaluate the full expression
for the $\calX_i$-dependent contributions to the Higgs mass 
in Eq.~(\ref{Higgsmass1}) via
\beq
  \widehat m_\phi^2(\rho,a)\Bigl|_\calX ~=~ \frac{1}{1+\rho a^2} \,A_\rho\,
          a^2 \frac{\partial}{\partial a} \,
          \biggl[   P(\rho a) - P(a) \biggr]~
\label{operator}
\eeq
where {\mbox{$A_\rho\equiv \rho/(\rho-1)$}}.
Given that {\mbox{$P(a)= P(1/a)$}}, the expression in Eq.~(\ref{operator}) for the Higgs
mass will be invariant
under {\mbox{$a\to 1/(\rho a)$}}.   
We also emphasize that the result of our calculation will be a manifestly finite 
quantity, as ensured by the presence of the regulator $\widehat \calG_\rho(a,\tau)$ in our 
integrand.   Therefore the calculation we shall be performing here is nothing but the direct
evaluation of an integral, with no additional regulators needed.

As in Sect.~\ref{sec3},
our first step is to recast Eq.~(\ref{mainint}) as an integral over the strip $\calS$
in Eq.~(\ref{Sdef}).
In order to do this, we first note that we can perform a Poisson resummation
of the expression for $Z_{\rm circ}(a,\tau)$ in Eq.~(\ref{Zcircdef}).
Indeed, we can resum either the winding modes or the momentum modes in Eq.~(\ref{Zcircdef}),
ultimately obtaining the two alternative expressions
\beqn
   Z_{\rm circ}(a,\tau) ~&=&~  a \,\sum_{j,k} \exp\left( -\frac{ \pi a^2}{\tau_2}\, |j+k\tau|^2\right)\nonumber\\ 
                   ~&=&~ \frac{1}{a}\, \sum_{j,k} \exp\left( -\frac{\pi }{a^2 \tau_2}\,|j+k\tau|^2 \right)~~~~
\label{newforms}
\eeqn
respectively.
Indeed, the existence of these two equivalent expressions for $Z_{\rm circ}$ is nothing but
a manifestation of the symmetry of $Z_{\rm circ}$ under {\mbox{$a\to 1/a$}}.  
Of course, each of these expressions independently retains
the {\mbox{$a\to 1/a$}} symmetry [since each is equal to $Z_{\rm circ}(a,\tau)$], but
this symmetry is no longer manifest.

In principle, we could now proceed using either of the two expressions in Eq.~(\ref{newforms}).
However, since we shall be most interested in the physics that emerges for {\mbox{$a\ll 1$}}, we shall find
it most useful to continue from the second expression in Eq.~(\ref{newforms}).
This is the expression in which the momentum modes within $Z_{\rm circ}$ are Poisson-resummed,
as appropriate when the compactification radius $a^{-1}$ is large.
Indeed, it is precisely through this resummation that we find {\mbox{$Z_{\rm circ}\sim a^{-1}$}} as {\mbox{$a\to 0$}}. 

Continuing from Eq.~(\ref{newforms}),
we next define the greatest common divisor {\mbox{$r\equiv \gcd(j,k)$}}, where {\mbox{$r=0$}} if {\mbox{$j=k=0$}},
where {\mbox{$r>0$}} in all other cases, and where {\mbox{$\gcd(0,k)\equiv |k|$}} for all $k$. 
With this definition, the second line of Eq.~(\ref{newforms}) becomes
\beqn
   Z_{\rm circ}(a,\tau) \,
                                                 &=&\, \frac{1}{a} + \frac{1}{a}\, \sum_{r=1}^\infty  \sum_{\substack{j,k\\ (j,k)=1}}
                      \exp\left( -\frac{ \pi r^2}{a^2 \tau_2}\, |j+k\tau|^2\right)\nonumber\\
                     &=&\, \frac{1}{a} + \frac{2}{a}\, \sum_{r=1}^\infty  \sum_{\substack{j,k\\ (j,k)=1\\ j+k>0}}
                      \exp\left( -\frac{ \pi r^2}{a^2 \tau_2}\, |j+k\tau|^2\right)~.\nonumber\\
\label{newforms2}
\eeqn
Note that in these expressions the new $(j,k)$ summations are over values of $j$ and $k$ which
are relatively prime.
In the first line of Eq.~(\ref{newforms2}) we have explicitly separated those contributions with {\mbox{$r=0$}} from those
with {\mbox{$r>0$}},
while in the second line we have further restricted our sum so that {\mbox{$j+k>0$}}
[thereby ensuring that if $(j,k)$ is included then $(-j,-k)$ is excluded, and vice versa].

Next, we observe that any modular transformation {\mbox{$\tau \to \tau'\equiv (A\tau+B)/(C\tau+D)$}}
sends {\mbox{$\tau_2 \to \tau'_2\equiv \tau_2/|C\tau+D|^2$}}.
We further note that $C$ and $D$ are relatively prime for any such modular transformation
(thanks to the constraint {\mbox{$AD-BC=1$}})
and that the set of modular transformations consisting of one representative for
each possible pair of relatively prime integers $(C,D)$ with {\mbox{$C+D>0$}}
are precisely those that fill out the coset {\mbox{$\Gamma_\infty\backslash \Gamma$}}.
(Indeed, the infinite number of possible choices for $A$ and $B$ in each case generate the distinct cosets.)
As a result,
when acting on $\calF$,
these modular transformations
fill out the strip $\calS$.
Thus,
multiplying Eq.~(\ref{newforms}) by $F(\tau)$, integrating over the fundamental domain $\calF$, 
and then utilizing the unfolding relation in Eq.~(\ref{unfold})
on the second term in the second line of Eq.~(\ref{newforms2})
yields the result 
\beq
         P(a) ~=~ P_1(a) + P_2(a) 
\eeq
where
\beqn
P_1(a) ~&\equiv &~ \frac{1}{a} \int_\calF \dmu\, F(\tau) \nonumber\\  
P_2(a) ~&\equiv &~ \frac{2}{a} \int_\calS \dmu \, F(\tau) \sum_{r=1}^\infty  e^{-\pi r^2/ (a^2 \tau_2)}~~\nonumber\\
       ~&=&~ \frac{2}{a} \int_0^\infty \frac{d\tau_2}{\tau_2^2} \, g(\tau_2) \sum_{r=1}^\infty  e^{-\pi r^2/(a^2 \tau_2)}~.~~~
\label{newforms3}
\eeqn
Indeed, $P_1(a)$ is nothing but our original integral $P(a)$ in the full {\mbox{$a\to 0$}} limit, wherein {\mbox{$Z_{\rm circ}\to a^{-1}$}}.

In principle, our goal at this stage is to 
evaluate $P_1(a)$ and $P_2(a)$.
Unfortunately, although the sum {\mbox{$P(a)\equiv P_1(a)+P_2(a)$}} leads to a finite result
for the Higgs mass in Eq.~(\ref{operator}),
the individual terms $P_{1,2}(a)$ do not;  instead they lead to expressions which each exhibit the original
logarithmic divergence associated with our unregulated Higgs mass.
Indeed, these logarithmic divergences only cancel in the Higgs-mass contributions coming from the sum $P(a)$.
For this reason, we shall now reshuffle our expressions for $P_1(a)$ and $P_2(a)$, producing
new quantities $P'_1(a)$ and $P'_2(a)$ such that each 
leads to an independently finite contribution to the Higgs mass. 
To do this, we recall that our Higgs-mass calculation yields
{\mbox{$g(\tau_2)\sim c_0 + c_1\tau_2$}} as {\mbox{$\tau_2\to\infty$}},
where $c_{0}$ and $c_1$ are given in Eq.~(\ref{c0c1}). 
We shall therefore define
\beqn
        P'_1(a) ~&\equiv &~  P_1(a) - \frac{1}{a} \int_t^\infty  \frac{d\tau_2}{\tau_2} \, c_1~\nonumber\\
        P'_2(a) ~&\equiv &~  P_2(a) + \frac{1}{a} \int_t^\infty  \frac{d\tau_2}{\tau_2} \, c_1~
\label{Pprimes}
\eeqn
where $t$ is an arbitrary finite parameter.
As we shall see, the extra terms in Eq.~(\ref{Pprimes}) have the net
effect of transferring this logarithmic divergence between the 
Higgs-mass contributions coming from these separate terms, 
thereby allowing these divergences to separately cancel. 
Indeed, for any finite $t$,
each of these new quantities $P'_i(a)$ leads to a finite contribution to the Higgs mass
[and $P'_1(a)$ will even be finite by itself].
Of course, with these extra terms,
the new quantities $P'_i(a)$ are no longer individually modular invariant.
However, modular invariance continues to be preserved for their sum, as required.
Likewise, although $P'_1(a)$ and $P'_2(a)$ will now depend on $t$,
all dependence on $t$ will cancel in their sum.

We emphasize that despite a superficial similarity to the non-minimal regulator, 
the act of passing from $\lbrace P_1(a),P_2(a)\rbrace$ to $\lbrace P'_1(a),P'_2(a)\rbrace$ is
not one in which we are regulating our Higgs mass by softening or eliminating a net divergence.
We are simply performing an algebraic reshuffling of terms, transferring 
a logarithmic divergence from one contributing expression to another.  Indeed, the only regulator 
in our Higgs-mass calculation remains the $\widehat \calG_\rho(a,\tau)$ function with which we started.

Having defined these quantities, we now begin by evaluating $P'_1(a)$.   However, upon comparing
Eq.~(\ref{Pprimes}) with Eq.~(\ref{I3}), we note that 
\beq
            P'_1(a) ~=~ \frac{1}{a}\, \widehat m_\phi^2(t)\bigl|_{\bQ}
\label{P1prime}
\eeq
where on the right side we are explicitly disregarding the $\Lambda$-term
(\ie, keeping only those terms that result from non-trivial $\bQ$-insertions).
Thus, even though we are not employing the non-minimal regulator 
in this calculation (and thus we do not interpret $t$ as corresponding to a mass scale),
we nevertheless find that $P'_1(a)$ by itself is algebraically identical to what we would have 
obtained for the $\bQ$-dependent contributions to the Higgs mass using the non-minimal regulator.
The same algebraic manipulations that took
us from Eq.~(\ref{I3})
to Eq.~(\ref{Zagierresult3}) and
ultimately Eq.~(\ref{nonminresult})
then yield
\beq 
         P'_1(a) ~=~          -  \frac{\calM^2}{2a} \, \biggl[
          \frac{\pi}{3} \, \Str\, \mathbbX_1
    +  (\, \zStr \,\mathbbX_2) \log 4\pi t e^{-\gamma} \biggr]~.~~
\label{Pprime2}
\eeq
                                                                    
We now evaluate $P'_2(a)$.
Given the form of $P_2(a)$ in Eq.~(\ref{newforms3}),
we shall begin our evaluation of $P'_2(a)$
by breaking $P_2(a)$ into three contributions:  those from 
massless string states 
charged under $\mathbbX_1$;
those from massless string states charged under $\mathbbX_2$;
and those from the massive string states charged under 
$\mathbbX_1$ and/or $\mathbbX_2$.
Note that only the second of these contributions to $P_2(a)$ is
divergent.  It is therefore within this contribution to $P'_2(a)$ that 
we shall absorb the extra divergent term 
in Eq.~(\ref{Pprimes}).  This will allow each of these three contributions
to take an explicitly finite form.

We can easily evaluate the first of these contributions:
\beqn
  P_2(a)\biggl|_{\substack{ \mathbbX_1\\ M=0}}  \!  
     &=& 
      \, -\frac{\calM^2}{a}  \,\int_0^\infty \frac{d\tau_2}{\tau_2^2} \,
              \, \zStr \, \mathbbX_1 \,\sum_{r=1}^\infty e^{-\pi  r^2 /(a^2 \tau_2)}\nonumber\\
     &=&  \, - \frac{\calM^2}{a} \, \, \zStr \,\mathbbX_1 \, \sum_{r=1}^\infty \, (\pi r^2/a^2)^{-1} \nonumber\\
     &=&  \, - \half a  \calM^2 \left( \frac{\pi}{3} \,\, \zStr \mathbbX_1\right)~.
\label{Pprimetwo}
\eeqn
                                         By contrast, evaluating the contribution to $P_2(a)$ from massless states charged under $\mathbbX_2$
is more subtle.
Including the extra logarithmically divergent term from Eq.~(\ref{Pprimes}),
we have
\beqn
  P'_2(a)\biggl|_{\substack{ \mathbbX_2\\ M=0}}  \!  
     &=& 
      \,- \frac{ \calM^2}{a}  \,\int_0^\infty \frac{d\tau_2}{\tau_2} \, 
              \, \zStr \, \mathbbX_2 \,\sum_{r=1}^\infty e^{-\pi r^2 /(a^2 \tau_2)}\nonumber\\
     && ~~~~~~~~ - \frac{\calM^2}{2a} \int_t^\infty  \frac{d\tau_2}{\tau_2} \, \, \zStr~\mathbbX_2 ~.\nonumber\\
\label{intermediate}
\eeqn
Each line of Eq.~(\ref{intermediate}) is individually logarithmically divergent, but their sum is not.
In order to isolate these divergences algebraically and then cancel them between these two terms, 
we can insert a factor of $\tau_2^s$ inside each integral, with {\mbox{$s\leq 0$}}.   
We then find explicitly that 
each term diverges as {\mbox{$s\to 0$}}, but that their sum remains finite as {\mbox{$s\to 0$}}.
Explicitly, we have
\beqn
   P'_2(a)\biggl|_{\substack{ \mathbbX_2\\ M=0}} 
    \!&=&\, - \frac{\calM^2}{a}  \,\int_0^\infty \frac{d\tau_2}{\tau_2} \,\tau_2^s\,  
              \, \zStr \, \mathbbX_2 \,\sum_{r=1}^\infty e^{-\pi r^2 /(a^2 \tau_2)}~\nonumber\\
     && ~~~~~~~~~~~ - \frac{\calM^2}{2a} \int_t^\infty  \frac{d\tau_2}{\tau_2} \,\tau_2^s\,  \, \zStr~\mathbbX_2 \nonumber\\
   = && \!\!\!\!\!\!  - \frac{\calM^2}{2a} \, \zStr\,\mathbbX_2\, \biggl[
                2 \pi^s a^{-2s} \Gamma(-s) \zeta(-2s) - \frac{t^s}{s} \biggr]~.\nonumber\\
\eeqn
Each of the terms inside the square brackets on the final line
has an expansion around {\mbox{$s=0$}} beginning with a leading simple pole $1/s$.
These then cancel, whereupon the {\mbox{$s\to 0$}} limit leaves behind
the net finite contribution
\beq
   P'_2(a)\biggl|_{\substack{ \mathbbX_2\\ M=0}} ~=~  
      \frac{\calM^2}{2a} \, \zStr\,\mathbbX_2\, 
                \log(4\pi a^2 t e^{-\gamma}) ~.
\label{secondresult}
\eeq
Indeed, although this result continues to depends on $t$,
this $t$-dependence ultimately cancels when this result is added to the result coming
from  Eq.~(\ref{Pprime2}).
Thus, as required, none of our results depend on the parameter $t$ that 
characterized our reshuffling of the logarithmic divergence in Eq.~(\ref{Pprimes}).

Finally, we turn to the contributions to $P_2(a)$ from the massive states 
charged under $\mathbbX_1$ and/or $\mathbbX_2$.
These are also finite, and are given by
\beqn
&& P_2(a) \biggl|_{M>0}  \!  
 = \, - \frac{\calM^2}{a} \sum_{r=1}^\infty
          \int_0^\infty \frac{d\tau_2}{\tau_2^2}\, \bigg\lbrack\nonumber\\
 &&~ ~~~~~~~~~~~~~~~\, \pStr\, (\mathbbX_1+ \tau_2\mathbbX_2) \, e^{-\pi \alpha' M^2 \tau_2 - \pi r^2/(a^2 \tau_2)}\biggr\rbrack\nonumber\\
 &&~~=  \, - \frac{2\calM^2}{a} \sum_{r=1}^\infty \, \biggl\lbrack
     \frac{1}{2\pi } \, \pStr\, \mathbbX_1 \, \left(\frac{ aM}{r\calM}\right)\, 
             K_1\left( \frac{r  M}{a\calM} \right)\nonumber\\
 &&~~~~~~~~~~~~ + \, \pStr\, \mathbbX_2 \, K_0\left( \frac{r M}{a\calM} \right)\biggr\rbrack~  
\label{Bessels}
\eeqn
where $K_\nu(z)$ is the modified Bessel function of the second kind. 
     
Thus, combining Eqs.~(\ref{Pprime2}), (\ref{Pprimetwo}), (\ref{secondresult}), and (\ref{Bessels}),
we obtain our final result
\beqn
P(a) && \,=\, -\half \calM^2 \Biggl\lbrace \nonumber\\ 
     ~~&& \phantom{+} 
          \, \zStr \mathbbX_1 \left\lbrack \frac{\pi}{3}(a+1/a)               \right\rbrack \nonumber\\
     ~~&& + \, \zStr \mathbbX_2 \left\lbrack - \frac{2}{ a} \log\,a\,               \right\rbrack \nonumber\\
     ~~&& + \, \pStr \mathbbX_1 \left\lbrack \frac{\pi}{3a} +  
  \frac{2}{\pi} \,\sum_{r=1}^\infty \left(\frac{M}{r\calM}\right)
          \,K_1\left( \frac{r  M}{a\calM} \right) 
  \right\rbrack ~~\nonumber\\
     ~~&& + \, \pStr \mathbbX_2 \left\lbrack \frac{4}{a}\, \sum_{r=1}^\infty  
            K_0\left(  \frac{ r  M}{a\calM} \right) \right\rbrack \Biggr\rbrace ~.
\label{finalPa}
\eeqn
Indeed, this result is exact for all $a$ and for any modular-invariant theory.

At no step in our calculation for the total $P(a)$ did we break the {\mbox{$a\to 1/a$}} symmetry.
It therefore remains true that {\mbox{$P(a)=P(1/a)$}}.
However, this symmetry is deeply hidden.
Indeed, our manipulations in deriving this result presupposed that $F(\tau)$ in Eq.~(\ref{mainint}) is modular invariant,
and this in turn provides tight (but not obvious) relative constraints on the supertraces which appear on each line
of Eq.~(\ref{finalPa}).   For example, given that modular transformations
mix massless and massive string modes, the supertraces over the massless modes in any modular-invariant
theory are non-trivially balanced against the supertraces over the massive modes --- especially when these supertraces are weighted
by extra factors of $M$ and the Bessel functions thereof.
We have also seen that $\mathbbX_1$ and $\mathbbX_2$ are modular completions of each other, and hence
their contributions are also mixed under modular transformations.
Thus, although $P(a)$ continues to be invariant under {\mbox{$a\to 1/a$}}, we no longer expect to see this explicitly
when our terms are organized in the manner presented in Eq.~(\ref{finalPa}).

Given this result for $P(a)$, we can now directly calculate $\widehat m_\phi^2(\rho,a)$ via Eq.~(\ref{operator}).
Our result is
\beqn
 && \widehat m_\phi^2(\rho,a)\Bigl|_\calX  \,=\, \frac{\calM^2}{1+\rho a^2} \Biggl\lbrace \nonumber\\ 
     && ~~\phantom{+} 
          \, \zStr \mathbbX_1 \left\lbrack - \frac{\pi}{6}\left(1+\rho a^2\right)               \right\rbrack \nonumber\\
     && ~+ \, \zStr \mathbbX_2 \left\lbrack  \log\, a - 1 - \frac{\log \rho}{\rho-1} 
               \right\rbrack \nonumber\\
     && ~+ \, \pStr \mathbbX_1 \, \biggl\lbrace - \frac{\pi}{6}  
     - \frac{1}{2\pi(\rho-1)} \left(\frac{M}{\calM}\right)^2  \times \nonumber\\
     && ~~~~~~~~~~~~~~~~~  
      \times \left\lbrack    \calK_0^{(0,1)}\!\left( \frac{M}{a\calM}\right) + \calK_2^{(0,1)}\!\left( \frac{M}{a\calM}\right)
      \right\rbrack \biggr\rbrace ~~~~~~\nonumber\\
     && ~+ \, \pStr \mathbbX_2 \, \biggl\lbrace 
     \frac{2}{\rho-1} \biggl \lbrack 
               \calK_0^{(0,1)}\!\left(\frac{M}{a\calM}\right)  
           - \frac{1}{\rho} \calK_1^{(1,2)}\!\left( \frac{M}{a\calM}\right) 
              \biggr\rbrack  \biggr\rbrace \Biggr\rbrace  \nonumber\\
                                                           \label{finalhiggsmassa}
\eeqn
where $\calK_\nu^{(n,p)} (z)$ are the combinations of Bessel functions in Eq.~(\ref{Besselcombos}).

\section{Evaluating the cosmological constant with the modular-invariant regulator:~
Explicit calculation\label{lambdaappendix}} 

In Appendix~\ref{higgsappendix}, 
we provided an explicit calculation of the first contribution 
to the total regulated Higgs mass listed  in Eq.~(\ref{twocontributions}). 
In this Appendix, with an eye towards evaluating the second contribution, 
we now provide an explicit calculation of $\widehat \Lambda(\rho,a)$.

In any tachyon-free modular-invariant theory, the 
one-loop cosmological constant $\Lambda$ is finite.   
As such, evaluating $\Lambda$ requires no regulator.
However,  we have seen
that $\Lambda$ also appears as a contributing term within our 
total result for the Higgs mass in Eq.~(\ref{relation1}), along with contributions stemming from the non-trivial $\calX_i$
insertions.
These latter contributions  are divergent, and thus the Higgs mass requires regularization.
However, when we regularize our Higgs mass, for consistency we must apply the same regulator 
to all of the terms that contribute to the Higgs mass, and this includes the
cosmological-constant term as well.
Accordingly, in this Appendix, we shall evaluate
the quantity $\widehat \Lambda(\rho,a)$ defined in Eq.~(\ref{Lambdahatdef})
             where $\calZ$ is the one-loop partition function of our theory with no charge insertions,
and express our result in terms of supertraces over only physical string states.
Indeed, our result for $\widehat \Lambda (\rho,a)$ can then be incorporated
alongside our result for 
$\widehat m_\phi^2 (\rho,a)\Bigl|_\calX$ from Appendix~\ref{higgsappendix}
in order to obtain a full expression for the regularized Higgs mass $\widehat m_\phi^2 (\mu)$. 

Our calculation will parallel the calculation presented in Appendix~\ref{higgsappendix}.~  In particular,
we shall begin by evaluating the core integral
\beq
       P(a) ~\equiv~ \int_\calF \dmu~ \calZ(\tau)\,Z_{\rm circ}(a,\tau)~,
\label{mainint2}
\eeq
which is the same as $P(a)$ in Eq.~(\ref{mainint}) except that we have
replaced {\mbox{$F(\tau)\to \calZ(\tau)$}} within the integrand of Eq.~(\ref{mainint}). 
We therefore now have
\beq
     g(\tau_2) ~=~ -\frac{\calM^4}{2} \, \tau_2^{-1} \biggl\lbrack
              \, \zStr {\bf 1} + \, \pStr e^{-\pi \alpha' M^2 \tau_2}\biggr\rbrack~.
\label{gtau2}
\eeq
Of course, 
\beq
            \, \zStr {\bf 1} ~=~ n_B - n_F~,
\label{nBnF}
\eeq
where $n_B$ and $n_F$ are respectively the numbers of physical massless bosonic and fermionic
degrees of freedom in the string spectrum.
Proceeding exactly as in Appendix~\ref{higgsappendix}, we can  
then separate $P(a)$ into two distinct contributions $P_1(a)$ and $P_2(a)$
as in Eq.~(\ref{newforms}), except with {\mbox{$F(\tau)\to \calZ(\tau)$}}.
However, unlike the situation in Appendix~\ref{higgsappendix}, there is no need
to transfer any divergences between these two terms.

Evaluating $P_1(a)$ is straightforward, yielding
\beq
   P_1(a) ~=~ \frac{1}{a} \,\Lambda ~=~  \frac{\calM^2}{24 a}  \,\Str\,M^2~
\label{P1cos}
\eeq
where in passing to the final expression we have followed the derivation in Eq.~(\ref{eq:lamlam}).
Evaluating $P_2(a)$ is also relatively straightforward.
The contribution to $P_2(a)$ from the massless states --- \ie,
from the first term in Eq.~(\ref{gtau2}) --- 
is given by
\beqn
  P_2(a)\biggl|_{M=0}  \!\! &=&\, 
      -  \frac{\calM^4}{a}\, 
       (n_B-n_F)\,
          \sum_{r=1}^\infty\,
       \int_0^\infty \frac{d\tau_2}{\tau_2^3} \, e^{-\pi r^2 /(a^2 \tau_2)}\nonumber\\   
  &=&\, 
      -  \frac{\calM^4}{a}\, 
       (n_B-n_F)\,
          \sum_{r=1}^\infty\,   \frac{a^4}{\pi^2 r^4 } \nonumber\\
  &=&\, -  \frac{\calM^4}{2}  \frac{\pi^2}{45}   
            \,(n_B-n_F) \,{a^3}~.
\label{P2cosa}
\eeqn
By contrast, the contribution to $P_2(a)$ from the massive states --- \ie,
from the second term in Eq.~(\ref{gtau2}) ---
is given by
\beqn
  && P_2(a)\biggl|_{M>0}\!  = \,
      -  \frac{\calM^4}{ a} 
       \, \pStr  \sum_{r=1}^\infty
       \int_0^\infty \!\frac{d\tau_2}{\tau_2^3}
               e^{- \pi \tau_2\alpha' M^2 -\pi r^2 /(a^2 \tau_2)}\nonumber\\   
  && ~~~~~~~~=\, 
  -  \frac{\calM^2}{2} \frac{a }{\pi^2 } 
       \, \pStr  \!\left\lbrack M^2 \,\sum_{r=1}^\infty
                \, \frac{1}{r^2} \,K_2\left(\frac{ r  M}{a\calM }\right)\right\rbrack\nonumber\\
\label{P2cosb}
\eeqn
where $K_2(z)$ is the order-two modified Bessel function of the second kind. 
Combining our results from Eqs.~(\ref{P1cos}), (\ref{P2cosa}), and (\ref{P2cosb})
then yields our final expression for $P(a)$:
\beqn
     P(a) \,&=&\,~   
    \frac{\calM^2}{24a} \, \Str\,M^2~ -  
 \frac{\calM^4}{2}  \frac{\pi^2}{45}\,(n_B-n_F)   \,{a^3}~\nonumber\\
  && \,-  \frac{\calM^2}{2} \frac{a}{\pi^2 } 
       \, \pStr  \!\left\lbrack M^2 \sum_{r=1}^\infty
                \frac{1}{r^2} K_2\left(\frac{  r M}{ a \calM}\right)\right\rbrack~.~~\nonumber\\
\label{Pacos}
\eeqn
Applying the operator in Eq.~(\ref{operator}),
we then find that our regulated cosmological constant $\widehat\Lambda(\rho,a)$ is given by
\beqn
  && \widehat\Lambda(\rho,a) ~=~ \frac{1}{1+\rho a^2} \, \Biggl\lbrace
    \frac{\calM^2}{24} \, \Str\,M^2 \nonumber\\
  && ~~~~   - \frac{\pi^2}{30} \,\rho ( 1+ \rho+\rho^2) \,(n_B-n_F)\,  (a\calM)^4  
~\nonumber\\
  && ~~~~ - \frac{1}{4\pi^2} \frac{\rho}{\rho-1} \,
   \, \pStr  
       M^4 \,\Biggl[ \calK_1^{(-1,0)}\!\left( \frac{M}{a\calM}\right) ~~~~~~\nonumber\\
  && ~~~~~~~~~~~~~ +      2 \rho \,\calK_2^{(-2,-1)}\!\left( \frac{M}{a\calM}\right)\nonumber\\
  && ~~~~~~~~~~~~~ +       \calK_3^{(-1,0)}\!\left( \frac{M}{a\calM}\right)
     \Biggr] ~  \Biggr\rbrace~
\label{lambdaresult}
\eeqn
where $\calK_\nu^{(n,p)}(z)$ are the Bessel-function combinations defined in Eq.~(\ref{Besselcombos}).

\bigskip
\bigskip

\bibliography{TheLiterature}
\end{document}